\documentclass[10pt,letterpaper]{article}
\usepackage[top=0.85in,left=2.75in,footskip=0.75in]{geometry}

\usepackage{amsmath,amssymb}

\usepackage{changepage}

\usepackage{textcomp,marvosym}

\usepackage{cite}

\usepackage{nameref,hyperref}

\usepackage[right]{lineno}

\usepackage[nopatch=eqnum]{microtype}
\DisableLigatures[f]{encoding = *, family = * }

\usepackage[table]{xcolor}

\usepackage{array}
\usepackage{tabularx}
\usepackage{comment}
\usepackage{subcaption}
\usepackage{wrapfig}
\usepackage{url}
\usepackage{lscape}

\usepackage{adjustbox}

\newcolumntype{+}{!{\vrule width 2pt}}

\newlength\savedwidth

\raggedright
\usepackage[aboveskip=1pt,labelfont=bf,labelsep=period,justification=raggedright,singlelinecheck=off]{caption}
\renewcommand{\figurename}{Fig}

\usepackage{pifont}
\usepackage{tikz}
\usetikzlibrary{automata, calc, fit}

\definecolor{color0}{HTML}{023eff}
\definecolor{color1}{HTML}{ff7c00}
\definecolor{color2}{HTML}{1ac938}
\definecolor{color3}{HTML}{e8000b}
\definecolor{color4}{HTML}{8b2be2}
\definecolor{color5}{HTML}{9f4800}
\definecolor{color6}{HTML}{f14cc1}
\definecolor{color7}{HTML}{a3a3a3}
\definecolor{color8}{HTML}{ffc400}
\definecolor{color9}{HTML}{00d7ff}

\newcolumntype{L}[1]{>{\raggedright\let\newline\\\arraybackslash\hspace{0pt}}m{#1}}
\newcolumntype{C}[1]{>{\centering\let\newline\\\arraybackslash\hspace{0pt}}m{#1}}
\newcolumntype{R}[1]{>{\raggedleft\let\newline\\\arraybackslash\hspace{0pt}}m{#1}}

\usepackage{colortbl}
\usepackage{xcolor}

\definecolor{high}{HTML}{ffffff}  
\definecolor{low}{HTML}{1ac938}  

\definecolor{highnmi}{HTML}{1ac938}  
\definecolor{lownmi}{HTML}{ffffff}  

\makeatletter
\renewcommand{\@biblabel}[1]{\quad#1.}
\makeatother

\usepackage{lastpage,fancyhdr,graphicx}
\usepackage{epstopdf}
\fancyheadoffset[L]{2.25in}
\fancyfootoffset[L]{2.25in}
\usepackage{array}
\newcommand{\PreserveBackslash}[1]{\let\temp=\\#1\let\\=\temp}
\newcolumntype{C}[1]{>{\PreserveBackslash\centering}p{#1}}
\newcolumntype{R}[1]{>{\PreserveBackslash\raggedleft}p{#1}}
\newcolumntype{L}[1]{>{\PreserveBackslash\raggedright}p{#1}}

\usepackage{siunitx}
\usepackage{booktabs, tabularx, threeparttable}

\usepackage{etoolbox}
\newrobustcmd\B{\DeclareFontSeriesDefault[rm]{bf}{b}\bfseries}   

\begin{document}
\vspace*{0.2in}

\begin{flushleft}
{\Large
\textbf\newline{Quantifying Group Fairness in Community Detection} 
}
\newline
\\
Elze de Vink,
Frank W. Takes,
Akrati Saxena\textsuperscript{*}


\bigskip
Leiden Institute of Advanced Computer Science, Leiden University, Leiden, The Netherlands
\\
\bigskip

%
%





* a.saxena@liacs.leidenuniv.nl

\end{flushleft}
\section*{Abstract}
Understanding community structures is crucial for analyzing networks, as nodes join communities that collectively shape large-scale networks. In real-world settings, the formation of communities is often impacted by several social factors, such as ethnicity, gender, wealth, or other attributes. These factors may introduce structural inequalities; for instance, real-world networks can have a few majority groups and many minority groups. Community detection algorithms, which identify communities based on network topology, may generate unfair outcomes if they fail to account for existing structural inequalities, particularly affecting underrepresented groups. In this work, we propose a set of novel group fairness metrics to assess the fairness of community detection methods. Additionally, we conduct a comparative evaluation of the most common community detection methods, analyzing the trade-off between performance and fairness. Experiments are performed on synthetic networks generated using LFR, ABCD, and HICH-BA benchmark models, as well as on real-world networks. Our results demonstrate that the fairness-performance trade-off varies widely across methods, with no single class of approaches consistently excelling in both aspects. We observe that Infomap and Significance methods are high-performing and fair with respect to different types of communities across most networks. The proposed metrics and findings provide valuable insights for designing fair and effective community detection algorithms.


\section*{Introduction}

Social networks are used to represent complex social systems where nodes represent individuals or entities, and the relationships between these entities, such as friendships, collaborations, or shared interests, are denoted by edges. In social network analysis, the objective is to uncover meaningful patterns, understand structural properties within networks, and analyze the dynamic processes taking place on these networks. A key concept in social network analysis is community detection, which identifies groups of nodes that are more densely connected internally and more loosely connected with the rest of the network. With that, a community is ``a group of nodes that have a higher likelihood of connecting to each other than to nodes from other communities” \cite{barabasi2014network}. Detecting communities is essential for understanding the functional organization of networks and improving applications such as recommendation systems \cite{kumar2024community, Saxena2022hm}, awareness spread by influence maximization \cite{chen2014cim}, anomaly detection \cite{chen2012community}, disrupting networks \cite{miller2018discovering, miller2022community} and disease outbreak modeling \cite{kitchovitch2011community}. Community detection algorithms take the structure of a social network as the input and output a partitioning of the nodes into communities~\cite{Fortunato2016}. In the literature, many different community detection algorithms have been proposed to identify meaningful clusters, which can reveal hidden structures driving real-world interactions \cite{Fortunato2010}. 
  
One important aspect of network analysis is understanding how the structure of a social network reflects inherent social inequalities~\cite{Saxena2024fairsna}. The formation of communities within these networks is influenced by factors such as ethnicity, gender, race, and socioeconomic status, leading to variations in network structure in terms of community size, density, and connectivity \cite{Saxena2024fairsna}. If these structural inequalities are not accounted for in the algorithm design phase, network analysis algorithms might produce biased outcomes, particularly disadvantaging minority groups. To mitigate bias and promote equitable outcomes, it is essential to integrate structural inequalities into the design of network analysis methods, ensuring fair treatment for all users and groups, regardless of their type, size, or any protected attribute. A fair community detection method should accurately identify all types of communities, whether small or large, dense or sparse, or having different connectivity in the network, while ensuring high-quality results. 
Community detection methods, which leverage network structure to identify communities, often struggle to accurately detect small or sparsely connected groups \cite{ghasemian2019evaluating}. This misclassification can further propagate bias in other downstream network analysis tasks, such as influence maximization \cite{Farnad2020}, influence minimization \cite{Saxena2023fairness}, link prediction \cite{Saxena2022hm, Saxena2022nodesim}, and centrality ranking \cite{Tsioutsiouliklis2020}, that utilize the community structure to ensure fairness. 

Ghasemian et al. \cite{ghasemian2019evaluating} analyzed 16 community detection methods and found significant variation, for example, the number of identified communities across different methods. However, their study did not explore the impact of these methods on communities of varying sizes and densities. While numerous metrics exist to evaluate the quality of detected communities \cite{Chakraborty2016}, there is currently no established metric to assess the fairness of a community detection method, particularly in relation to measuring its bias against minority groups. Despite the extensive literature on community detection methods and evaluation metrics \cite{fortunato2016community}, fairness remains an underexplored aspect in community detection, lacking clear definitions and comprehensive evaluation frameworks. Yet, its importance is evident in ensuring fair network analysis. 

In this work, we introduce a set of group fairness metrics $(\Phi)$ to assess the fairness of community detection methods. The objective of these metrics is that given as input the ground truth communities (community labels of nodes), the communities identified by a detection method, a network dataset, and fairness criterion, output a fairness score based on the specified community property. This indicates the extent to which the employed community detection method provided a fair partitioning of the network given the chosen fairness criterion, such as community size, density, or connectivity. In \cite{de2024group}, we discussed the group fairness metric. 
Our approach begins by matching the predicted communities to the ground truth communities. In particular, we introduce three measures, called FCCN, F1, and FCCE, to compute fairness for each ground truth community, which will be subsequently used to compute group fairness score $(\Phi)$. 

In the experiments, we first extensively analyze the behavior of the proposed metric. Next, we conduct a comparative analysis of existing community detection methods, examining the performance-fairness trade-off to determine whether high-performing methods exhibit biases, i.e., we assess the extent to which they are fair. We measure this fairness based on three structural properties of communities - (i) size, (ii) density, and (iii) conductance. Here, size refers to the total number of nodes in a community, density represents the ratio of internal edges to possible internal edges, and conductance measures the fraction of a community’s edge volume that connects external nodes. We categorize community detection methods based on their algorithmic approach into six classes: (i) Optimization, (ii) Spectral, (iii) Propagation, (iv) Dynamics, (v) Representation Learning, and (vi) Probabilistic. Experiments are performed on synthetic benchmark models, including LFR \cite{Lancichinetti2008}, ABCD \cite{Kamiński2021}, and HICH-BA \cite{Saxena2023fairness} models and real-world networks. The performance of community detection methods is measured using NMI \cite{Fred2003}, RMI \cite{Newman2020}, ARI~\cite{Hubert1985}, NF1 \cite{Rossetti2017}, and PF1 \cite{Rossetti2016} evaluation metrics. Our experiments reveal that no single class of community detection methods consistently outperforms others. The performance-fairness trade-off varies significantly across methods as well as on networks. The analysis highlights that some of the best-performing and fairest approaches including Infomap \cite{Rosvall2008}, RSC-V \cite{Zhang2018}, RSC-K \cite{Zhang2018}, Significance \cite{Traag2013}, Walktrap \cite{Pons2005}, SBM \cite{Peixoto2014}, and SBM-Nested \cite{Peixoto2014a} methods.  

\vspace{2mm}
The main contributions of our work are mentioned below.
\begin{itemize}
    \item The paper introduces a novel set of group fairness measures, denoted as $\Phi$, to evaluate the fairness of community detection methods. These measures quantify how fairly a community detection method identifies communities of varying structural properties, e.g., size, density, and conductance.
    \item Extensive experiments are conducted to study the fairness-performance trade-off of 24 community detection methods on synthetic benchmark models (LFR, ABCD, HICH-BA) and real-world networks.
    \item Through empirical analysis, the paper highlights the suitability of different methods for different types of networks. We recommend the Infomap and Significance Community detection methods as they achieve high fairness-performance trade-offs on different networks.
\end{itemize}

The paper is structured as follows. First, we review related work on evaluation metrics for detected communities and algorithmic fairness in community detection. Next, we introduce the proposed group fairness metric, followed by a description of the experimental setup. We then present the empirical analysis and insights, summarizing the most important findings and giving an outline of potential future research directions, followed by the conclusion.

\section*{Related Work}

In this section, we discuss the existing literature on evaluation metrics and algorithmic fairness for community detection.

\subsection*{Evaluation Metrics}

Community detection consists of two phases: identifying meaningful community structures by means of a community detection algorithm and evaluating the quality and relevance of the detected communities. Here, we first discuss metrics for evaluating detected communities without relying on ground truth labels, which are also used as optimization criteria for community detection algorithms. Next, we discuss metrics to assess the quality of the identified communities.

\subsubsection*{Metrics for Community Detection}

Community detection methods aim to identify meaningful group structures within a network by optimizing quality metrics. These measures compute the goodness of the community on the basis of the connectivity of nodes and the network structure. Such quality metrics can be categorized into four main types: (i) internal connectivity-based, (ii) external connectivity-based, (iii) internal and external connectivity-based, and (iv) network model-based \cite{Chakraborty2016}.

Internal connectivity-based metrics assess the quality of identified communities using the structure within a community. These include Internal Density, Edge Inside, Average Degree, Fraction Over Median Degree, and Triangle Participation Ratio \cite{radicchi2004defining}. They measure factors such as edge density, internal connectivity, node degree distribution, and the prevalence of triangular motifs. For instance, Internal Density measures the density of edges within the community by comparing the actual number of internal edges to the total possible internal edges. Edge Inside counts the total number of internal edges within the community. Overall, these measures analyze how tightly knit nodes are within the identified community compared to the rest of the network.

External connectivity-based metrics evaluate how a community interacts with the rest of the network. Important measures falling under this category include the Cut Ratio \cite{wei1989towards}, which calculates the fraction of outgoing edges relative to all possible edges, and Expansion \cite{radicchi2004defining}, which measures the number of external edges from the community divided by the size of the community. 
The next class of metrics considers both the internal and external connections of nodes from the community's perspective. Common metrics include Conductance \cite{shi2000normalized}, Normalized cut \cite{shi2000normalized}, and Maximum-ODF (Out Degree Fraction), Average-ODF, and Flake-ODF \cite{flake2000efficient}. 

Network model-based metrics, such as modularity, compare actual community structures against a randomized null model to determine the strength of detected communities. Modularity \cite{Newman2006a, Newman2004} evaluates community quality by comparing the actual number of internal edges to the expected number in a random graph with the same degree distribution. The higher modularity values indicate stronger community structures. However, modularity optimization faces challenges, including the resolution limit \cite{fortunato2007resolution}, which prevents the detection of smaller communities with varying levels of interconnectedness, and the degeneracy problem \cite{good2010performance}, where multiple distinct community structures yield similar modularity values. To address these issues, various modifications have been proposed, such as modularity density \cite{li2008quantitative}, modularity intensity \cite{sun2013maximizing}, Adaptive scale modularity \cite{van2014axioms}, Community Score \cite{pizzuti2008ga}, SPart \cite{chira2012evolutionary}, Permanence \cite{chakraborty2014permanence}, and Significance \cite{Traag2013}. A variety of community detection algorithms have been developed that optimize these measures, aiming to detect better communities \cite{Blondel2008, chakraborty2013constant, Traag2019, Bonald2018, Traag2013}.

\subsubsection*{Ground truth-based Validation Metrics} 

The performance of a community detection method is evaluated by comparing the detected communities with a ground truth structure. 
To compare the detected and ground truth communities, several metrics from the data mining clustering literature have been adapted and reformulated to incorporate network-specific information \cite{berkhin2006survey}.

Mutual Information (MI) \cite{danon2005comparing}, derived from information theory, measures how much one partition tells us about the other one. Normalized Mutual Information (NMI) \cite{Fred2003} refines MI by incorporating the entropy of the respective community structures, while Reduced Mutual Information (RMI) \cite{Newman2020} addresses a flaw in MI by ensuring a value of zero when the predicted partition consists of $n$ communities (each having one node), indicating no meaningful structure. 
Purity \cite{lin2005foundations} assigns each detected community to the most frequent ground truth label within it. A Purity score of 1 indicates a perfect match. However, Purity is asymmetric, meaning $Purity(C, P)$ and $Purity(P, C)$ are not necessarily equal, where $C$ and $P$ are the set of ground truth and predicted communities. The former, commonly referred to as ``Purity," is more widely used, while the latter ($Purity(P, C)$) is known as ``Inverse Purity" \cite{artiles2007semeval}. The F-Measure \cite{artiles2007semeval} overcomes this limitation by computing the harmonic mean of Purity and Inverse Purity.

An alternative way to define a community is as a collection of pairwise decisions for nodes in a network \cite{hubert1985comparing}. Two nodes are considered part of the same community if they share the ground truth label. The Rand Index (RI) evaluates accuracy by measuring the proportion of correctly assigned node pairs: true positives (TP) and true negatives (TN) indicate correct assignments, while false positives (FP) and false negatives (FN) represent errors. A true positive (TP) occurs when two nodes belonging to the same ground truth community are correctly grouped within the same detected community. Similarly, we can compute other values of the confusion matrix. RI reflects the overall alignment between detected and ground truth communities. However, it has limitations, which the Adjusted Rand Index (ARI) \cite{Hubert1985} addresses by reducing sensitivity to the number of communities. 
Recent metrics in this class include Variation of Information (VI) \cite{Meila2007}, Edit Distance \cite{aynaud2010static}, NF1 \cite{Rossetti2017}, and PF1 \cite{Rossetti2016}. In our study, we use a selection of the most commonly used metrics from different categories to evaluate the performance of community detection methods.

\subsection*{Fairness-aware Community Detection} 

Fairness-aware network analysis aims to ensure that algorithms used for analyzing social networks produce unbiased and equitable outcomes for all users and groups \cite{Saxena2024fairsna}. The evolution of social networks is impacted by several factors, such as ethnicity, gender, or socioeconomic status, which leads to structural inequalities. Fairness-agnostic methods, such as community detection, link prediction, and influence maximization, often disproportionately favor majority groups and reinforce existing structural inequalities in networks. Fairness-aware network analysis introduces techniques to mitigate structural inequalities by incorporating fairness constraints, redefining evaluation metrics, and adjusting algorithmic decisions to ensure equitable representation and treatment across different social groups. In recent years, fairness-aware methods have been proposed for several downstream network analysis tasks, such as link prediction \cite{Saxena2022hm, Saxena2022nodesim}, centrality ranking \cite{Tsioutsiouliklis2021}, influence maximization \cite{Stoica2019, feng2023influence}, and influence minimization \cite{Saxena2023fairness}. However, fairness in community detection is still underexplored.

Community detection aims to uncover structural patterns in networks, but no single algorithm is universally optimal across all inputs, as stated by the No Free Lunch theorem for community detection \cite{peel2017ground}. Ghasemian et al. \cite{ghasemian2019evaluating} analyzed 16 community detection algorithms on a benchmark corpus of 572 diverse real-world networks to examine their over- and underfitting behaviors. The findings reveal that (i) algorithms vary significantly in the number and composition of communities detected, (ii) similar algorithms cluster together based on their outputs, (iii) performance differences impact link-based learning tasks, and (iv) no algorithm consistently outperforms others across all networks. Probabilistic and non-probabilistic methods exhibit distinct behaviors, with spectral techniques producing more similar results to each other than to other approaches. The study highlights the importance of evaluating community detection algorithms across a wide range of networks, as results from small-scale studies may not generalize.

The detectability of communities is a crucial factor in ensuring fairness in community detection. Prior research has established detectability thresholds~\cite{Decelle2011, Fortunato2016, Nadakuditi2012} that define conditions under which communities become undetectable. Radicchi~\cite{Radicchi2013} demonstrated that degree heterogeneity enables modularity-based community detection algorithms to recover network community structures accurately. However, in complex networks, such as those generated by the LFR and ABCD benchmark models, the existence of a well-defined detectability threshold remains uncertain~\cite{Fortunato2016}. If some specific community properties hinder detectability, it may introduce bias in community detection outcomes, disproportionately affecting specific groups and leading to unfair representations in social network analysis.

Mehrabi et al.~\cite{Mehrabi2019} highlight that community detection algorithms, particularly those optimizing modularity, tend to exclude low-degree nodes. To address this issue, the authors proposed the Communities with Lowly-connected Attributed Nodes (CLAN) method, designed for networks with attributed nodes. CLAN incorporates a supervised learning step that reassigns attributed nodes from smaller predicted communities into larger ones. However, this approach assumes that smaller communities are not meaningful and should be merged for downstream tasks, potentially leading to the dissolution of actual minority communities rather than their correct identification. While the study does not explicitly define fairness, it aims to introduce a method for mitigating an observed bias in community detection algorithms.

Manolis et al.~\cite{Manolis2024} introduce two fairness metrics for communities: balance fairness and modularity fairness. Their analysis focuses on networks with two disjoint groups of nodes, represented as blue and red, where red nodes constitute the protected group. Balance fairness quantifies the deviation of the fraction of red nodes in a community from their overall fraction in the entire network. Modularity fairness evaluates how well red and blue nodes are connected within a community, using modularity as a measure of group connectedness. These metrics assess whether the protected group is adequately represented and integrated within each community. Through experiments on synthetic networks, the study finds that group size imbalance has the most significant impact on both fairness metrics.

However, these methods do not fully capture the underlying definition of communities and the mesoscale connectivity of networks. Communities in social networks emerge based on human behavior and connection patterns, and enforcing proportional group representation in each detected community may not align with real-world community structures. Consequently, the proposed fairness definitions resemble node clustering rather than true community detection and may be more suitable for applications requiring equitable clustering rather than structural community identification.

\section*{The Proposed Group Fairness Metric ($\Phi$)} 
 
To calculate the proposed fairness metrics, we first map the ground truth communities with the identified communities. We then evaluate the community-wise fairness using the three introduced metrics, and finally, community-wise fairness scores are used to determine the overall fairness of a community detection method. The detailed 3-step methodology is outlined below, followed by an analysis of the proposed metric's behavior. 

\subsection*{1. Community Mapping}

Consider a network $G = (V,E)$ with $m$ ground truth communities, denoted as $C=\{c_1, c_2,..., c_{m}\}$. A community detection method applied to $G$, produces a set of $k$ predicted communities, represented as $P=\{p_1, p_2, ..., p_{k}\}$. To evaluate potential bias in a community detection method, it is crucial to measure the quality of each identified ground truth community. 
This is accomplished by mapping each ground truth community to the most relevant predicted community. 

The mapping process is done using the following steps iteratively until at least one ground truth and one predicted community remains unmapped:
\begin{enumerate}
        \item Compute the Jaccard similarity for each pair of ground truth and predicted communities as follows: 
        \begin{equation*}
            \label{eq:jaccard_sim}
            J(c_i, p_j) = \frac{|c_i \cap p_j|}{|c_i \cup p_j|}, \forall \; i \; \& \; j
        \end{equation*}
        \item Select the pair with the highest similarity score and map the corresponding ground truth and predicted communities. If multiple pairs have the same highest score, the tie is broken by randomly selecting a ground truth and predicted community pair for mapping.
    \end{enumerate}
If any ground truth community remains unmapped after this process, it is considered completely misclassified and is mapped to an empty set.

\subsection*{2. Community-wise Performance Metrics}

We introduce the following three metrics to assess how well one ground truth community is captured by its corresponding predicted community, considering both node membership and structural connectivity through edges. 

\begin{enumerate}
    \item \textbf{Fraction of Correctly Classified Nodes (FCCN):} A straightforward way to evaluate how well a predicted community ($p_j$) represents the ground truth community ($c_i$) is by measuring the fraction of ground truth nodes correctly captured within the predicted community. This metric, referred to as FCCN, is calculated as follows:
        \begin{equation}
            \textit{FCCN}(c_i, p_j) = \frac{|c_i \cap p_j|}{|c_i|}
        \end{equation}
        
    \item  \textbf{F1 Score}: The FCCN primarily considers the overlap of nodes between the ground truth and predicted communities but does not penalize the presence of extra nodes in the predicted community. To address this limitation, we introduce the F1 score, inspired by the F1 score used in machine learning \cite{Chinchor1992}. It is computed as follows:
        \begin{equation}\label{eq:f1}
            F1(c_i, p_j) = \frac{2 |c_i \cap p_j|}{|c_i| + |p_j|}
        \end{equation}
        
    \item \textbf{Fraction of Correctly Classified Edges (FCCE):} Community structure is primarily driven by its edges, making it essential to evaluate a community detection method based on how well it preserves intra-community connections. To capture this aspect, we introduce the FCCE metric, which measures the proportion of ground truth community edges present in the corresponding predicted community. It is defined as:
    \begin{equation}
            FCCE(c_i, p_j) = \frac{|E(c_i) \cap E(p_j)|}{|E(c_i)|}
    \end{equation}
    where $E(c_i)$ represents the set of intra-community edges in the ground truth community $c_i$, and it is computed as: $E(c_i) = \{(u, v) \in E \mid u \in c_i \text{ and } v \in c_i \}$.
\end{enumerate}

\subsection*{3. Group Fairness Metric ($\Phi$)}

\begin{figure}[h!]
  \begin{center}
    \includegraphics[width=7cm]{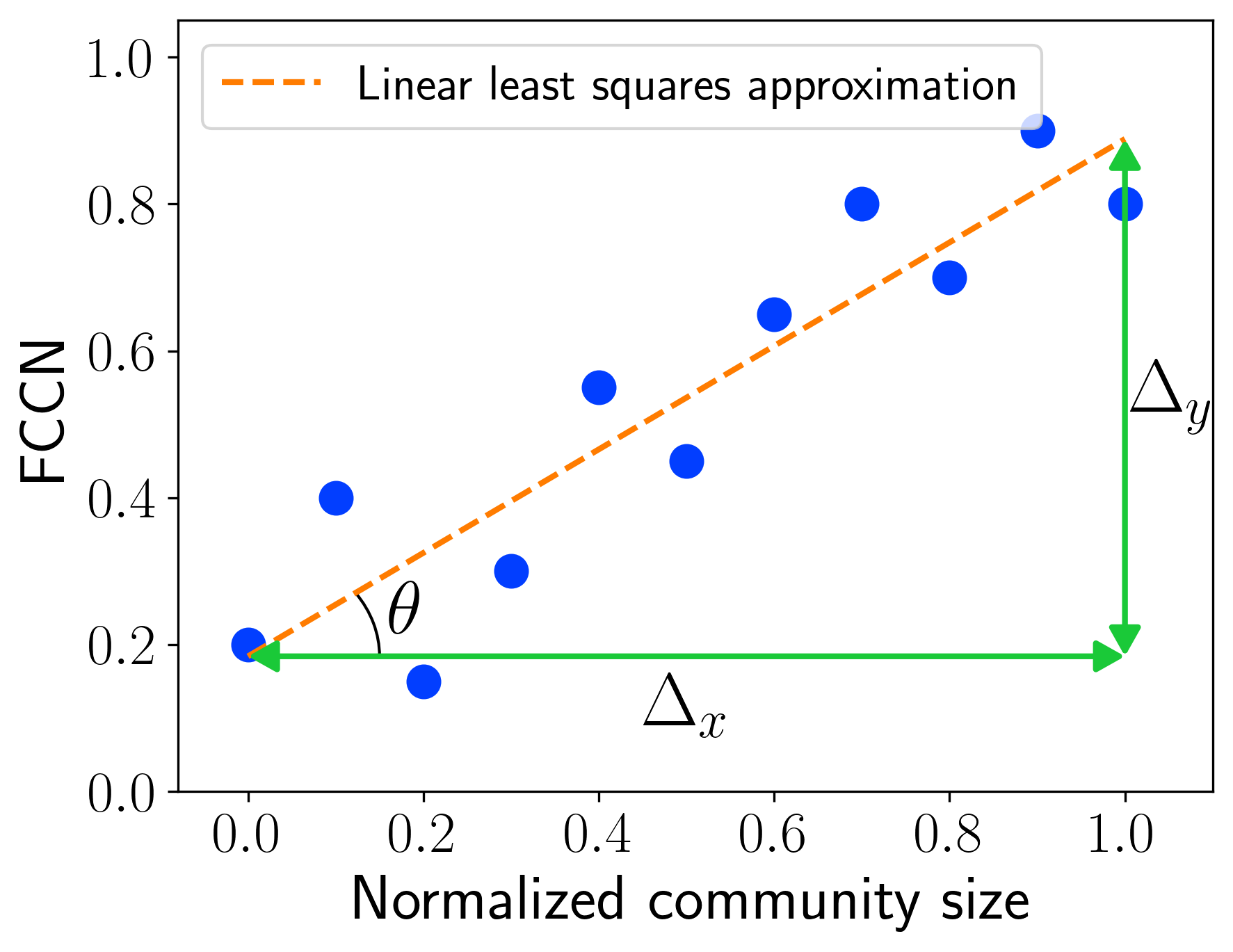}
  \end{center}
  \caption{FCCN vs. normalized community size for each community on a small sample network. The best-fit line shows the trend of a community detection method.}
  \label{fig:fairness-example}
\end{figure}

Our goal is to investigate whether a given community detection method favors or exhibits a bias toward communities with specific characteristics, concretely, size, density, and conductance. We do so using the previously defined community-wise performance metrics. Our approach starts by analyzing the relationship between these performance scores and the attribute value of each community. For instance, Fig.~\ref{fig:fairness-example} illustrates FCCN versus normalized community size for a sample network, revealing that larger communities tend to be identified more accurately than smaller ones. To ensure fair comparisons across different networks, we normalize community attribute values between 0 and 1 by applying min-max scaling. 

To compute the group fairness metric ($\Phi$), we fit a linear regression line using least squares approximation~\cite{Hastie2009} on the proposed community-wise fairness metrics (represented by $F^*$) versus normalized community property ($p$). For example, in Fig.~\ref{fig:fairness-example}, the dashed line represents the best fit linear line for the community-wise fairness metric (FCCN) versus normalized community size. The group fairness metric is defined as the slope of this regression line, which ranges in $(-1, 1)$. The regression line provides the changes in the x-axis ($\Delta x$) and y-axis ($\Delta y$), and the angle $\theta$ is computed using the arctangent function as follows:
\begin{equation*}
    \theta = \textit{arctan}\left(\frac{\Delta y}{\Delta x}\right)
\end{equation*}
To get the fairness metric value ranging in $(-1,1)$, we multiply the angle with $\frac{2}{\pi}$ as the arctangent angle is in radians within $(-\frac{\pi}{2}, \frac{\pi}{2})$. Finally, the fairness of a community detection method, with respect to a community-wise fairness metric (F*) and community property (p), is computed as:
    \begin{equation}
        \label{eq:angle}
        \Phi_p^{F*} = \frac{2}{\pi}\textit{arctan}\left(\frac{\Delta y}{\Delta x}\right) = \frac{2}{\pi}\textit{arctan}\left({\Delta y}\right)
    \end{equation}
The x-axis values are normalized using min-max scaling, meaning that $\Delta x =1$, and therefore, it can be excluded from the calculation. Referring to the example in Fig.~\ref{fig:fairness-example}, where $\Delta y = 0.70$, the fairness score is computed as follows:  
\begin{equation*}
    \Phi_{size}^{FCCN} = \frac{2}{\pi}\cdot \textit{arctan}(0.70) \approx 0.39
\end{equation*}

This value represents the fairness score for the community-wise metric (FCCN) with respect to community size.
 
The proposed metric $(\Phi_p^{F*})$ ranges from $-1$ to $1$, where a value of $0$ (corresponding to a straight best-fit line) indicates a fair result, meaning that all communities are identified either equally well or equally poorly. Negative values suggest that the community detection method favors communities with lower property values $(p)$, while positive values indicate a bias toward communities with higher property values $(p)$.

\subsection*{Analyzing Metric Behavior}\label{sec:metric-behavior}

\begin{figure}[h!]
        \centering
        \includegraphics[width=7cm]{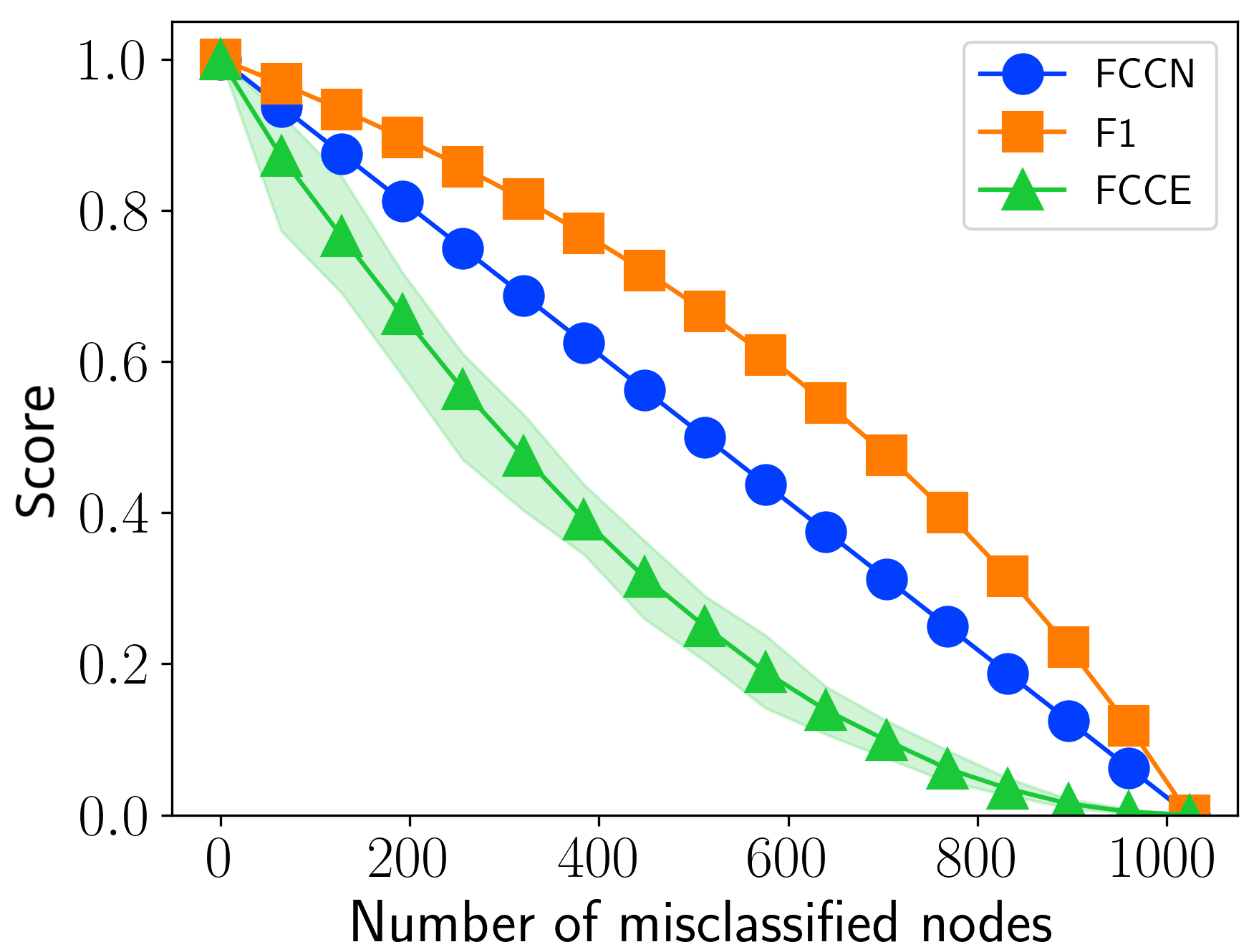}
        \caption{Community-wise fairness metric score as the number of misclassified nodes in the predicted community increases. The plot shows the average FCCE values over 20 iterations, along with the highest and lowest recorded values at each point.}
        \label{fig:behavior-mapped}
    \end{figure}

In this section, we analyze how the proposed fairness measures respond to node misclassification. To do this, we evaluate different fairness metrics under various levels of node misclassification to understand its impact on community fairness. We construct a HICH-BA network \cite {Saxena2023fairness} consisting of a single community with 1,024 nodes and $\sim$90k edges. The misclassification process gradually removes the nodes from this community, starting from 0 to 1,024. Initially, the classification is entirely accurate, with the predicted community perfectly aligning with the ground truth community. However, as misclassification increases, the prediction deviates by progressively removing nodes. Figure~\ref{fig:behavior-mapped} shows how the fairness metric scores change as the number of misclassified nodes increases over time. For FCCE, the figure highlights the range between the highest and lowest values observed across 20 repetitions of misclassification, with the average score marked. 

\begin{figure}[h!]
\centering
    \begin{subfigure}[b]{0.4\textwidth}            
            \centering\includegraphics[width=\textwidth]{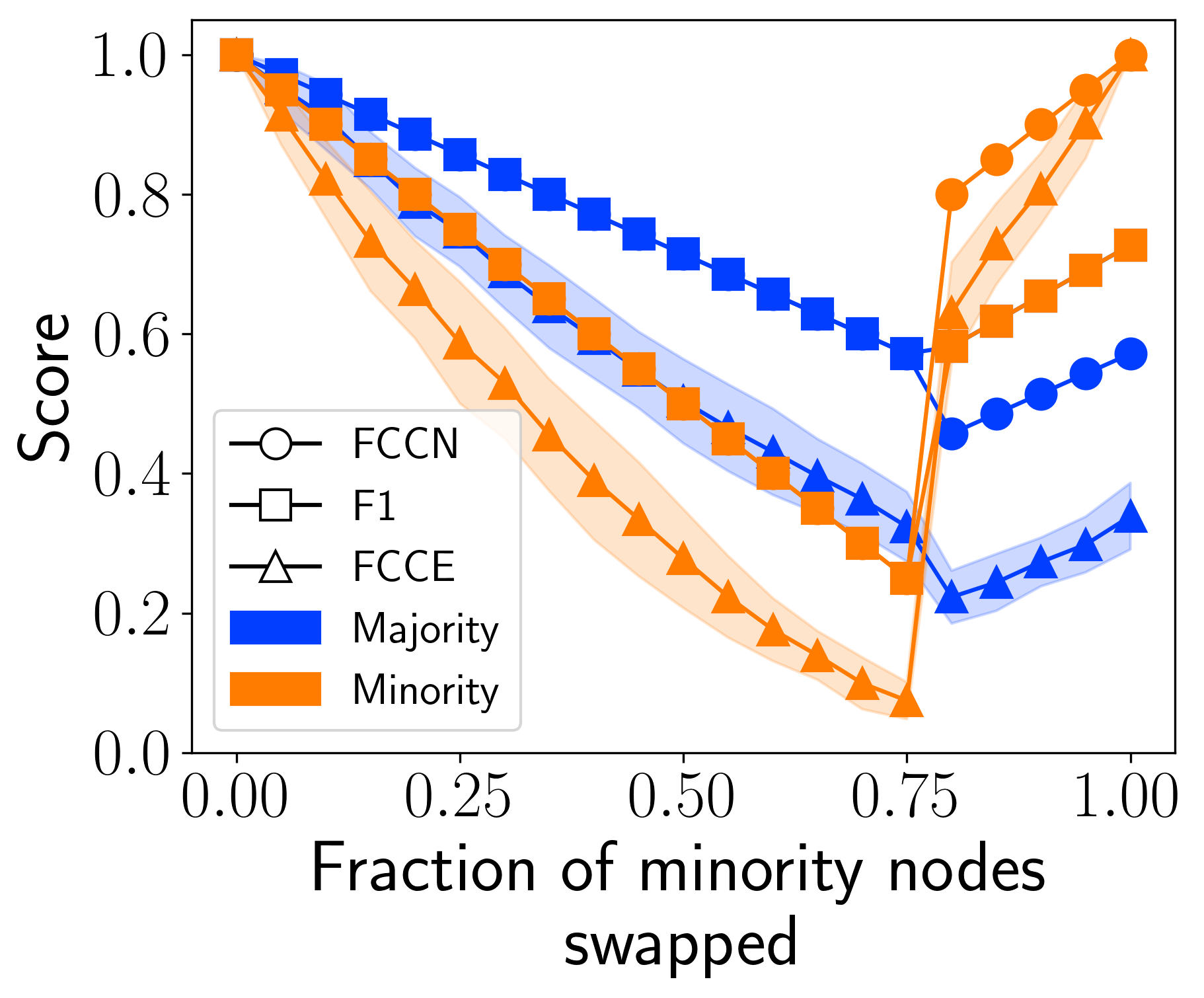}\\
            (a)
    \end{subfigure} \quad 
    \begin{subfigure}[b]{0.42\textwidth}
            \centering
            \includegraphics[width=\textwidth]{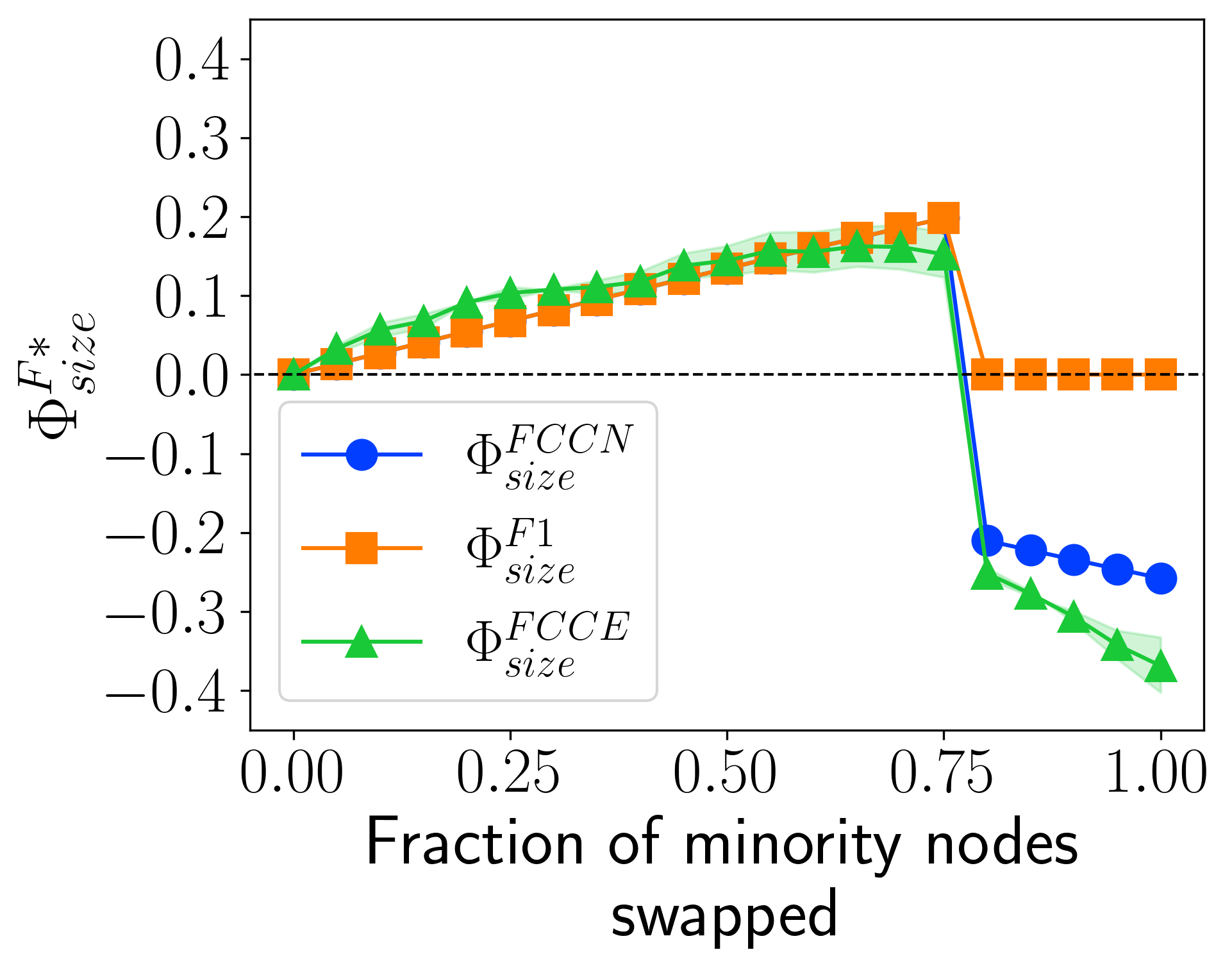}\\
            (b)
    \end{subfigure}
    \caption{Analysis of (a) the behavior of community-wise performance metrics and (b) group fairness on a HICH-BA network having both minority and majority communities.} 
    \label{fig_metric_behavior}
\end{figure}

To observe the behavior of the proposed metric in networks where misclassified nodes can be reassigned to different communities, we generate a HICH-BA network \cite{Saxena2023fairness} with a homophily factor of 0.9. The network consists of two communities: a majority group with 70 nodes and a minority group with 40 nodes, connected by approximately 900 edges. Initially, the predicted communities align perfectly with the ground truth communities. To introduce misclassification, nodes are progressively swapped between the minority and majority communities, ranging from 0 to 40. The resulting predicted and ground truth communities are then mapped for evaluation. Figure~\ref{fig_metric_behavior} presents the fairness scores for individual communities (left) and group fairness (right) as a function of the number of swapped nodes. Since the FCCE score depends on which specific nodes are reassigned, the figure reports the average value along with variations observed over 20 iterations. Due to the homophilic nature of the network, FCCE values tend to be lower than those of FCCN and F1.

As an equal number of nodes are exchanged between the minority and majority communities, community-wise fairness remains lower for the minority group compared to the majority until a critical threshold ($\sim$0.75). Beyond this point, the mapping between majority and minority communities is switched, leading to an increase in fairness for the minority group relative to the majority. A similar trend is observed in $\Phi^{F*}_{size}$, which initially favors the majority but, after the mapping transition, either favors the minority or remains neutral. Additionally, $\Phi^{F1}_{size}$  is fair after this threshold, as it accounts for nodes in the predicted communities that do not appear in the ground truth. It is important to note that this example represents an extreme scenario that may not commonly occur in real-world settings.

The proposed fairness metric ($\Phi$) provides insights into biases, allowing community detection methods to be assessed not only based on overall performance but also in terms of fairness across communities with varying sizes, densities, and conductance.

\section*{Experimental Setup}

This section outlines the experimental setup, detailing the community detection methods used in the analysis, the datasets (both synthetic and real-world), and the metrics for evaluating the quality of the identified communities, focusing on the performance-fairness trade-off.

\subsection*{Community Detection Methods}

We analyze 24 community detection methods based on their performance and fairness using our proposed group fairness metric, $\Phi$. Table~\ref{tab:community_detection_table} classifies these methods into six classes according to their approach to community partitioning. These approaches are briefly explained below.

\begin{enumerate}
    \item \textbf{Optimization} methods optimize a quality function to assess partition or community quality. Most of the methods in this category use the modularity function \cite{Newman2006a, Newman2004}, except significance \cite{Traag2013}. Given that modularity optimization is NP-hard, these methods typically use heuristic techniques.

    \item \textbf{Dynamics} methods infer community structure through network traversal, often using random walks. The intuition is that random walks tend to remain within communities, as they are more densely connected than the rest of the graph. For instance, Walktrap \cite{Pons2005} proposes a similarity measure based on random walks, which effectively captures dense subgraphs in sparse graphs.
    
    \item \textbf{Spectral} methods create partitions based on spectral properties of matrices describing the network, such as the adjacency matrix or the Laplacian matrix~\cite{Fortunato2016}. These methods analyze the eigenvalues and eigenvectors of these matrices, which provide insights into the network structure. For example, Spectral Clustering~\cite{Higham2007} makes use of the Fiedler vector, the second smallest eigenvalue, to construct communities. 
    
    \item \textbf{Propagation} methods initially assign a community label to all nodes and then iteratively update nodes' community label based on neighboring nodes, aiming for a stable configuration that reflects the community structure. The Label Propagation Algorithm (LPA) \cite{Cordasco2011} was the first such method, updating the nodes' label according to the majority of neighbors, with ties resolved randomly. The speed and scalability of the LPA method make it well-suited for large networks and serve as the foundation for various extensions, including FLPA \cite{Traag2023}, LLPA \cite{Hu2016}, LPA-MNI \cite{Li2021}, WSSLPA \cite{Malhotra2021}, and DCC \cite{das2022dcc}.

    \item \textbf{Representation Learning} based methods first generate a network embedding that is a latent representation of the network in a low-dimensional space, and then apply clustering algorithms, such as k-means, to create a partition of the network \cite{arya2022node}.

    \item \textbf{Probabilistic} methods approach community detection by modeling the network as being generated by an underlying probabilistic process. These models estimate the likelihood that each node belongs to a certain community based on the observed edges in the network. The methods falling under this category include the Expectation-Maximization~\cite{Newman2007}, Stochastic Block Model~\cite{Peixoto2014}, and SBM - Nested~\cite{Peixoto2014a}.
    
\end{enumerate}

\begin{table}[t]
    \centering
    \caption{Overview of community detection methods used in experimentation from 6 classes.}
    \label{tab:community_detection_table}
    \begin{tabularx}{\linewidth}{@{}|X|X|X|@{}}
        \hline \multicolumn{1}{|c|}{Optimization} & \multicolumn{1}{c|}{Spectral Properties} & \multicolumn{1}{c|}{Representation Learning} \\ \hline
        \begin{tabular}[c]{@{}l@{}}
            $\bullet$~Clauset-Newman-Moore \\\hspace{9.91664pt}Algorithm (CNM)~\cite{Clauset2004} \\
            $\bullet$~Combo~\cite{Sobolevsky2014} \\ 
            $\bullet$~Leiden~\cite{Traag2019} \\ 
            $\bullet$~Louvain~\cite{Blondel2008} \\ 
            $\bullet$~Paris~\cite{Bonald2018} \\ 
            $\bullet$~Reichardt-Bornholdt - \\\hspace{9.91664pt}configuration null model \\\hspace{9.91664pt}(RB-C)~\cite{Reichardt2006} \\ 
            $\bullet$~Reichardt-Bornholdt - \\\hspace{9.91664pt}Erdős-Rényi null model \\\hspace{9.91664pt}(RB-ER)~\cite{Reichardt2006} \\
            $\bullet$~Significance~\cite{Traag2013}
        \end{tabular}  & 
        \begin{tabular}[c]{@{}l@{}}
            $\bullet$~Eigenvector~\cite{Newman2006} \\
            $\bullet$~Regularized Spectral \\\hspace{9.91664pt}Clustering with k-means \\\hspace{9.91664pt}(RSC-K)~\cite{Zhang2018} \\
            $\bullet$~RSC sklearn Spectral \\\hspace{9.91664pt}Embedding (RCS-SSE)\\\hspace{9.91664pt}\cite{Zhang2018} \\
            $\bullet$~RSC - Vanilla (RSC-V)\\\hspace{9.91664pt}\cite{Zhang2018} \\
            $\bullet$~Spectral Clustering~\cite{Higham2007} \\ \\ \\ \\
        \end{tabular} &
        \begin{tabular}[c]{@{}l@{}}
            $\bullet$~Deepwalk~\cite{Perozzi2014} \\
            $\bullet$~Fairwalk~\cite{Rahman2019} \\
            $\bullet$~Node2Vec~\cite{Grover2016} \\ \\ \\ \\ \\ \\ \\ \\ \\ \\ \\
        \end{tabular} \\ \hline
    \end{tabularx} \\
    
    \vspace{.1cm}
    
    \begin{tabularx}{\linewidth}{@{}|X|X|X|@{}}
        \hline \multicolumn{1}{|c|}{Dynamics} & \multicolumn{1}{c|}{Propagation} & \multicolumn{1}{c|}{Probabilistic} \\ \hline
        \begin{tabular}[c]{@{}l@{}}
            $\bullet$~Infomap~\cite{Rosvall2008} \\
            $\bullet$~Spinglass~\cite{Reichardt2006} \\ 
            $\bullet$~Walktrap~\cite{Pons2005} \\ \\ \\
        \end{tabular} &
        \begin{tabular}[c]{@{}l@{}}
            $\bullet$~Fluid~\cite{Parés2017} \\ 
            $\bullet$~Label Propagation~\cite{Cordasco2011} \\ \\ \\ \\
        \end{tabular} &
        \begin{tabular}[c]{@{}l@{}}
            $\bullet$~Expectation-Maximization \\\hspace{9.91664pt}(EM)~\cite{Newman2007} \\ 
            $\bullet$~Stochastic Block Model \\\hspace{9.91664pt}(SBM)~\cite{Peixoto2014} \\
            $\bullet$~SBM - Nested~\cite{Peixoto2014a}
        \end{tabular} \\ \hline
    \end{tabularx}
    
\end{table}


\subsection*{Network datasets}

To perform our experiments, we use synthetic benchmark network-generating models and real-world networks, which are discussed below.

\subsubsection*{Synthetic Networks}

We use the following benchmark models to generate synthetic networks:

\begin{enumerate}
    \item \textbf{LFR Benchmark Model:} The Lancichinetti–Fortunato–Radicchi (LFR) benchmark~\cite{Lancichinetti2008} is a widely used method for generating synthetic networks with an inherent community structure. It improves upon the Girvan-Newman benchmark~\cite{Girvan2002} by incorporating power-law distributions for both degree and community sizes, making the generated networks more representative of real-world structures~\cite{Arenas2003, Stegehuis2016}. A key feature of the LFR benchmark is the mixing parameter $\mu$, which determines the fraction of intercommunity edges. When $\mu=0$, all edges are confined within communities, whereas $\mu=1$ results in no intra-community edges. Other configurable parameters in the LFR model include the number of nodes, average and maximum degree, minimum and maximum community size, and power law distribution exponent of community sizes ($\tau$), allowing for flexible network generation. 

    For our experiments, we primarily set the parameter values used by Lancichinetti et al.~\cite{Lancichinetti2011}. Specifically, we generate networks having 10,000 nodes with other parameters set as- the average degree $\textit{$deg_{avg}$}=20$, the maximum degree $deg_{max}=100$, the minimum community size $|c_{min}|=20$, the power-law exponent of the degree distribution $\tau_1=2$, and the power-law exponent of the community size distribution as $\tau_2=2.5$. The mixing parameter is varied as $\mu \in [0.2, 0.4, 0.6]$ to generate different types of networks.

\item \textbf{ABCD Model:} The Artificial Benchmark for Community Detection (ABCD) model~\cite{Kamiński2021} is similar to the LFR benchmark but offers improvements in scalability and interpretability of the mixing parameter. Like LFR, it generates networks where both the degree and community size distributions follow power laws. However, ABCD runs approximately 100 times faster. ABCD introduces $\xi$ as its mixing parameter, providing a more intuitive measure of community strength. When $\xi=0$, all edges remain within communities, similar to LFR. However, when $\xi=1$, edges are randomly distributed throughout the network. In contrast to LFR, where $\mu=1$ results in zero intra-community edges. Crucially, ABCD maintains intra-community edges proportional to community size.

To ensure comparability, we configure the ABCD model parameters to closely align with the graphs generated by the LFR benchmark. Therefore, the corresponding $\xi$ values are determined using the global method from~\cite{Kamiński2021}. While $\xi$ plays a similar role to $\mu$ in LFR, it differs slightly as $\xi \in [0.201, 0.402, 0.603]$. 

\item \textbf{HICH-BA Model:} HIgh Clustering Homophily Barab\'asi-Albert (HICH-BA)~\cite{Saxena2023fairness} is a recently proposed extension to the homophily BA model~\cite{Karimi2018}. It was specifically designed to generate homophilic networks with controlled clustering coefficient, density, and community sizes. The HICH-BA model takes the following parameters:
    \begin{itemize}
        \item $n$: Total number of nodes in the network.
        \item $r$: This is a list containing the likelihood of assigning a node to each community. The $i$-th element in $r$ shows the likelihood of adding a node to the $i_{th}$ community $c_i$. 
        \item $h$: This is the homophily factor.
        \item $p_N$: At each iteration, a node is added to the network with probability $p_N$; otherwise, an edge is added. 
        \item $p_t$: Probability to form a close triad connection. 
        \item $p_{PA}$: Probability for a new edge to be placed using preferential attachment. 
    \end{itemize}
    HICH-BA iteratively adds nodes and edges to the network based on probabilities provided by the user. HICH-BA allows for a custom community size distribution by setting $r$. We set $r$ to create two network cases: (i) multiple minority communities with one majority community (MMin), where the parameter $r$ was set as [0.005, 0.005, 0.005, 0.01, 0.01, 0.01, 0.02, 0.02, 0.02, 0.9], and (ii) multiple majority communities (MMaj), where $r$ is set as [0.003, 0.003, 0.003, 0.03, 0.03, 0.03, 0.3, 0.3, 0.3]. The other parameters are same for both cases: $n=10,000$, $h = 0.9$, $p_N = 0.1$, $p_T = 0.3$, and $p_{PA} = 0.8$. 

\end{enumerate}

Table \ref{tab:synthetic_networks} summarizes the average parameters of the generated networks using different benchmark models. In our experiments, we generate ten networks for each configuration using the chosen generative model. All results are shown by computing the average and standard deviation of the scores obtained across these networks.

\begin{table}[h!]
\centering
\resizebox{\textwidth}{!}{
\begin{tabular}{|l|*{8}{r|}}
    \hline \rule{0pt}{.45cm}Dataset name & \multicolumn{1}{c|}{$|V|$} & \multicolumn{1}{c|}{$|E|$} & \multicolumn{1}{c|}{$\textit{deg}_{avg}$} & \multicolumn{1}{c|}{$\textit{deg}_{max}$} & \multicolumn{1}{c|}{$\textit{deg}_{min}$} & \multicolumn{1}{c|}{$|C|$} & \multicolumn{1}{c|}{$|c_{max}|$} & \multicolumn{1}{c|}{$|c_{min}|$} \\ \hline 
    LFR, $\mu=0.2$ & 10,000 & 133,388.6 & 26.7 & 109.8 & 8 & 283.2 & 98.2 & 20 \\
    LFR, $\mu=0.4$ & 10,000 & 136,803.6 & 27.4 & 110.8 & 8 & 279.0 & 98.8 & 20 \\
    LFR, $\mu=0.6$ & 10,000 & 138,811.0 & 27.8 & 114.8 & 8 & 283.8 & 98.2 & 20 \\
    ABCD, $\xi=0.2$ & 10,000 & 105,197.0 & 21.0 & 100.2 & 8 & 267.8 & 115.2 & 20 \\
    ABCD, $\xi=0.4$ & 10,000 & 105,530.0 & 21.1 & 100.4 & 8 & 267.8 & 116.0 & 20 \\
    ABCD, $\xi=0.6$ & 10,000 & 104,462.2 & 20.9 & 100.2 & 8 & 259.2 & 117.4 & 20 \\
    HICH-BA MMaj & 10,000 & 98,054.4 & 19.6 & 1021.4 & 1.0 & 9 & 3038.8 & 27.4 \\
    HICH-BA MMin & 10,000 & 92,597.4 & 18.5 & 1283.6 & 1.0 & 10 & 8964.6 & 42.0 \\ \hline
\end{tabular}
}
\caption{Synthetic dataset summary: the values are the average of all networks of a given type. $|V|$: number of nodes, $|E|$: number of nodes, $\textit{deg}_{avg}$: average degree, $\textit{deg}_{max}$: maximum degree, $\textit{deg}_{min}$: minimum degree, $|C|$: number of communities, $|c_{max}|$: size of the largest community, $|c_{min}|$: size of the smallest community.}
\label{tab:synthetic_networks}
\end{table}

\begin{figure}[t]
\captionsetup[subfigure]{justification=centering}
\centering
\begin{subfigure}[c]{0.05\textwidth}
\caption*{\rotatebox{90}{(a) LFR}}%
\end{subfigure}%
\begin{minipage}[c]{0.9\textwidth}
\begin{subfigure}[t]{0.28\textwidth}
\includegraphics[height=3.3cm]{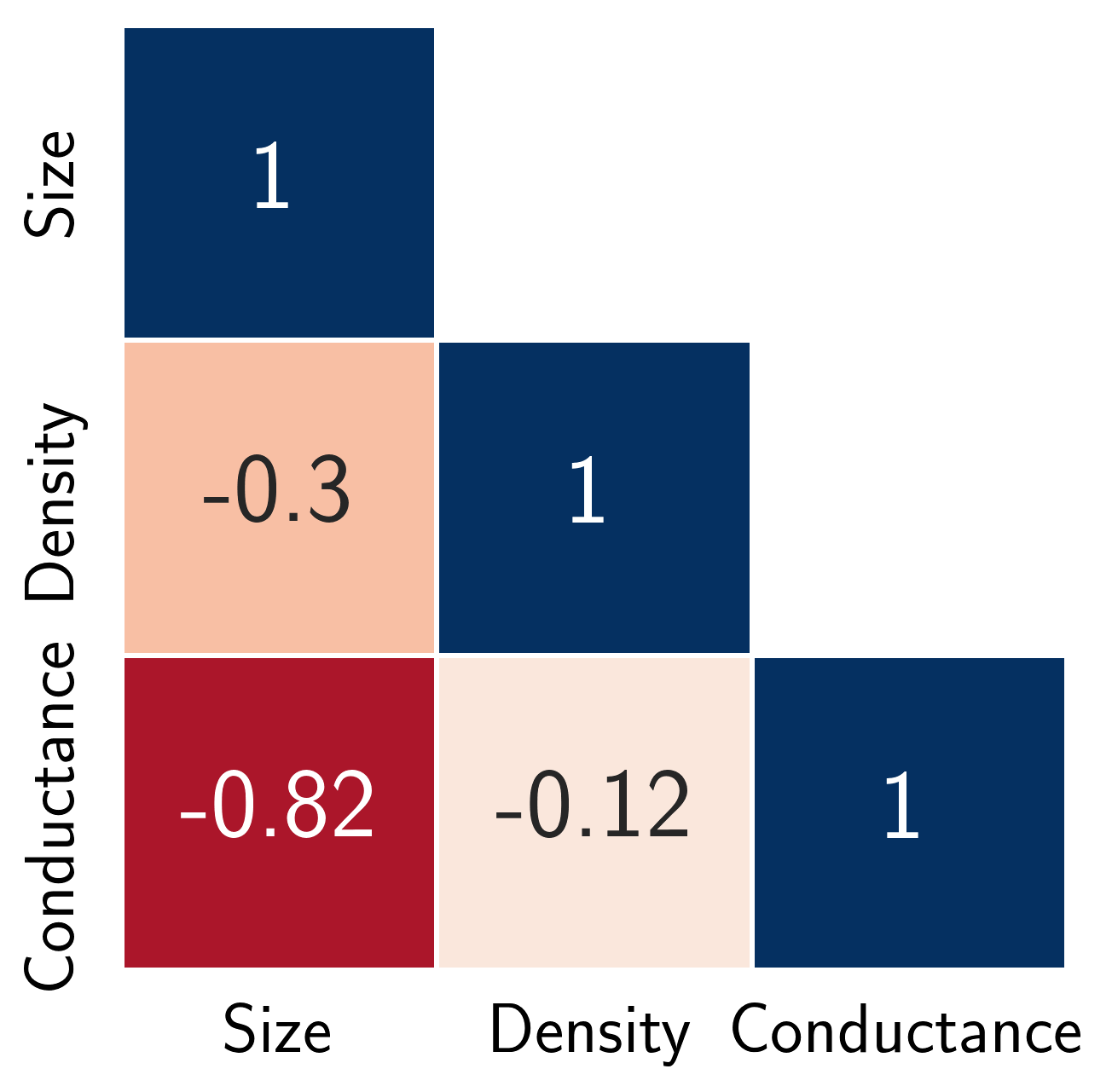}\caption*{$\mu=0.2$}\label{fig:corr_LFR_mu2}
\end{subfigure}
\begin{subfigure}[t]{0.28\textwidth}
\includegraphics[height=3.3cm]{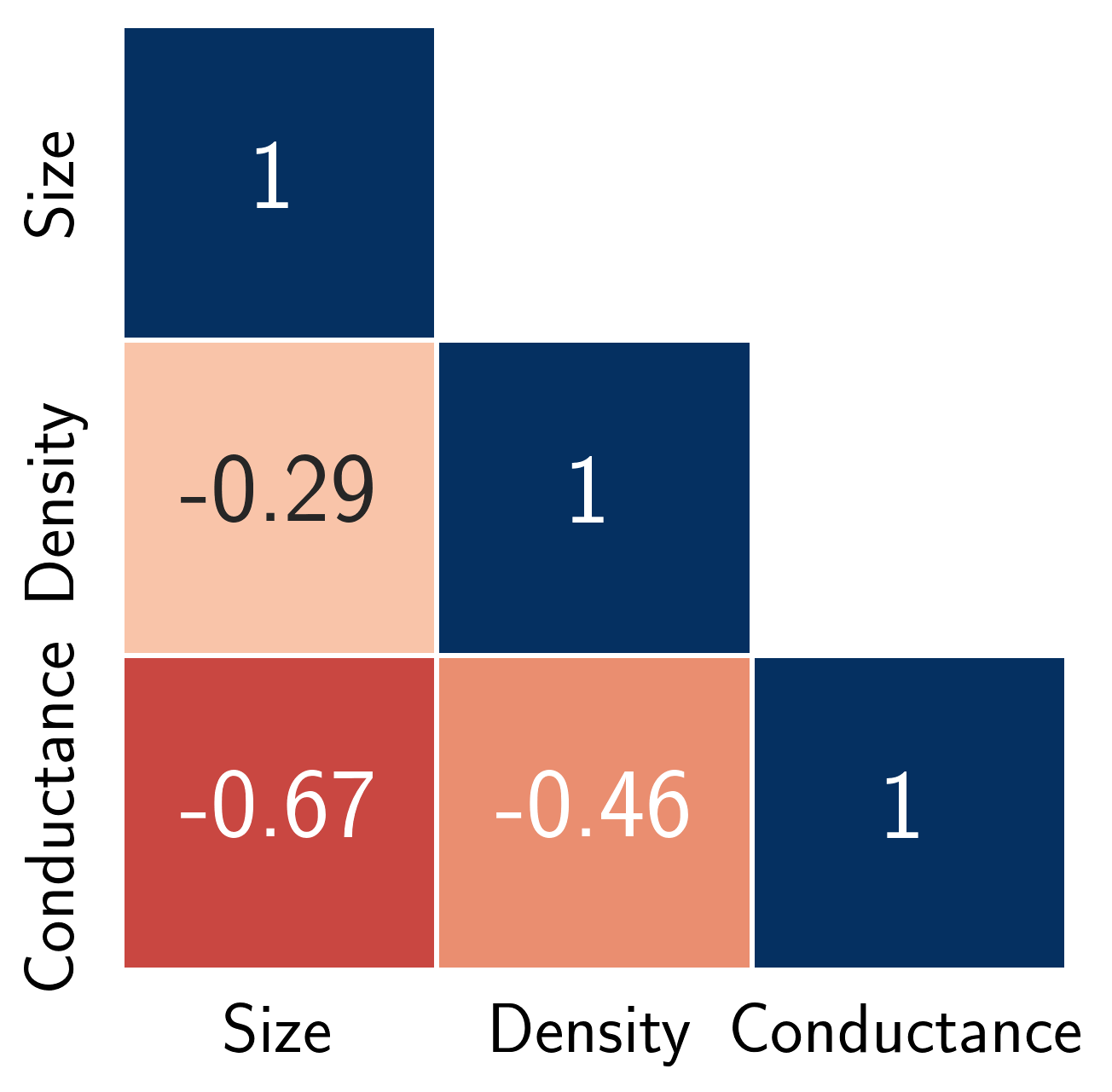}\caption*{$\mu=0.4$}\label{fig:corr_LFR_mu2}
\end{subfigure}
\begin{subfigure}[t]{0.33\textwidth}
\centering
\includegraphics[height=3.3cm]{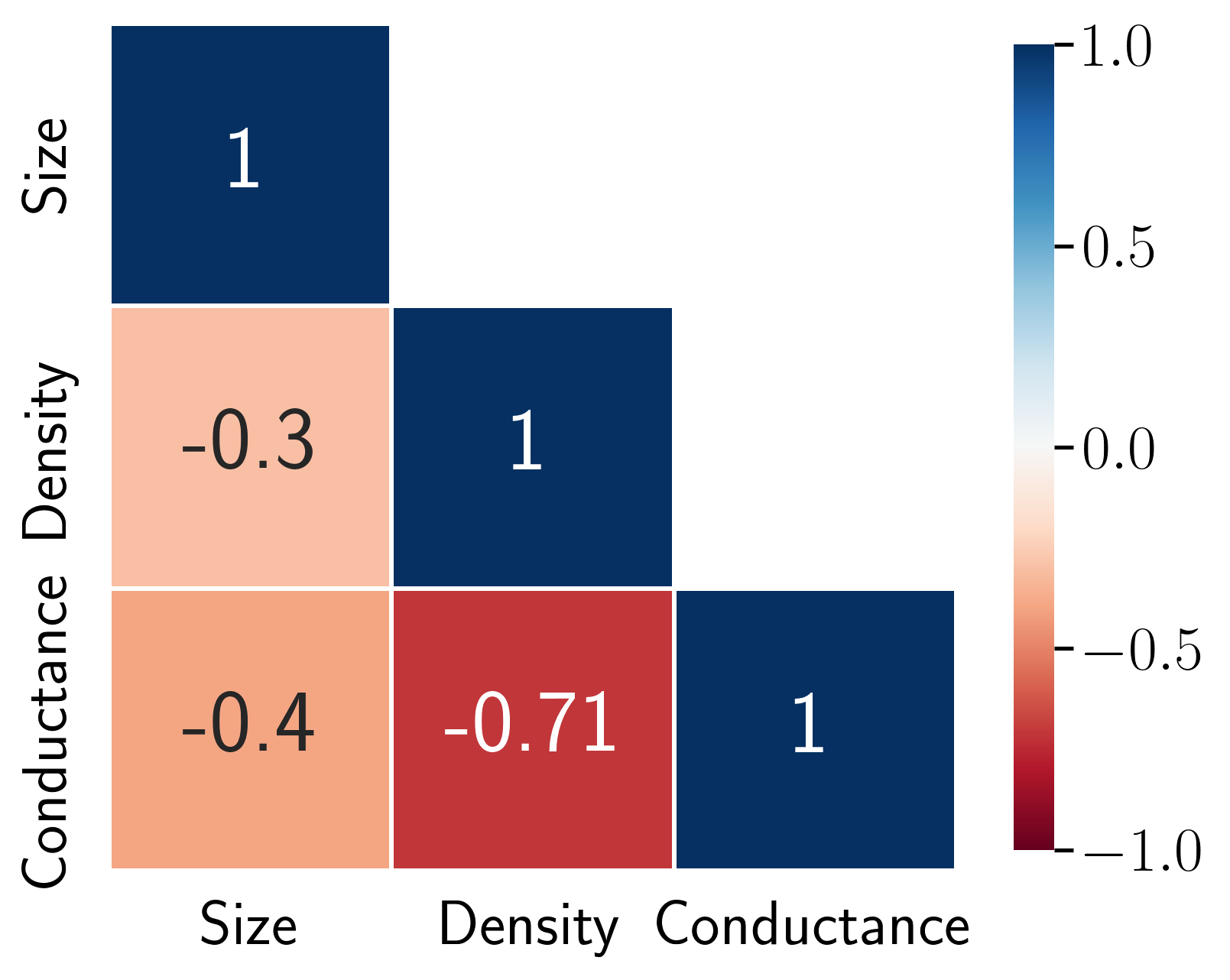}\caption*{$\mu=0.6$}\label{fig:corr_LFR_mu2}
\end{subfigure}
\end{minipage}
\\
\vspace{2mm}
\begin{subfigure}[c]{0.05\textwidth}
\caption*{\rotatebox{90}{(b) ABCD}}%
\end{subfigure}%
\begin{minipage}[c]{0.9\textwidth}
\begin{subfigure}[t]{0.28\textwidth}
\includegraphics[height=3.3cm]{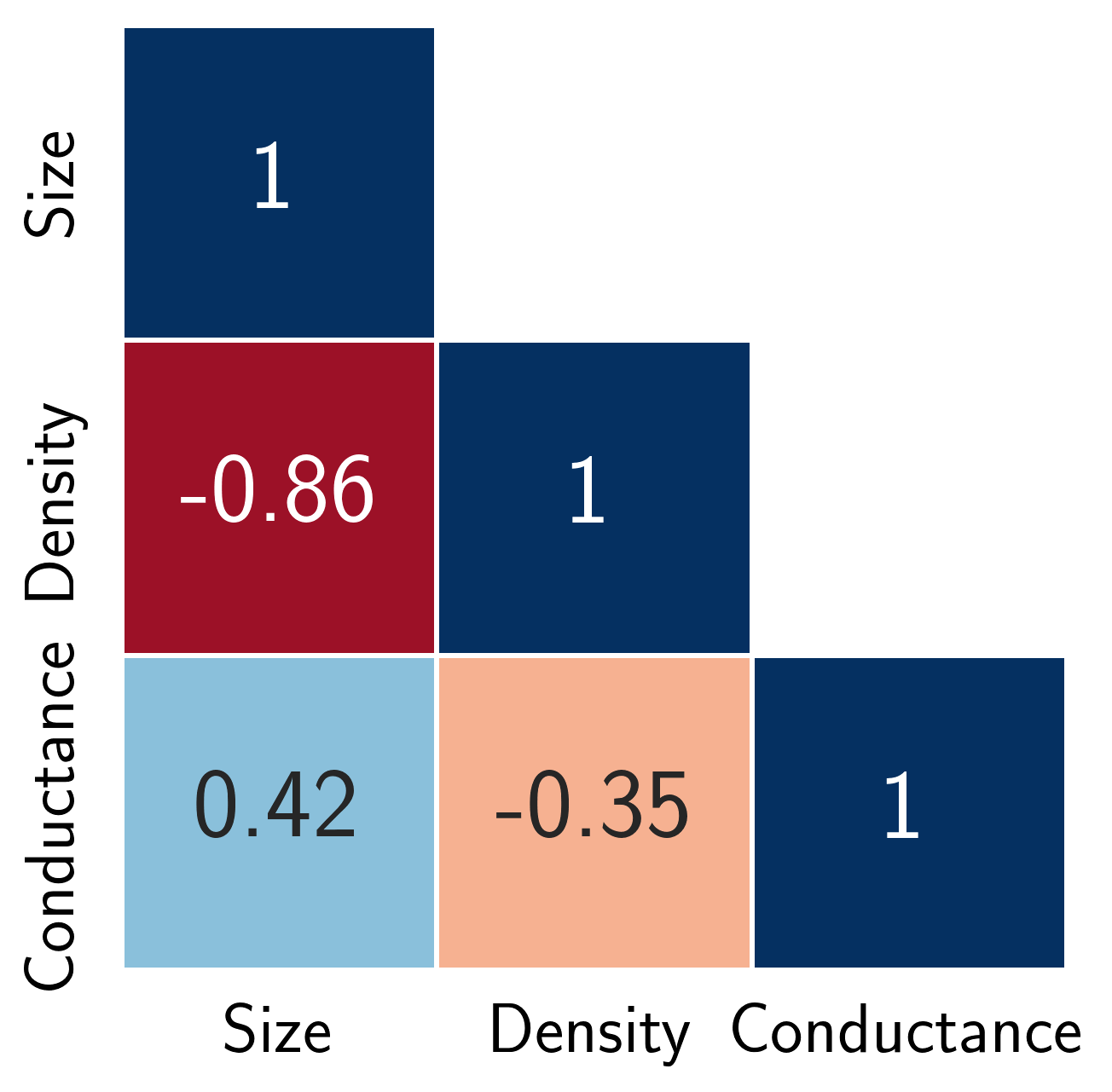}\caption*{$\xi=0.2$}\label{fig:corr_LFR_mu2}
\end{subfigure}
\begin{subfigure}[t]{0.28\textwidth}
\includegraphics[height=3.3cm]{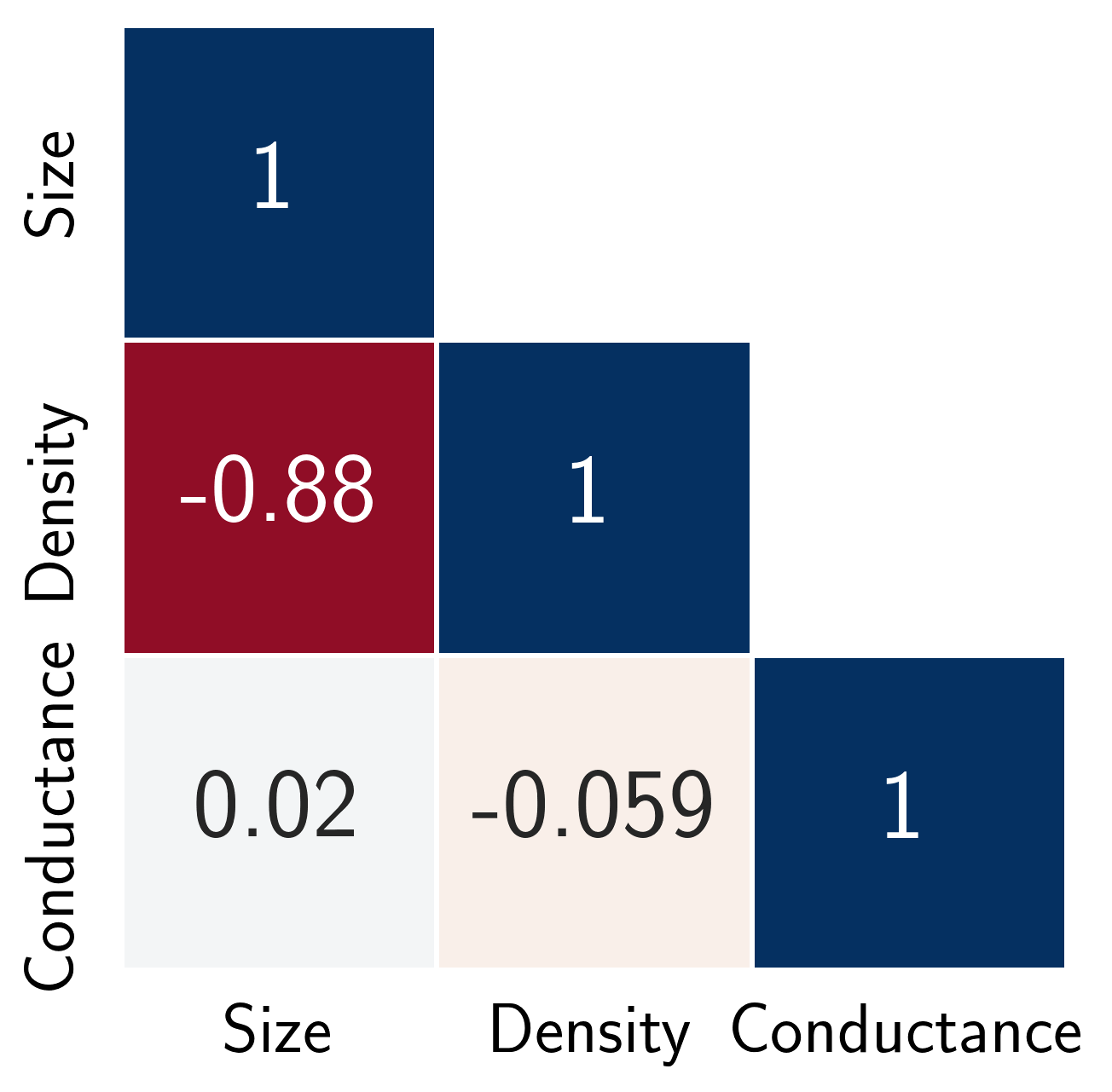}\caption*{$\xi=0.4$}\label{fig:corr_LFR_mu2}
\end{subfigure}
\begin{subfigure}[t]{0.33\textwidth}
\centering
\includegraphics[height=3.3cm]{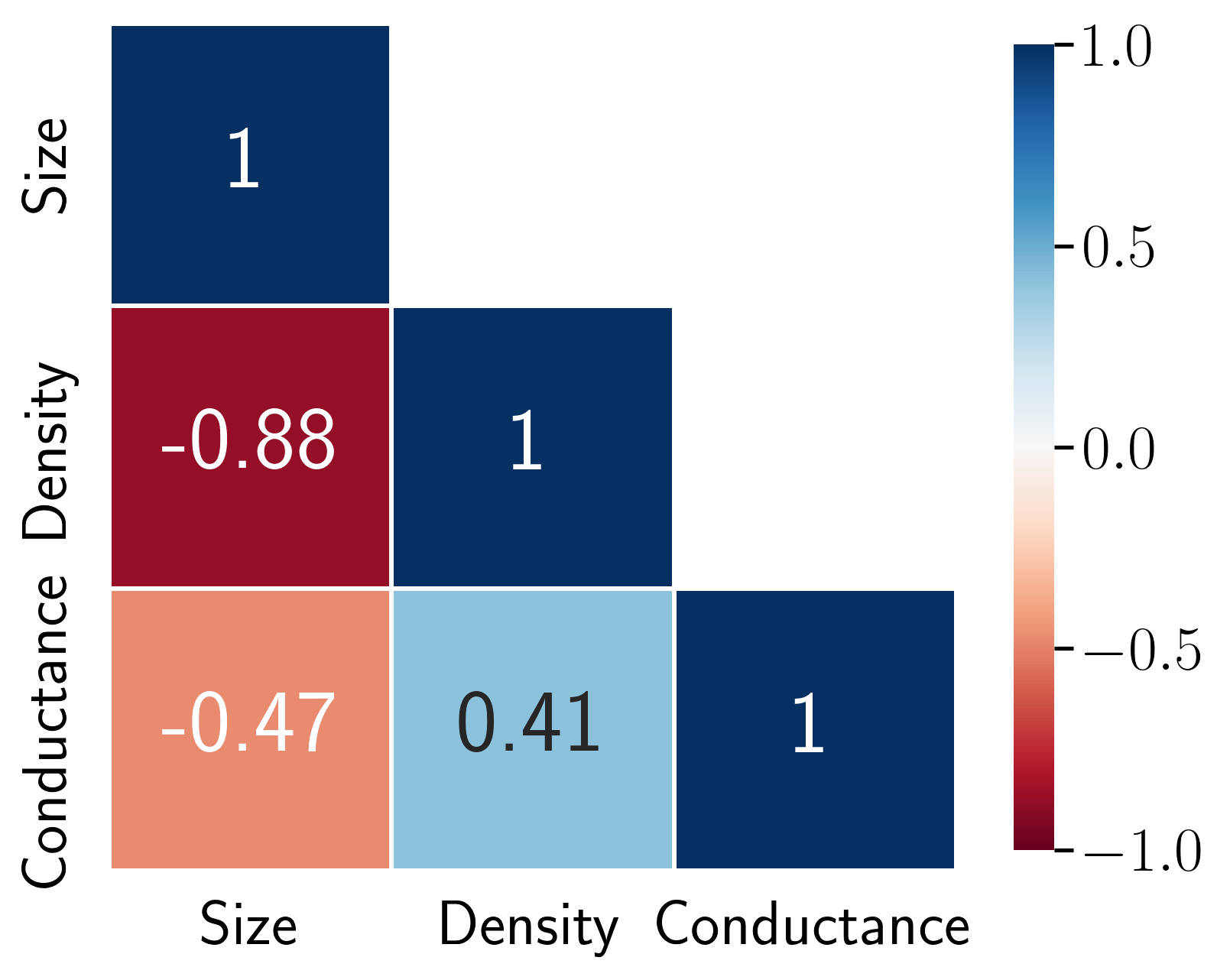}\caption*{$\xi=0.6$}\label{fig:corr_LFR_mu2}
\end{subfigure}
\end{minipage}
\\
\vspace{2mm}
\begin{subfigure}[c]{0.05\textwidth}
\caption*{\rotatebox{90}{(c) HICH-BA}}%
\end{subfigure}%
\begin{minipage}[c]{0.9\textwidth}
\centering
\begin{subfigure}[t]{0.28\textwidth}
\includegraphics[height=3.3cm]{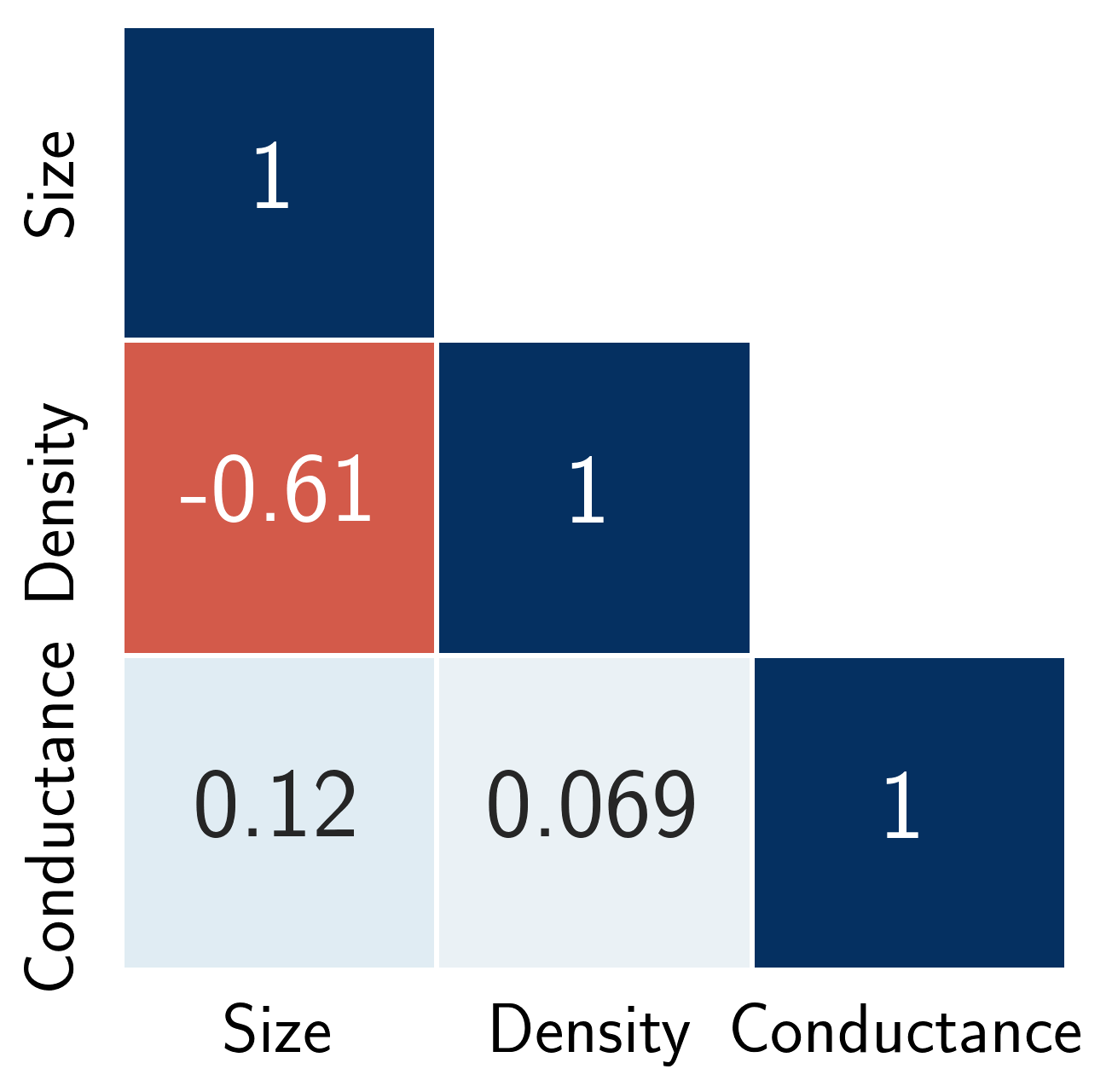}\caption*{$MMaj$}\label{fig:corr_LFR_mu2}
\end{subfigure}\quad\quad\quad\quad\quad
\begin{subfigure}[t]{0.28\textwidth}
\includegraphics[height=3.3cm]{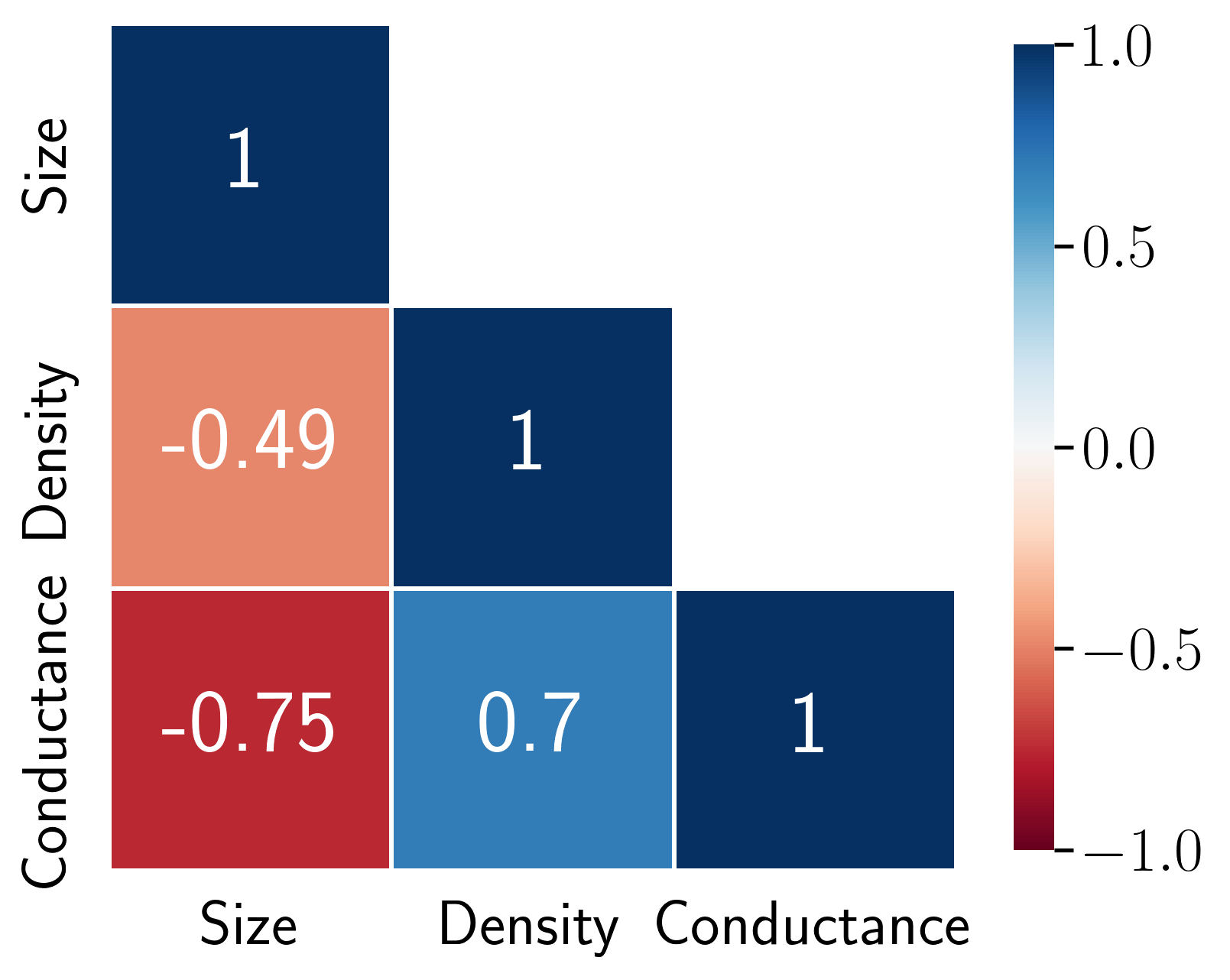}\caption*{$MMin$}\label{fig:corr_LFR_mu2}
\end{subfigure}
\end{minipage}
\vspace{2mm}
\caption{correlation between community properties — size, density, and conductance, in LFR, ABCD, and HICH-BA networks.}\label{fig:corr_matrix} 
\end{figure}

\vspace{5mm}
\textbf{Structural Properties of Networks.} The networks generated using different benchmark models have distinct structural characteristics, reflecting various aspects of real-world networks. Therefore, it is crucial to analyze the structural properties of these networks, which are later used to assess fairness, including community size, density, and conductance, to gain a deeper understanding of their internal and external connectivity. To achieve this, we generate networks using different models and study the correlation between community size, density, and conductance using the Pearson correlation coefficient. Figure~\ref{fig:corr_matrix} presents these correlations for LFR, ABCD, and HICH-BA models, providing insights into the interplay of community connectivity across different network types.  

For LFR networks (shown in Figure~\ref{fig:corr_matrix} (a)), there is a strong negative correlation between size and conductance at $\mu=0.2$, indicating that larger communities tend to have lower conductance. As $\mu$ increases, this correlation weakens, while a new relationship emerges between density and conductance, showing that the dense communities have lower conductance. In contrast, ABCD networks (Figure~\ref{fig:corr_matrix} (b)) consistently show a negative correlation between density and size across all $\xi$ values. Unlike LFR networks, large communities in ABCD networks have high conductance at low $\mu$ values, suggesting they are well separated. Additionally, the correlation between density and conductance differs across LFR and ABCD models, with dense communities in ABCD being more distinctly separated for high $\xi$ values. For HICH-BA networks (Figure~\ref{fig:corr_matrix} (c)), structural properties depend heavily on the network's composition, particularly whether multiple majority or minority groups exist. In MMaj and MMin networks, density and size are negatively correlated. However, in MMin networks, conductance also has a strongly negative correlation with size and a positive correlation with density.

\subsubsection*{Real-World Networks}
We use the following real-world networks; summarized in Table~\ref{tab:real-world data}.

\begin{enumerate}
    \item \textbf{Polbooks}~\cite{Krebs}: This network represents the co-purchasing of books on US politics, with data collected around the 2004 presidential election. The books are grouped into three communities based on political affiliation: conservative, liberal, and non-partisan.
    
    \item \textbf{Football}~\cite{Girvan2002}: The US college (American) football network represents the regular-season games of the 2000 NCAA Division I-A football matches. Nodes correspond to football teams, while edges indicate matches played between them. The communities in the network are defined based on the 12 conferences to which the teams belong.
    
    
    \item \textbf{Eu-core}~\cite{Leskovec2014, Yin2017}: This communication network is constructed from emails exchanged between employees of a European research center, where communities correspond to the departments employees belong to. The network is converted into an undirected network, and the largest connected component comprising 98\% of the nodes is considered. 
\end{enumerate}

\begin{table}[h!]
\centering
\begin{tabular}{|l|*{8}{r|}}
    \hline \rule{0pt}{.45cm}Dataset name & \multicolumn{1}{c|}{$|V|$} & \multicolumn{1}{c|}{$|E|$} & \multicolumn{1}{c|}{$\textit{deg}_{avg}$} & \multicolumn{1}{c|}{$\textit{deg}_{max}$} & \multicolumn{1}{c|}{$|C|$} & \multicolumn{1}{c|}{$|c_{max}|$} & \multicolumn{1}{c|}{$|c_{min}|$} \\ \hline 
    Polbooks~\cite{Krebs} & 105 & 441 & 8.40 & 25 & 3 & 49 & 13 \\
    Football~\cite{Girvan2002} & 115 & 613 & 10.66 & 12 & 12 & 13 & 5 \\
    Eu-core~\cite{Leskovec2014, Yin2017} & 986 & 16,687 & 33.85 & 347 & 42 & 107 & 1 \\ \hline
\end{tabular}
\caption{Characteristics of real-world datasets - $|V|$: number of nodes, $|E|$: number of edges, $deg_{avg}$: average degree, $deg_{max}$: maximum degree, $|C|$: total number of communities, $|c_{max}|$: size of the largest community, and $|c_{min}|$: size of the smallest community.}
\label{tab:real-world data}
\end{table}

All networks used in our experiments are undirected, unweighted, and connected with non-overlapping communities.

\subsection*{Evaluation Metrics}\label{qualitymetrics} 

We use the following metrics to measure the quality of the identified communities as compared to the ground truth. Let us assume that $G=(V,E)$ is the given network, $C$ is the set of ground truth communities, defined as $C=\{c_1, c_2,..., c_{m}\}$, and $P$ is the set of predicted communities $P=\{p_1, p_2, ..., p_{k}\}$. 

\begin{enumerate}
    \item \textbf{Normalized Mutual Information (NMI):}
    NMI~\cite{Fred2003} is, as the name suggests, a normalized variant of mutual information (MI). The MI is computed as

    \begin{equation*}
        MI= S(C) + S(P) - S(C, P)
    \end{equation*}
    
    where $S(C)$ represents the Shannon entropy of the clustering size distribution, while $S(C, P)$ denotes the Shannon entropy of the joint clustering size distribution. Now, NMI is computed as:

    \begin{equation*}
        NMI(C, P) = \frac{MI(C, P)}{\sqrt{H(C) H(P)}}
    \end{equation*}

    where $H(C)$ and $H(P)$ are the entropies of the ground truth and predicted partitions, respectively.

    \item \textbf{Reduced Mutual Information (RMI):}
    The RMI~\cite{Newman2020} was proposed to address a flaw with MI that the measure should return a value of $0$ when the predicted partition consists of $n$ communities (each having one node), indicating no meaningful structure. However, instead of $0$, MI returns $H(C)$~\cite{Newman2020}. RMI addresses this by adding a correction term and is defined as:
    \begin{equation*}
        \textit{RMI}(C,P) = MI(C, P) - \log \frac{\Omega(a,b)}{n} 
    \end{equation*}
    where $\Omega(a,b)$ represents the number of contingency tables with row and column sums equal to $a=\{|c_i|\}$ and $b=\{|p_j|\}$, respectively. We use the RMI method proposed in \cite{jerdee2024mutual}, which provides an improved approach for encoding contingency tables, and the upper limit of the computed values is bounded by 1
    
    \item \textbf{Adjusted Rand Index (ARI):} The ARI~\cite{Hubert1985} is a chance-adjusted version of the Rand Index (RI). RI is given by:
    \begin{equation*}\label{eq:RI}
        RI(C, P) = \frac{TP + TN}{TP + FP + FN + TN}
    \end{equation*}
    Here, TP is the number of true positives, TN is the number of true negatives, FP is the number of false positives, and FN is the number of false negatives. These terms are defined as follows: 
    \begin{itemize}
        \item TP: The number of node pairs that belong to the same community in both $C$ and $P$.
        \item TN: The number of node pairs that belong to a different community in both $C$ and $P$. 
        \item FP: The number of node pairs that are in a different community in $C$ and in the same community in $P$.
        \item FN: The number of node pairs that are in the same community in $C$ and in a different community in $P$.
    \end{itemize}

    Hubert et al.~\cite{Hubert1985} formulated a way to adjust for chance in any measure $M$ as follows:
    \begin{equation*}
        \textit{adjusted}\ M = \frac{M - E(M)}{M_{\textit{max}} - E(M)}
    \end{equation*}
    where $E(M)$ is the expected value for some null model. Hubert formulated that if partitions are generated randomly, the expected number of pairs in a community intersection $c_i \cap p_j$ is given by:
    \begin{equation*}
        E\binom{|c_i\cap p_j|}{2} = \binom{|c_i|}{2} \binom{|p_j|}{2} \bigg/ \binom{N}{2}
    \end{equation*}
    Using these definitions, the ARI is computed as:
    \begin{equation*}
        ARI(C, P) = \frac{\sum_{ij} \left(\genfrac{}{}{0pt}{0}{|c_i\cap p_j|}{2}\right) - \sum_i \left(\genfrac{}{}{0pt}{0}{|c_i|}{2}\right) \sum_j \left(\genfrac{}{}{0pt}{0}{|p_j|}{2}\right) \bigg/ \left(\genfrac{}{}{0pt}{0}{N}{2}\right)}
        {\frac{1}{2} \left( \sum_i \left(\genfrac{}{}{0pt}{0}{|c_i|}{2}\right) + \sum_j \left(\genfrac{}{}{0pt}{0}{|p_j|}{2}\right) \right) - \sum_i \left(\genfrac{}{}{0pt}{0}{|c_i|}{2}\right) \sum_j \left(\genfrac{}{}{0pt}{0}{|p_j|}{2}\right) \bigg/ \Bigl(\genfrac{}{}{0pt}{0}{N}{2}\Bigr)}
    \end{equation*}
    ARI has an upper bound of 1, which occurs when the predicted partition perfectly matches the ground truth partition. Its lower bound is -1, with negative values indicating that the similarity between the two partitions is lower than what would be expected from randomly assigned partitions.

    \item \textbf{Average F1 Score (PF1):} PF1, introduced by Rossetti et al. \cite{Rossetti2016}, evaluates the quality of predicted communities by mapping them to ground truth communities based on the highest label overlap. Multiple predicted communities can map to the same ground truth community. Once mapped, the similarity is assessed using the F1-score, i.e., the harmonic mean of recall and precision. The averaged F1-score provides a basis for comparing partitions. To differentiate it from our proposed measure, we referred to it as PF1.

    \item \textbf{Normalized F1 Score (NF1):} Rossetti et al.~\cite{Rossetti2017} refined PF1 by introducing NF1, addressing cases where some ground truth communities remain unmapped. If a ground truth community’s label is not the most frequent in any predicted community, it is excluded from mapping. The set of mapped ground truth communities is denoted as $C_{id}$. Two key measures are introduced: coverage, the fraction of ground truth communities that are mapped ($\textit{coverage} = \frac{|C_{id}|}{|C|}$) and redundancy, the ratio of predicted communities to mapped ground truth communities ($\textit{redundancy} = \frac{|P|}{|C_{id}|}$). NF1 normalizes PF1 by incorporating both measures and is computed as:
    \begin{equation*}
        NF1 = \frac{\textit{PF1}\cdot \textit{coverage}}{\textit{redundancy}}
    \end{equation*}
\end{enumerate}


\subsection*{Experimental details} 

LFR networks are generated using NetworkX’s Python library~\cite{Hagberg2008}. The code for ABCD \cite{Kamiński2021} and HICH-BA \cite{Saxena2023fairness} is available on GitHub, with links provided in their respective papers. 

Most community detection methods are implemented using the CDlib Python library~\cite{Rossetti2019}, which we use with default parameters unless specific settings are required. Community detection methods that require the number of communities as input include RSC-K, RSC-SSE, RSC-V, Spectral Clustering, DeepWalk, FairWalk, Node2Vec, Fluid, and EM. Since the ground truth number of communities is typically unknown, these methods have an advantage over those that must infer it. 

\section*{Results}

In this section, we discuss the fairness of community detection methods with respect to community size, density, and conductance versus the quality of identified communities.

\subsection*{Analysis on LFR networks}

\subsubsection*{Fairness-Performance trade-off versus Community Size}

\begin{figure}[t]
\centering
\begin{subfigure}[b]{0.98\textwidth}            
    \includegraphics[width=\textwidth]{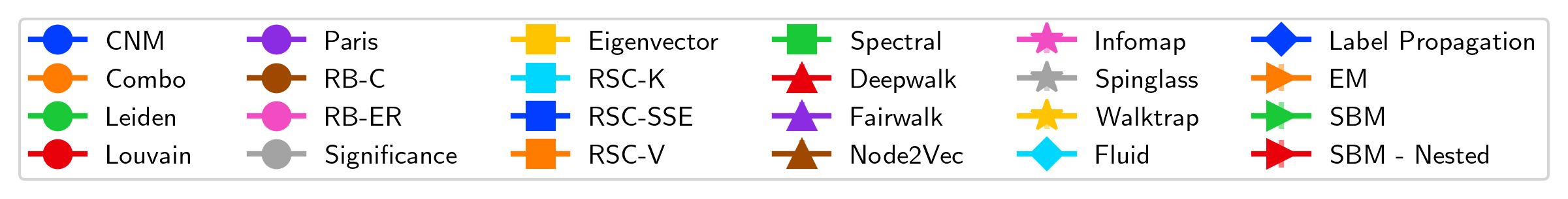}
\end{subfigure}\\
\begin{subfigure}[c]{0.05\textwidth}
\caption*{\rotatebox{90}{$\mu=0.2$}}
\end{subfigure}%
\begin{minipage}[c]{0.95\textwidth}
\includegraphics[width=0.31\textwidth]{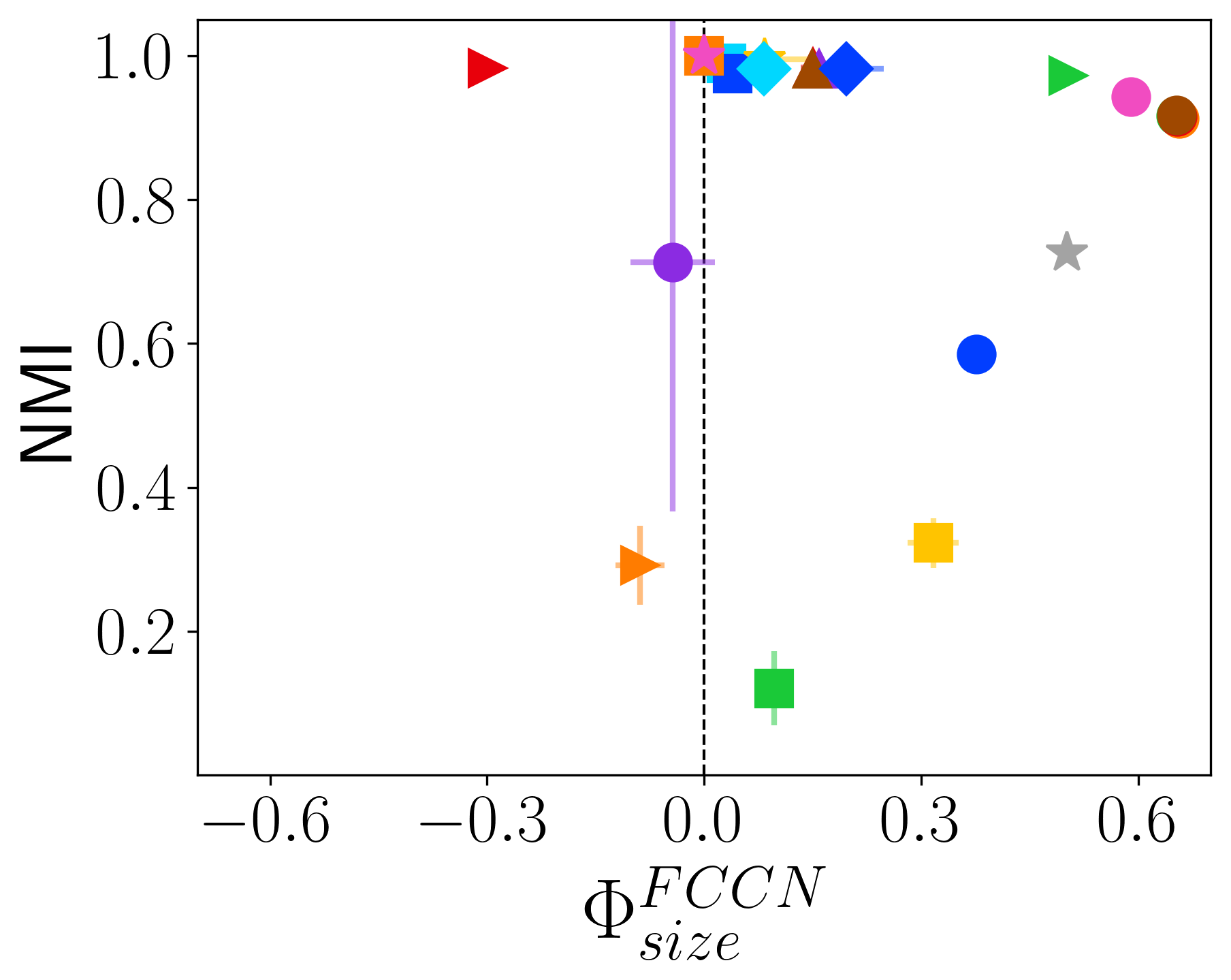}\quad
\includegraphics[width=0.31\textwidth]{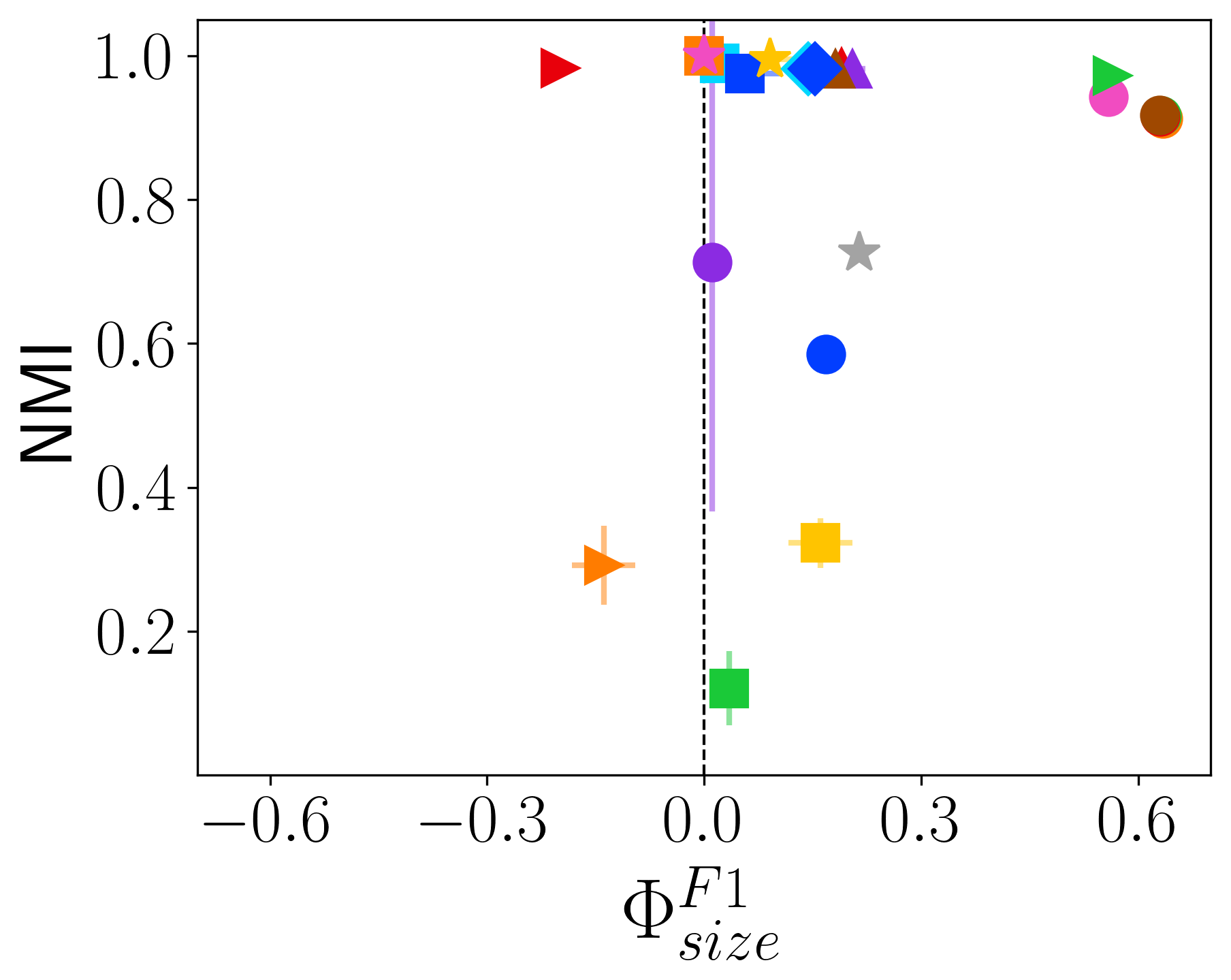}\quad
\includegraphics[width=0.31\textwidth]{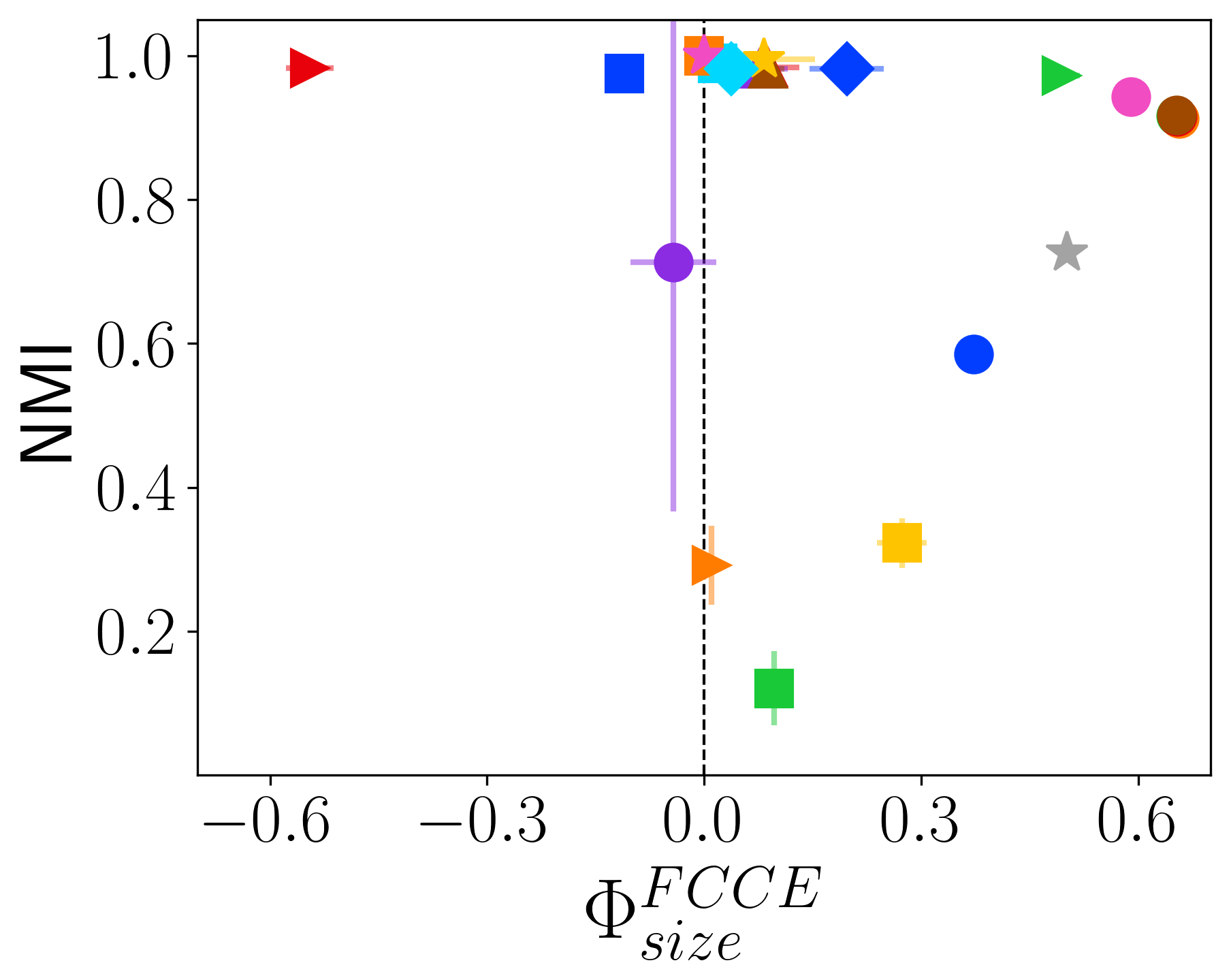}
\end{minipage}
\\
\begin{subfigure}[c]{0.05\textwidth}
\caption*{\rotatebox{90}{$\mu=0.4$}}
\end{subfigure}%
\begin{minipage}[c]{0.95\textwidth}
\includegraphics[width=0.31\textwidth]{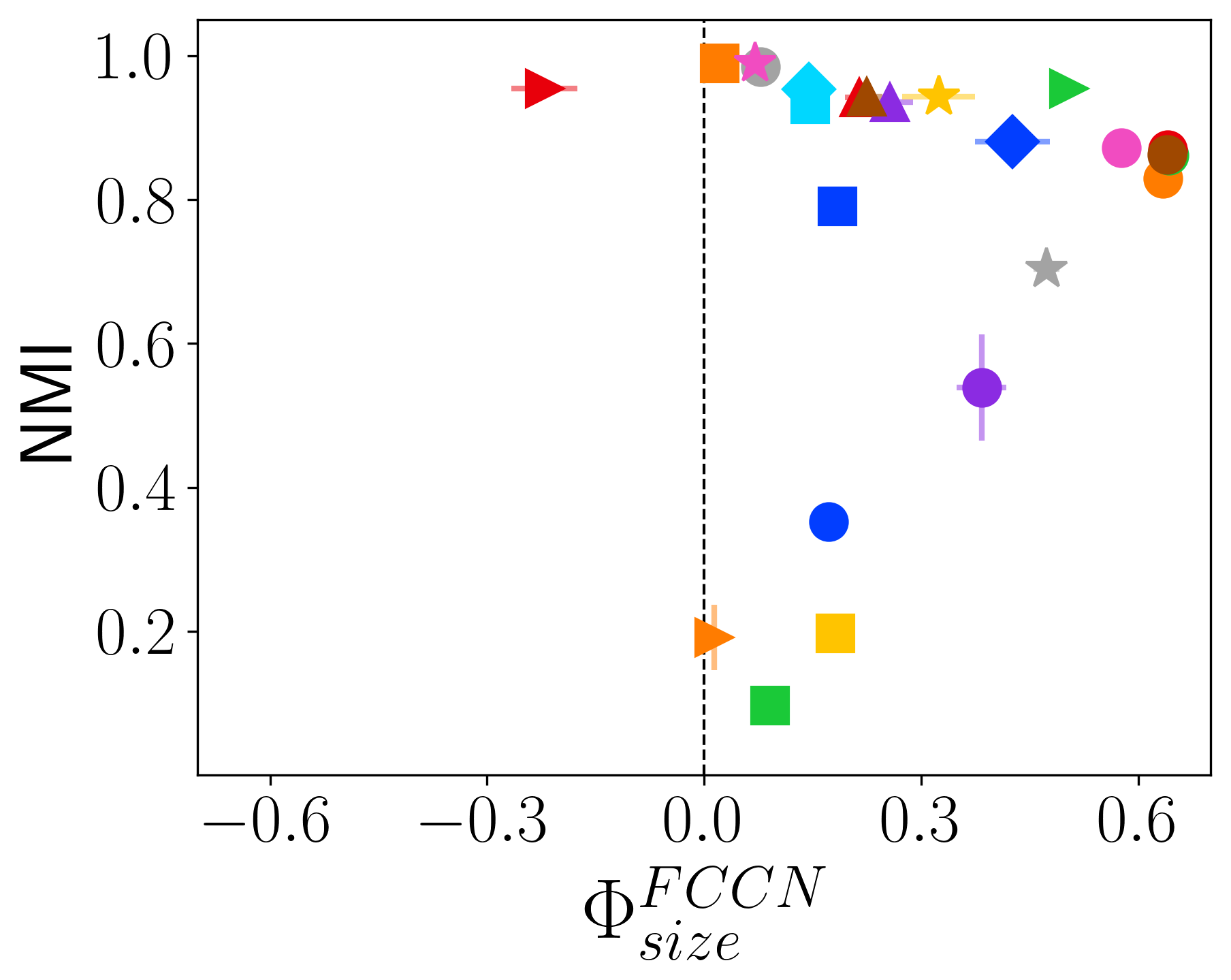}\quad
\includegraphics[width=0.31\textwidth]{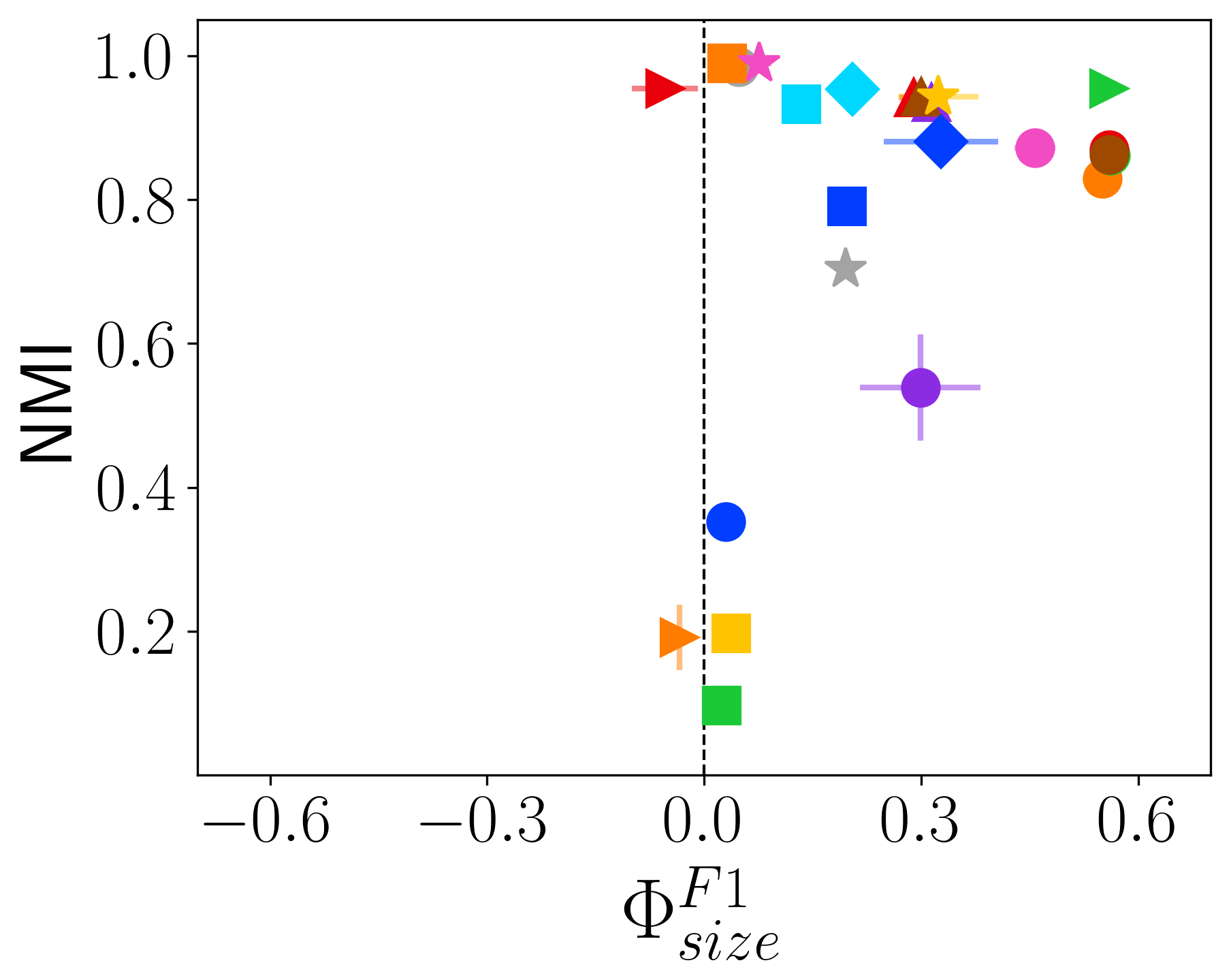}\quad
\includegraphics[width=0.31\textwidth]{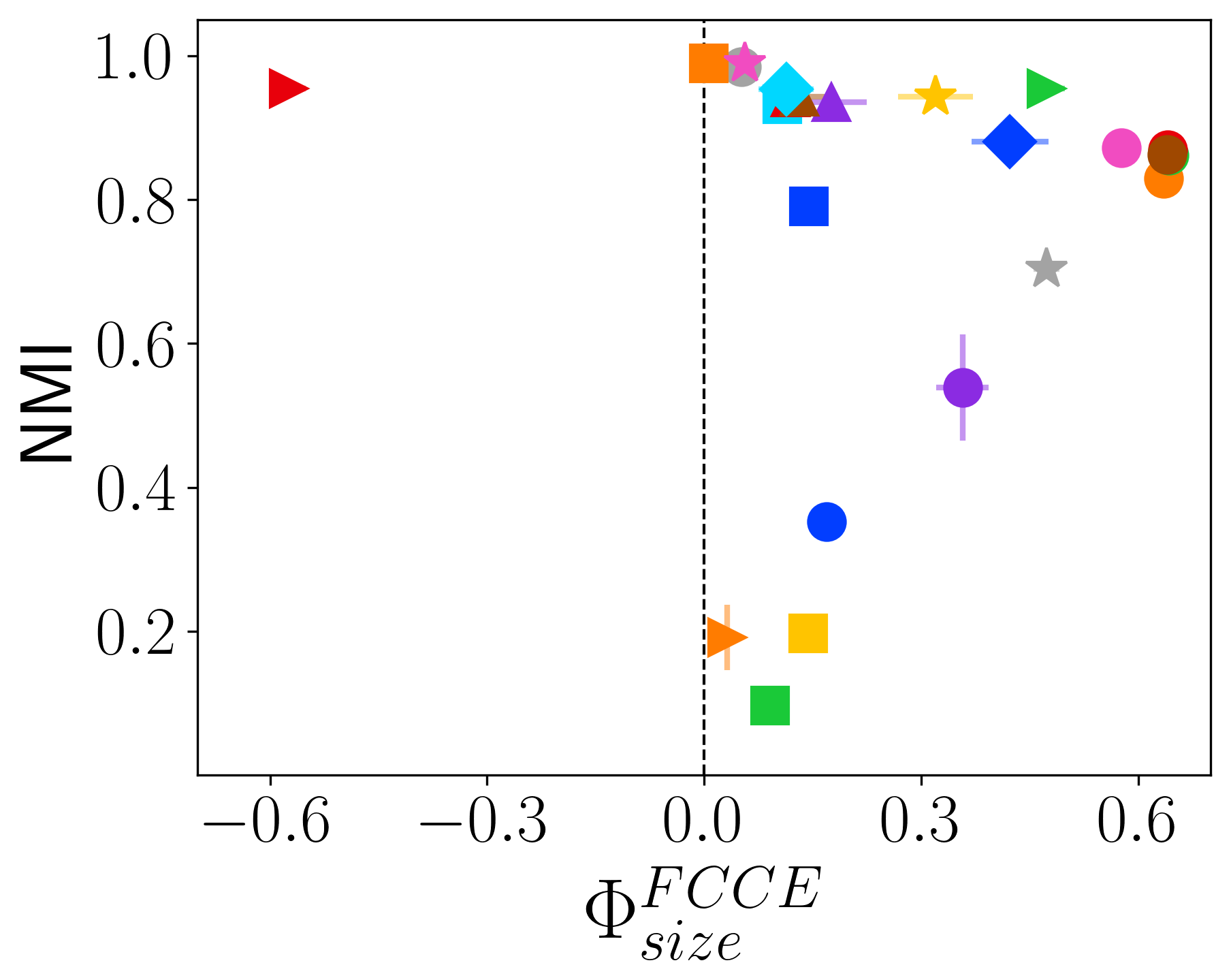}
\end{minipage}
\\
\begin{subfigure}[c]{0.05\textwidth}
\caption*{\rotatebox{90}{$\mu=0.6$}}
\end{subfigure}%
\begin{minipage}[c]{0.95\textwidth}
\includegraphics[width=0.31\textwidth]{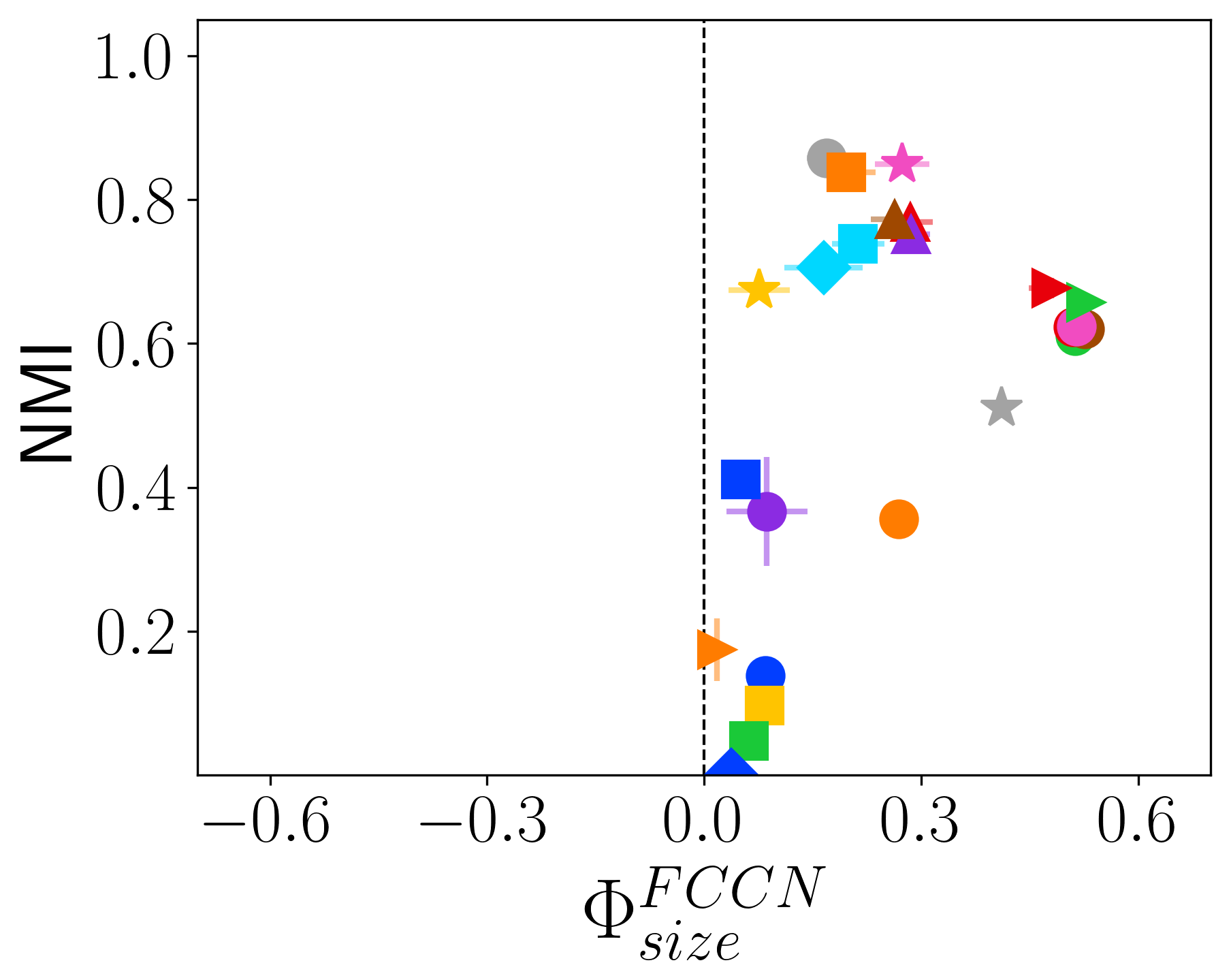}\quad
\includegraphics[width=0.31\textwidth]{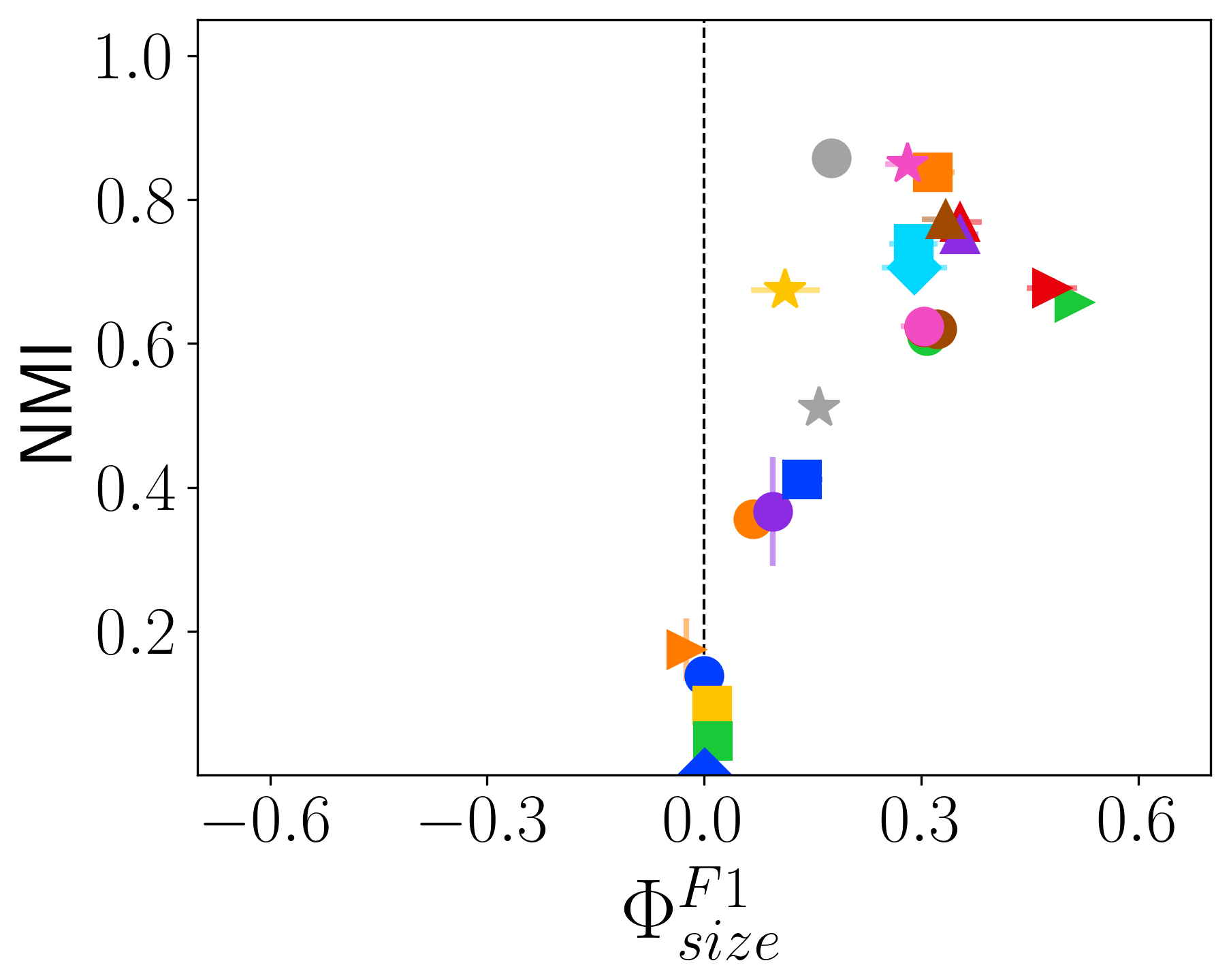}\quad
\includegraphics[width=0.31\textwidth]{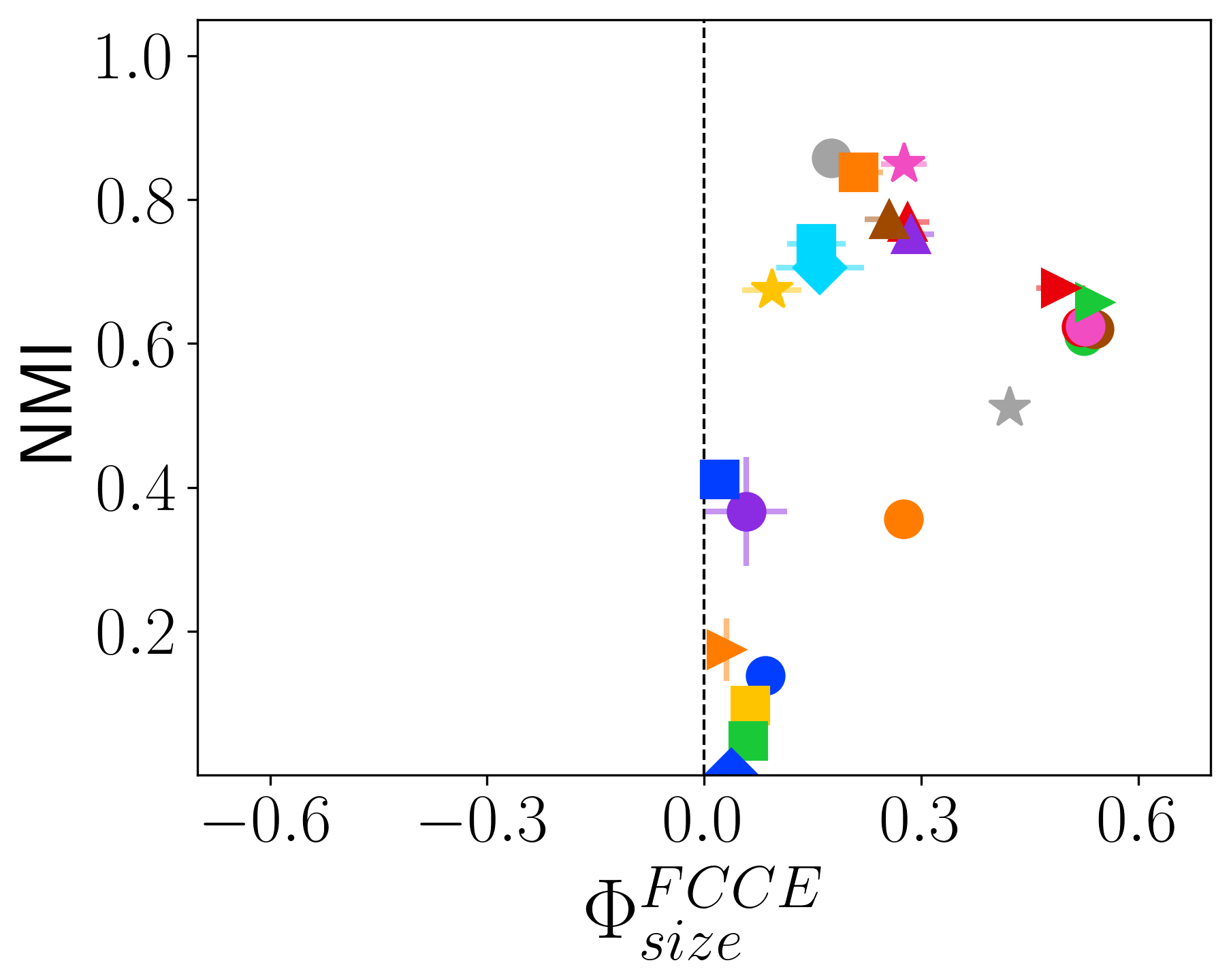}
\end{minipage}
\caption{NMI vs. fairness of community detection methods with respect to community size for LFR networks of 10,000 nodes having different $\mu$ values.}\label{lfr_phi_vs_size} 
\end{figure}

We begin by evaluating how effectively different community detection methods identify communities of varying sizes. Figure \ref{lfr_phi_vs_size} presents the relationship between NMI and three group fairness metrics ($\Phi^{FCCN}_{size}$, $\Phi^{F1}_{size}$, and $\Phi^{FCCE}_{size}$) for LFR networks with mixing parameters $\mu = 0.2$, $0.4$, and $0.6$. To conduct these experiments, we generate 10 LFR networks for each configuration and apply community detection methods. Community detection methods that produce different results across executions are run ten times, and we report the overall average and standard deviation.

For $\mu=0.2$,  most community detection methods, except for EM, Paris, SBM-nested, and RSC-SSE, tend to favor larger communities across all fairness metrics $F*$. Methods that achieve both high fairness and good quality community include RSC-K, RSC-V, Infomap, Walktrap, and Significance, all of which exhibit near-optimal NMI ($\sim$ 1). In contrast, SBM-Nested effectively detects smaller communities (negative $\Phi^{F*}_{size}$) due to its hierarchical nature, which allows it to accurately identify smaller communities at lower levels. As the mixing parameter $\mu$ increases to 0.4, SBM-Nested continues to favor smaller communities. However, when communities become highly interconnected ($\mu = 0.6$), no method perfectly identifies small groups. With high interconnectivity, all methods show a stronger bias toward larger communities, and the fair methods tend to have very low NMI, indicating that they struggle equally across all community sizes. Notably, across all types of networks, there is no observed correlation between fairness and the performance of community detection methods.

\subsubsection*{Fairness-Performance trade-off versus Community Density}

\begin{figure}[t]
\centering
\begin{subfigure}[b]{0.98\textwidth}            
    \includegraphics[width=\textwidth]{figures/legend_ncol6.png}
\end{subfigure}\\
\begin{subfigure}[c]{0.05\textwidth}
\caption*{\rotatebox{90}{$\mu=0.2$}}
\end{subfigure}%
\begin{minipage}[c]{0.95\textwidth}
\includegraphics[width=0.31\textwidth]{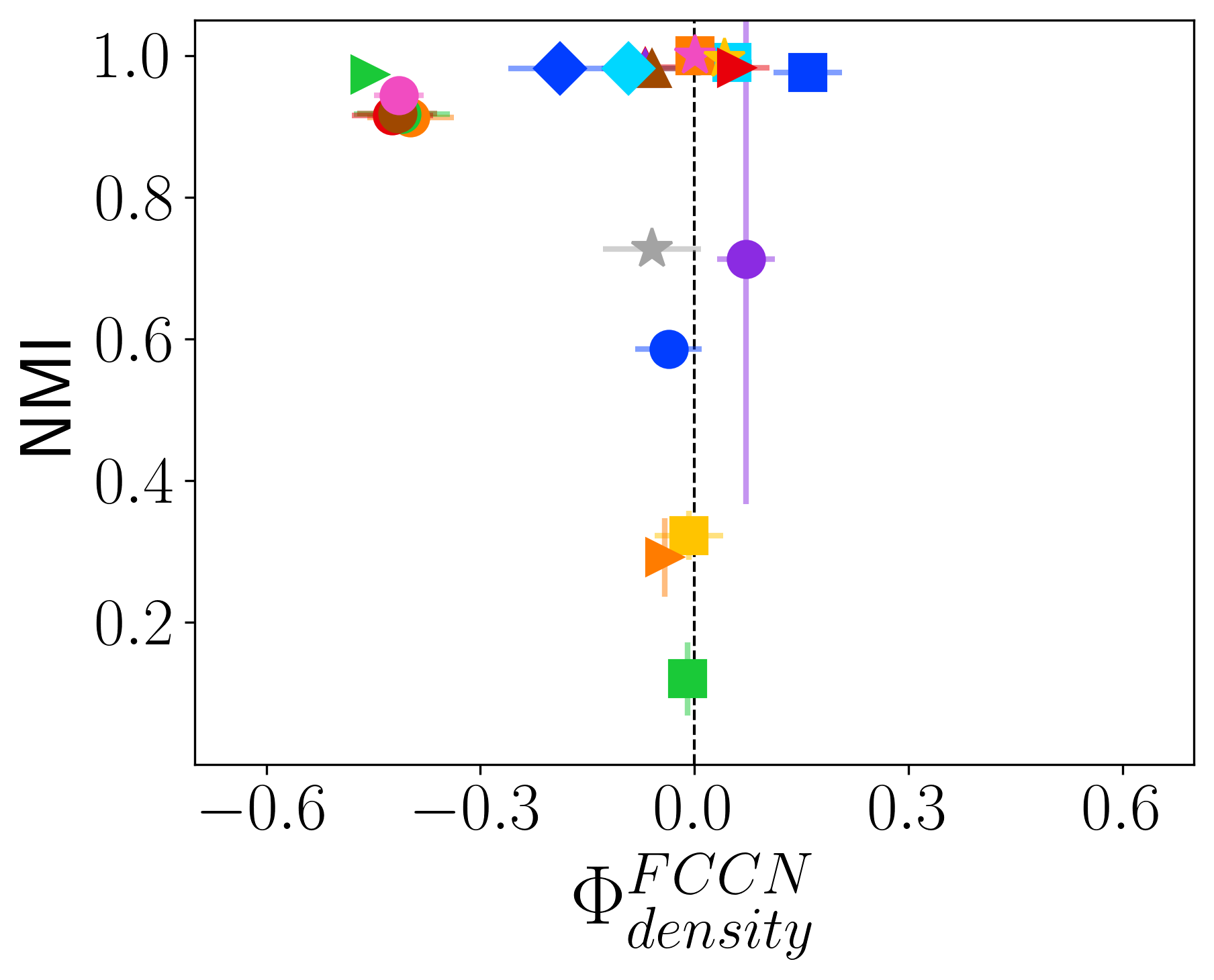}\quad
\includegraphics[width=0.31\textwidth]{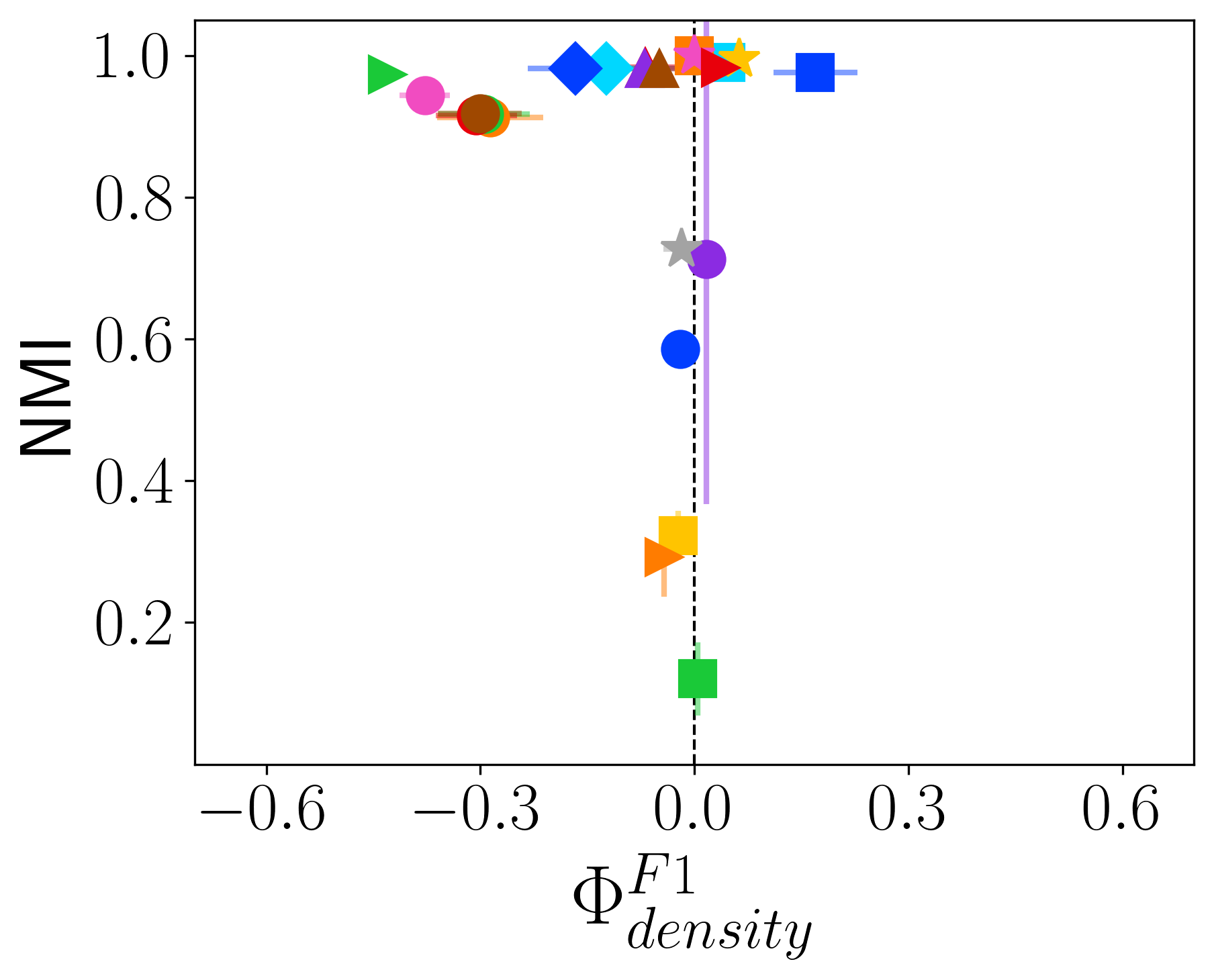}\quad
\includegraphics[width=0.31\textwidth]{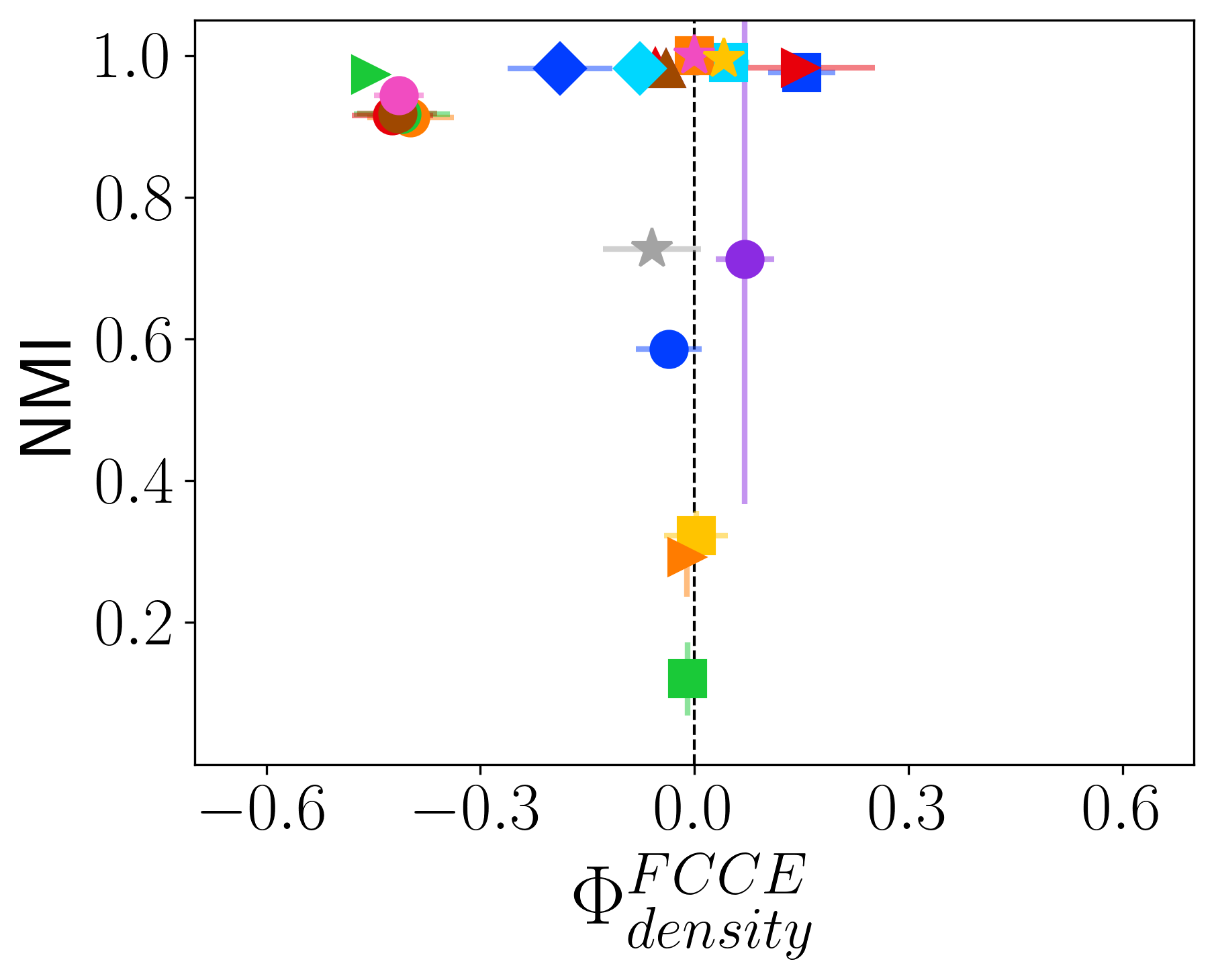}
\end{minipage}
\\
\begin{subfigure}[c]{0.05\textwidth}
\caption*{\rotatebox{90}{$\mu=0.4$}}
\end{subfigure}%
\begin{minipage}[c]{0.95\textwidth}
\includegraphics[width=0.31\textwidth]{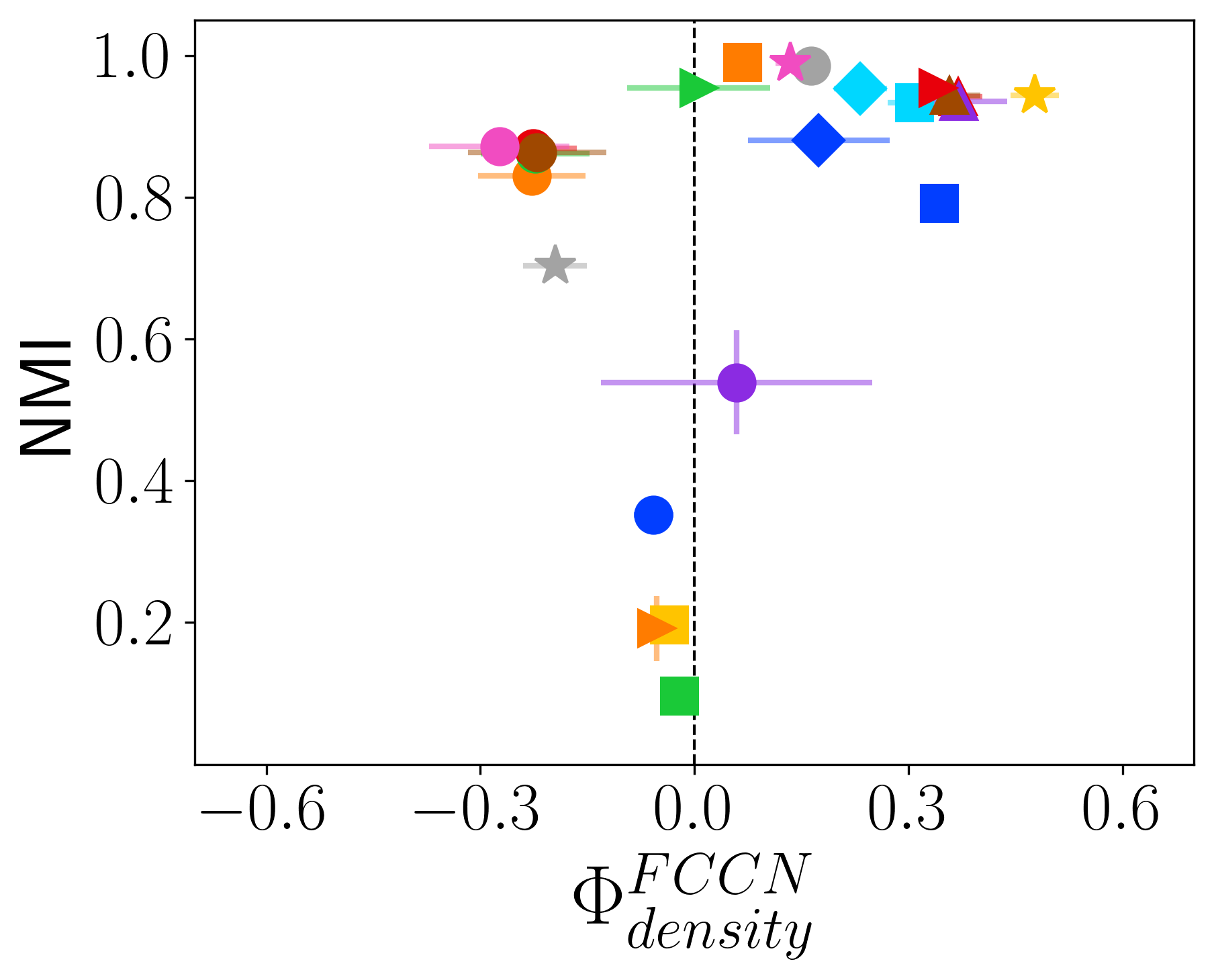}\quad
\includegraphics[width=0.31\textwidth]{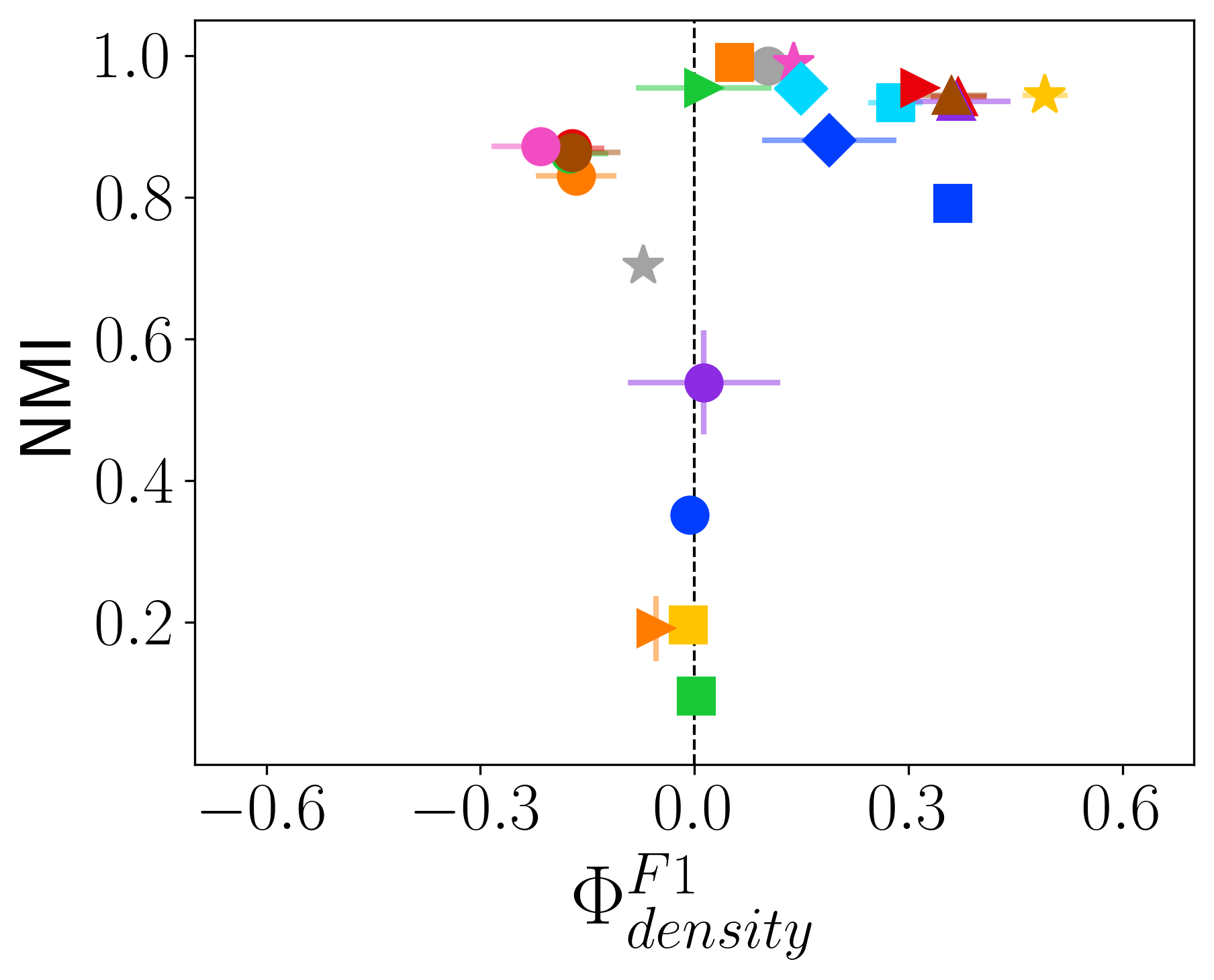}\quad
\includegraphics[width=0.31\textwidth]{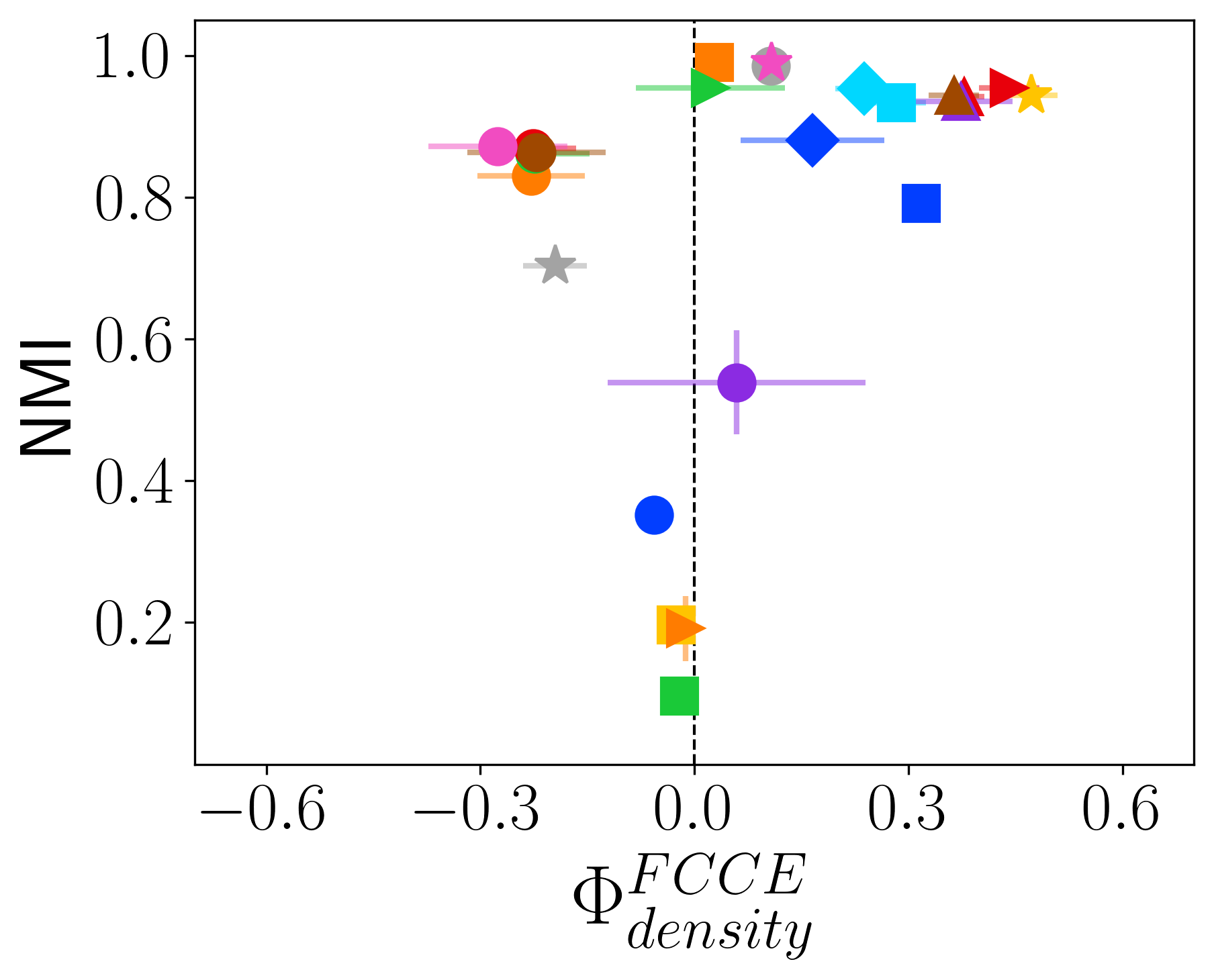}
\end{minipage}
\\
\begin{subfigure}[c]{0.05\textwidth}
\caption*{\rotatebox{90}{$\mu=0.6$}}
\end{subfigure}%
\begin{minipage}[c]{0.95\textwidth}
\includegraphics[width=0.31\textwidth]{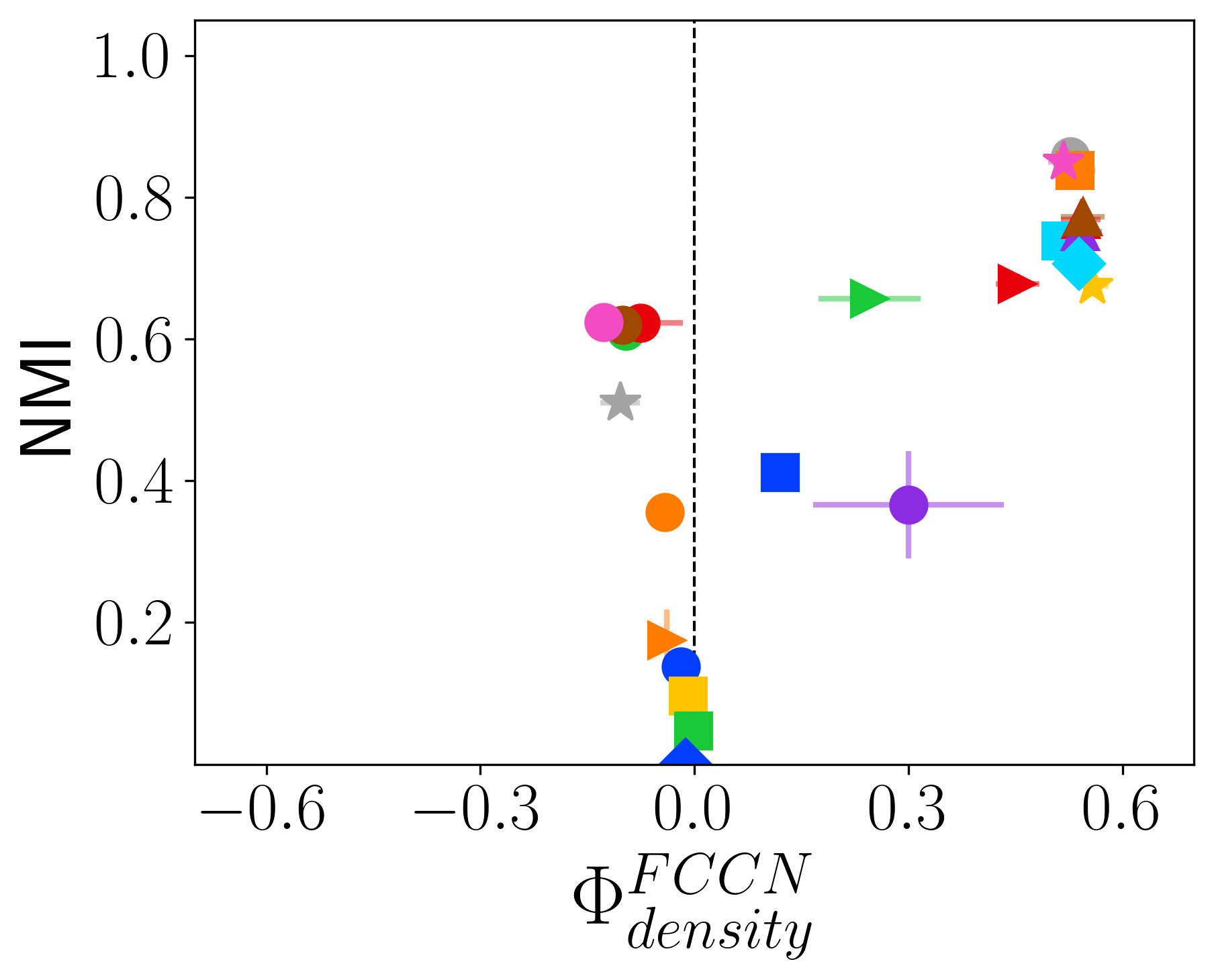}\quad
\includegraphics[width=0.31\textwidth]{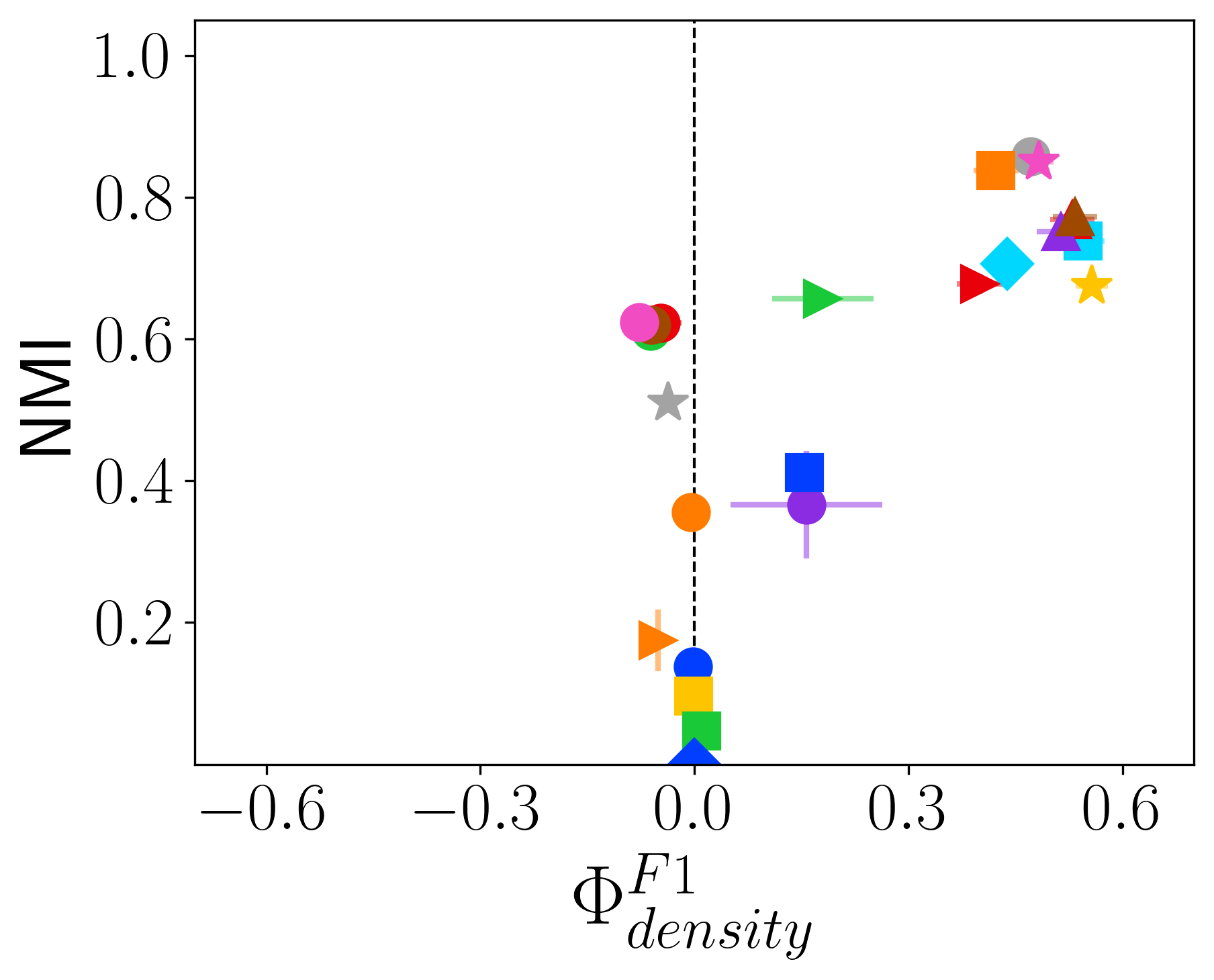}\quad
\includegraphics[width=0.31\textwidth]{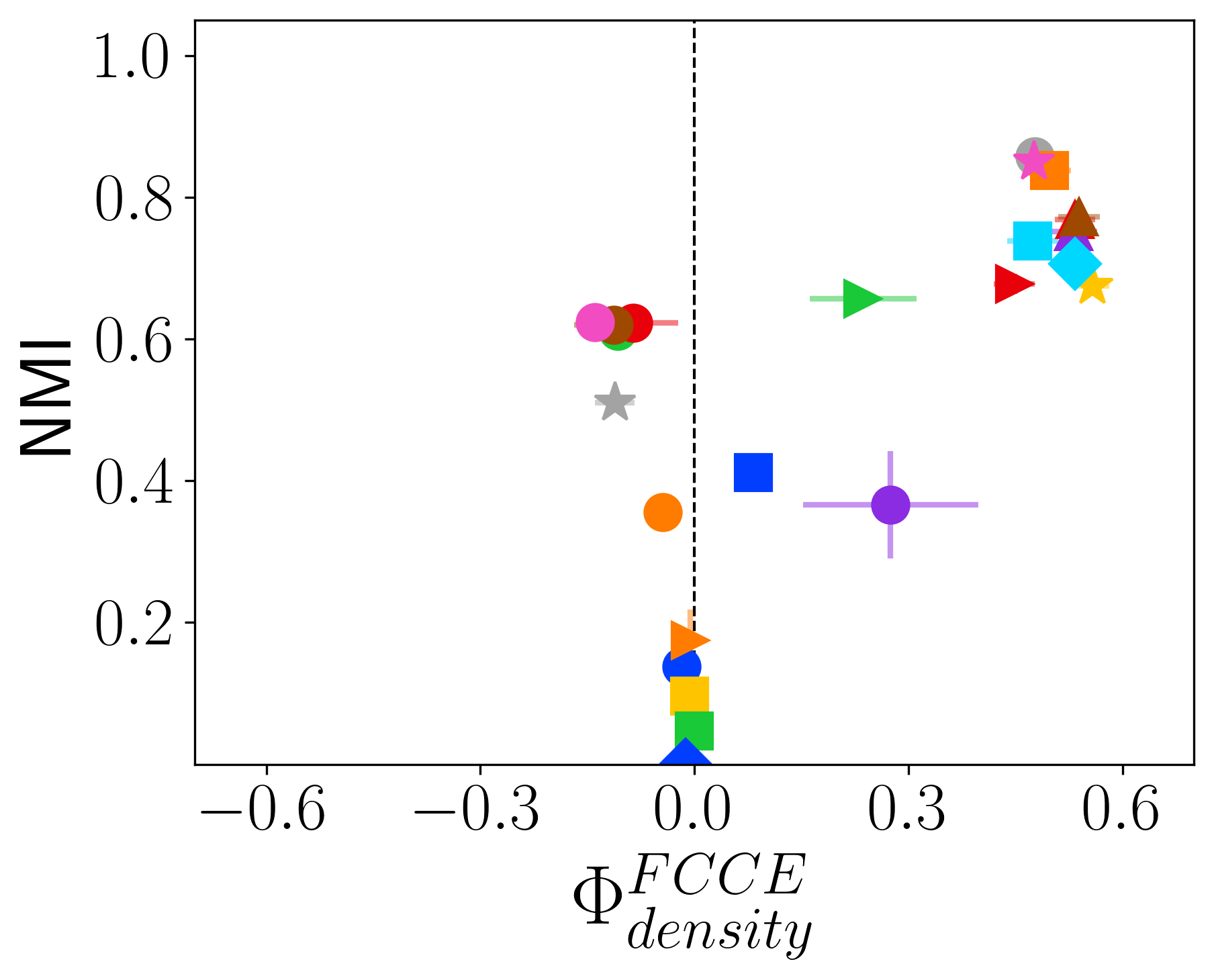}
\end{minipage}
\caption{NMI vs. fairness of community detection methods with respect to community density for LFR networks of 10,000 nodes having different $\mu$ values.}\label{lfr_phi_vs_dens} 
\end{figure}

We further examine the performance of community detection methods in identifying communities with varying densities. Figure~\ref{lfr_phi_vs_dens} shows that when the mixing parameter is low, community detection methods tend to detect sparse communities more effectively than denser ones. However, as $\mu$ increases and inter-community edges become more prevalent, this pattern shifts. At $\mu=0.4,$ and $0.6$, methods including Leiden, Louvain, RB-C, RB-ER, and Spinglass, have high NMI but negative $\Phi^{F*}_{density}$. These methods predict fewer communities than the ground truth, which are often mapped to low-density ground truth communities. At $\mu=0.6$, methods with high NMI also perform better at detecting denser communities. For instance, the Significance method predicts significantly more communities than the ground truth, effectively identifying dense structures while fragmenting sparser ones in its prediction.

\subsubsection*{Fairness Performance trade-off versus Community Conductance}

\begin{figure}[t]
\centering
\begin{subfigure}[b]{0.98\textwidth}            
    \includegraphics[width=\textwidth]{figures/legend_ncol6.png}
\end{subfigure}\\
\begin{subfigure}[c]{0.05\textwidth}
\caption*{\rotatebox{90}{$\mu=0.2$}}
\end{subfigure}%
\begin{minipage}[c]{0.95\textwidth}
\includegraphics[width=0.31\textwidth]{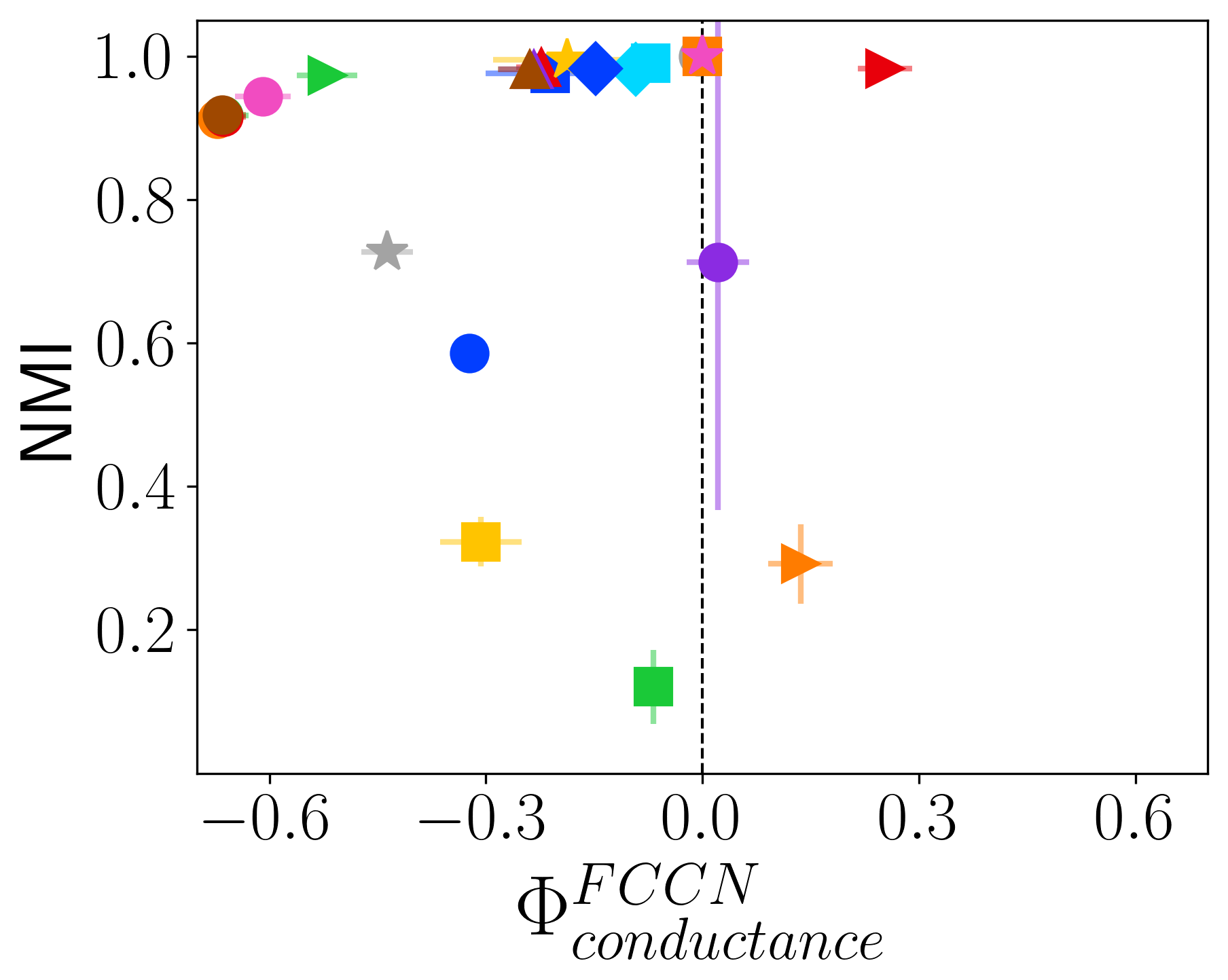}\quad
\includegraphics[width=0.31\textwidth]{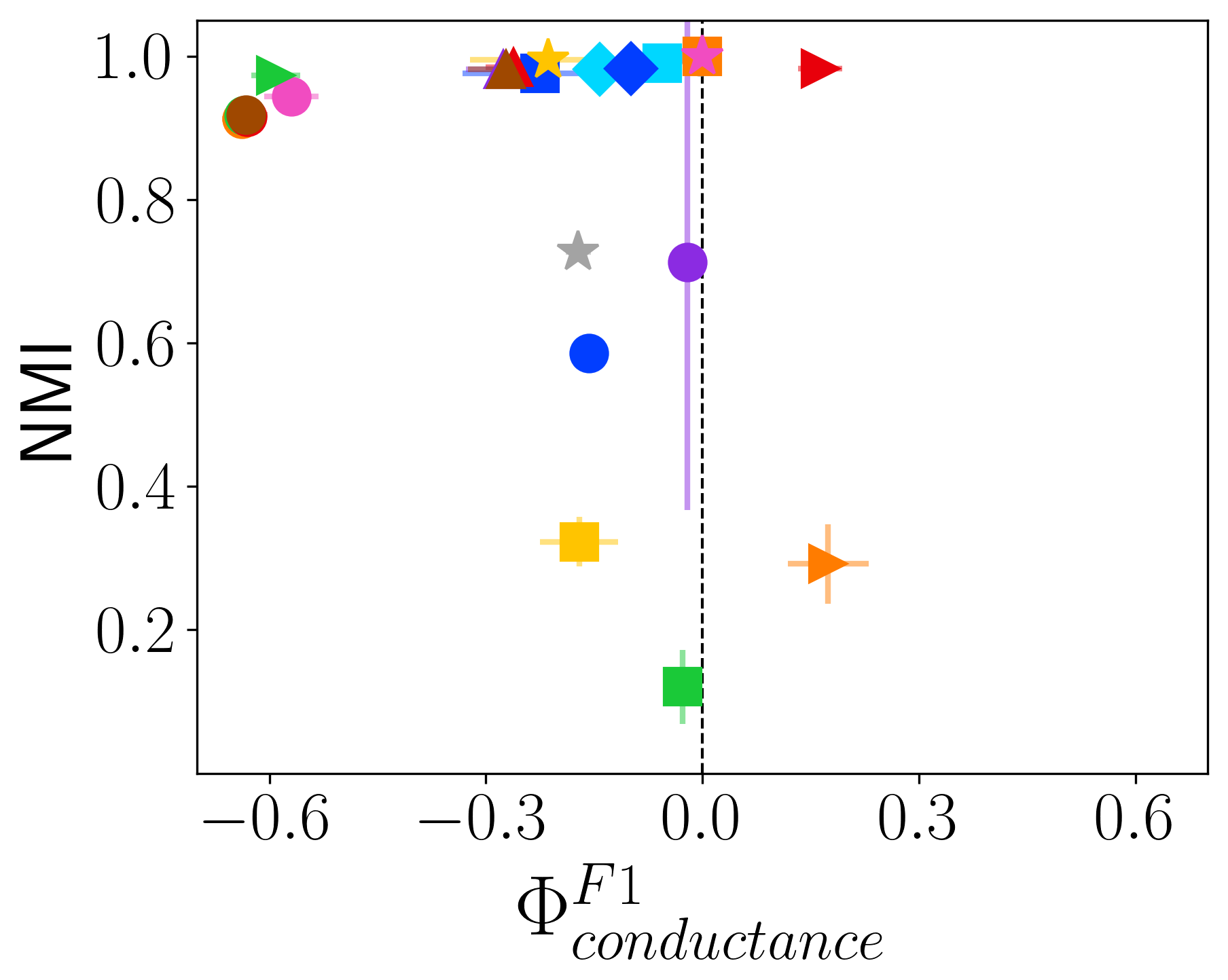}\quad
\includegraphics[width=0.31\textwidth]{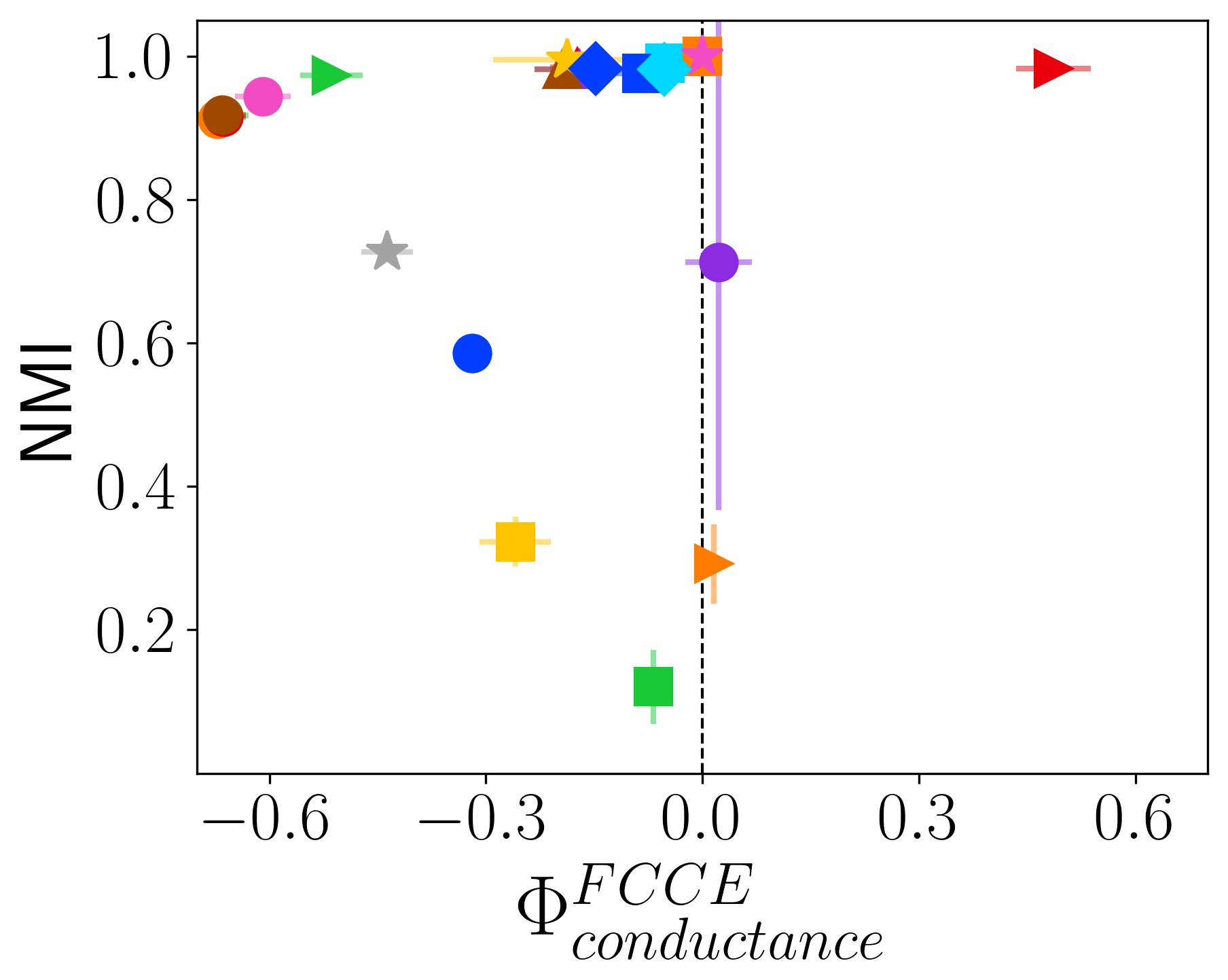}
\end{minipage}
\\
\begin{subfigure}[c]{0.05\textwidth}
\caption*{\rotatebox{90}{$\mu=0.4$}}
\end{subfigure}%
\begin{minipage}[c]{0.95\textwidth}
\includegraphics[width=0.31\textwidth]{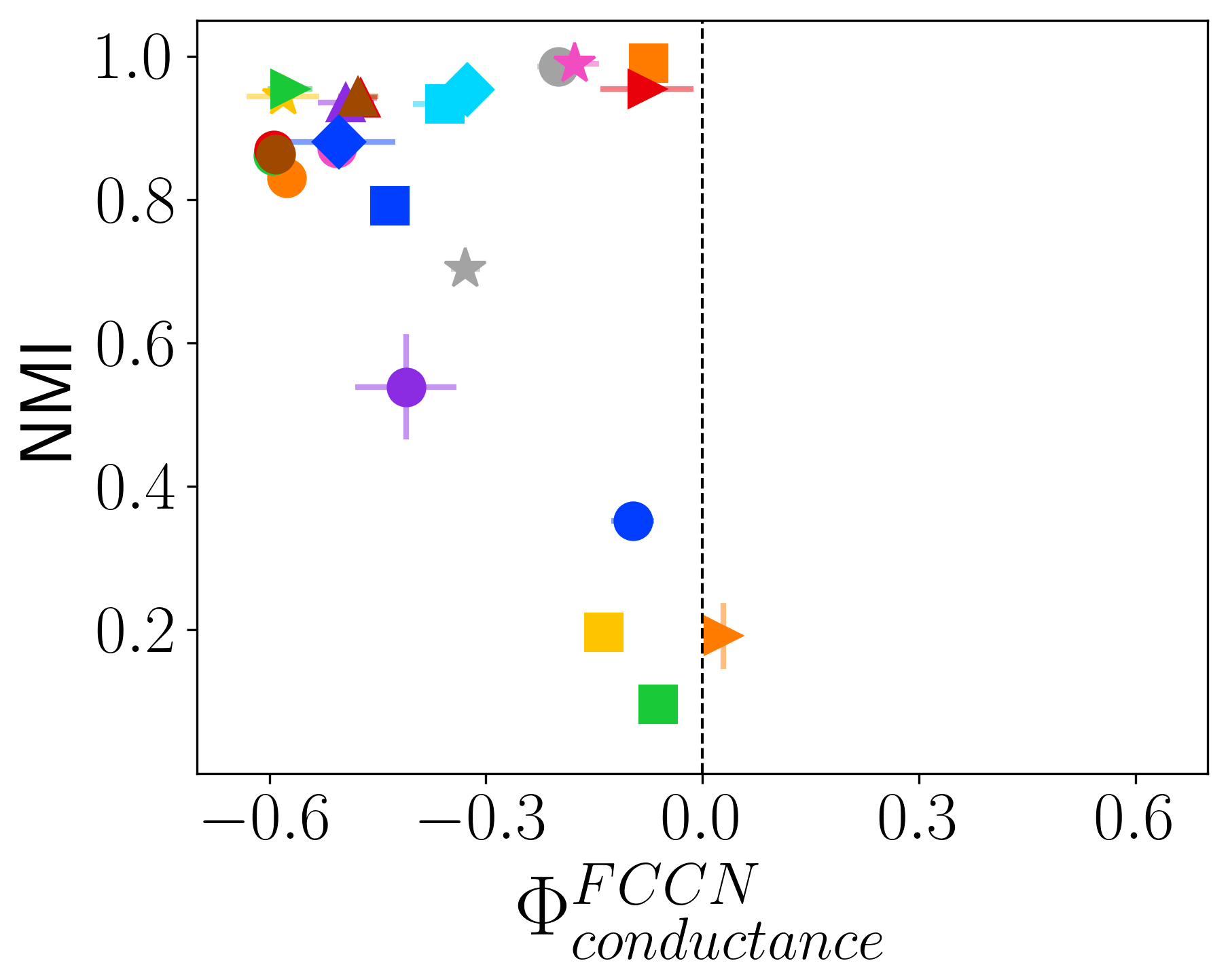}\quad
\includegraphics[width=0.31\textwidth]{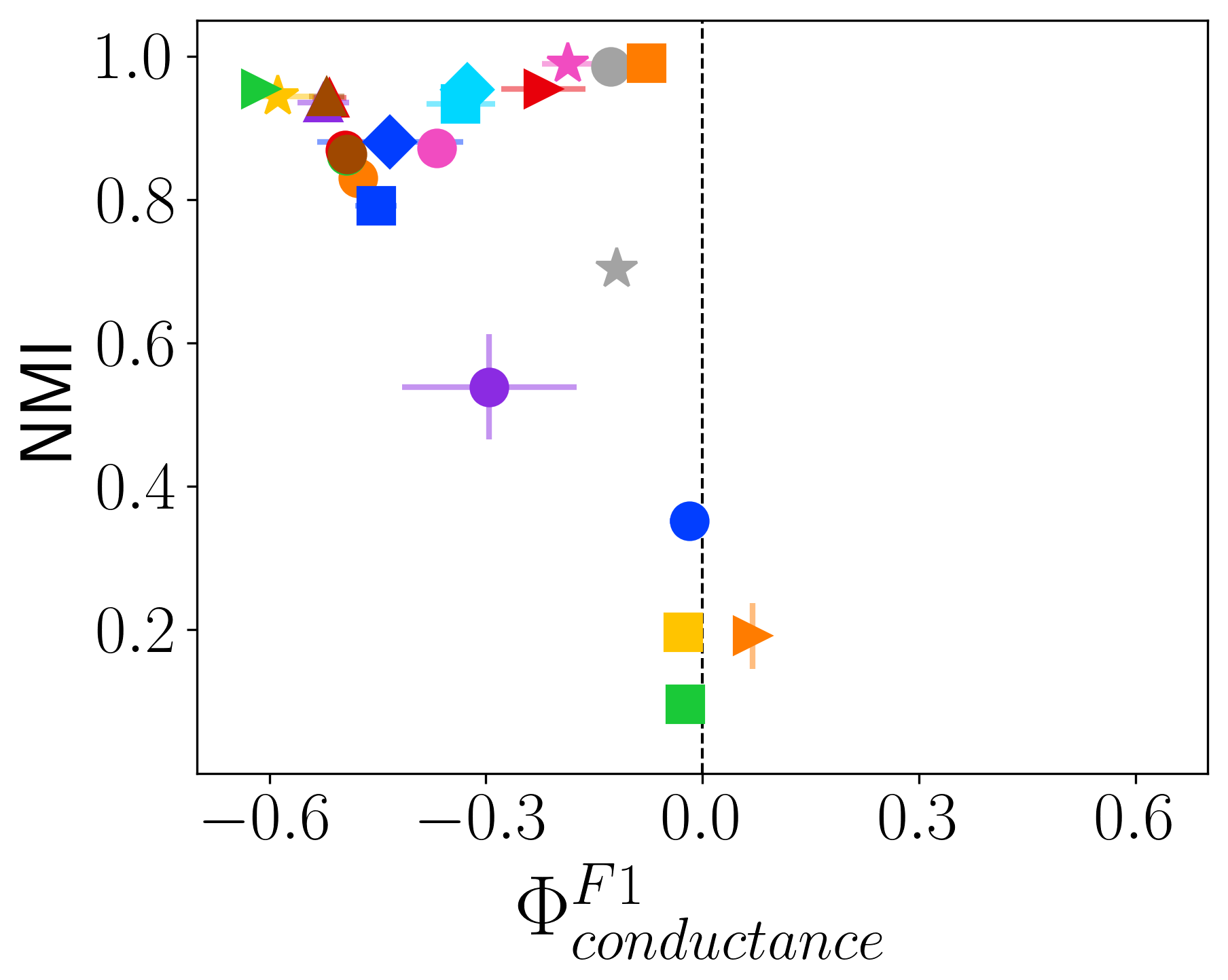}\quad
\includegraphics[width=0.31\textwidth]{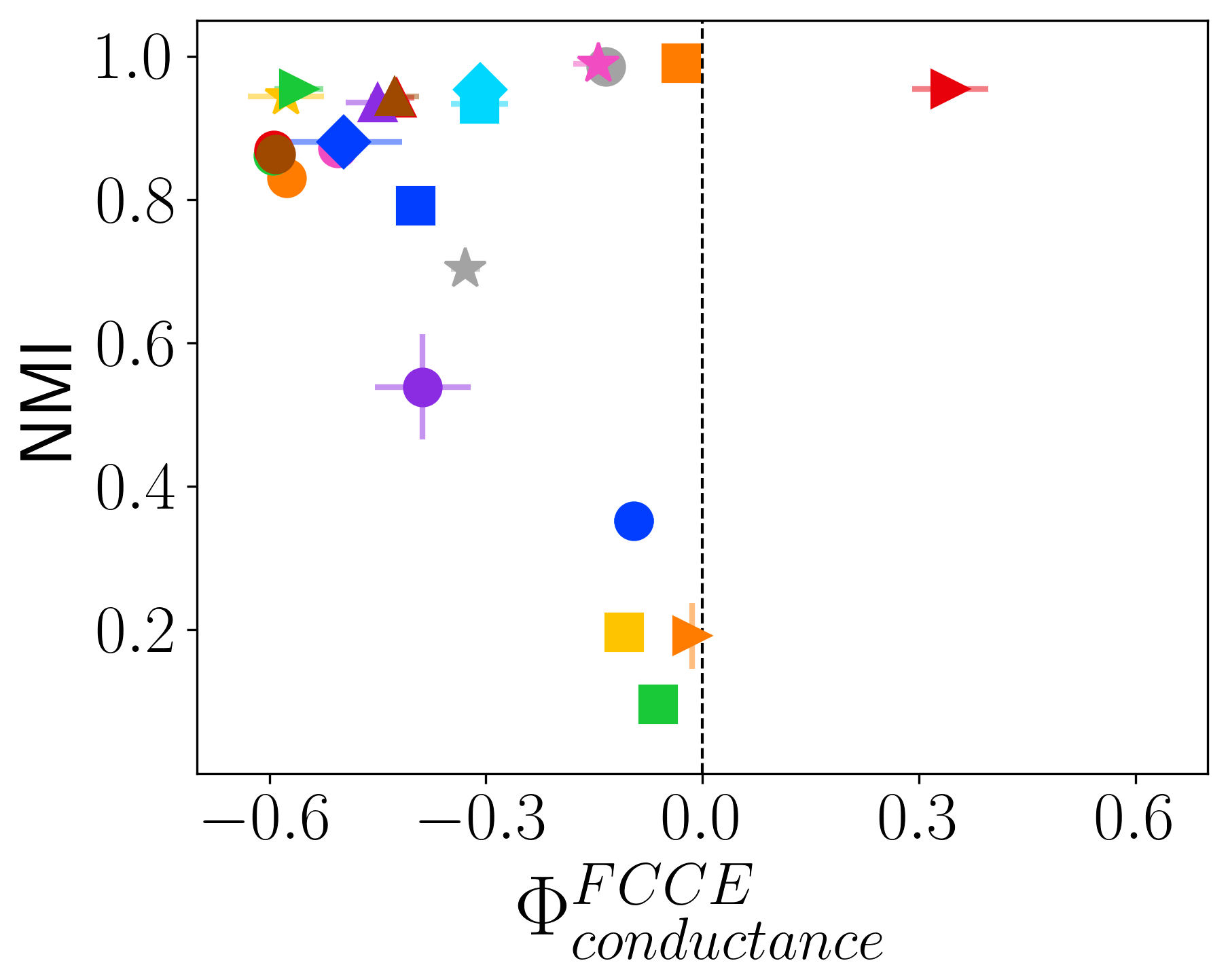}
\end{minipage}
\\
\begin{subfigure}[c]{0.05\textwidth}
\caption*{\rotatebox{90}{$\mu=0.6$}}
\end{subfigure}%
\begin{minipage}[c]{0.95\textwidth}
\includegraphics[width=0.31\textwidth]{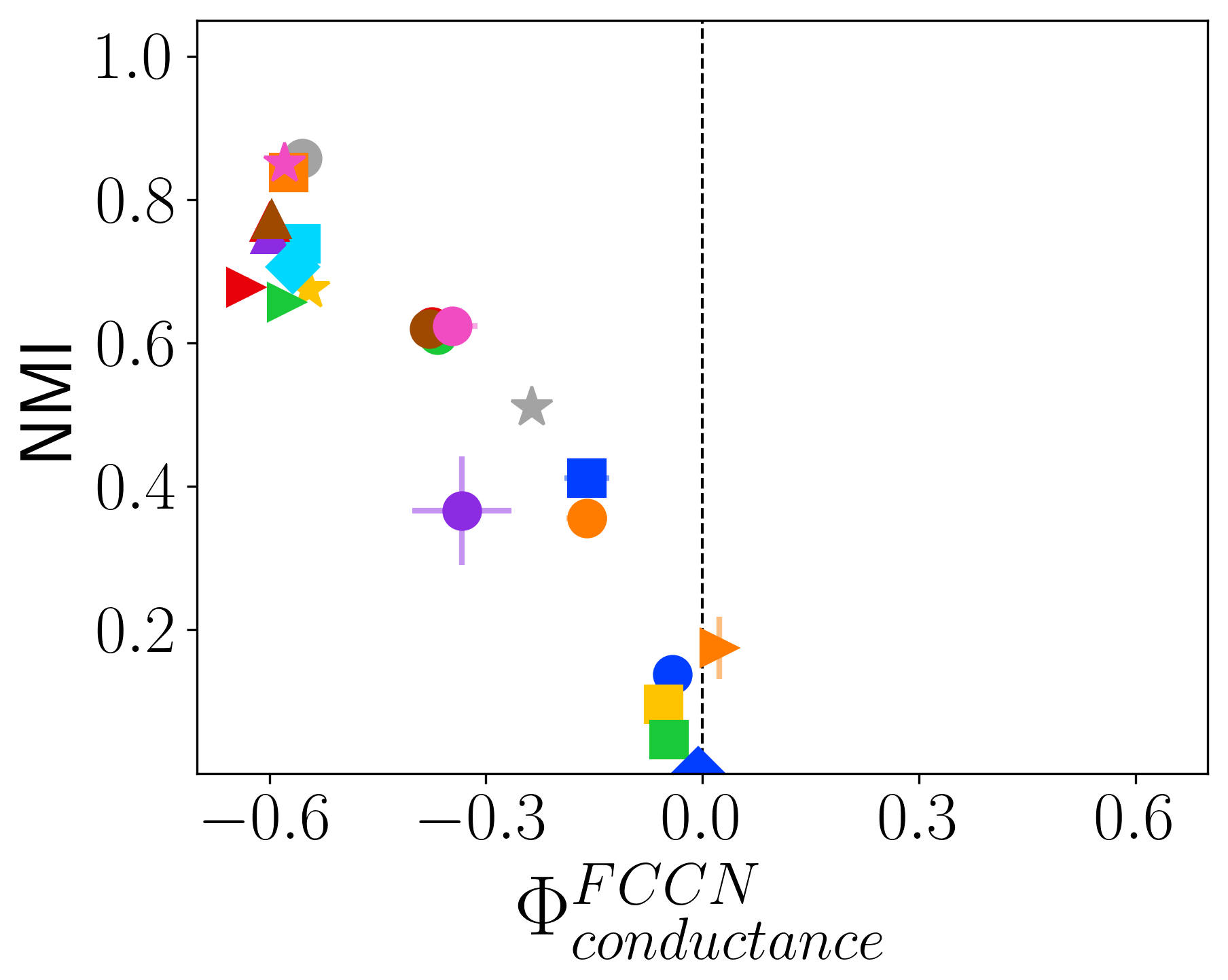}\quad
\includegraphics[width=0.31\textwidth]{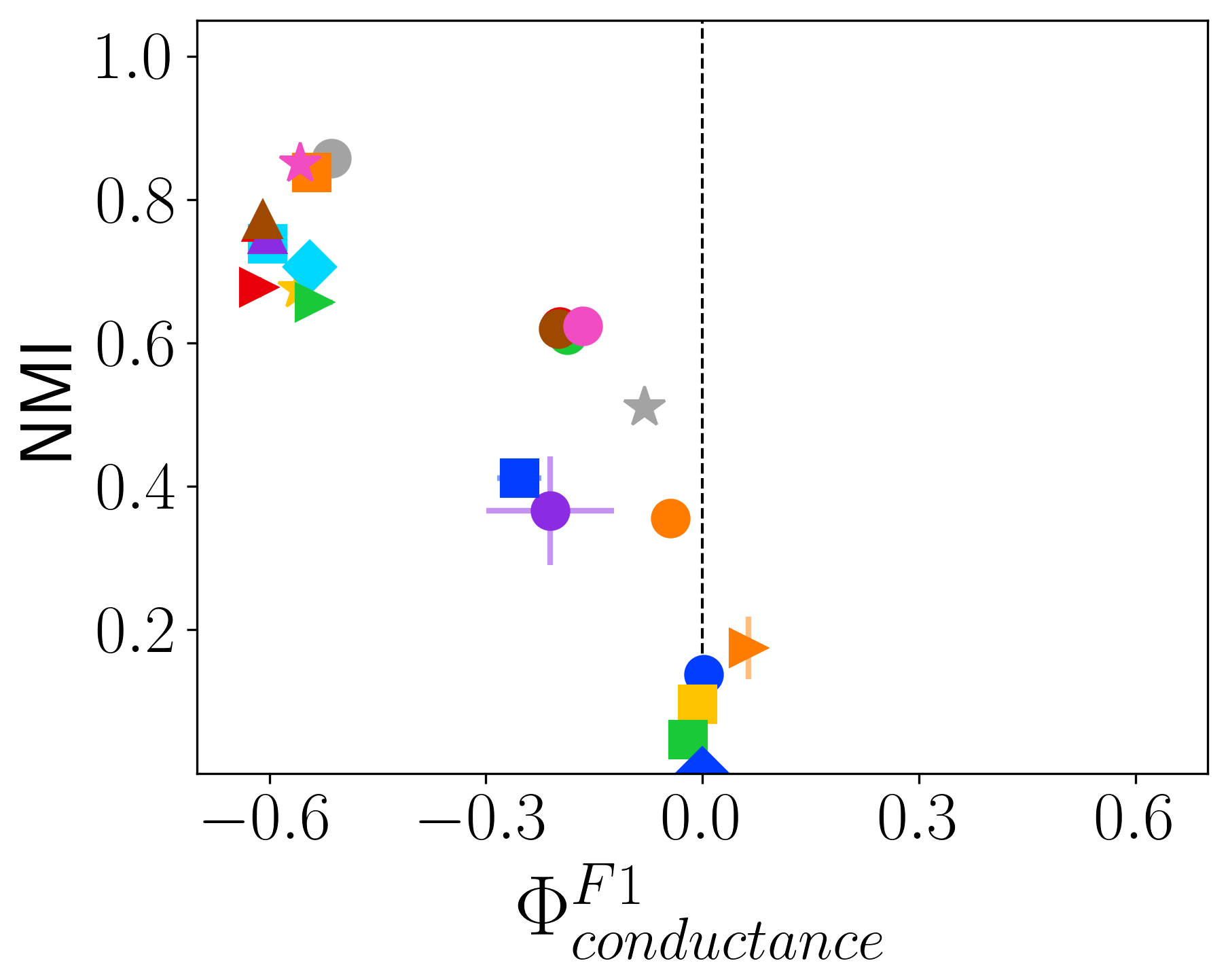}\quad
\includegraphics[width=0.31\textwidth]{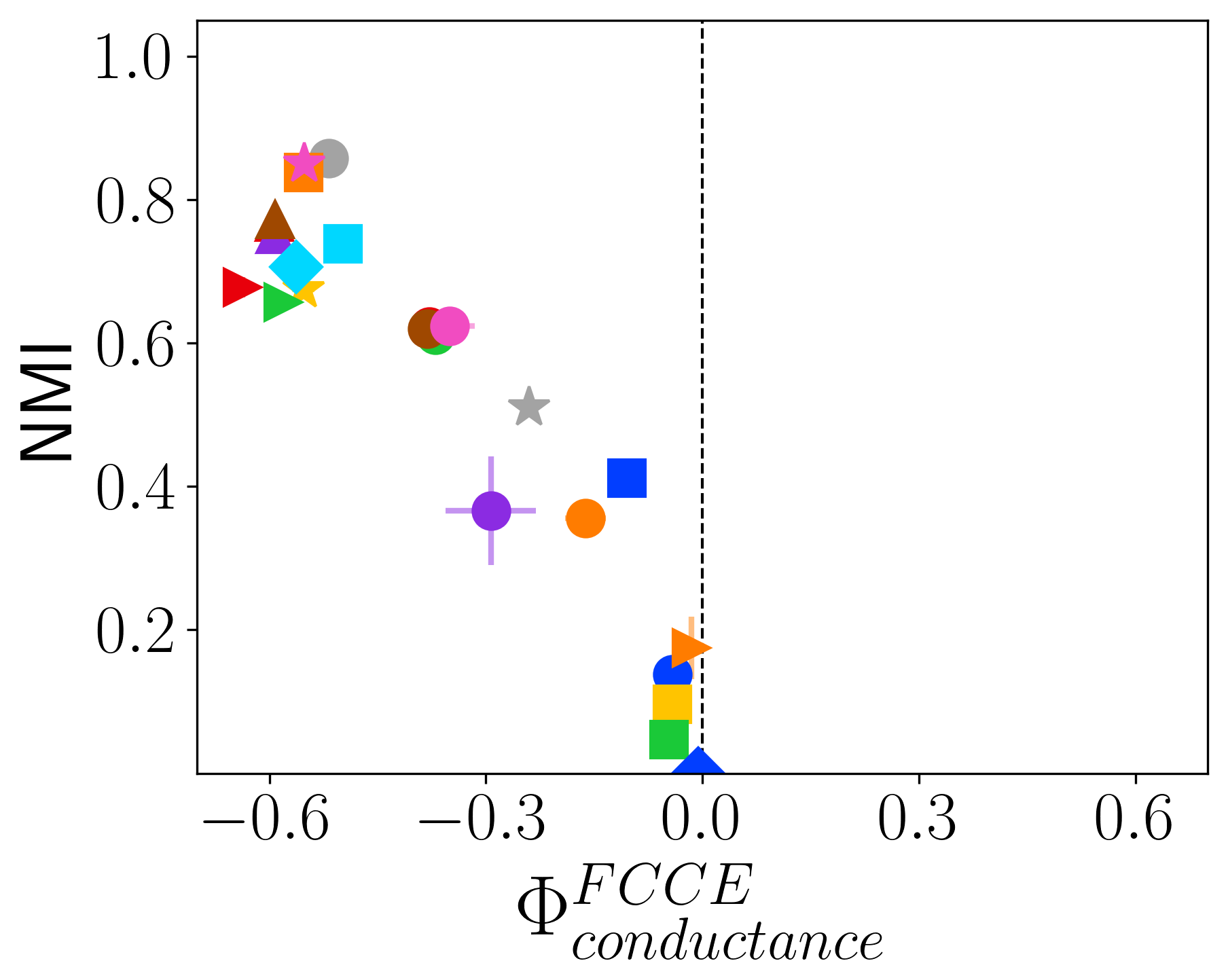}
\end{minipage}
\caption{NMI vs. fairness of community detection methods with respect to community conductance for LFR networks of 10,000 nodes having different $\mu$ values.}\label{lfr_phi_vs_cond} 
\end{figure}

Figure~\ref{lfr_phi_vs_cond} illustrates NMI versus fairness concerning community conductance. Most community detection methods tend to favor communities with lower conductance, which are more separated from other communities. This bias intensifies as $\mu$ increases, leading to lower $\Phi^{F*}_{conductance}$ scores across all fairness metrics. At medium mixing levels ($\mu=0.4$), methods such as RSC-V, SBM-Nested, Infomap, and Significance achieve both high fairness and good-quality communities. However, at $\mu=0.6$, a linear relationship emerges between $\Phi^{F*}_{conductance}$ and NMI, where methods that prioritize fairness tend to have poor predictive performance, while those with higher accuracy show a stronger bias.

\subsection*{Analysis on ABCD networks}

\subsubsection*{Fairness-Performance trade-off versus Community Size}

\begin{figure}[t]
\centering
\begin{subfigure}[b]{0.98\textwidth}            
    \includegraphics[width=\textwidth]{figures/legend_ncol6.png}
\end{subfigure}\\
\begin{subfigure}[c]{0.05\textwidth}
\caption*{\rotatebox{90}{$\xi=0.2$}}
\end{subfigure}%
\begin{minipage}[c]{0.95\textwidth}
\includegraphics[width=0.31\textwidth]{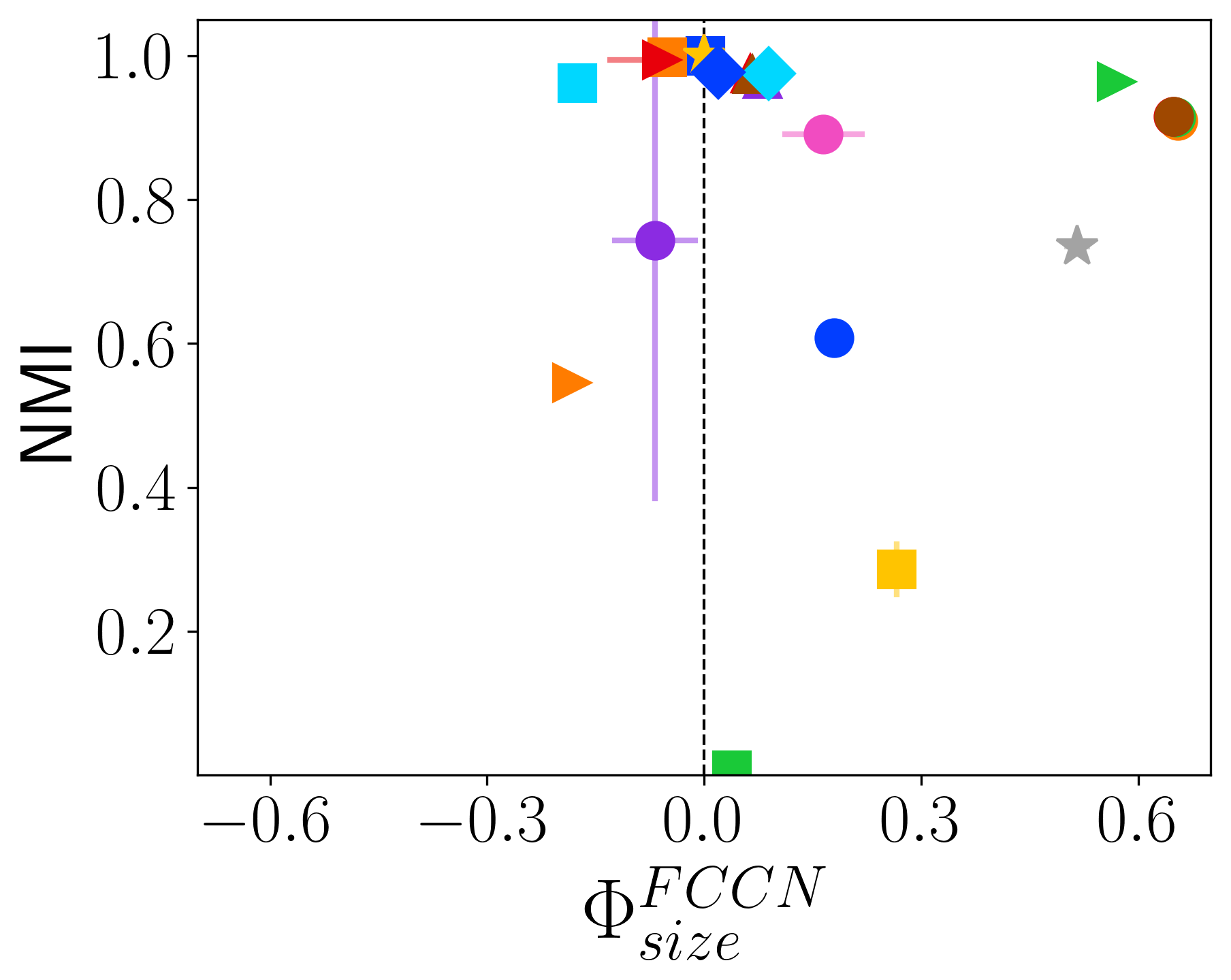}\quad
\includegraphics[width=0.31\textwidth]{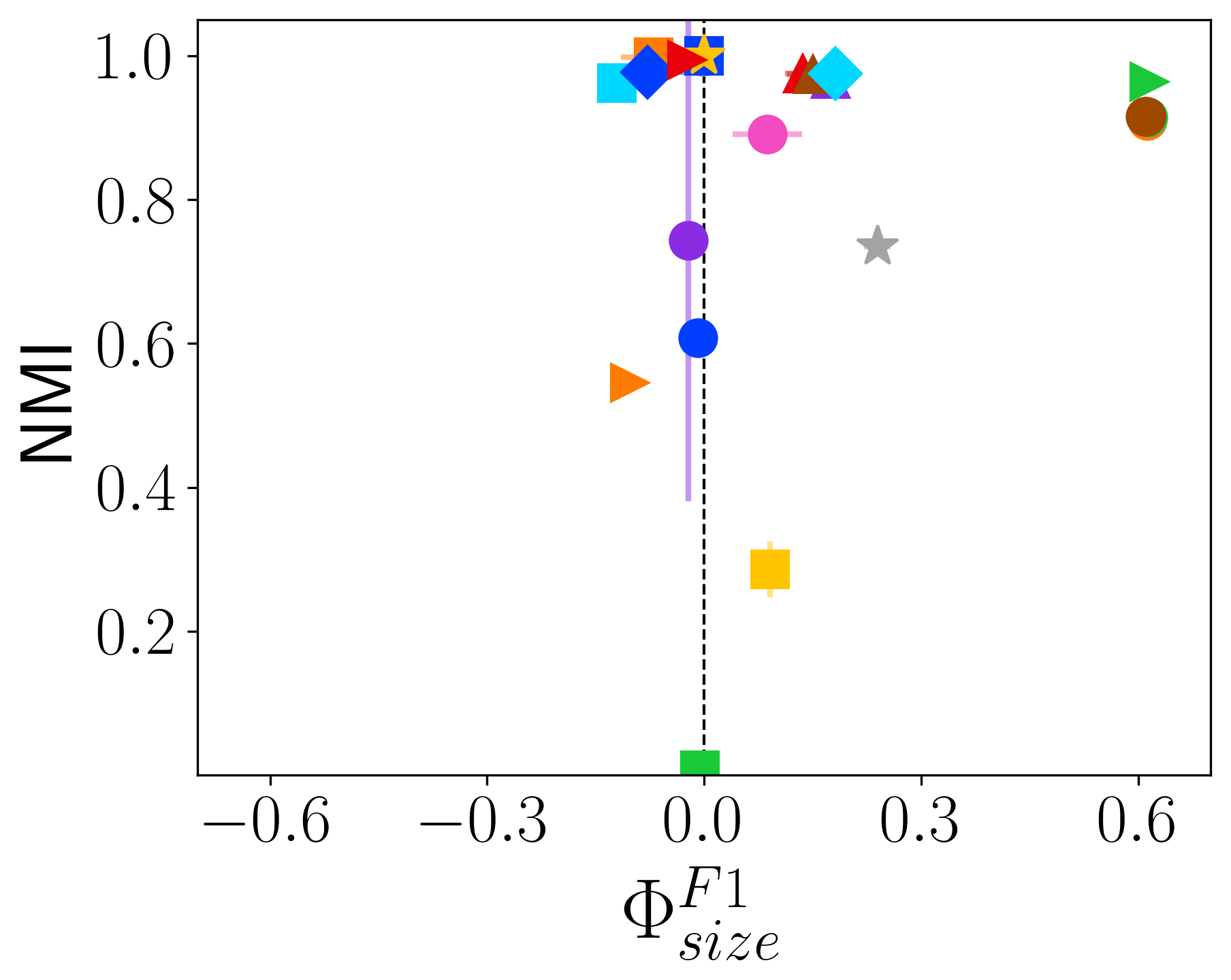}\quad
\includegraphics[width=0.31\textwidth]{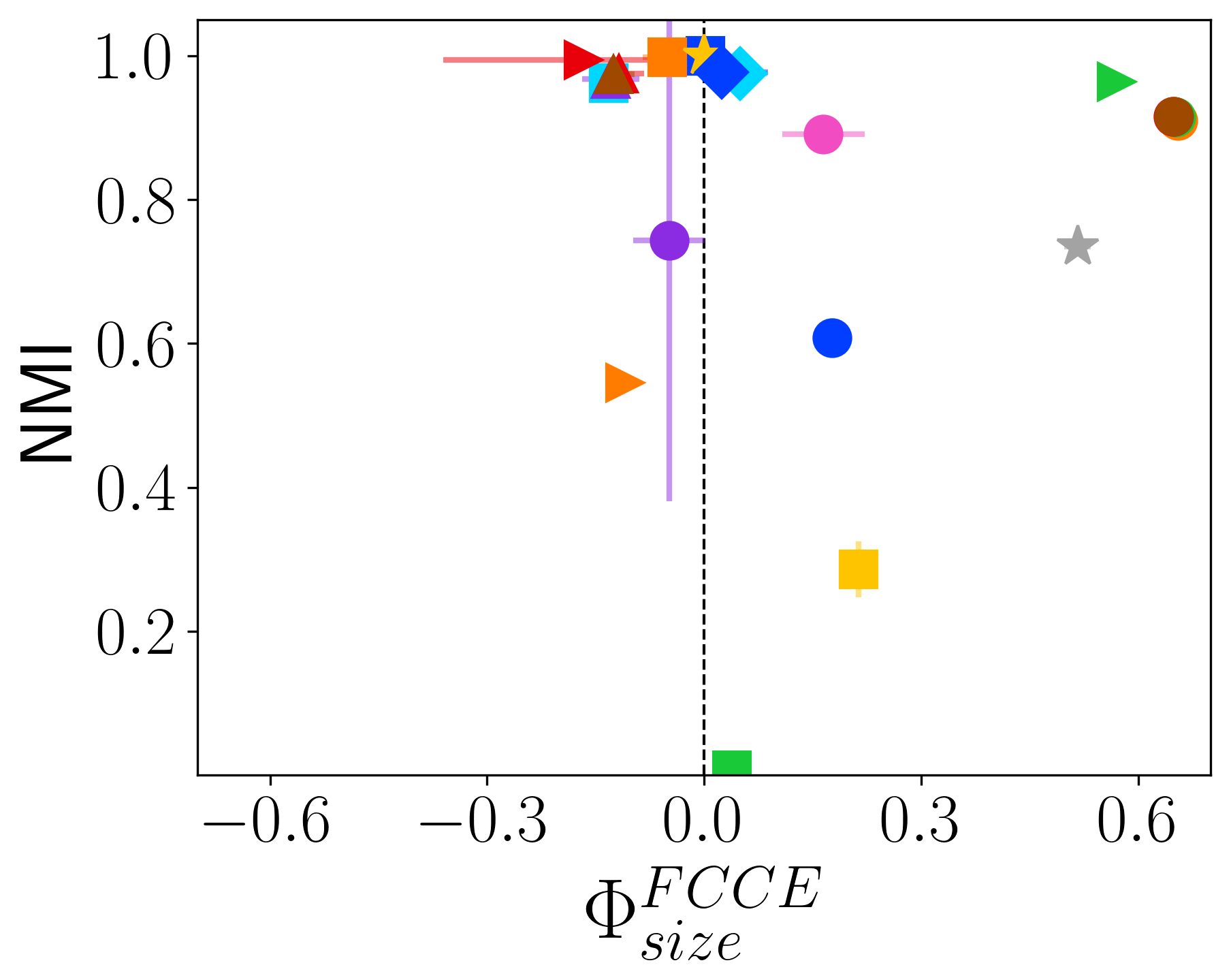}
\end{minipage}
\\
\begin{subfigure}[c]{0.05\textwidth}
\caption*{\rotatebox{90}{$\xi=0.4$}}
\end{subfigure}%
\begin{minipage}[c]{0.95\textwidth}
\includegraphics[width=0.31\textwidth]{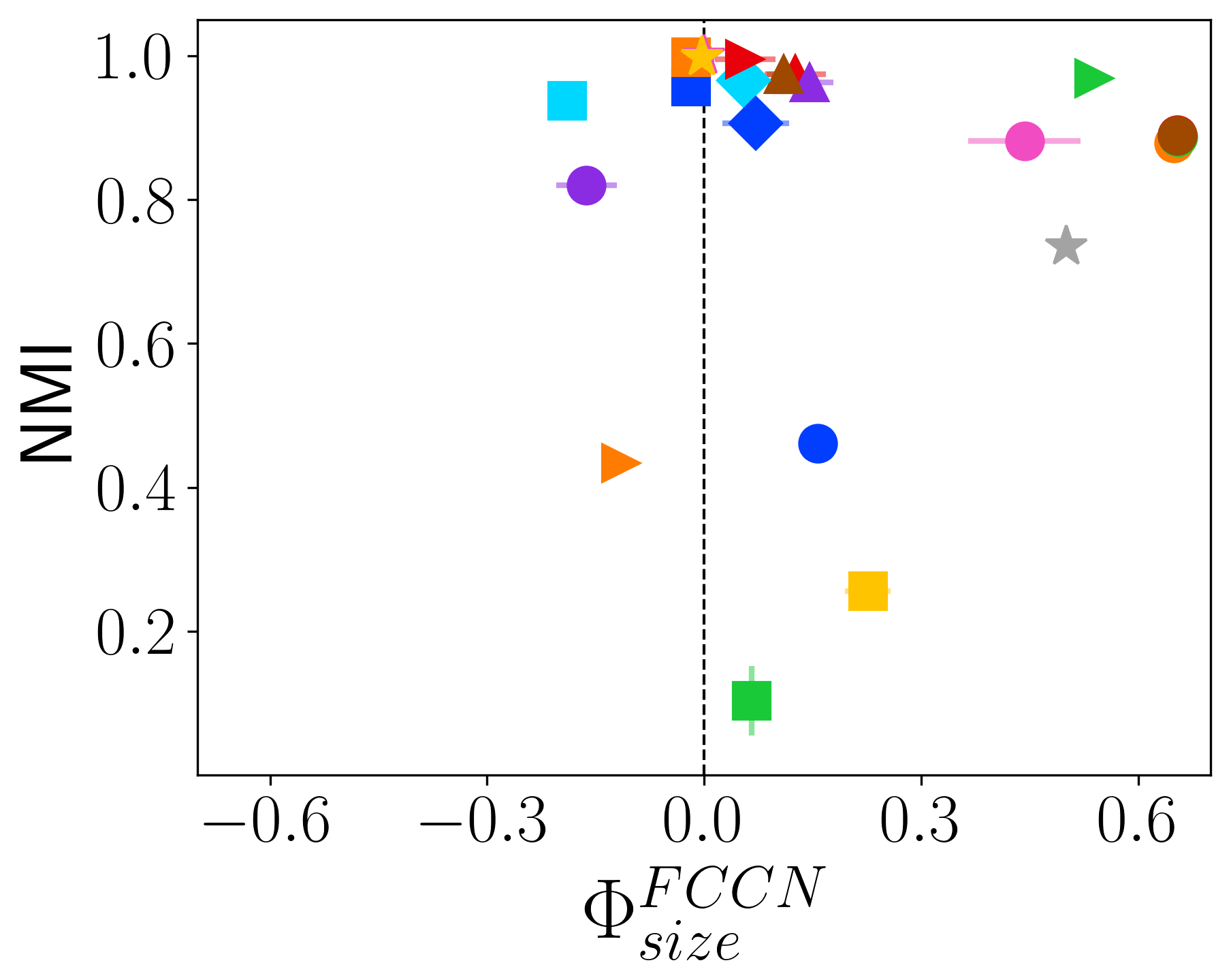}\quad
\includegraphics[width=0.31\textwidth]{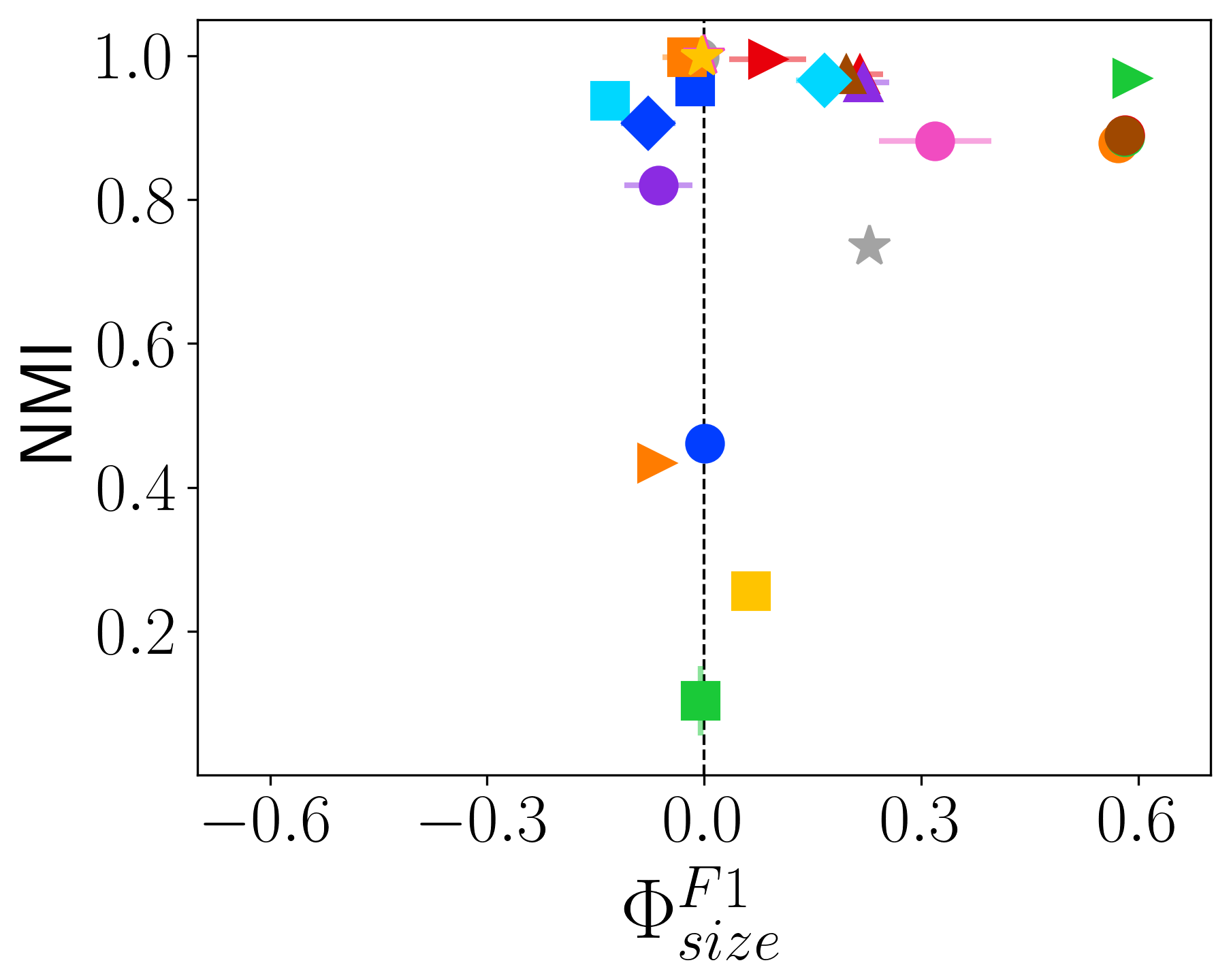}\quad
\includegraphics[width=0.31\textwidth]{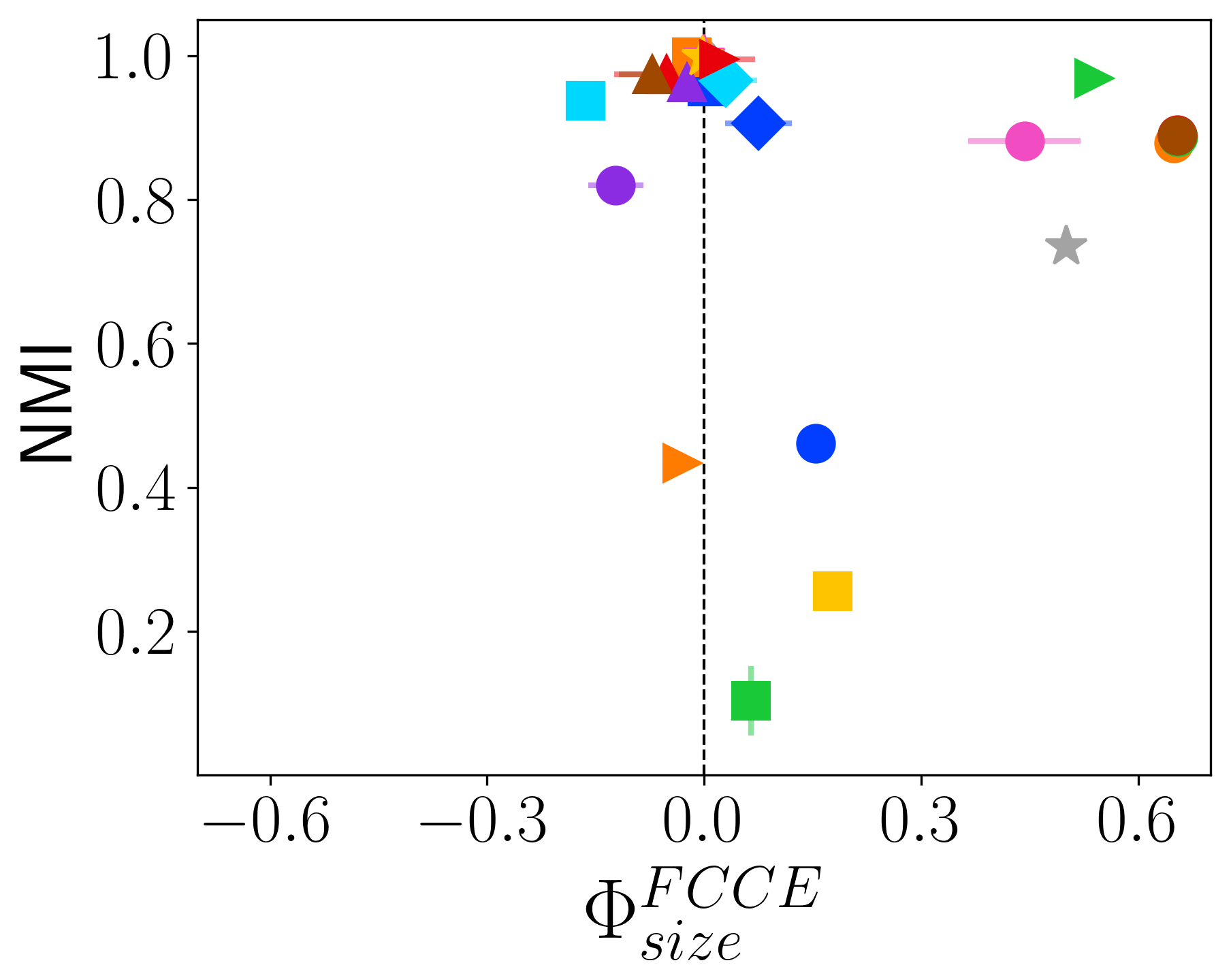}
\end{minipage}
\\
\begin{subfigure}[c]{0.05\textwidth}
\caption*{\rotatebox{90}{$\xi=0.6$}}
\end{subfigure}%
\begin{minipage}[c]{0.95\textwidth}
\includegraphics[width=0.31\textwidth]{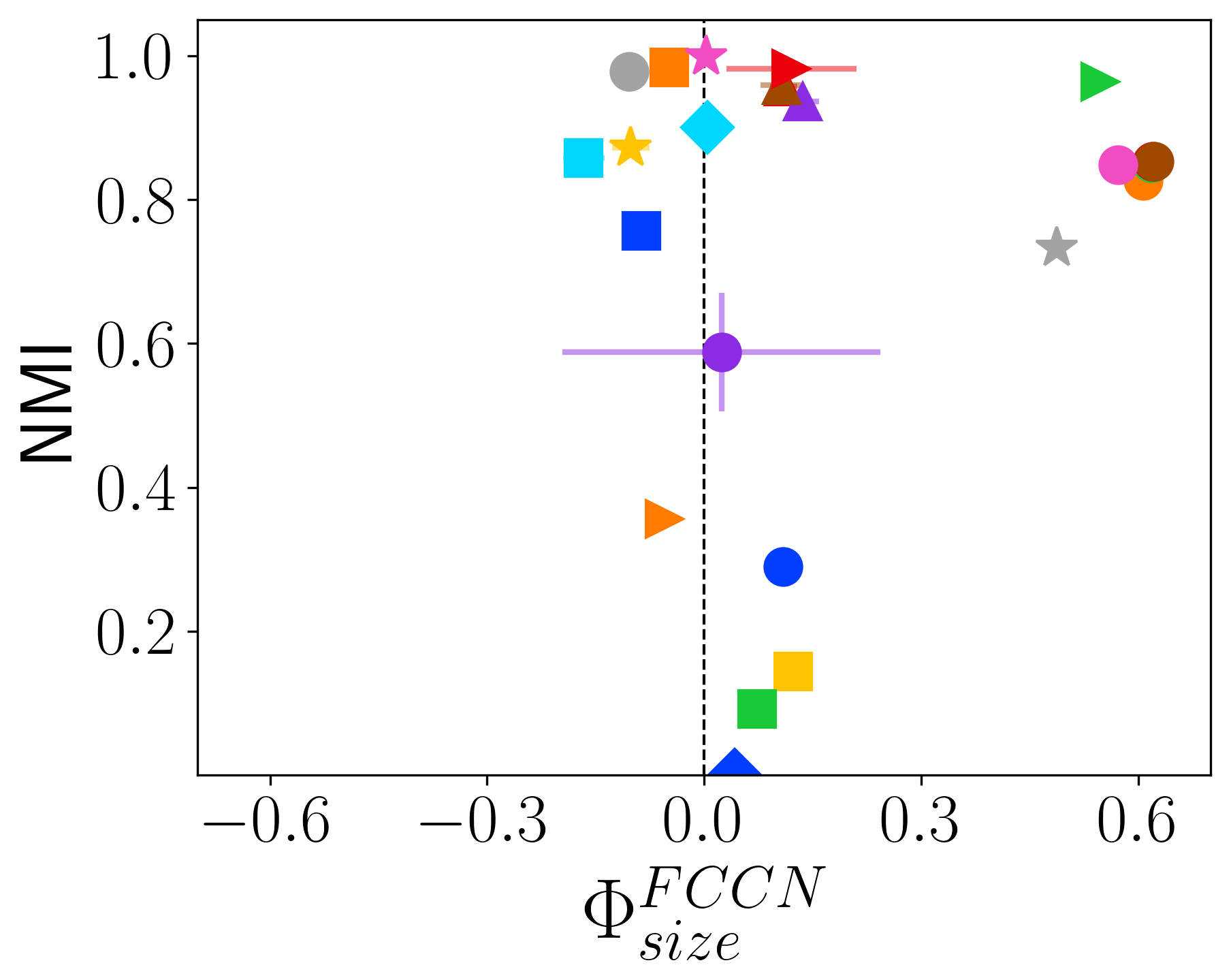}\quad
\includegraphics[width=0.31\textwidth]{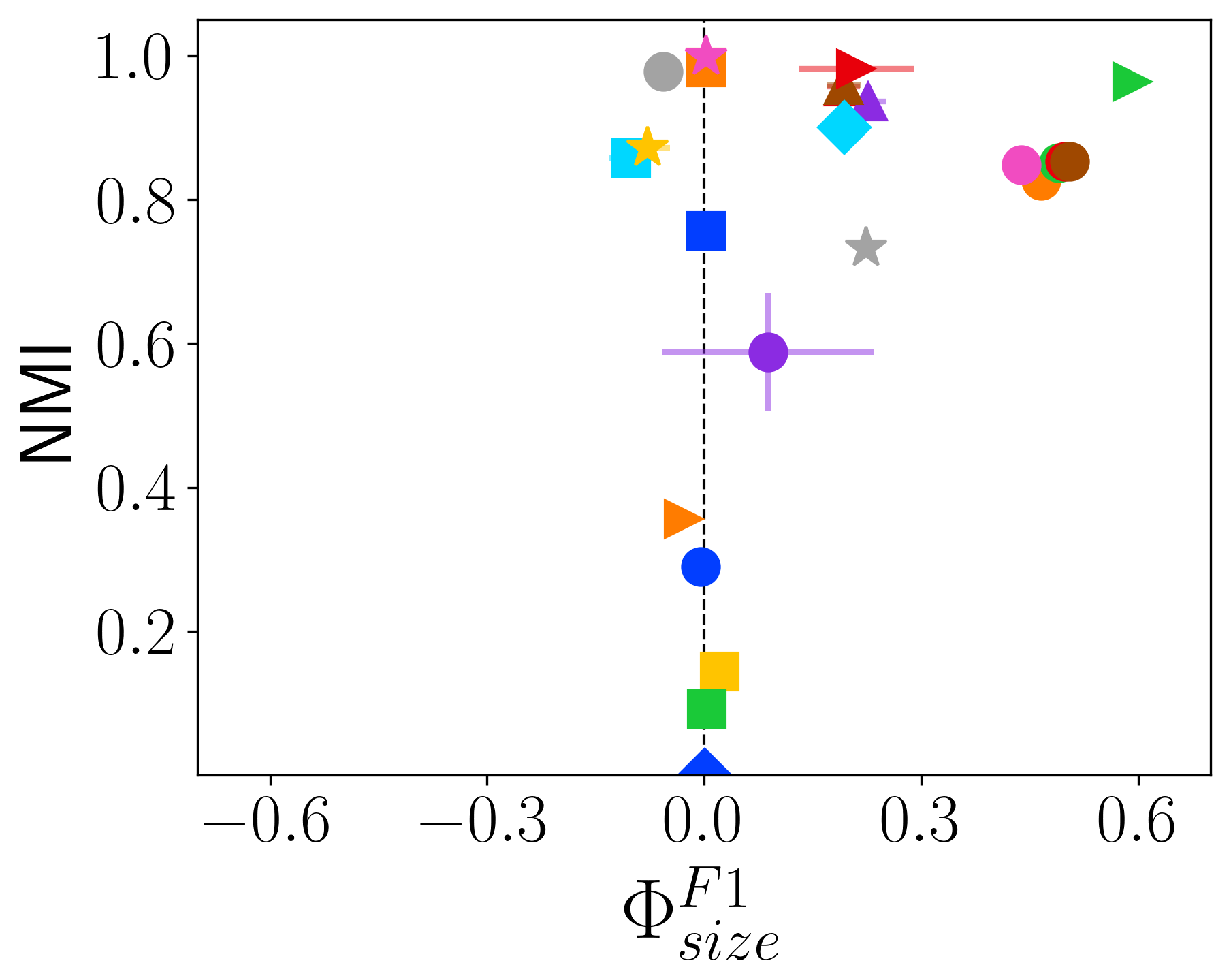}\quad
\includegraphics[width=0.31\textwidth]{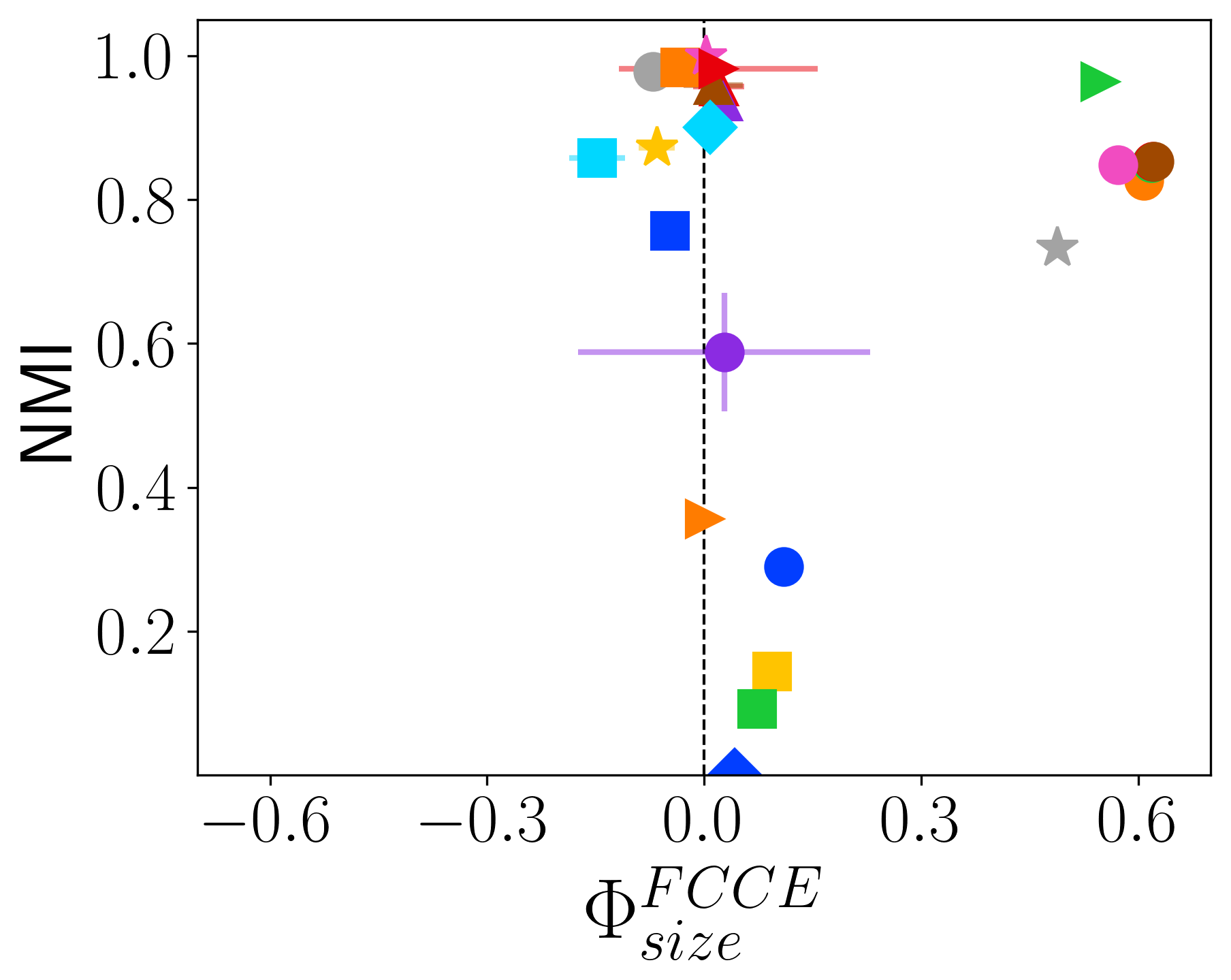}
\end{minipage}
\caption{NMI vs. fairness of community detection methods with respect to community size for ABCD networks of 10,000 nodes having different $\xi$ values.}\label{ABCD_phi_vs_size} 
\end{figure}

Figure~\ref{ABCD_phi_vs_size} shows the trade-off between Normalized Mutual Information (NMI) and fairness $\Phi_{size}^{F*}$ for various community detection methods applied to ABCD networks of 10,000 nodes under different values of $\xi$ (0.2, 0.4, and 0.6). The community detection methods perform better on ABCD even when $\xi=0.6$ compared to LFR networks with high $\mu$ values, as ABCD networks better replicate communities even with higher interconnectivity. However, as $\xi$ increases (from top to bottom in the figure), the distribution of fairness values becomes more dispersed, suggesting that different methods handle fairness differently at varying levels of structural mixing. The NMI values tend to decline as $xi$ increases for most of the community detection methods, as expected. 

For $\xi=0.2$, community detection Methods, including SIgnificance, RSC-SSE, Spectral, Infomap, Walktrap, and Label Propagation, are fair. However, as $\xi$ increases, Spectral tends to be less fair, and RSC-V emerges to be fair. For $\xi=0.4$ and $0.6$, the communities identified using the SBM-Nested method show a high value for performance metrics; however, the fairness is lower as the method favors large size, high conductance, and low-density communities. For $\xi=0.6$, Representation Learning-based methods (DeepWalk, Node2Vec, Fairwalk) show fairness with respect to conductance and fairly identify communities of all conductance. Fairwalk is a fairness-aware representation learning-based method; it still does not fairly identify communities of all types as well as the performance with respect to different validation metrics is lower as compared to many other methods. 

\subsubsection*{Fairness-Performance trade-off versus Community Density}

\begin{figure}[t]
\centering
\begin{subfigure}[b]{0.98\textwidth}            
    \includegraphics[width=\textwidth]{figures/legend_ncol6.png}
\end{subfigure}\\
\begin{subfigure}[c]{0.05\textwidth}
\caption*{\rotatebox{90}{$\xi=0.2$}}
\end{subfigure}%
\begin{minipage}[c]{0.95\textwidth}
\includegraphics[width=0.31\textwidth]{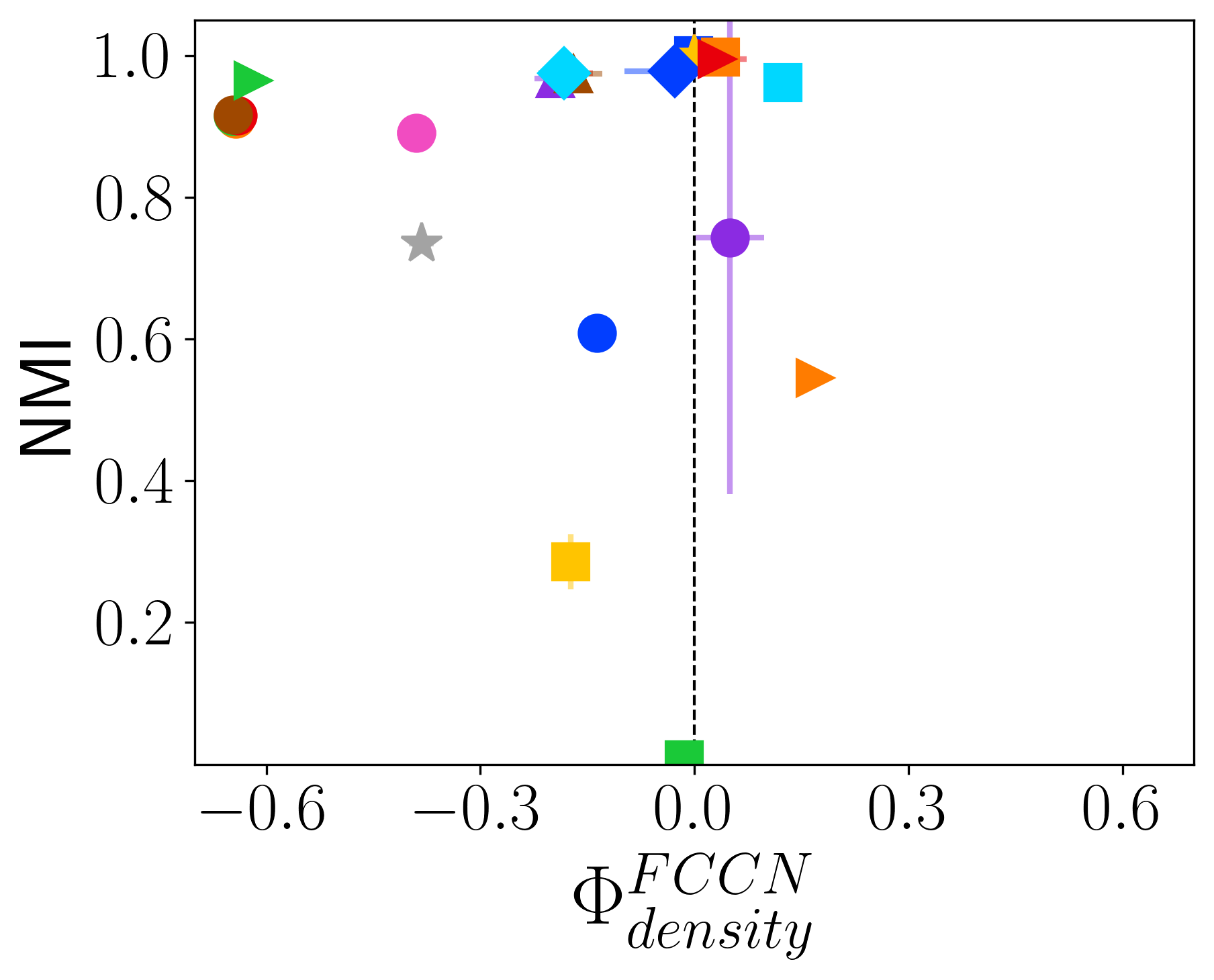}\quad
\includegraphics[width=0.31\textwidth]{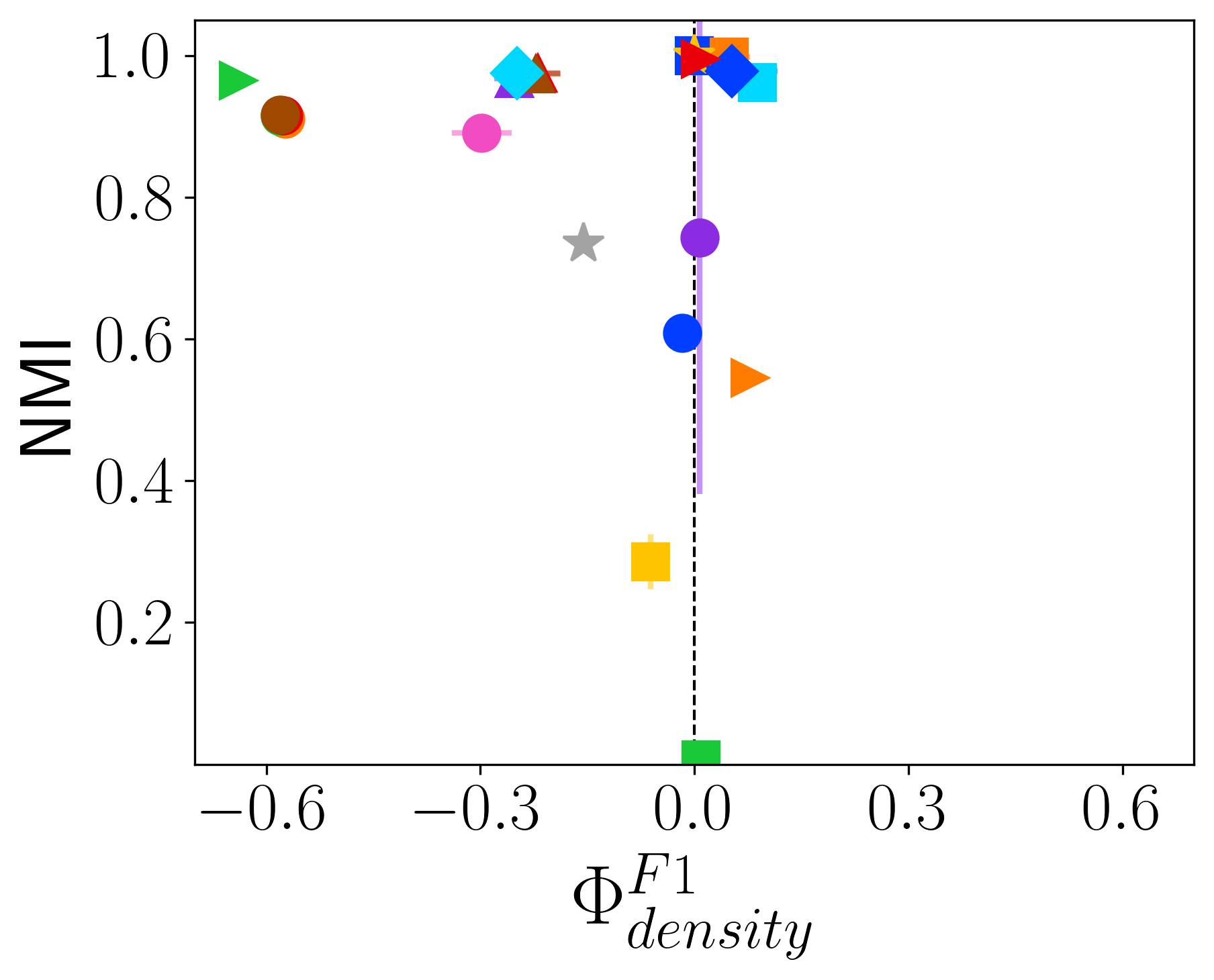}\quad
\includegraphics[width=0.31\textwidth]{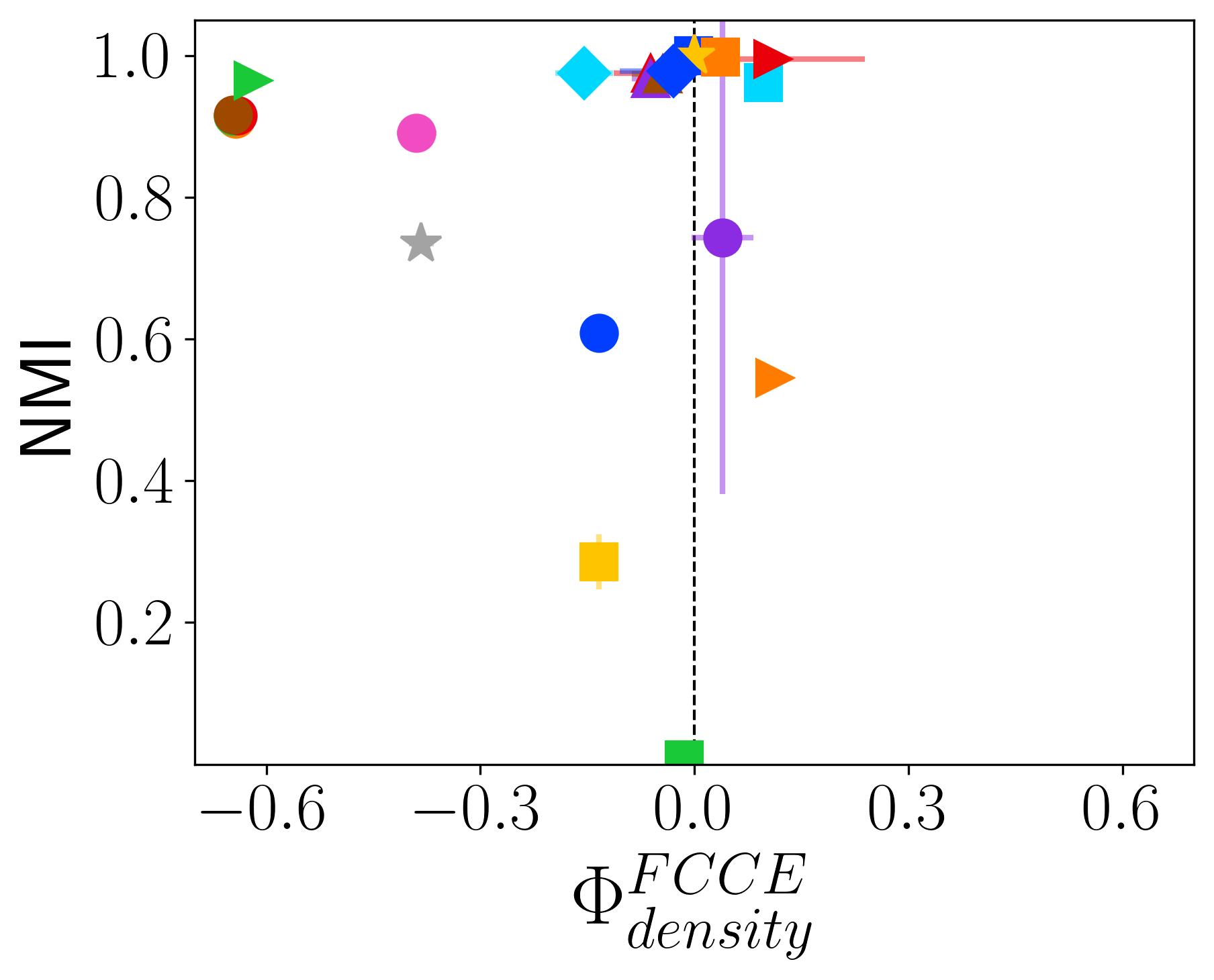}
\end{minipage}
\\
\begin{subfigure}[c]{0.05\textwidth}
\caption*{\rotatebox{90}{$\xi=0.4$}}
\end{subfigure}%
\begin{minipage}[c]{0.95\textwidth}
\includegraphics[width=0.31\textwidth]{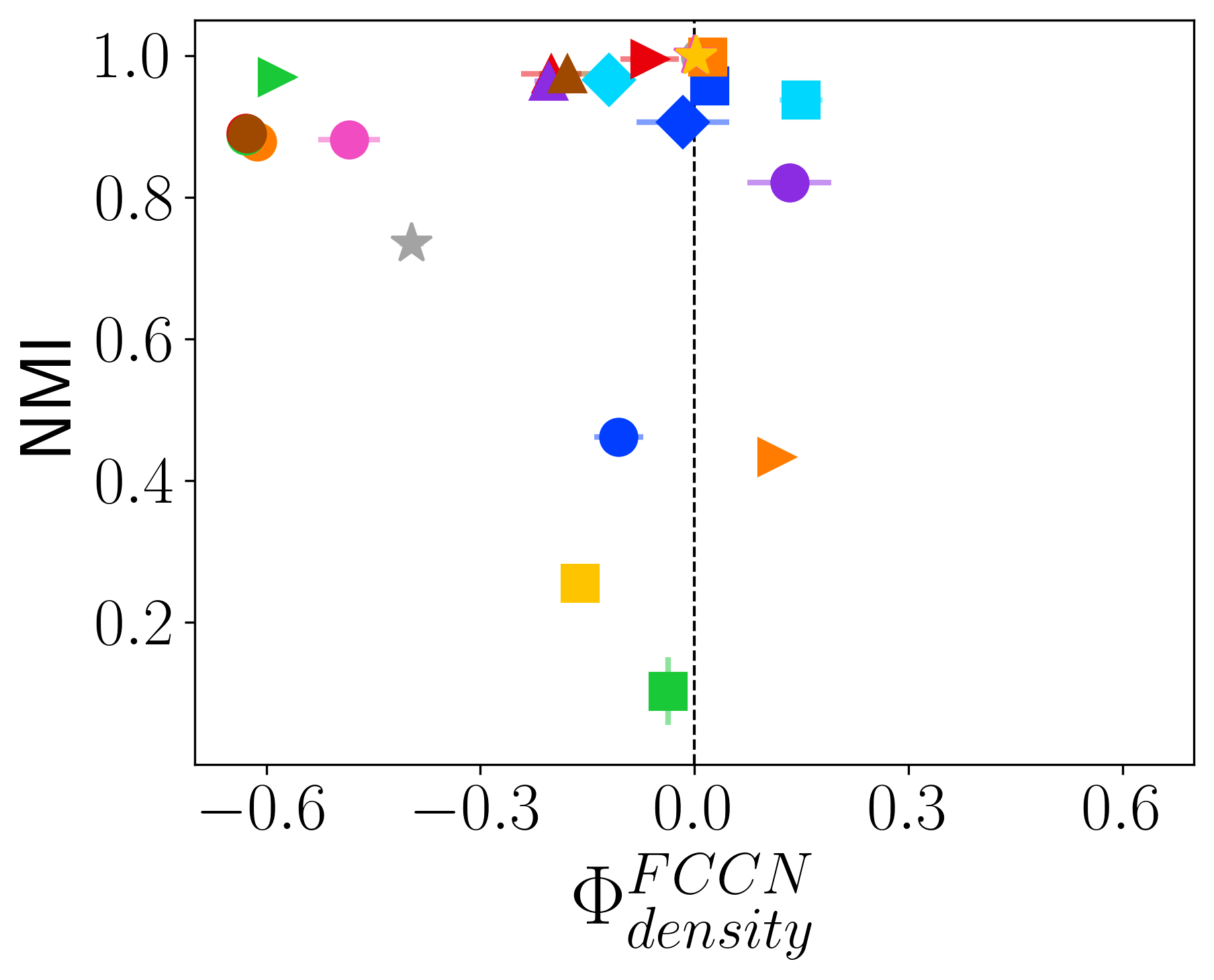}\quad
\includegraphics[width=0.31\textwidth]{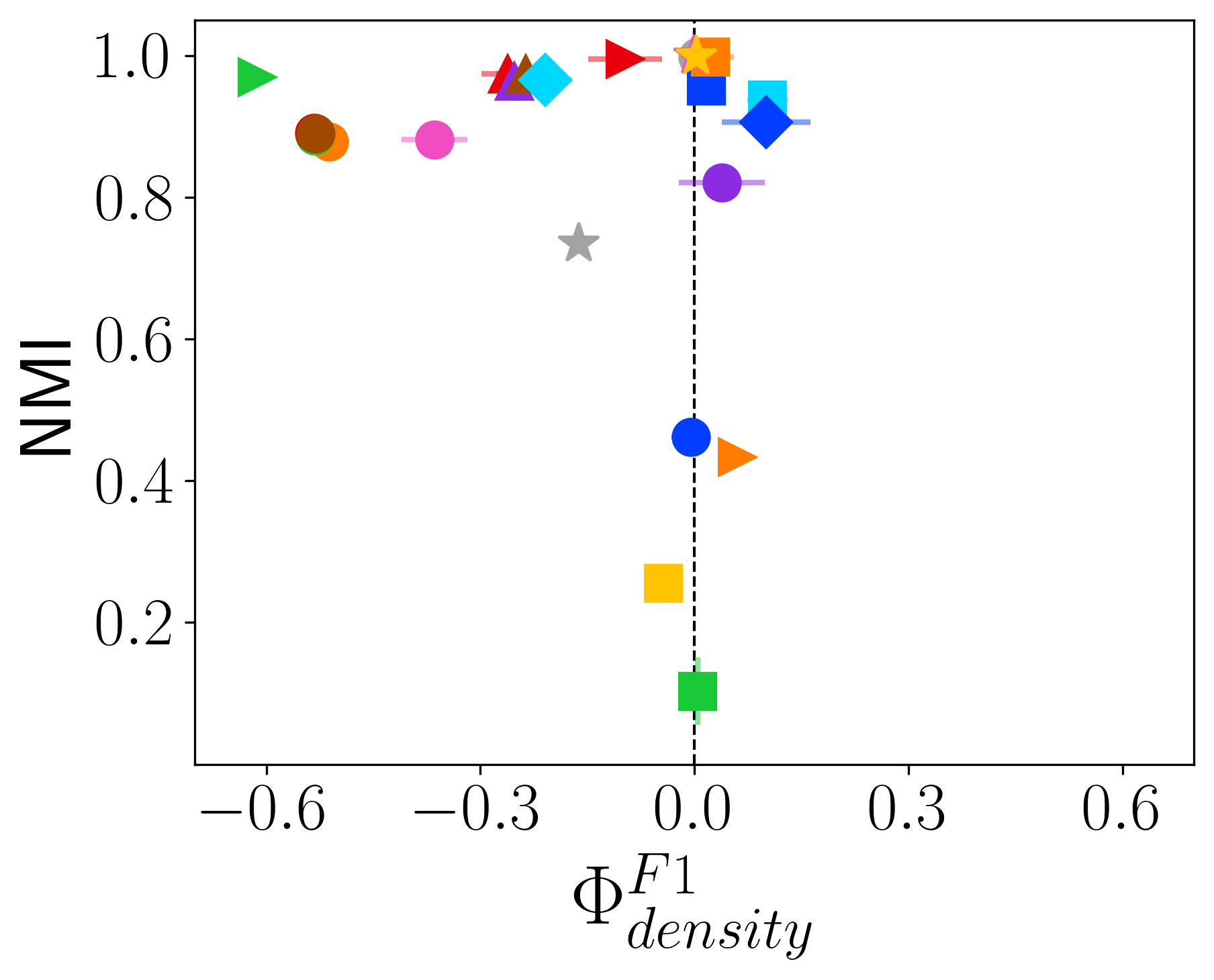}\quad
\includegraphics[width=0.31\textwidth]{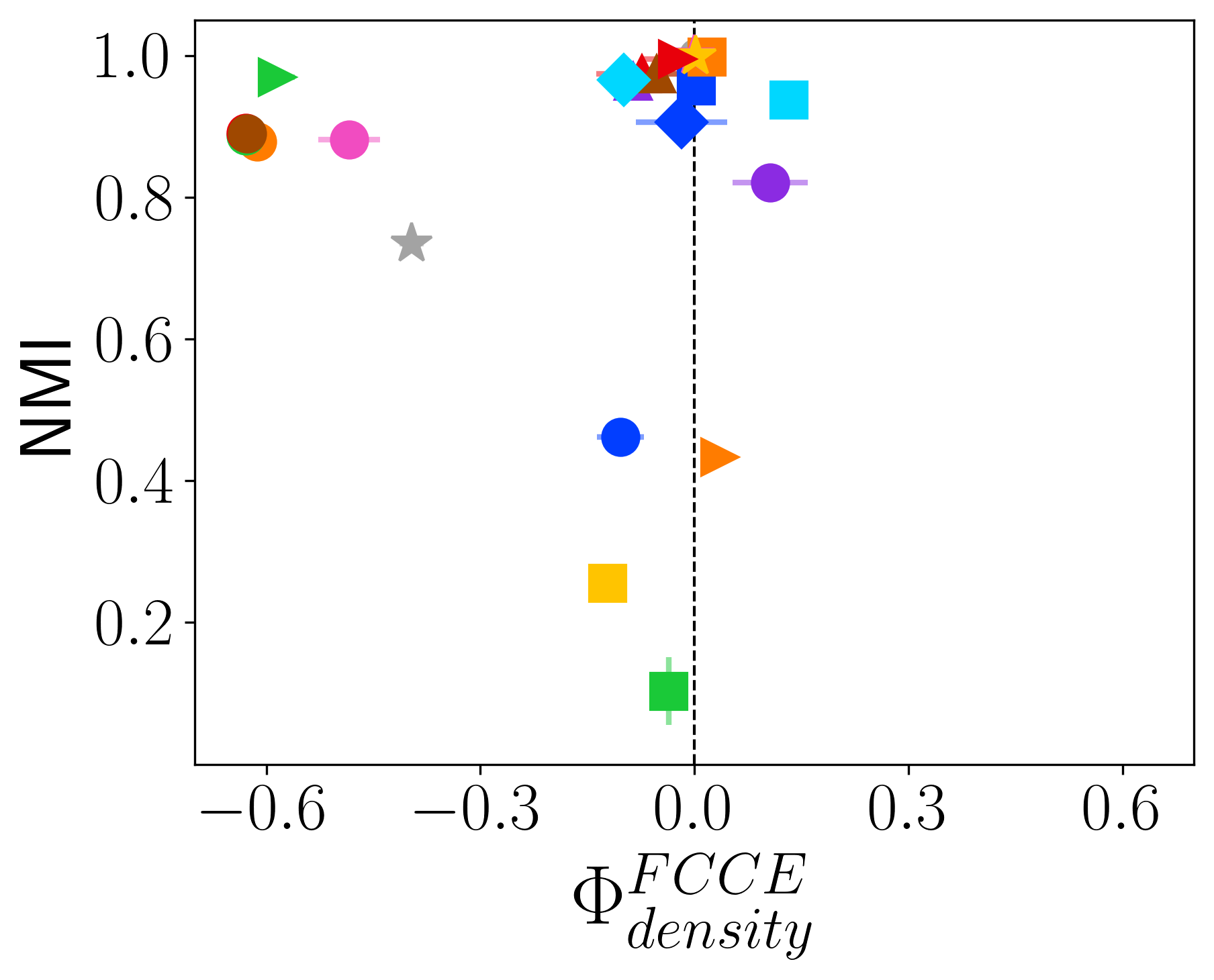}
\end{minipage}
\\
\begin{subfigure}[c]{0.05\textwidth}
\caption*{\rotatebox{90}{$\xi=0.6$}}
\end{subfigure}%
\begin{minipage}[c]{0.95\textwidth}
\includegraphics[width=0.31\textwidth]{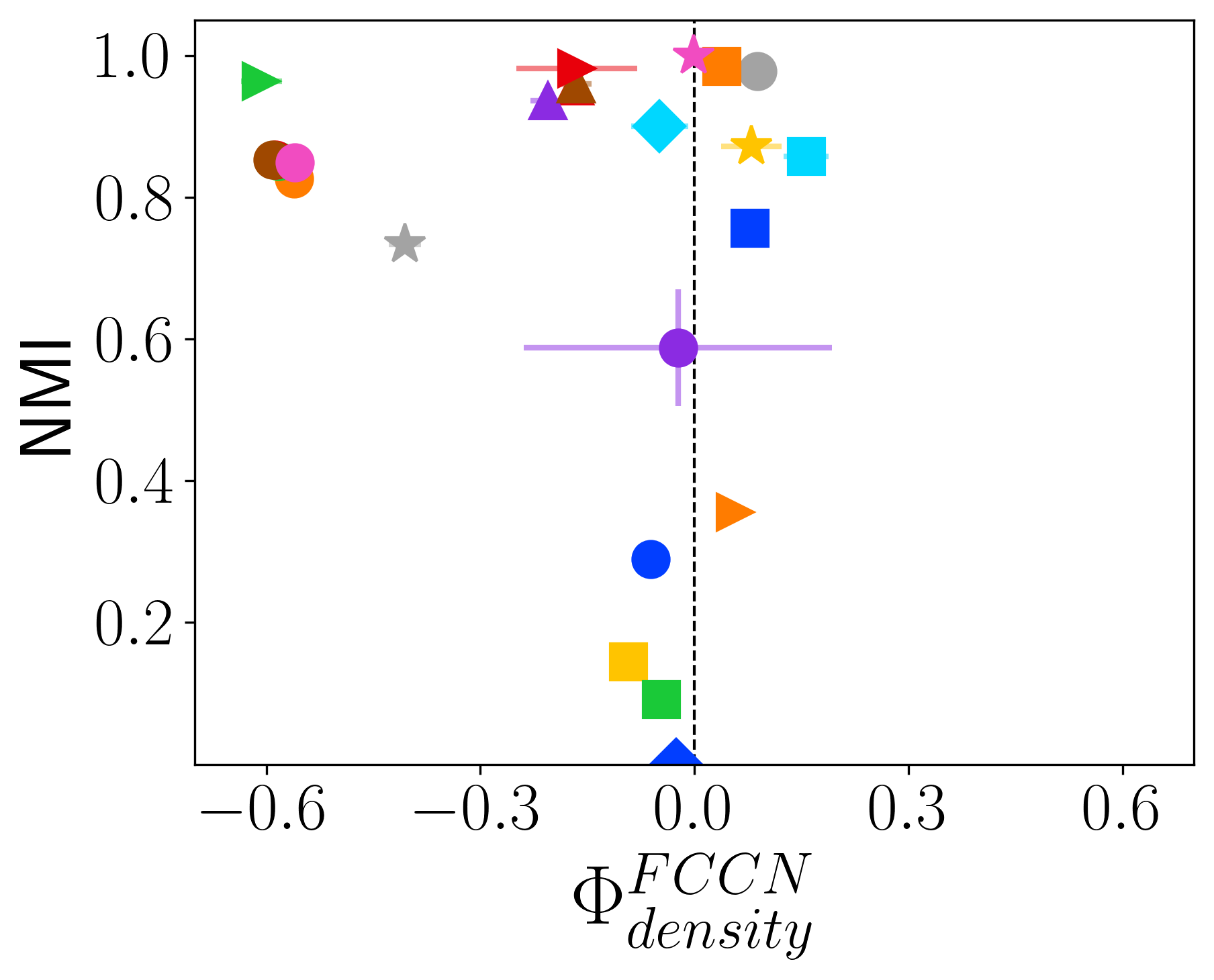}\quad
\includegraphics[width=0.31\textwidth]{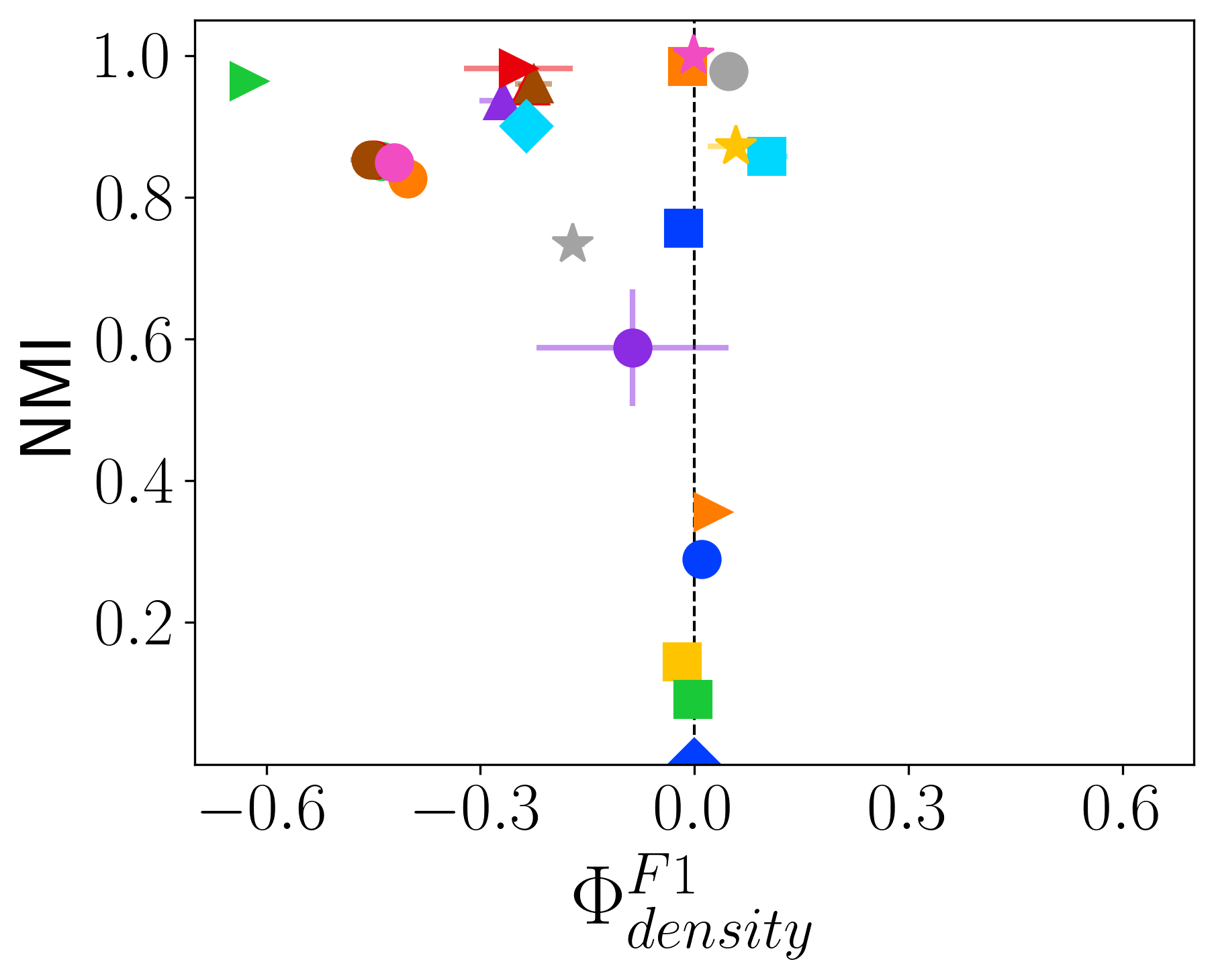}\quad
\includegraphics[width=0.31\textwidth]{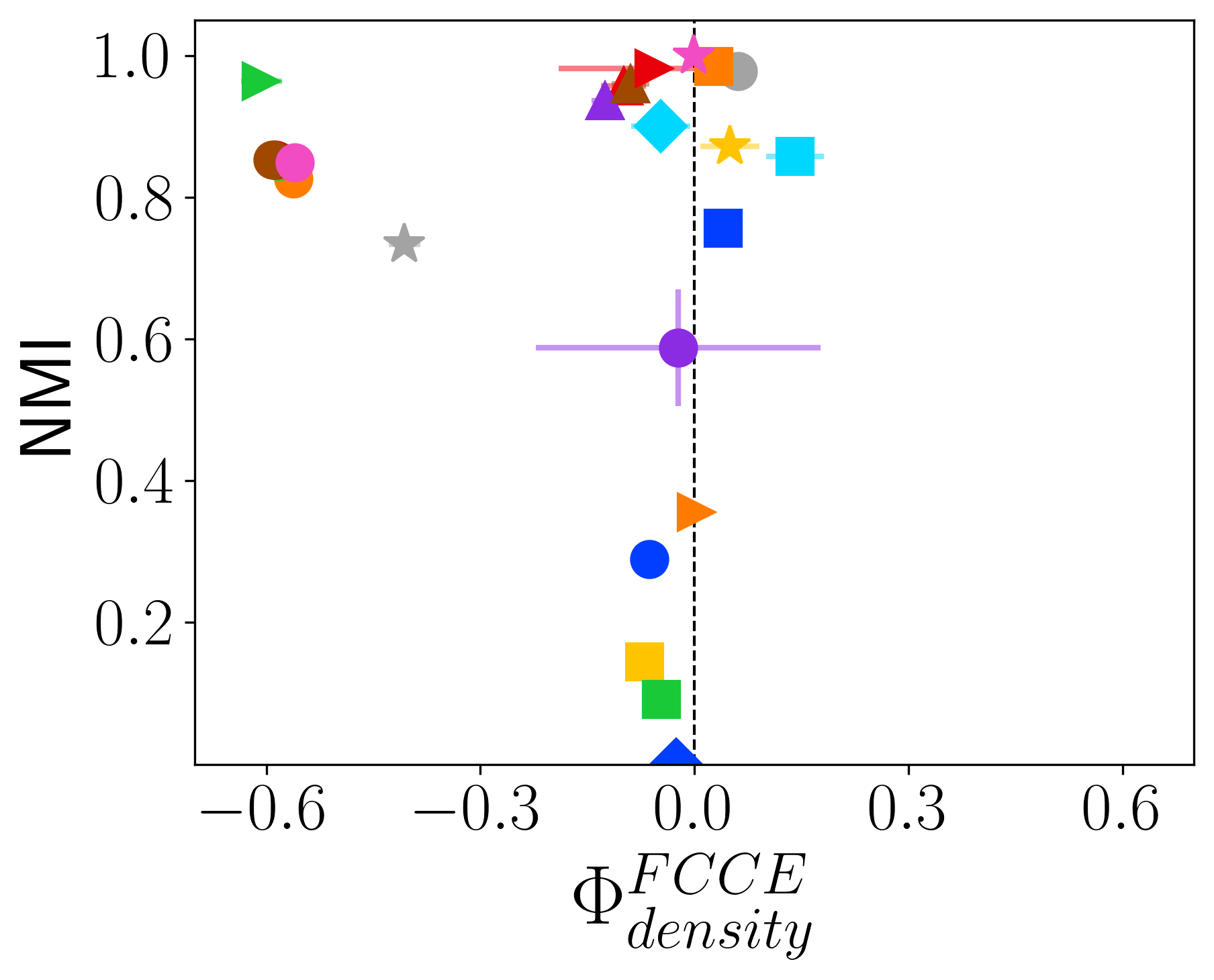}
\end{minipage}
\caption{NMI vs. fairness of community detection methods with respect to community density for ABCD networks of 10,000 nodes having different $\xi$ values.}\label{ABCD_phi_vs_dens} 
\end{figure}

Figure~\ref{ABCD_phi_vs_dens} shows the fairness of community detection methods with respect to density $\Phi^{F*}_{\textit{density}}$ on ABCD networks. The results are quite similar to what we observed for community size, as discussed earlier. However, there are a few key distinctions to note.  
The Paris method fairly identifies communities of different densities as shown by $\Phi^{F1}_{\textit{density}}$. However, as $\xi$ increases to $0.6$, Paris becomes fair with respect to $\Phi^{FCCN}_{\textit{density}}$ and $\Phi^{FCCE}_{\textit{density}}$. This occurs because Paris effectively detects large, sparse communities but also assigns a significant number of nodes from other communities to these larger groups, which is captured by $\Phi^{F1}_{\textit{density}}$. 
Similarly, the Spectral method consistently detects a comparable number of communities across all ABCD networks. However, its fairness decreases in terms of $\Phi^{FCCN}_{\textit{density}}$ and $\Phi^{FCCE}_{\textit{density}}$, while increasing for $\Phi^{F1}_{\textit{density}}$. This is because it tends to identify large, sparse communities while merging other community nodes into them, which is reflected in the measure.
 
Overall, most community detection methods favor lower-density communities rather than higher-density ones, particularly concerning $\Phi^{F1}_{\textit{density}}$. Methods such as Combo, Leiden, Louvain, RB-C, RB-ER, and SBM, consistently favor lower-density communities across all values of $\xi$. Unlike the findings for LFR networks, fairness does not necessarily improve as the mixing parameter increases, i.e., due to the structural connectivity of ABCD networks. 

\subsubsection*{Fairness Performance trade-off versus Community Conductance}

\begin{figure}[t]
\centering
\begin{subfigure}[b]{0.98\textwidth}            
    \includegraphics[width=\textwidth]{figures/legend_ncol6.png}
\end{subfigure}\\
\begin{subfigure}[c]{0.05\textwidth}
\caption*{\rotatebox{90}{$\xi=0.2$}}
\end{subfigure}%
\begin{minipage}[c]{0.95\textwidth}
\includegraphics[width=0.31\textwidth]{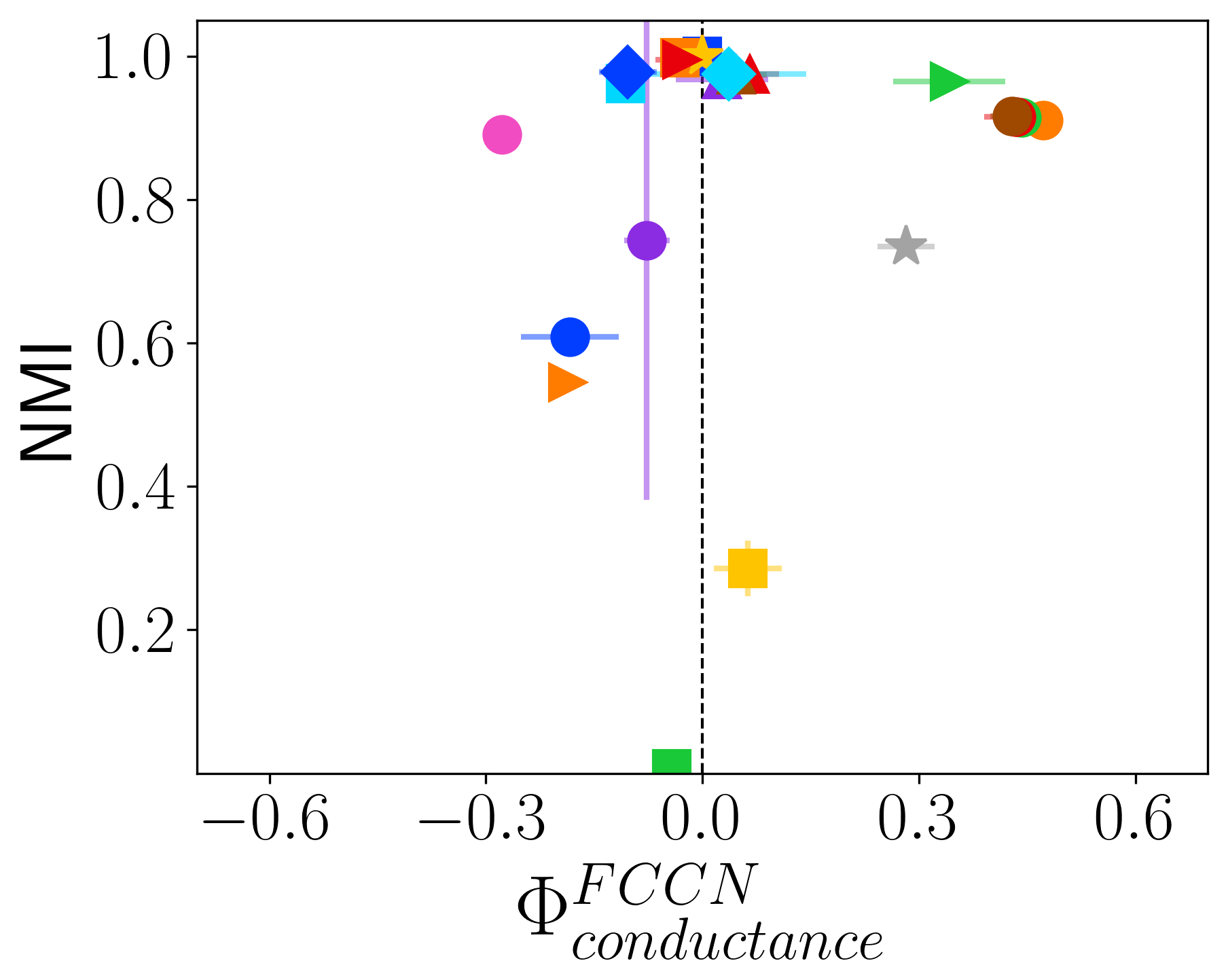}\quad
\includegraphics[width=0.31\textwidth]{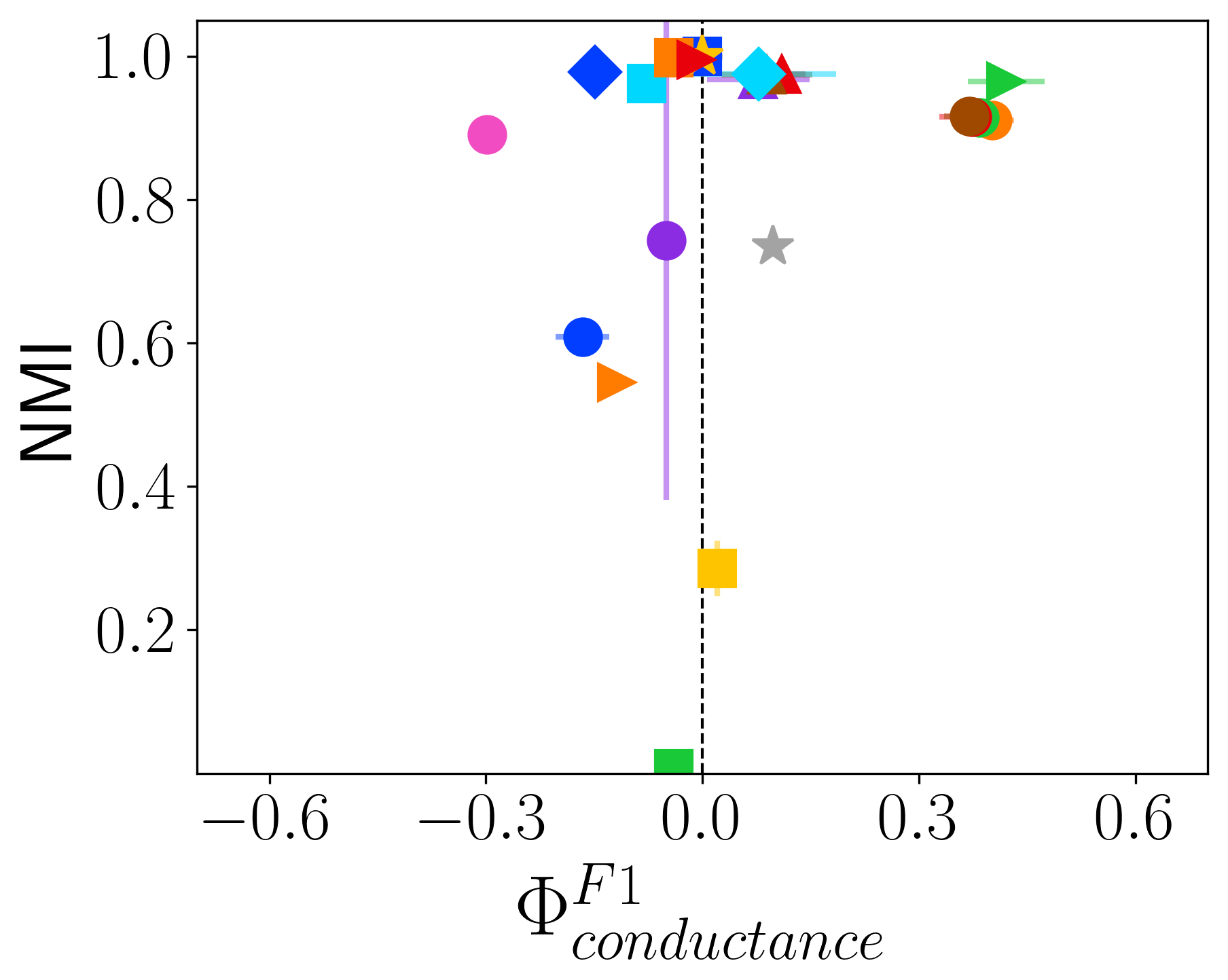}\quad
\includegraphics[width=0.31\textwidth]{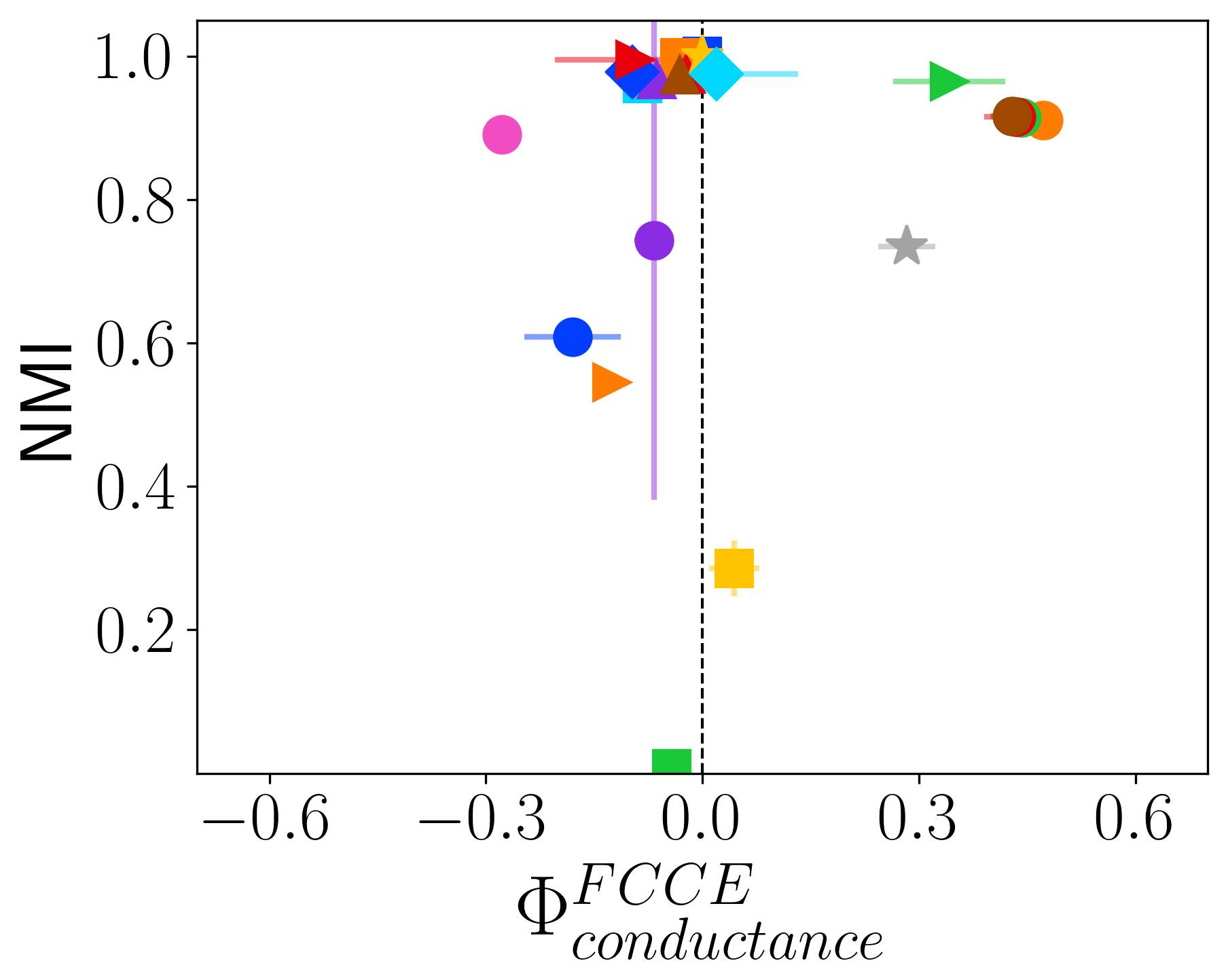}
\end{minipage}
\\
\begin{subfigure}[c]{0.05\textwidth}
\caption*{\rotatebox{90}{$\xi=0.4$}}
\end{subfigure}%
\begin{minipage}[c]{0.95\textwidth}
\includegraphics[width=0.31\textwidth]{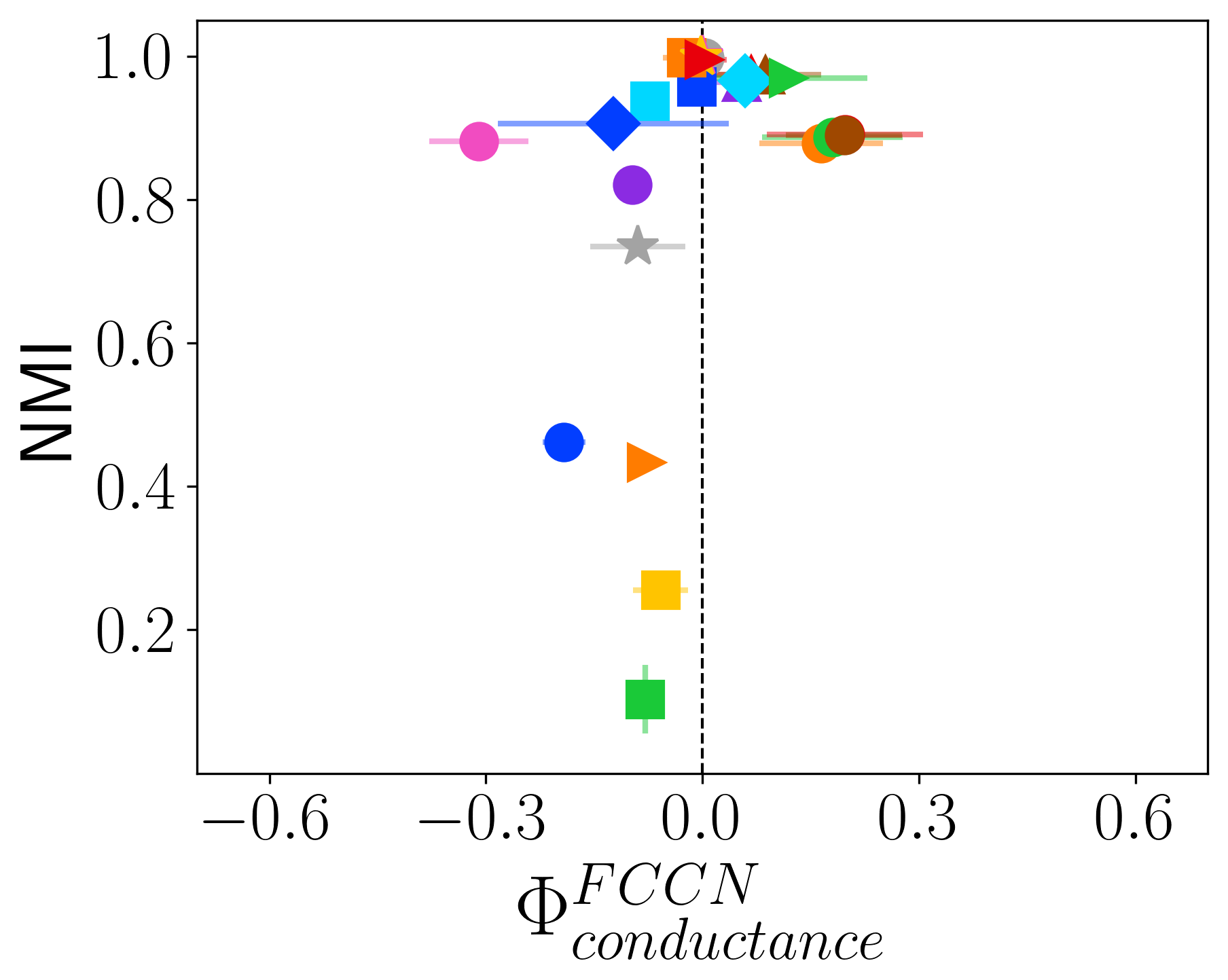}\quad
\includegraphics[width=0.31\textwidth]{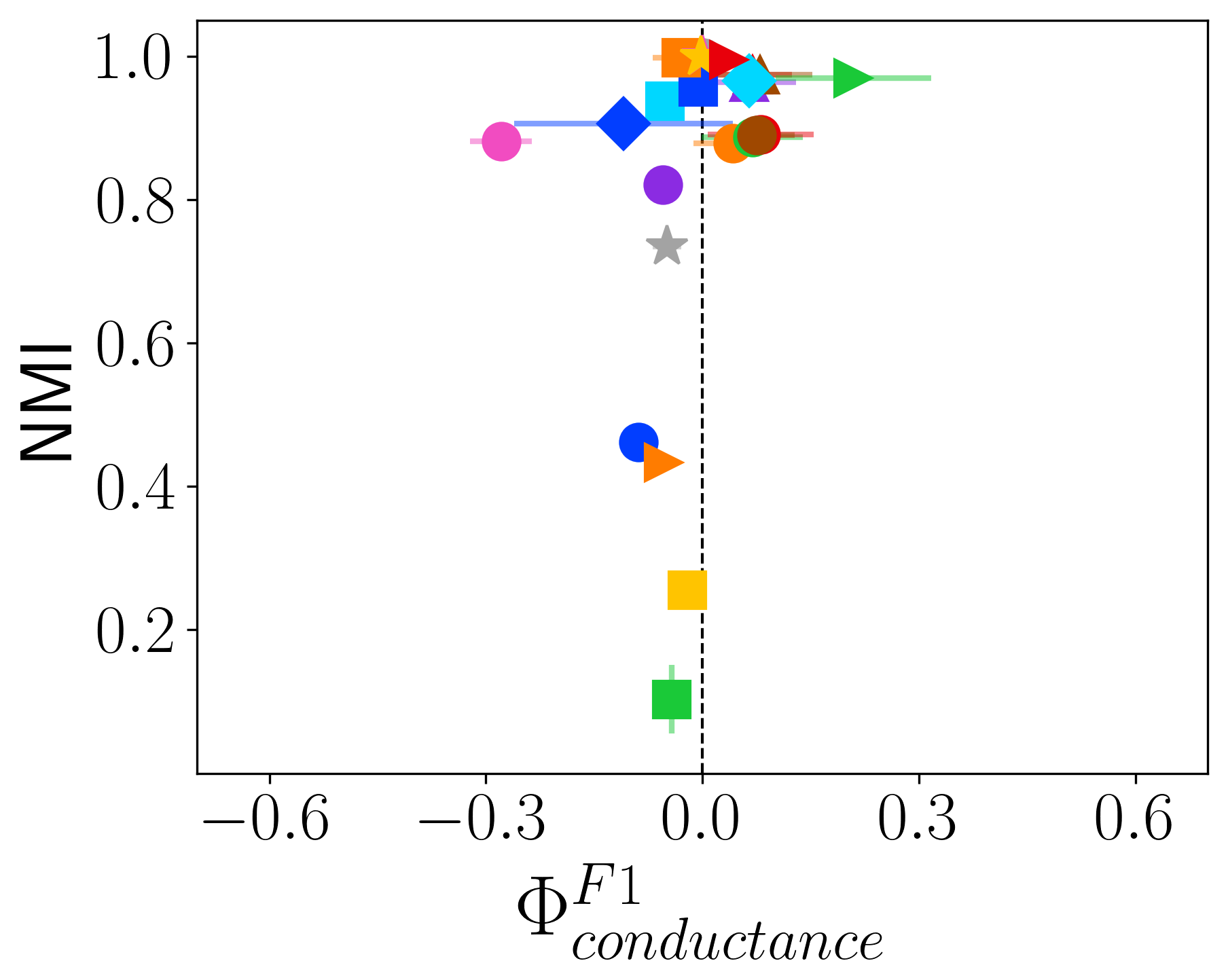}\quad
\includegraphics[width=0.31\textwidth]{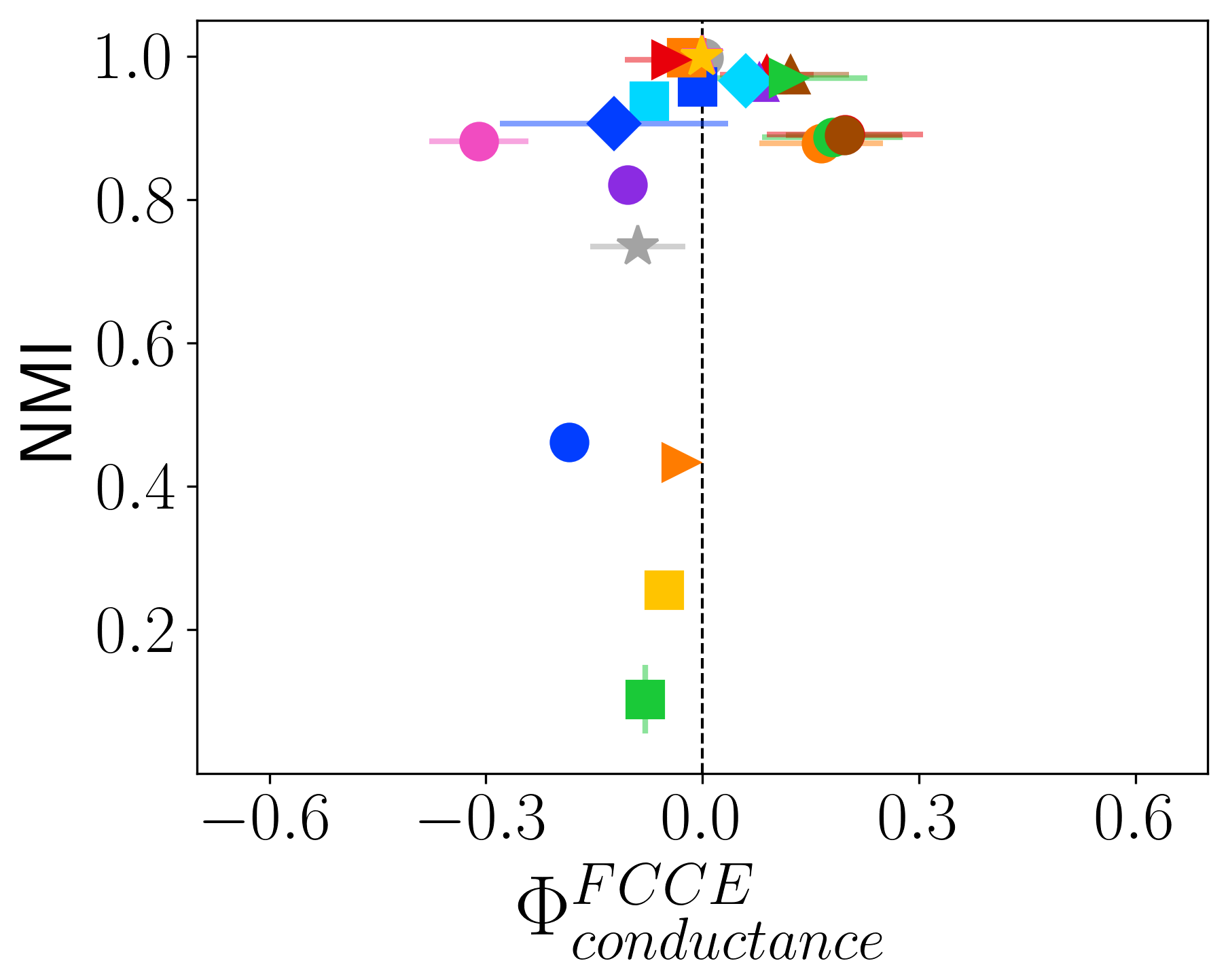}
\end{minipage}
\\
\begin{subfigure}[c]{0.05\textwidth}
\caption*{\rotatebox{90}{$\xi=0.6$}}
\end{subfigure}%
\begin{minipage}[c]{0.95\textwidth}
\includegraphics[width=0.31\textwidth]{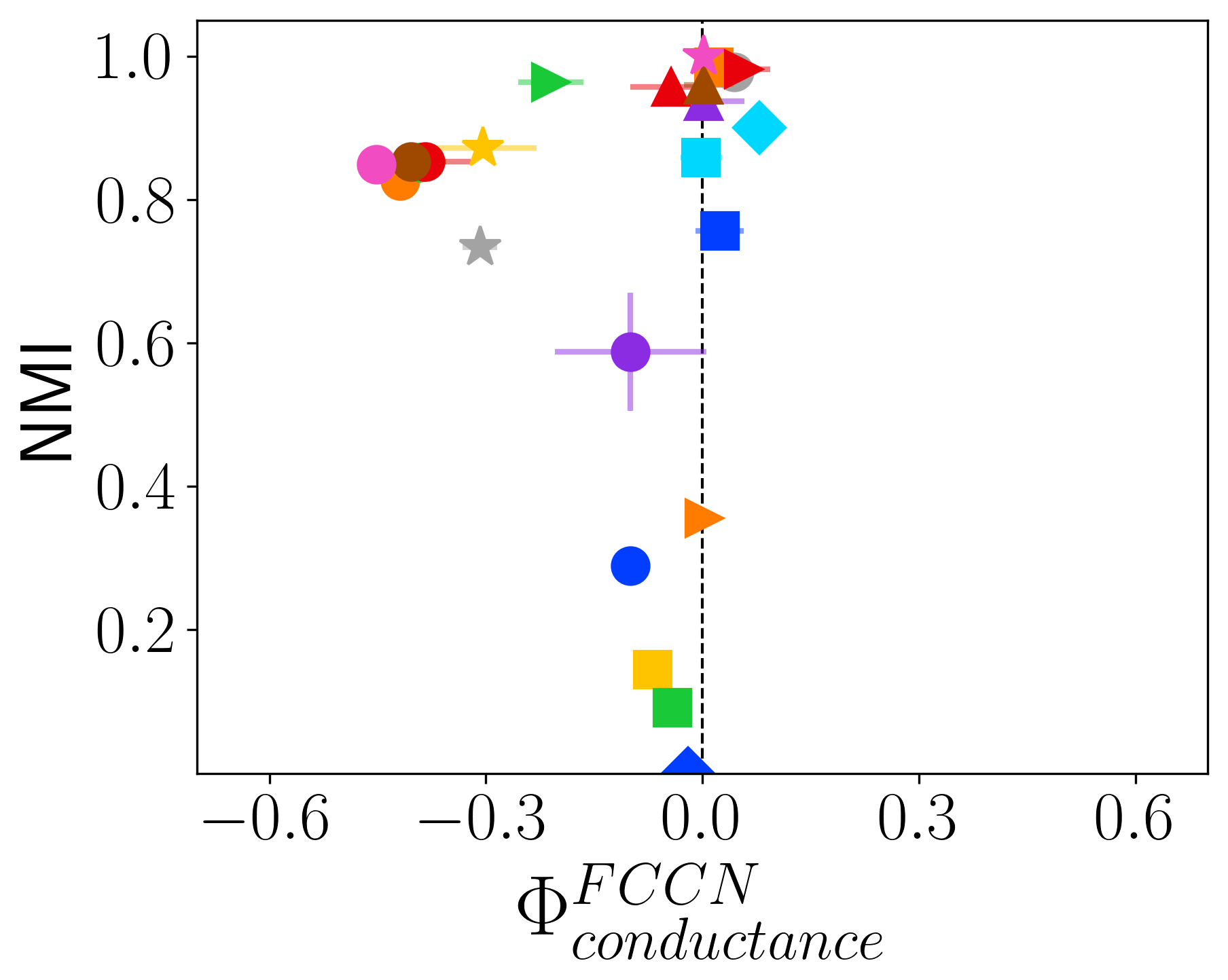}\quad
\includegraphics[width=0.31\textwidth]{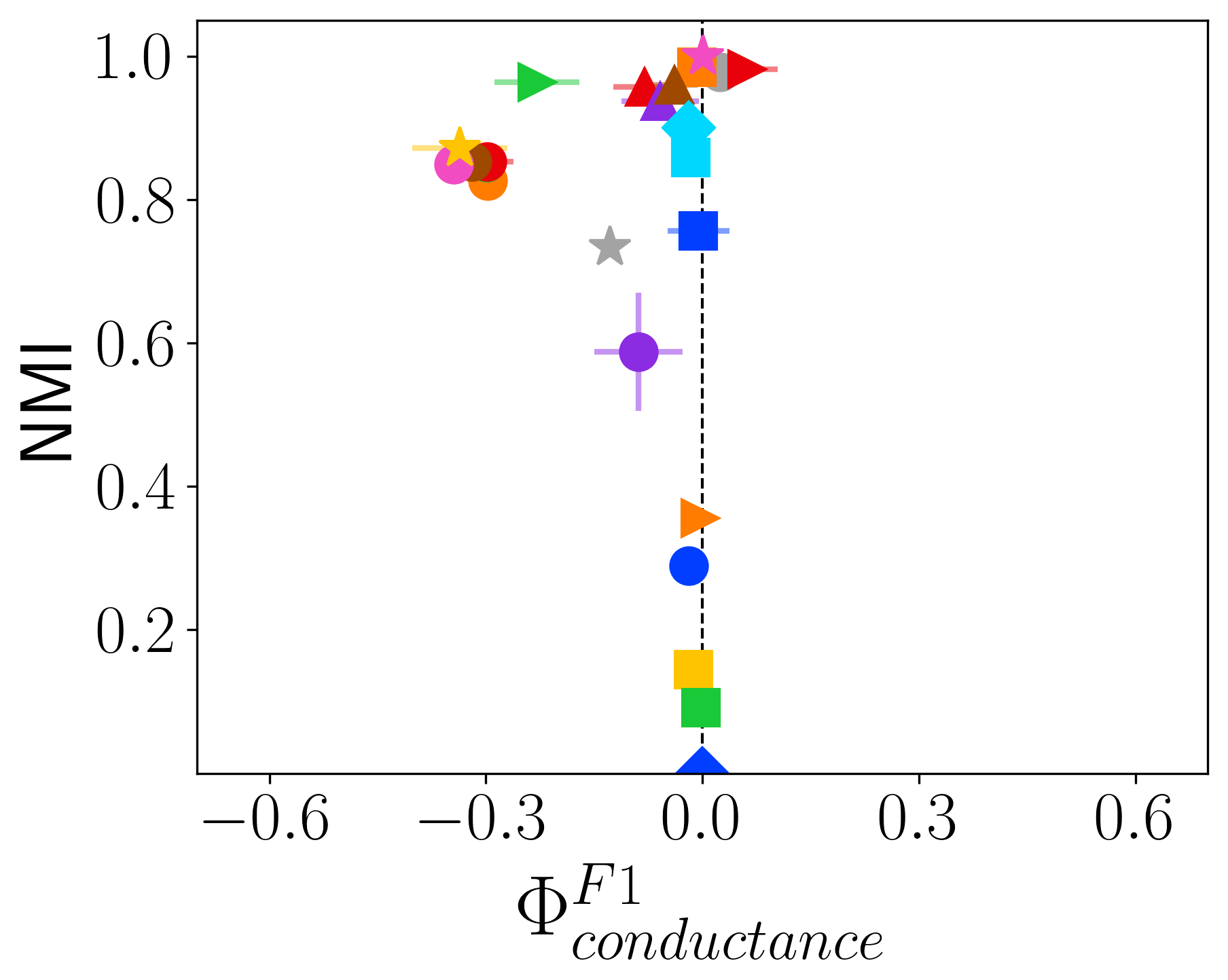}\quad
\includegraphics[width=0.31\textwidth]{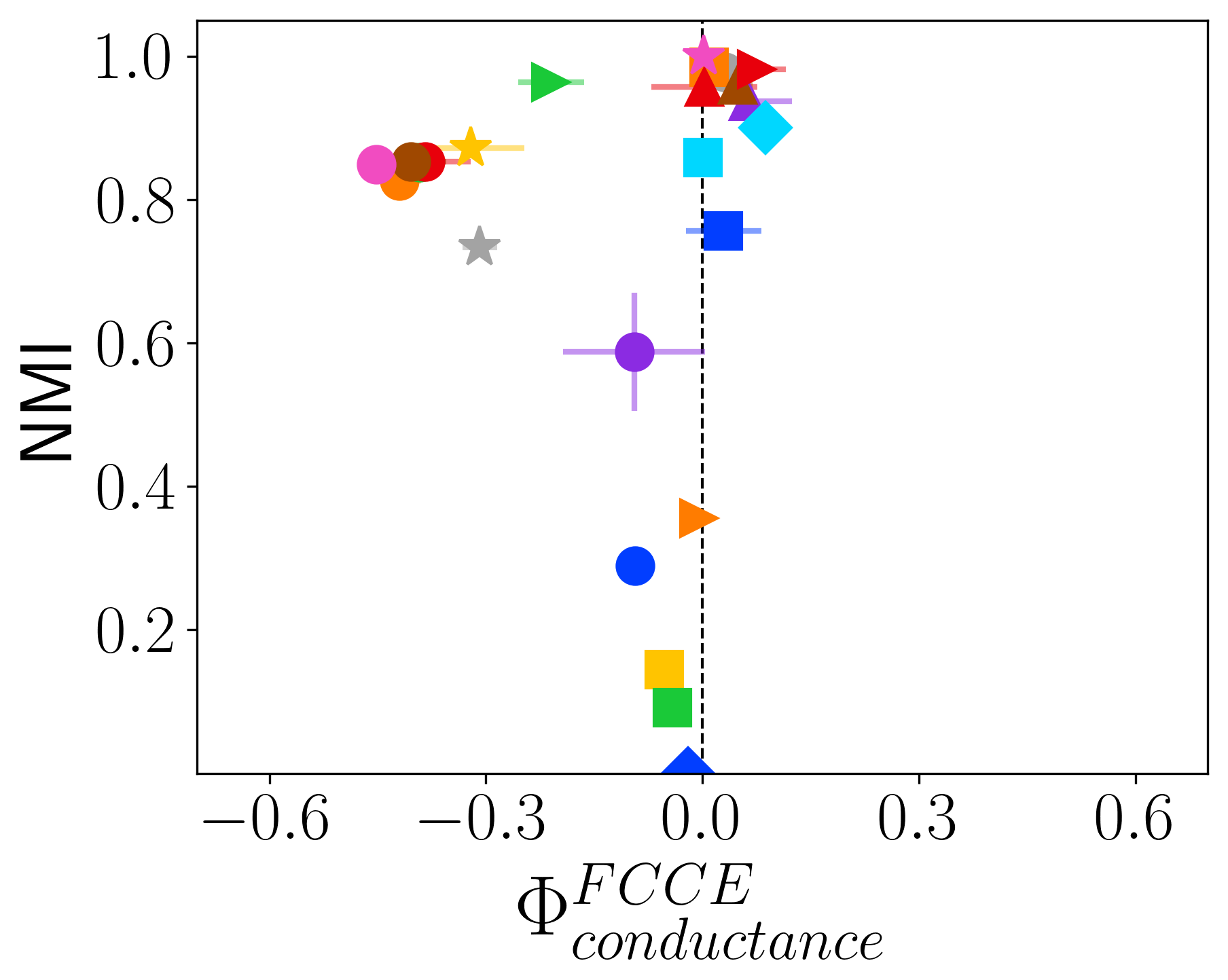}
\end{minipage}
\caption{NMI vs. fairness of community detection methods with respect to community conductance for ABCD networks of 10,000 nodes having different $\xi$ values.}\label{ABCD_phi_vs_cond} 
\end{figure}

Figure~\ref{ABCD_phi_vs_cond} presents the fairness of community detection methods with respect to conductance on ABCD networks. As observed previously, methods that achieve both high fairness and good community quality include Significance, RSC-SSE, Infomap, and Walktrap, which are also in this case. Another group of methods that demonstrates slightly lower fairness and community quality but still performs relatively well include FairWalk, Node2Vec, Spectral, RSC-V, SBM-Nested, Fluid, and EigenVector. 

An important observation is that as $\xi$ increases, methods, including Combo, Leiden, Louvain, RB-C, and SBM, deviate from other community detection methods in terms of fairness. At $\xi=0.2$, these methods tend to favor high-conductance communities. At $\xi=0.4$, their fairness improves, but by $\xi=0.6$, they begin favoring low-conductance communities. This shift occurs because, in ABCD networks, the correlation between conductance and community size varies significantly with $\xi$ (see Figure~\ref{fig:corr_matrix}). As $\xi$ increases, these methods detect fewer communities, which are often mapped to larger, low-conductance ground truth communities. Additionally, we observe that community detection methods that require the number of communities as input have lower fairness. While they effectively identify high-conductance communities, they do not correctly group nodes from lower-conductance communities, leading to discrepancies from the ground truth.

\subsection*{Analysis on HICH-BA networks}

We construct two homophilic networks: MMaj, which consists of multiple majority groups (large-sized communities), and MMin, which consists of multiple minority groups (small-sized communities). In both networks, smaller communities have higher density. However, the correlation of size and density with conductance differs significantly between them. Next, we analyze the fairness and performance of various community detection methods on these networks.

\begin{figure}[]
\centering
\begin{subfigure}[b]{0.98\textwidth}            
    \includegraphics[width=\textwidth]{figures/legend_ncol6.png}
\end{subfigure}\\
\begin{subfigure}[c]{0.05\textwidth}
\caption*{\rotatebox{90}{$MMaj$}}
\end{subfigure}%
\begin{minipage}[c]{0.95\textwidth}
\includegraphics[width=0.31\textwidth]{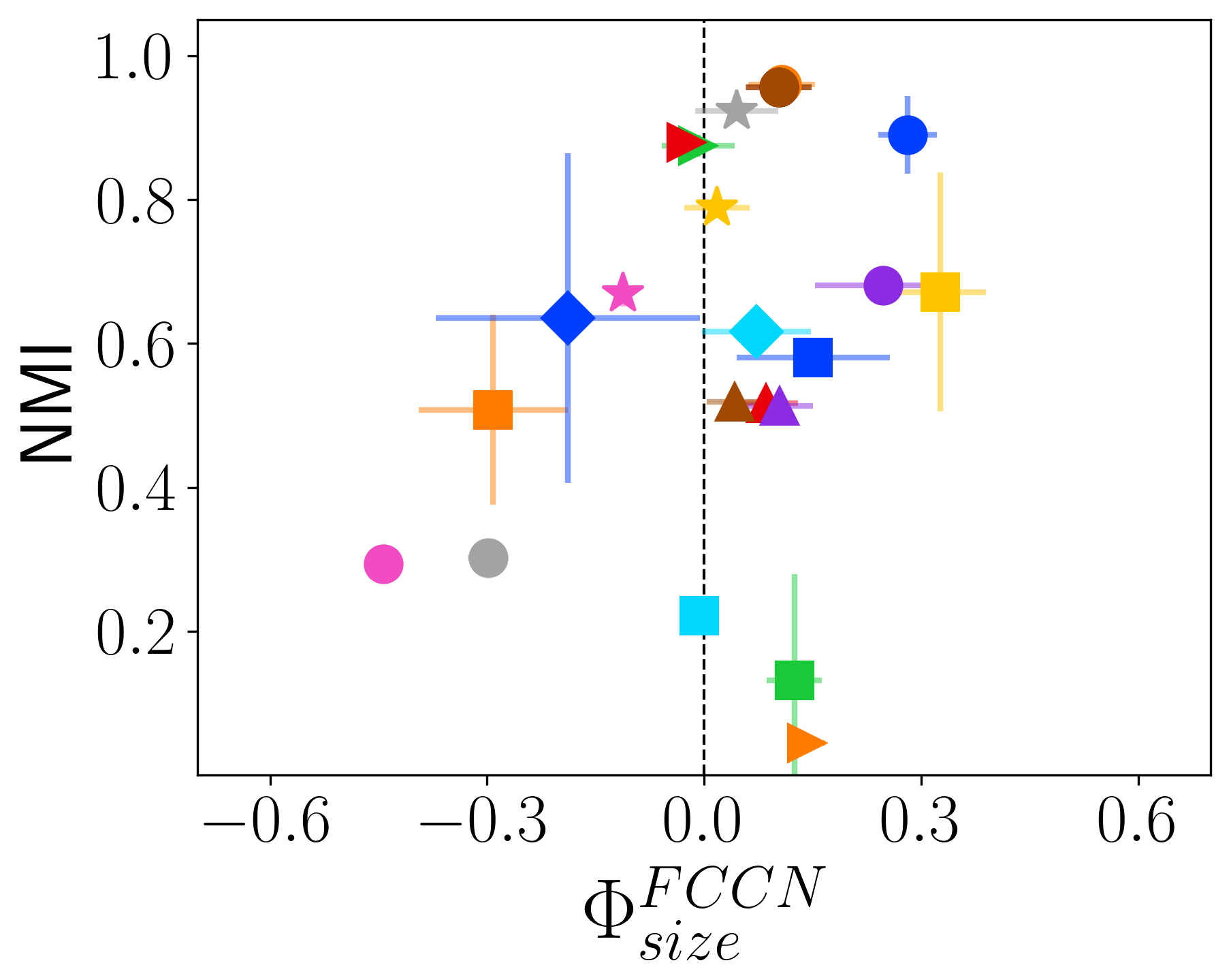}\quad
\includegraphics[width=0.31\textwidth]{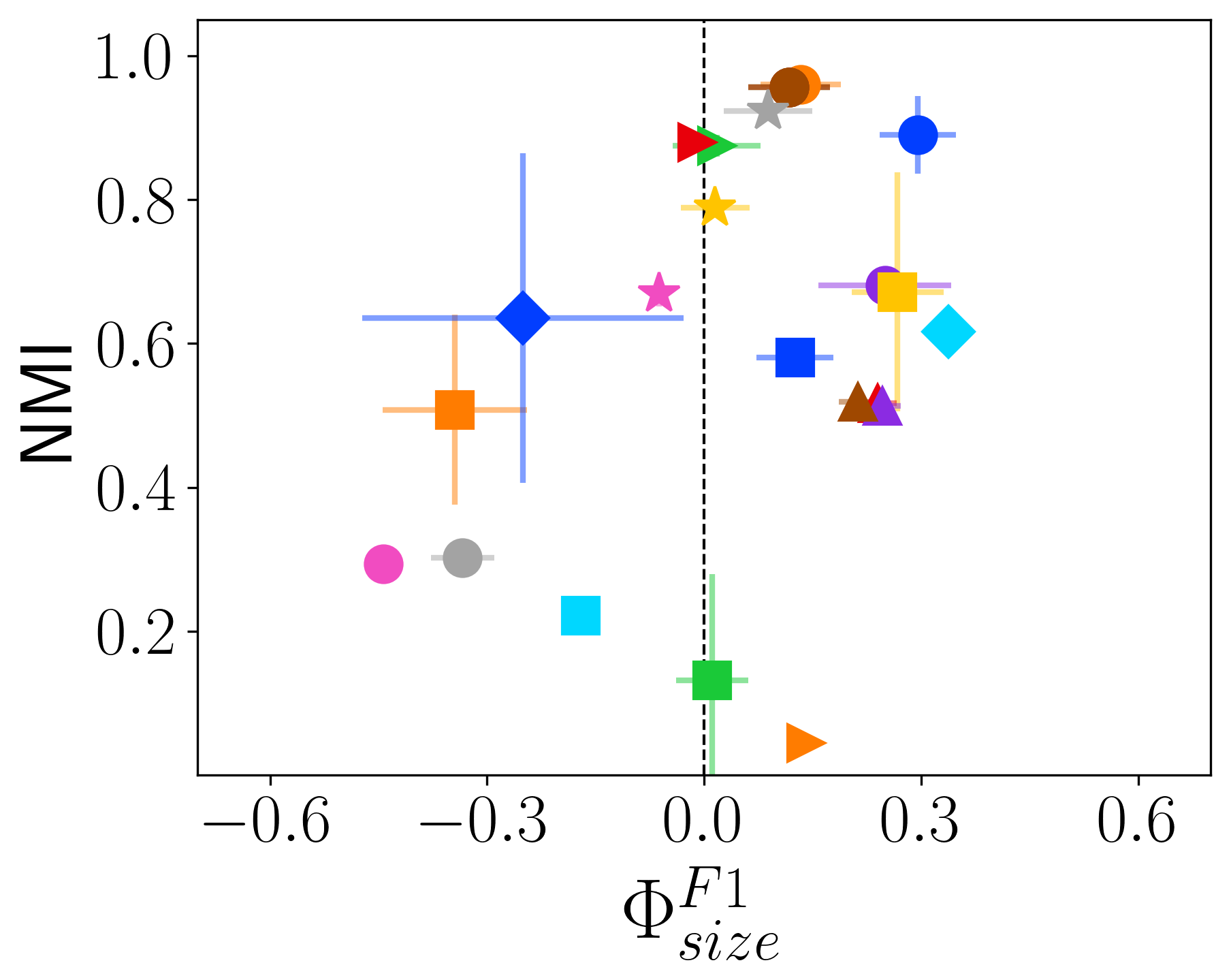}\quad
\includegraphics[width=0.31\textwidth]{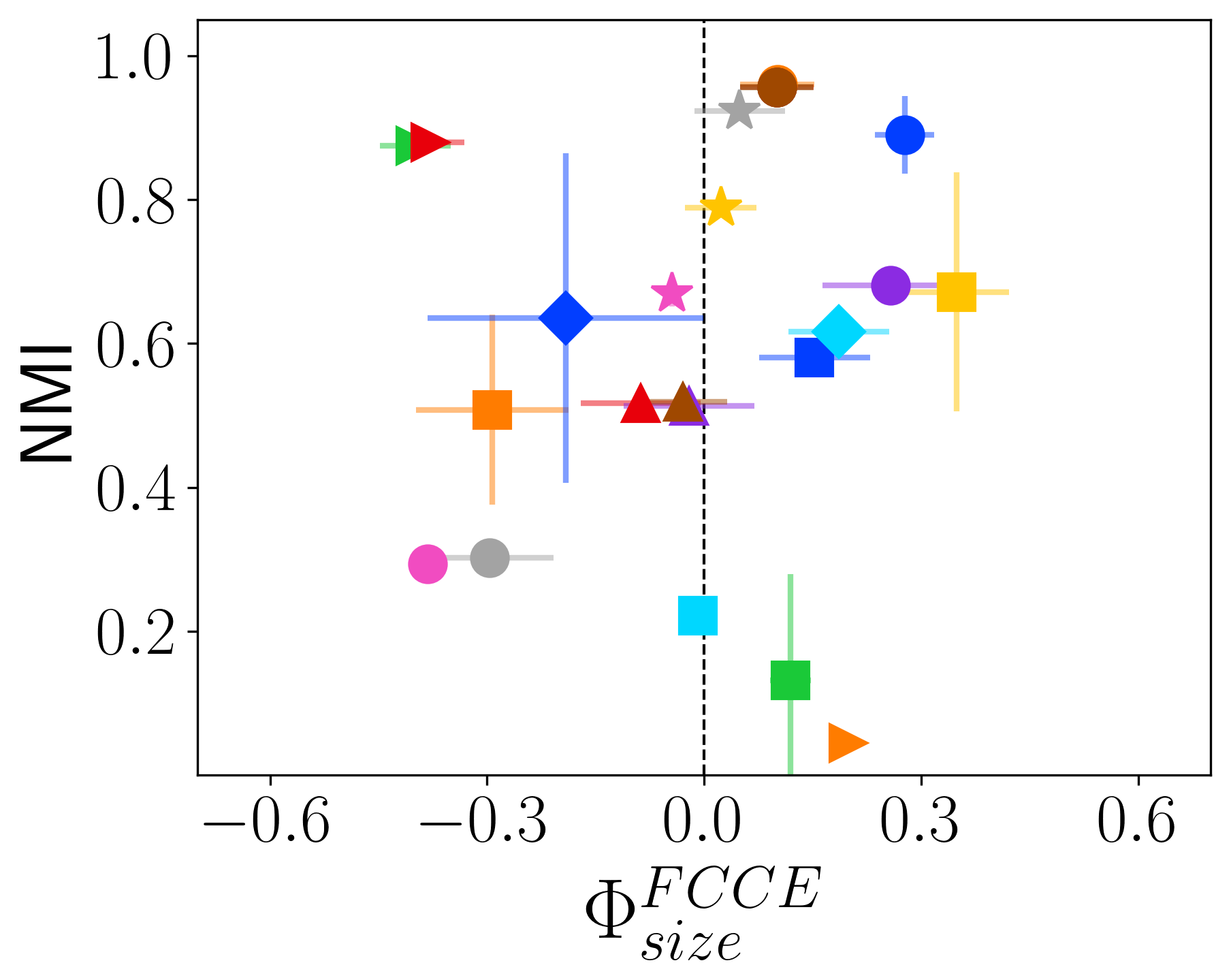}
\end{minipage}
\\
\begin{subfigure}[c]{0.05\textwidth}
\caption*{\rotatebox{90}{$MMin$}}
\end{subfigure}%
\begin{minipage}[c]{0.95\textwidth}
\includegraphics[width=0.31\textwidth]{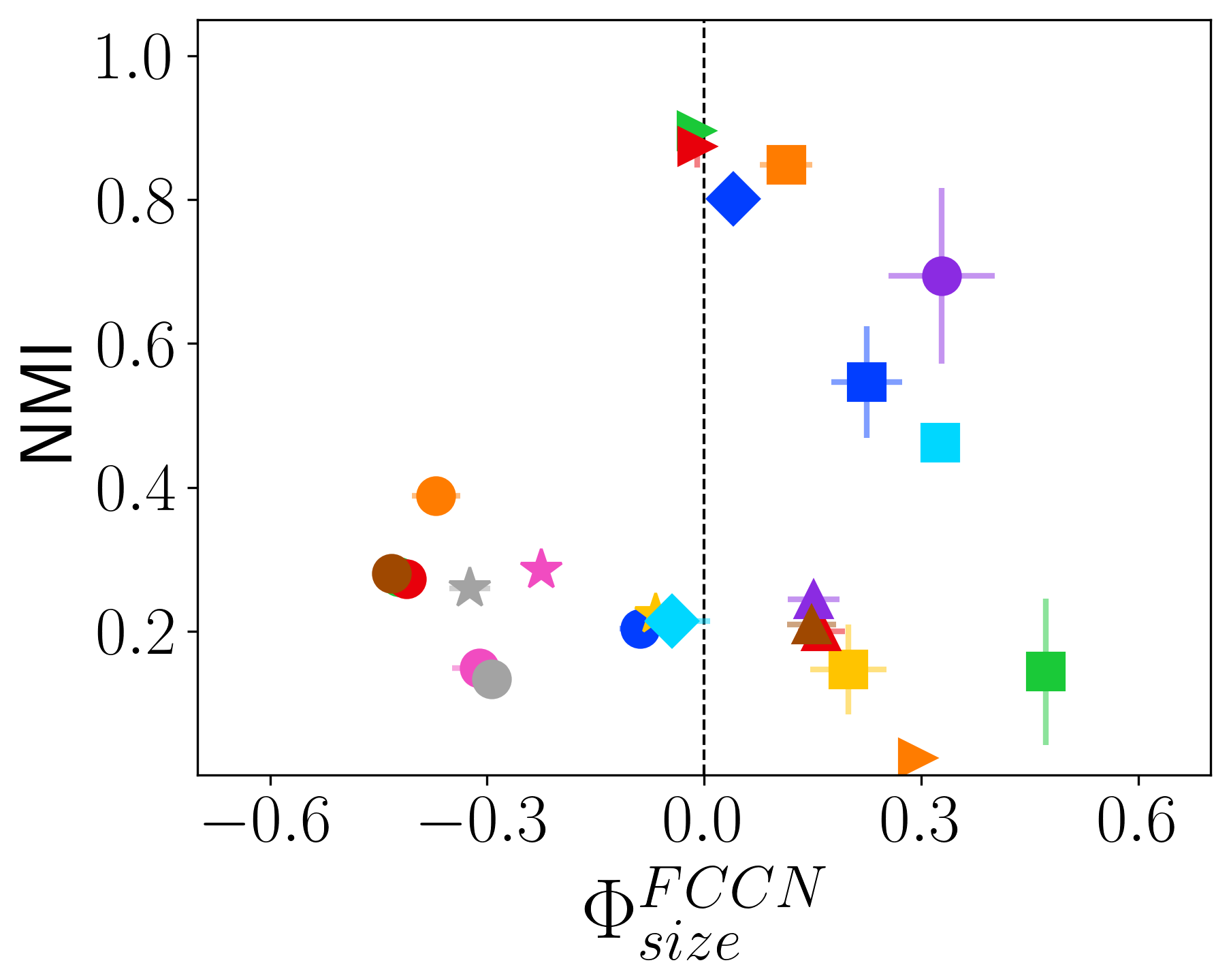}\quad
\includegraphics[width=0.31\textwidth]{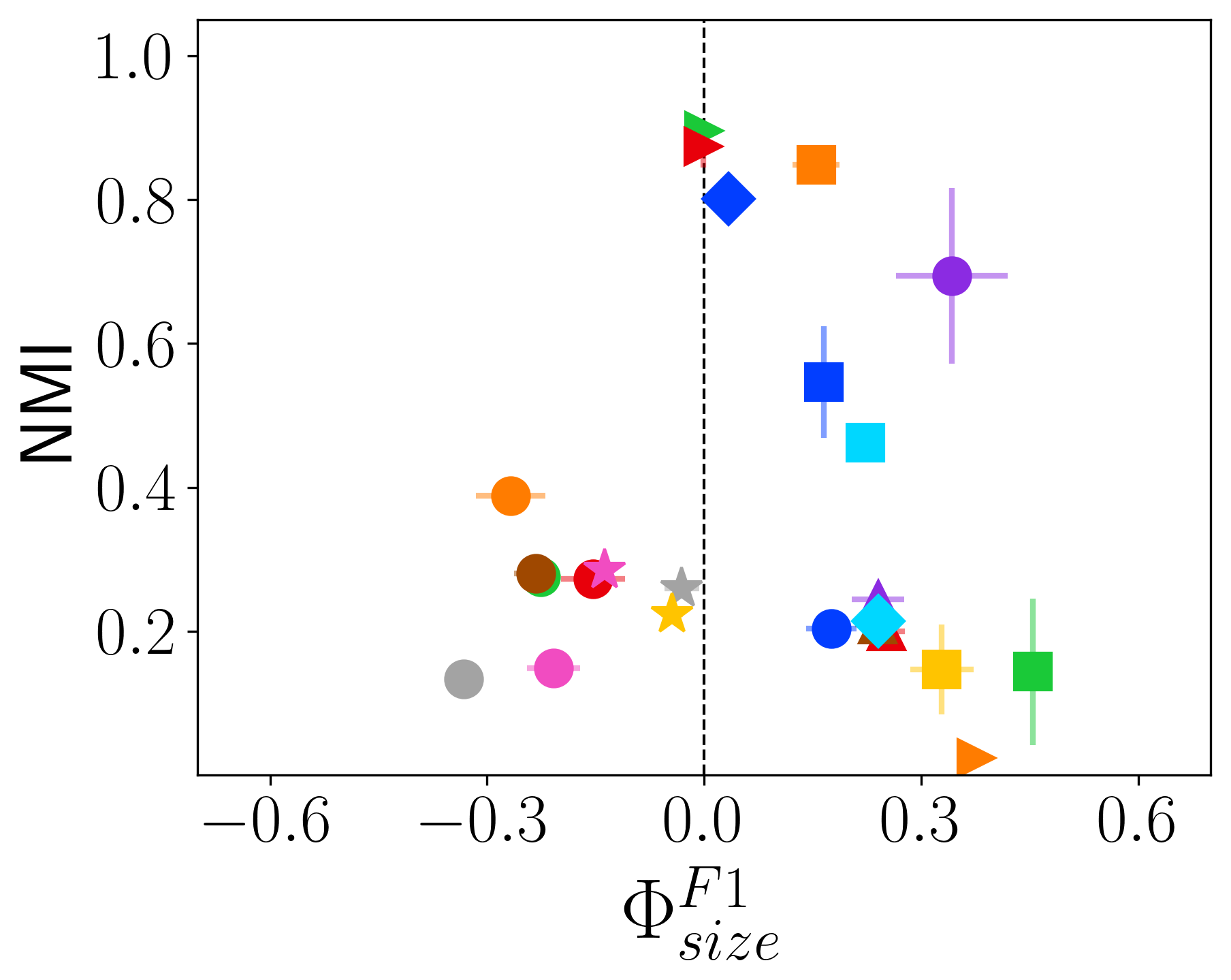}\quad
\includegraphics[width=0.31\textwidth]{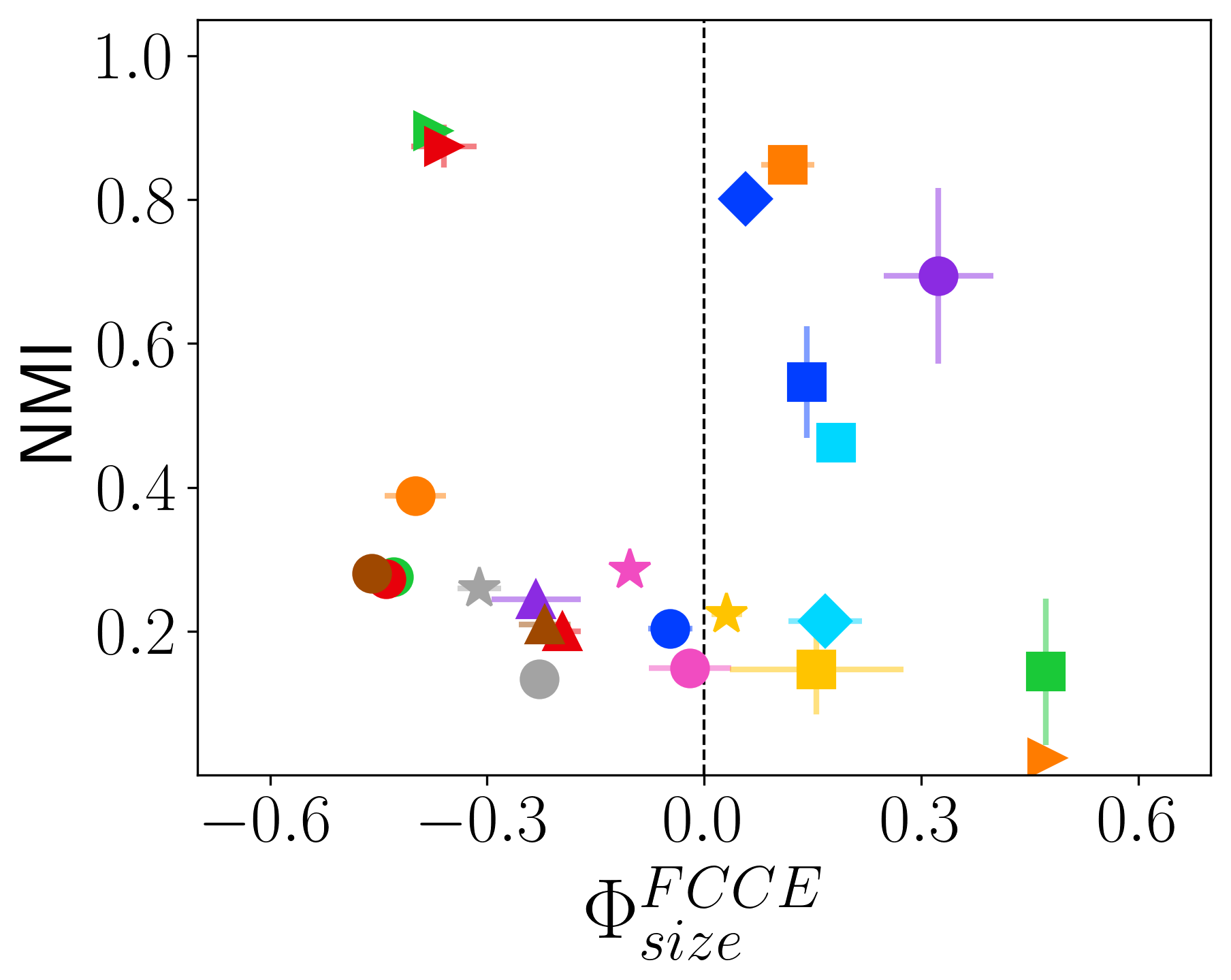}
\end{minipage}
\caption{NMI vs. fairness of community detection methods with respect to community size for HICH-BA networks of 10,000 nodes having (i) MMaj network having multiple majority communities and (ii) MMin network having multiple minority communities}\label{hichba_phi_vs_size} 
\end{figure}

\begin{figure}[]
\centering
\begin{subfigure}[b]{0.98\textwidth}            
    \includegraphics[width=\textwidth]{figures/legend_ncol6.png}
\end{subfigure}\\
\begin{subfigure}[c]{0.05\textwidth}
\caption*{\rotatebox{90}{$MMaj$}}
\end{subfigure}%
\begin{minipage}[c]{0.95\textwidth}
\includegraphics[width=0.31\textwidth]{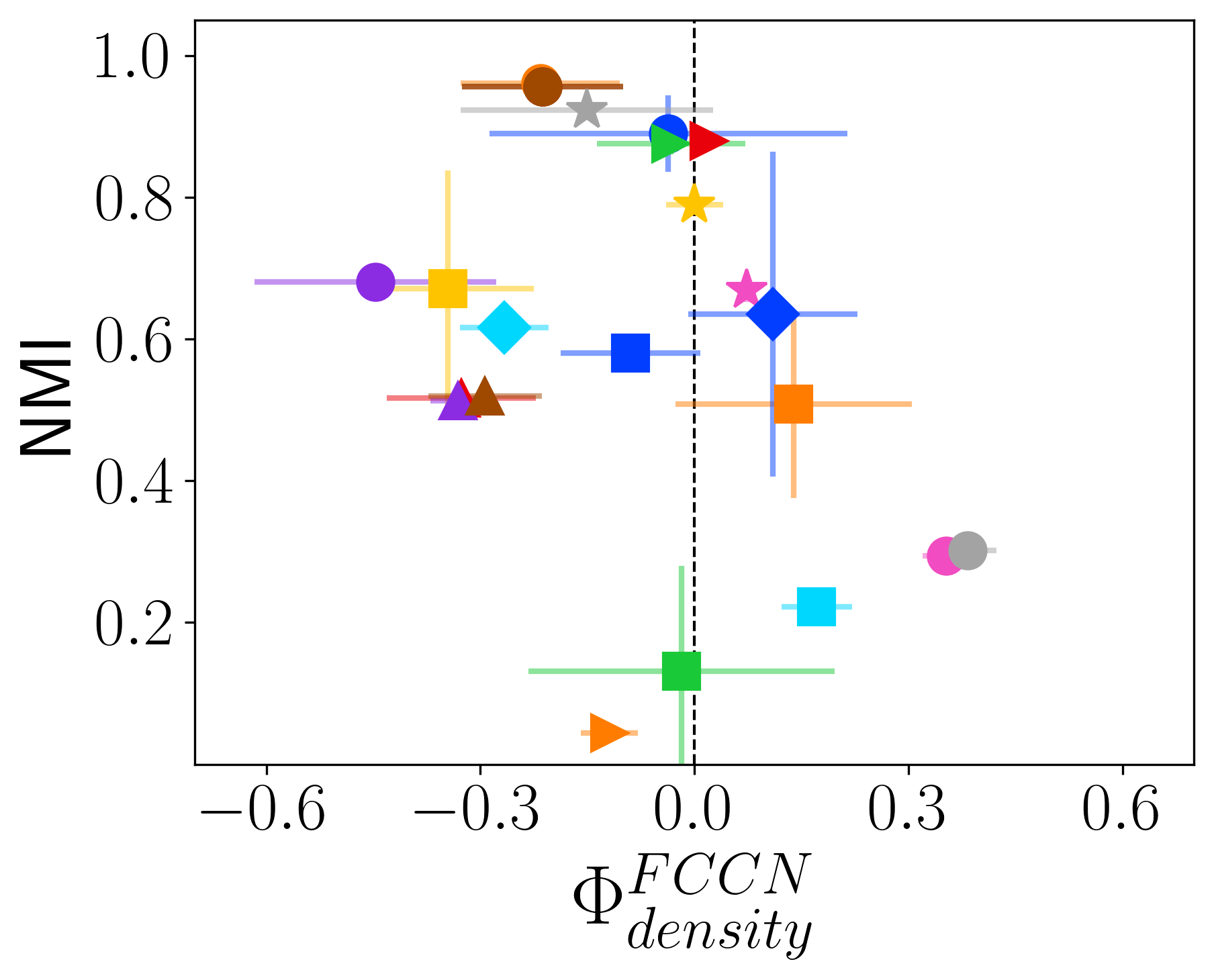}\quad
\includegraphics[width=0.31\textwidth]{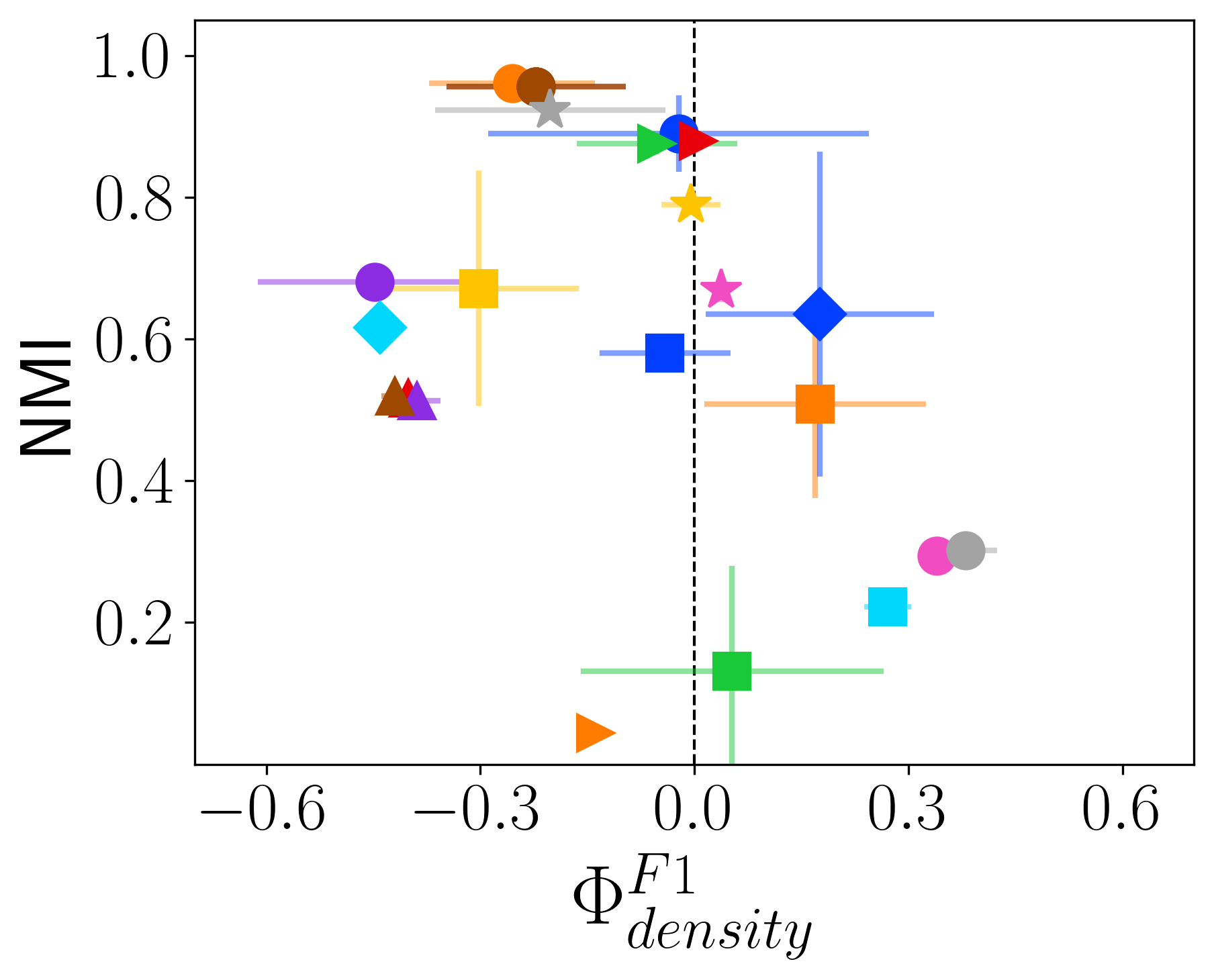}\quad
\includegraphics[width=0.31\textwidth]{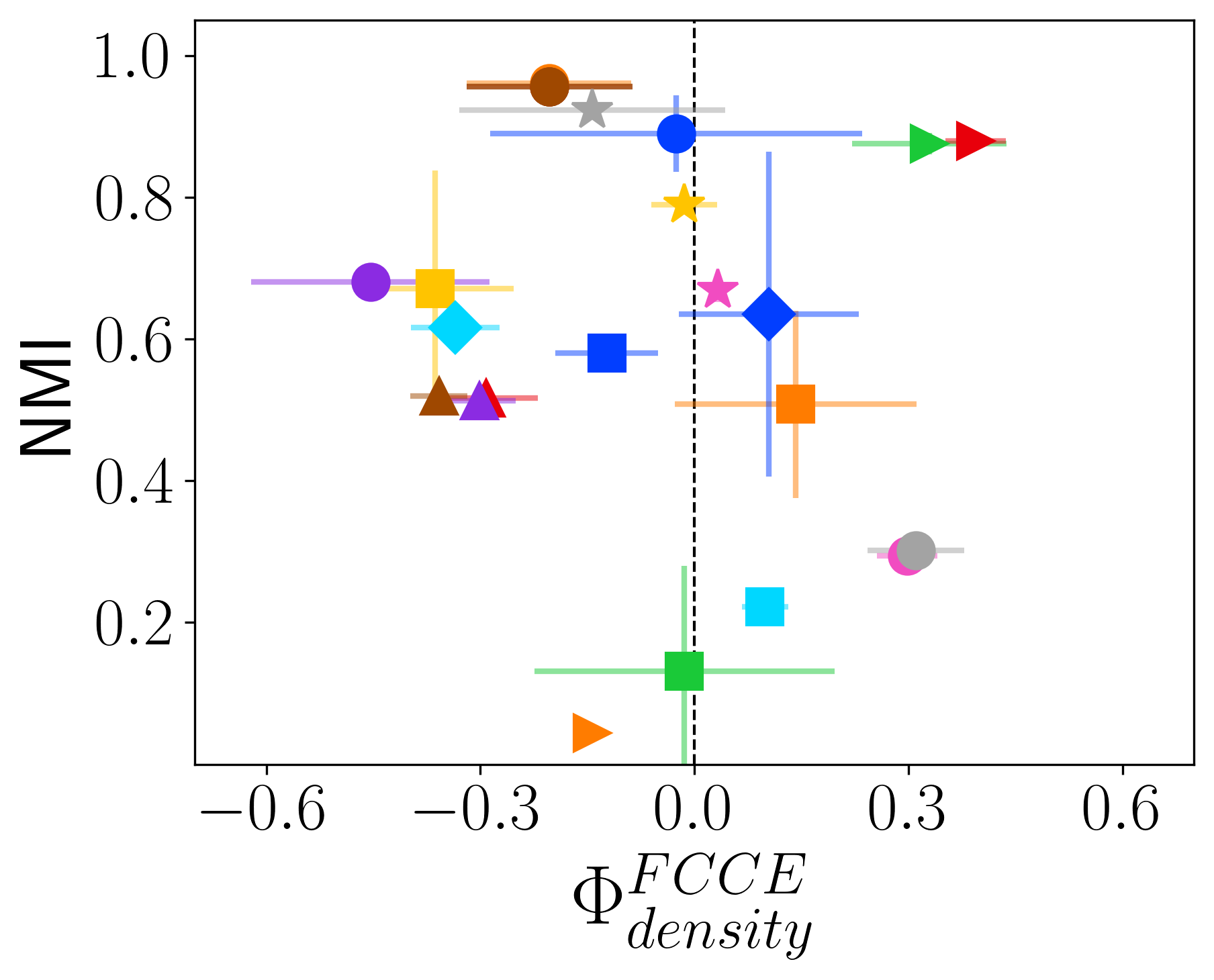}
\end{minipage}
\\
\begin{subfigure}[c]{0.05\textwidth}
\caption*{\rotatebox{90}{$MMin$}}
\end{subfigure}%
\begin{minipage}[c]{0.95\textwidth}
\includegraphics[width=0.31\textwidth]{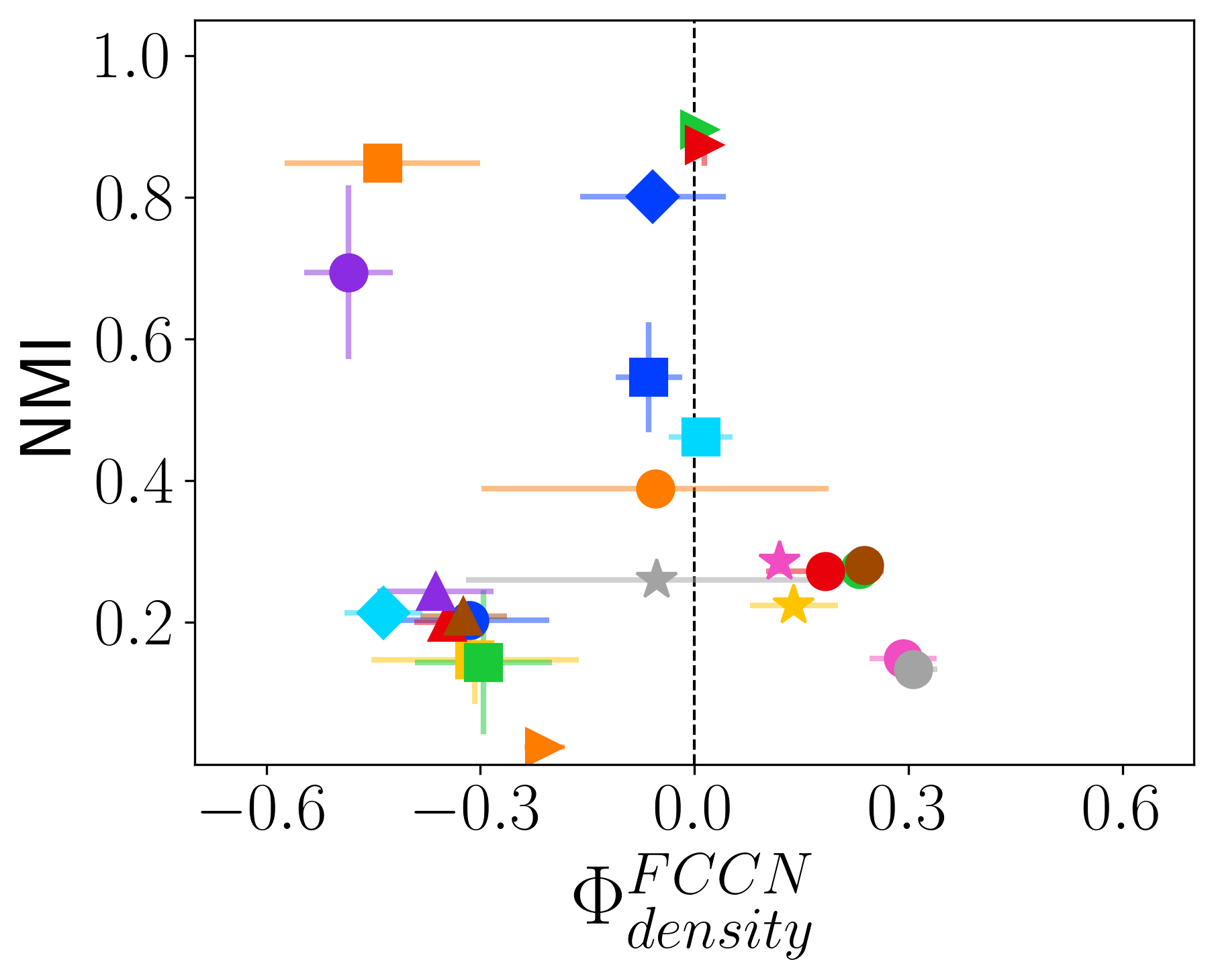}\quad
\includegraphics[width=0.31\textwidth]{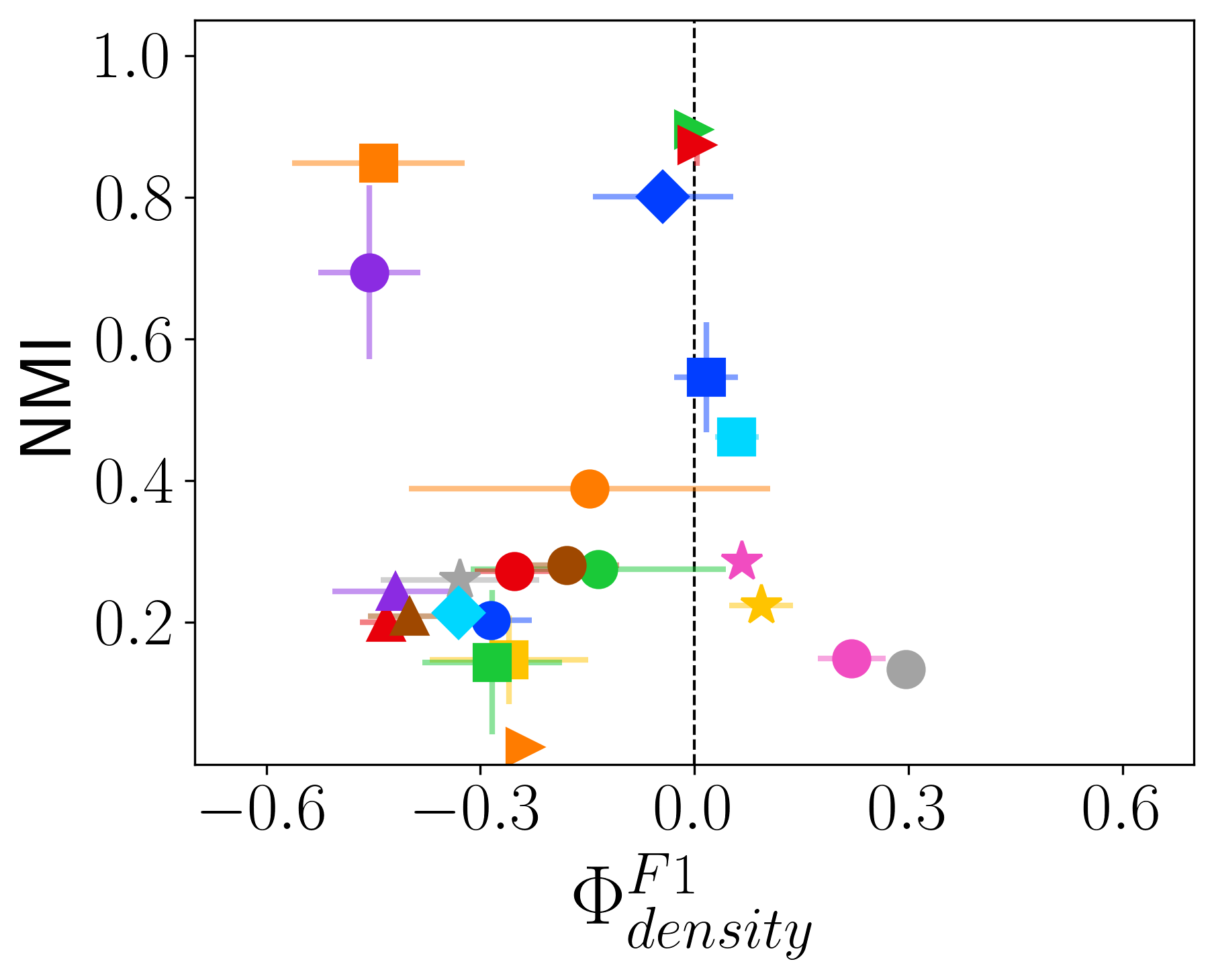}\quad
\includegraphics[width=0.31\textwidth]{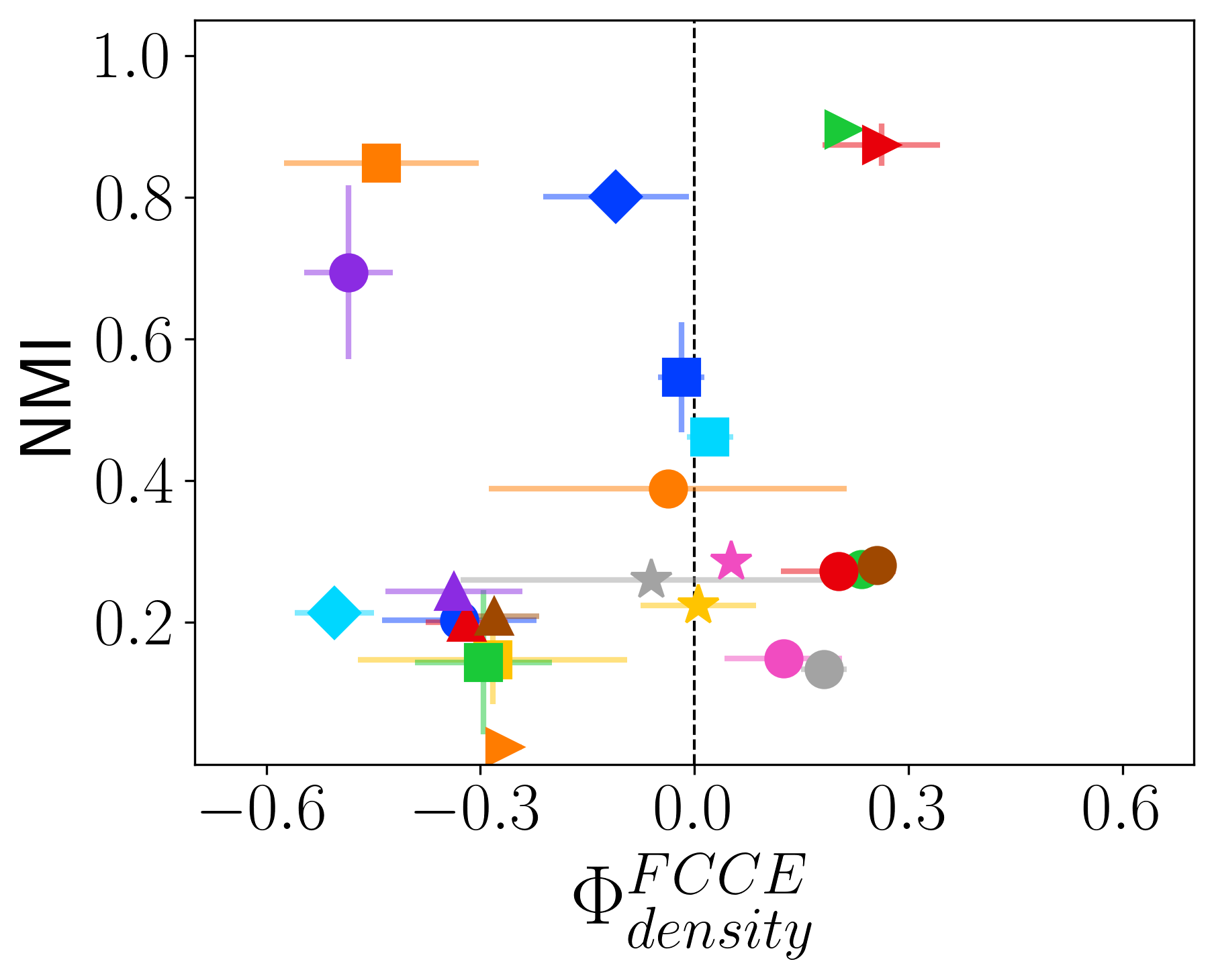}
\end{minipage}
\caption{NMI vs. fairness of community detection methods with respect to community density for HICH-BA networks of 10,000 nodes having (i) MMaj network having multiple majority communities and (ii) MMin network having multiple minority communities}\label{hichba_phi_vs_dens} 
\end{figure}

\begin{figure}[]
\centering
\begin{subfigure}[b]{0.98\textwidth}          
\includegraphics[width=\textwidth]{figures/legend_ncol6.png}
\end{subfigure}\\
\begin{subfigure}[c]{0.05\textwidth}
\caption*{\rotatebox{90}{$MMaj$}}
\end{subfigure}%
\begin{minipage}[c]{0.95\textwidth}
\includegraphics[width=0.31\textwidth]{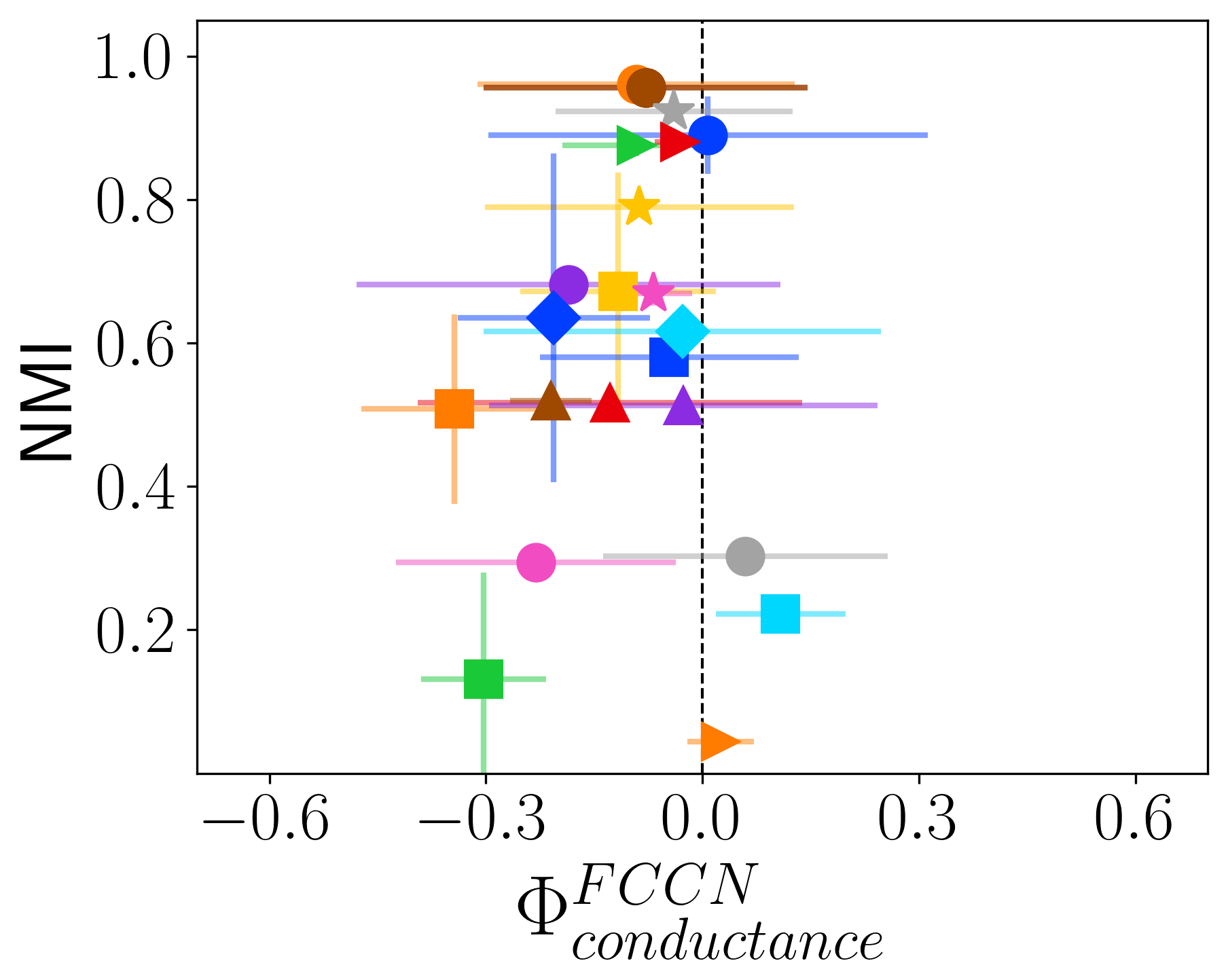}\quad
\includegraphics[width=0.31\textwidth]{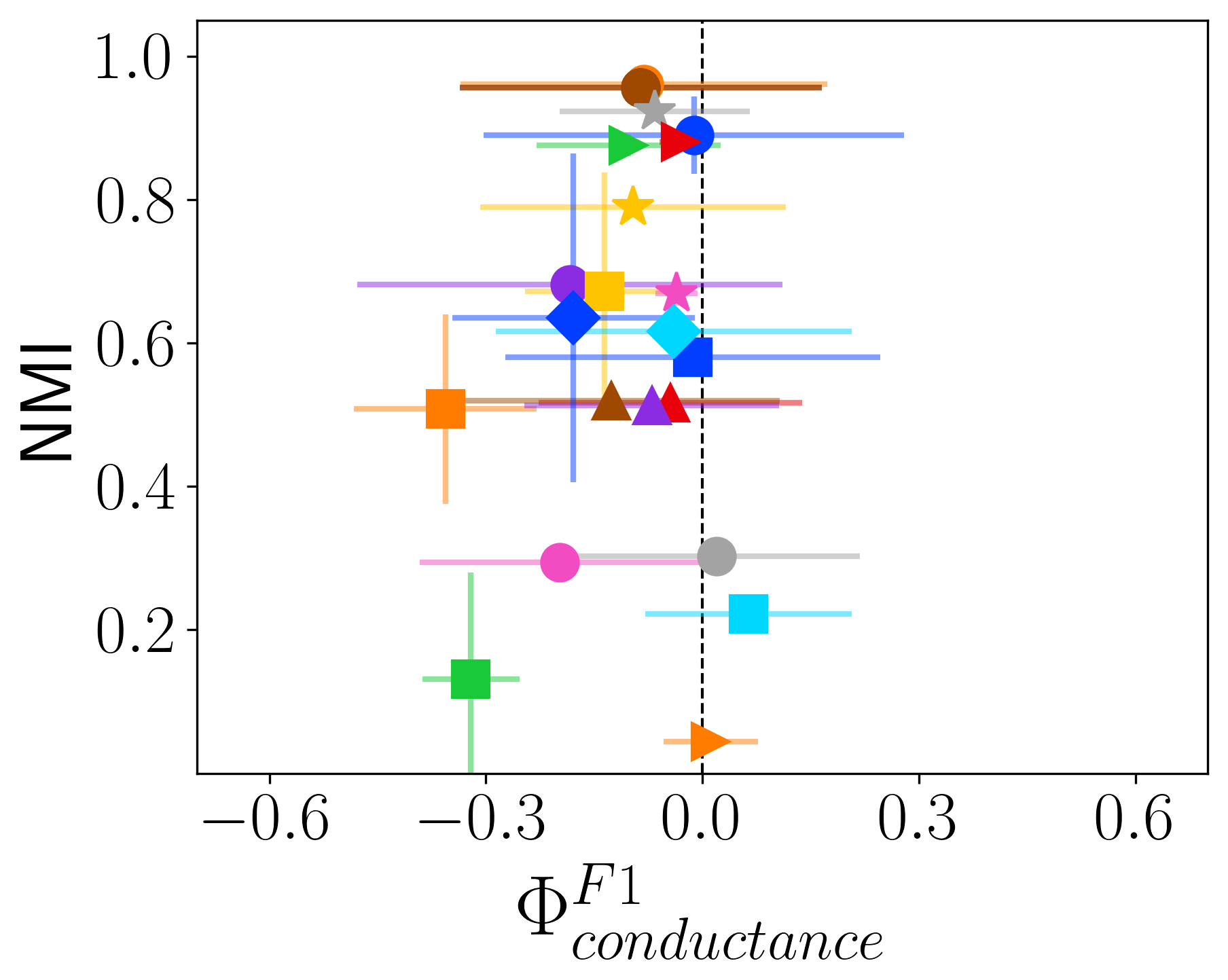}\quad
\includegraphics[width=0.31\textwidth]{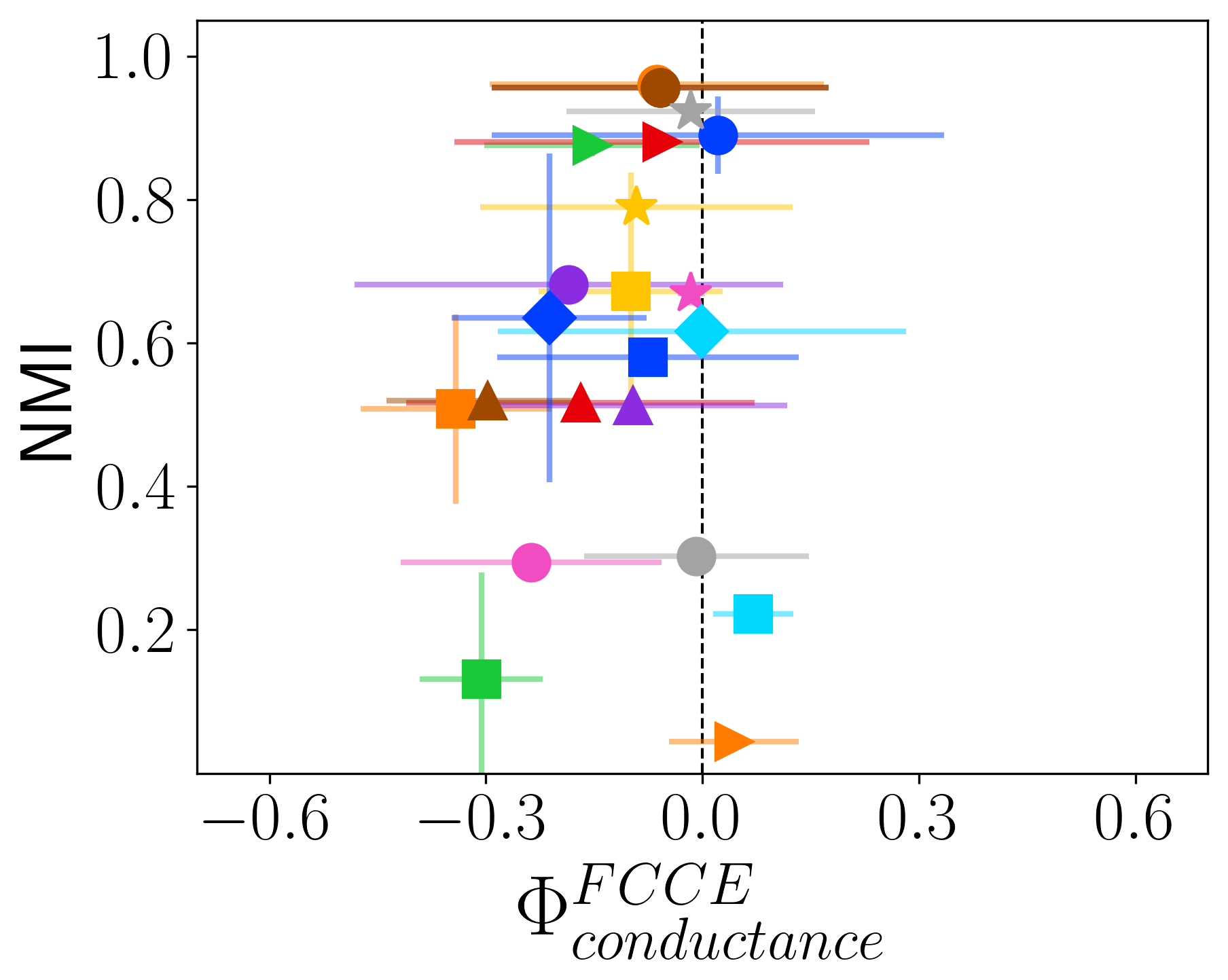}
\end{minipage}
\\
\begin{subfigure}[c]{0.05\textwidth}
\caption*{\rotatebox{90}{$MMin$}}
\end{subfigure}%
\begin{minipage}[c]{0.95\textwidth}
\includegraphics[width=0.31\textwidth]{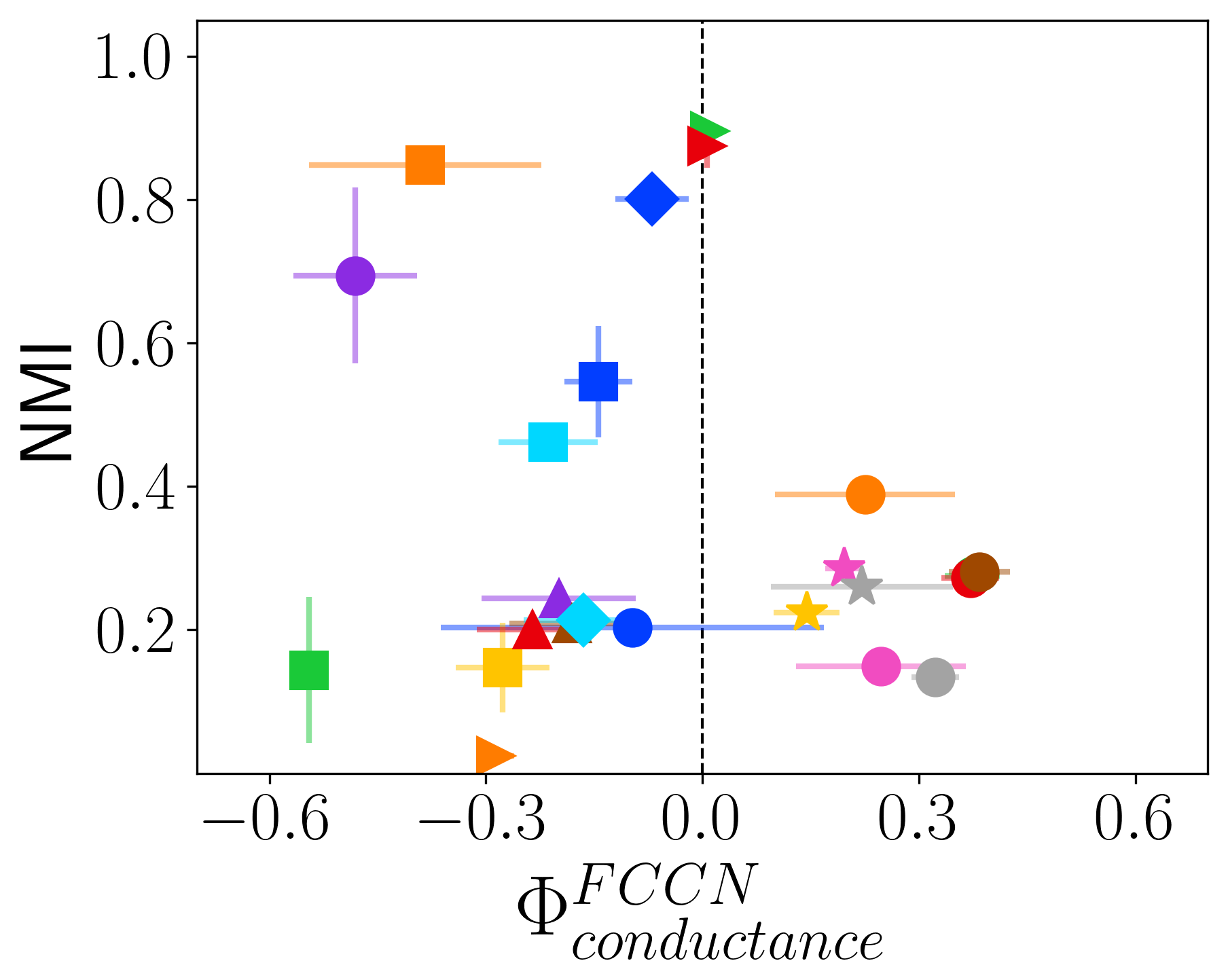}\quad
\includegraphics[width=0.31\textwidth]{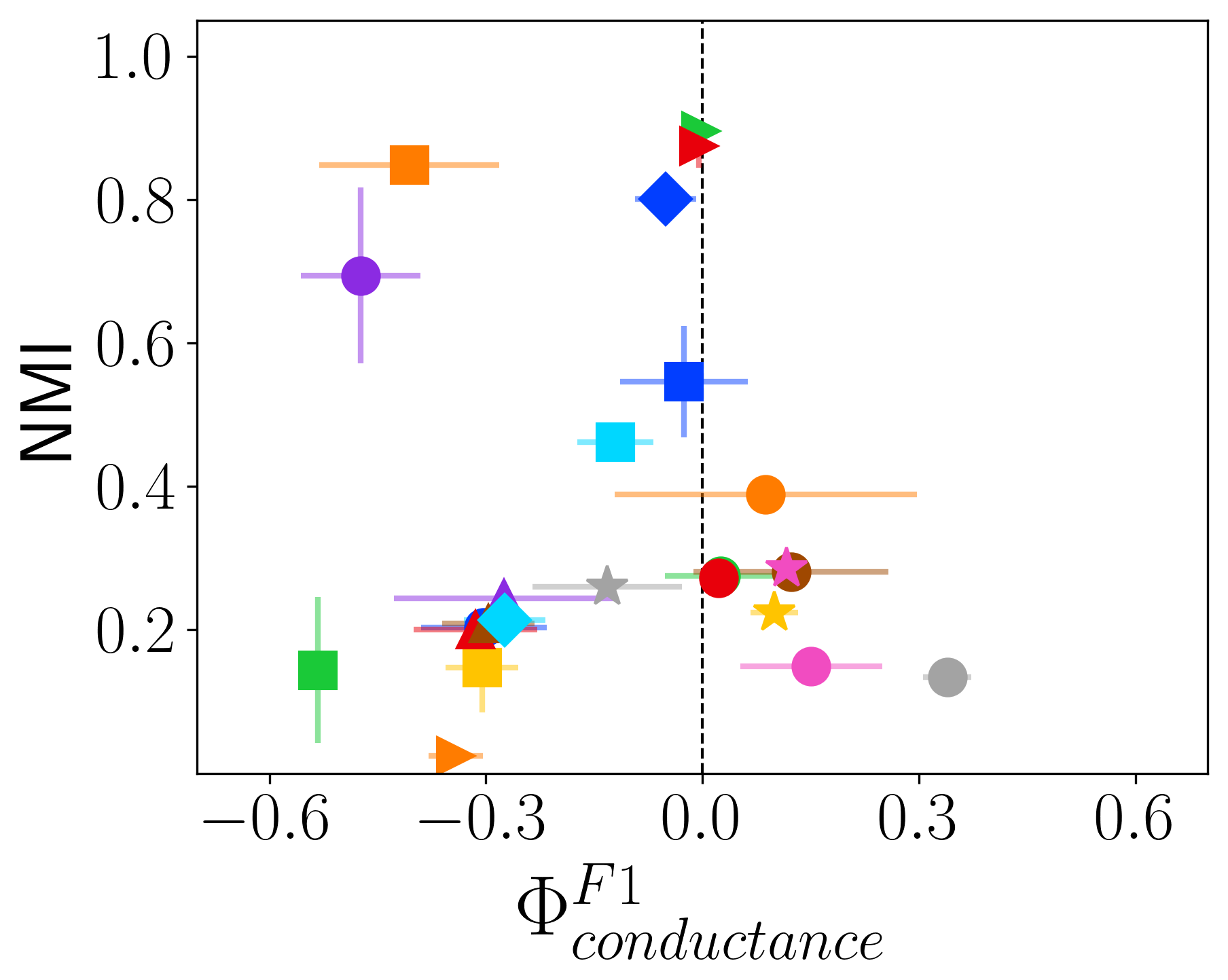}\quad
\includegraphics[width=0.31\textwidth]{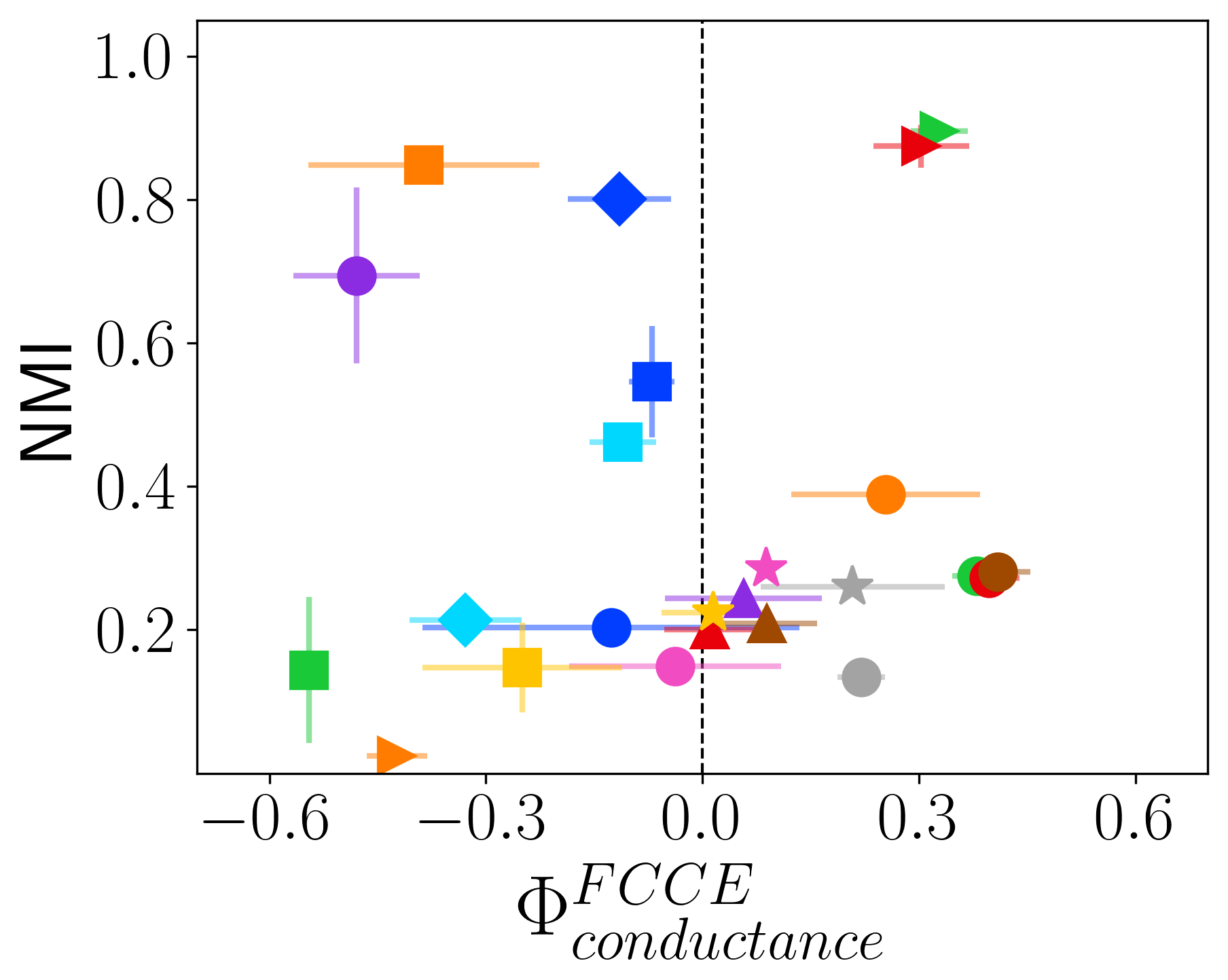}
\end{minipage}
\caption{NMI vs. fairness of community detection methods with respect to community conductance for HICH-BA networks of 10,000 nodes having (i) MMaj network having multiple majority communities and (ii) MMin network having multiple minority communities}\label{hichba_phi_vs_cond} 
\end{figure}

MMaj and MMin represent two extreme types of network structures. A key observation is that the RB-ER and Significance methods predict an excessive number of communities (over 3,000) in both networks. This suggests that when small-sized communities are present, these methods also tend to fragment larger communities into numerous smaller ones. Most community detection methods are not inherently designed to handle such extreme cases. Nevertheless, we briefly discuss their results and the fairness-performance trade-off of different methods. Figures \ref{hichba_phi_vs_size}, \ref{hichba_phi_vs_dens}, and \ref{hichba_phi_vs_cond} show NMI versus fairness of different community detection methods with respect to size, density, and conductance, respectively.

Community detection methods that achieve relatively high NMI scores include Combo, Leiden, Louvain, RB-C, Spinglass, SBM, and SBM-Nested. These methods also predict a reasonable number of communities, maintaining good performance. However, on MMin networks, most community detection methods do not achieve high NMI values. This is influenced by their ability to detect the largest ground truth community, which most of the methods are not able to identify properly. Methods that perform well in terms of NMI on MMin networks include Paris, RSC-V, Label Propagation, SBM, and SBM-Nested. Notably, SBM and SBM-Nested methods are fair across all types of communities. 

A particularly interesting observation is that Walktrap and Infomap perform fairly and maintain high performance across different LFR and ABCD Networks. These methods maintain fairness on HICH-BA networks; however, the quality of the detected communities is poor. This is because both methods overestimate the number of communities, leading to lower-quality results across varying community properties.

\subsection*{Real-world Networks}

\begin{figure}[t]
\centering
\begin{subfigure}[b]{0.93\textwidth}            
    \includegraphics[width=\textwidth]{figures/legend_ncol6.png}
\end{subfigure}\\
\begin{subfigure}[c]{0.05\textwidth}
\caption*{\rotatebox{90}{Polbooks}}
\end{subfigure}%
\begin{minipage}[c]{0.95\textwidth}
\includegraphics[width=0.3\textwidth]{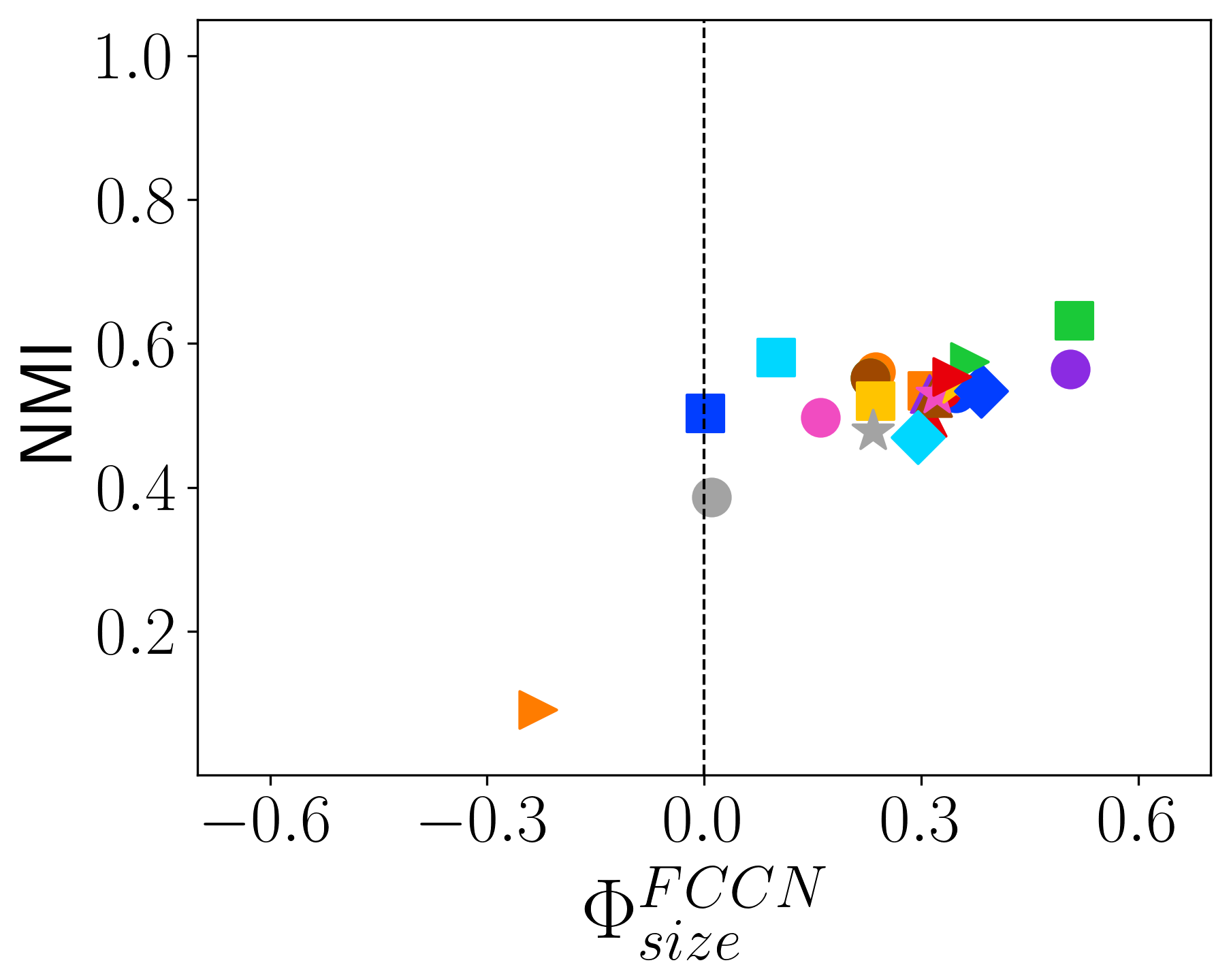}\quad
\includegraphics[width=0.3\textwidth]{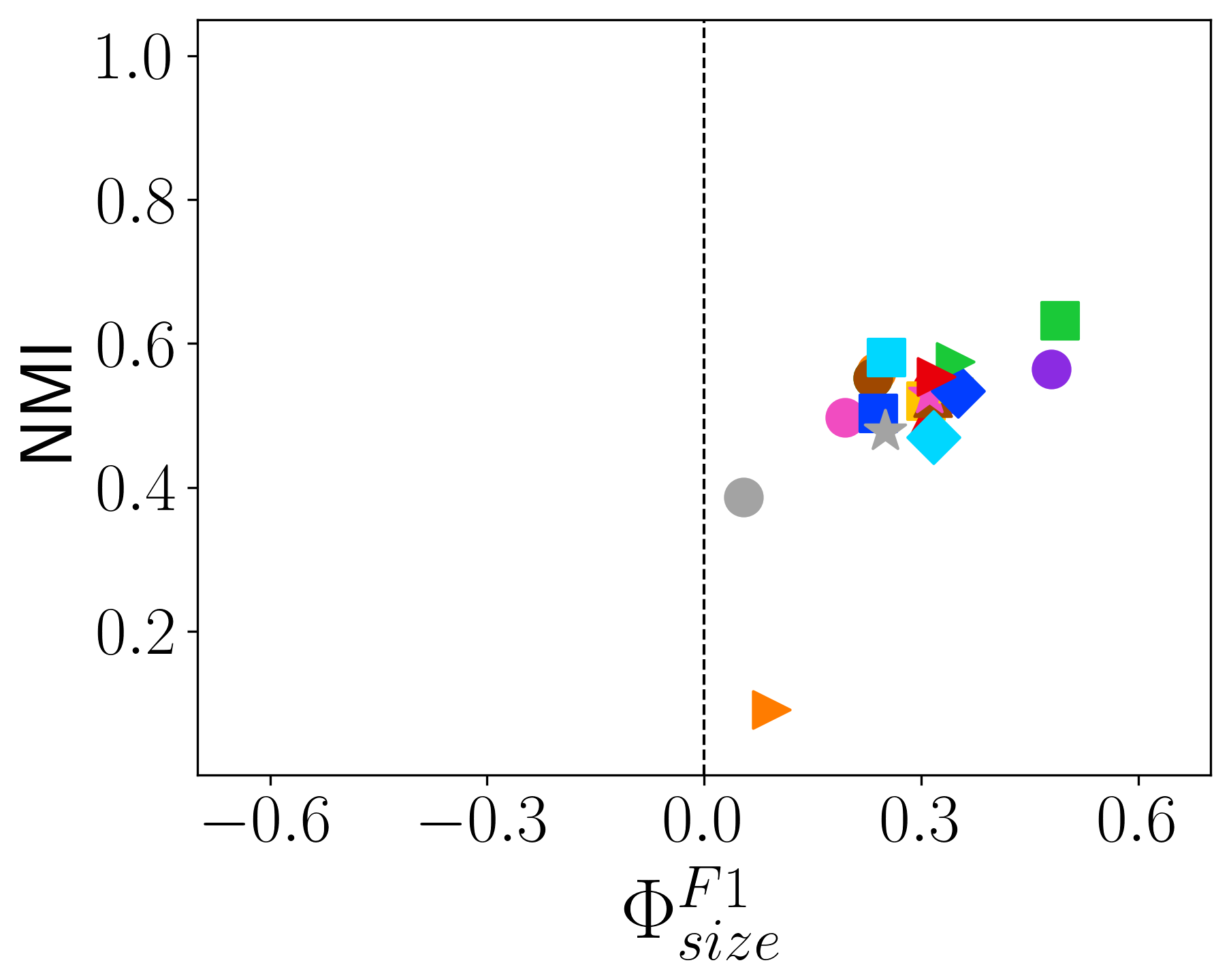}\quad
\includegraphics[width=0.3\textwidth]{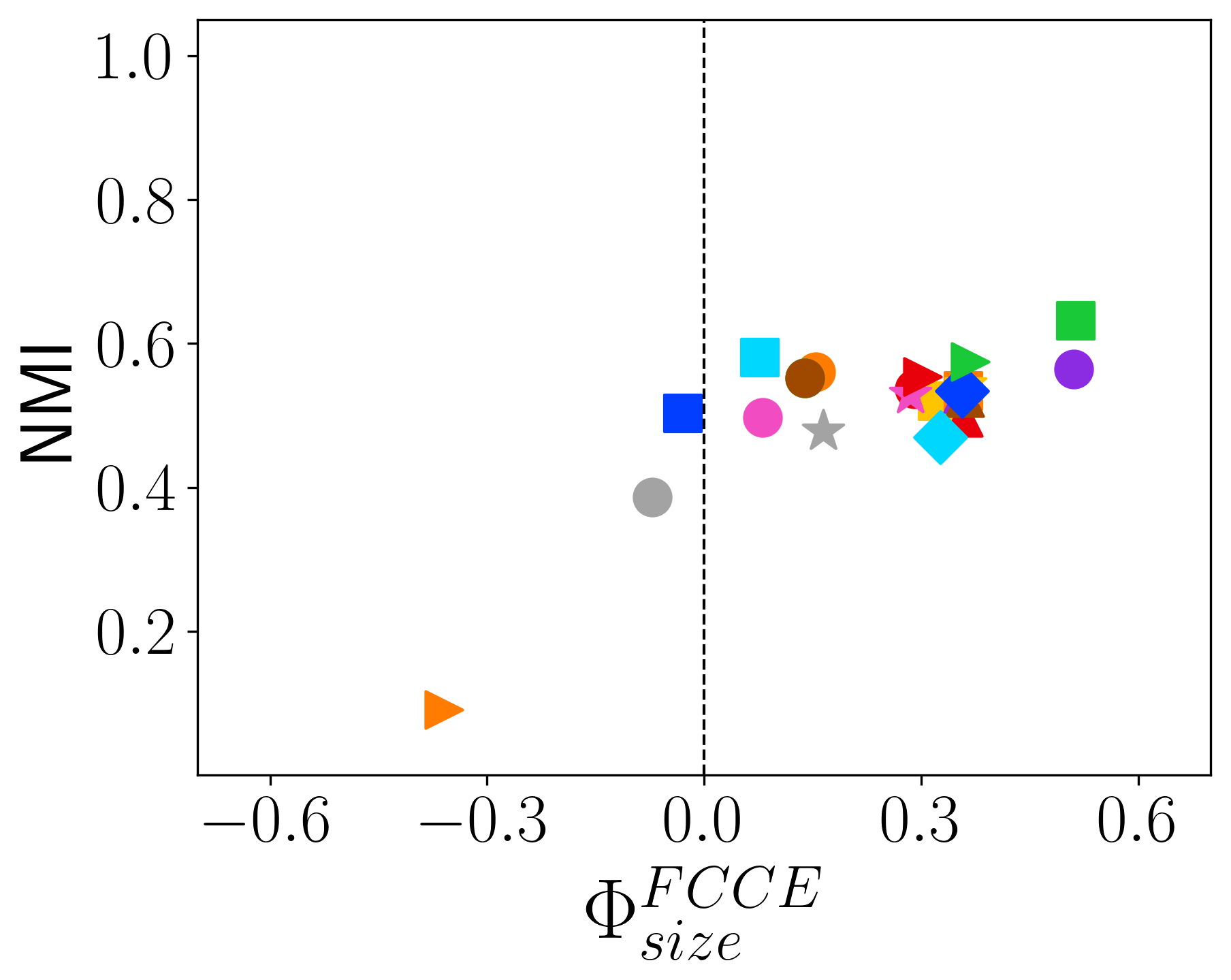}
\end{minipage}
\\
\begin{subfigure}[c]{0.05\textwidth}
\caption*{\rotatebox{90}{Football}}
\end{subfigure}%
\begin{minipage}[c]{0.95\textwidth}
\includegraphics[width=0.3\textwidth]{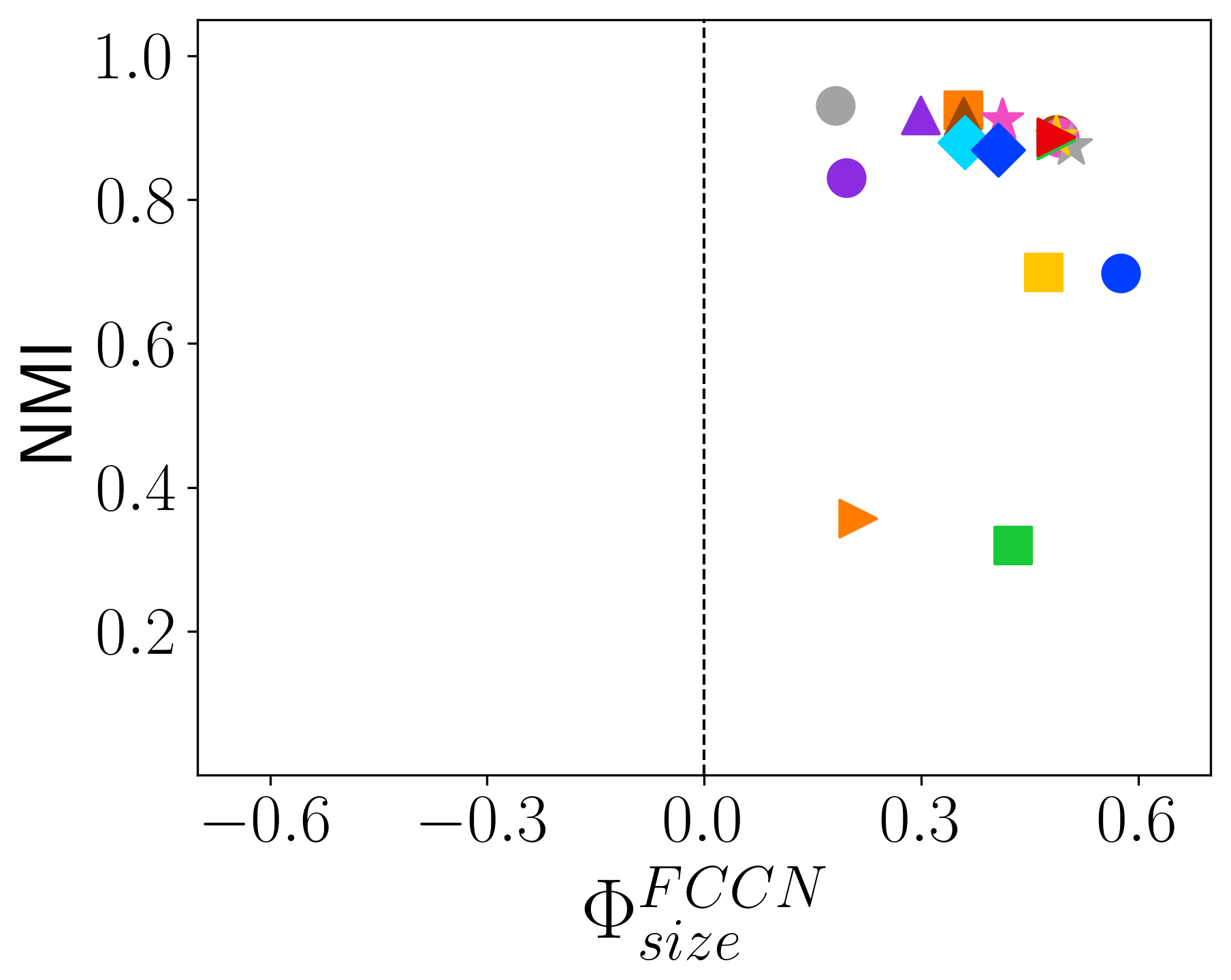}\quad
\includegraphics[width=0.3\textwidth]{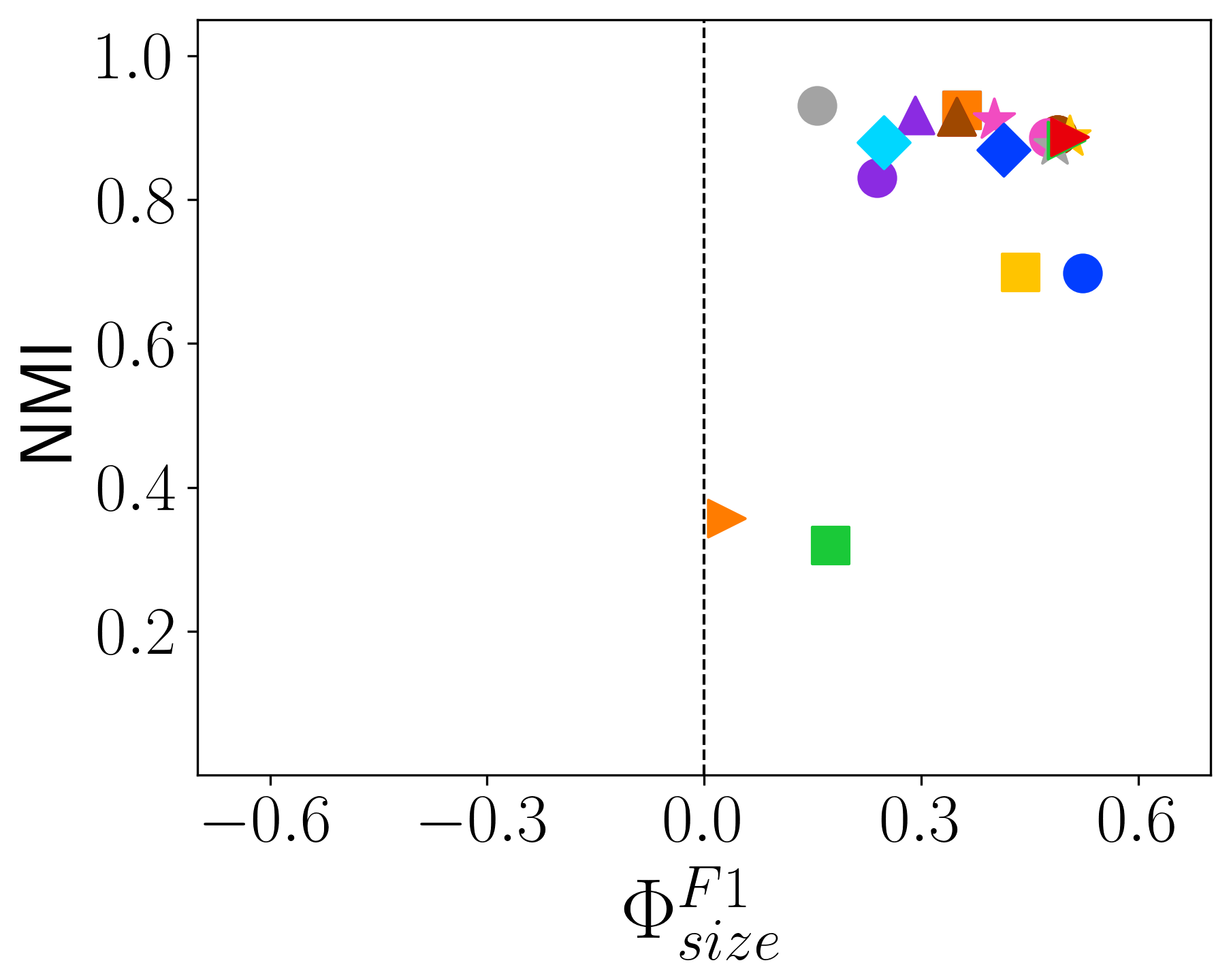}\quad
\includegraphics[width=0.3\textwidth]{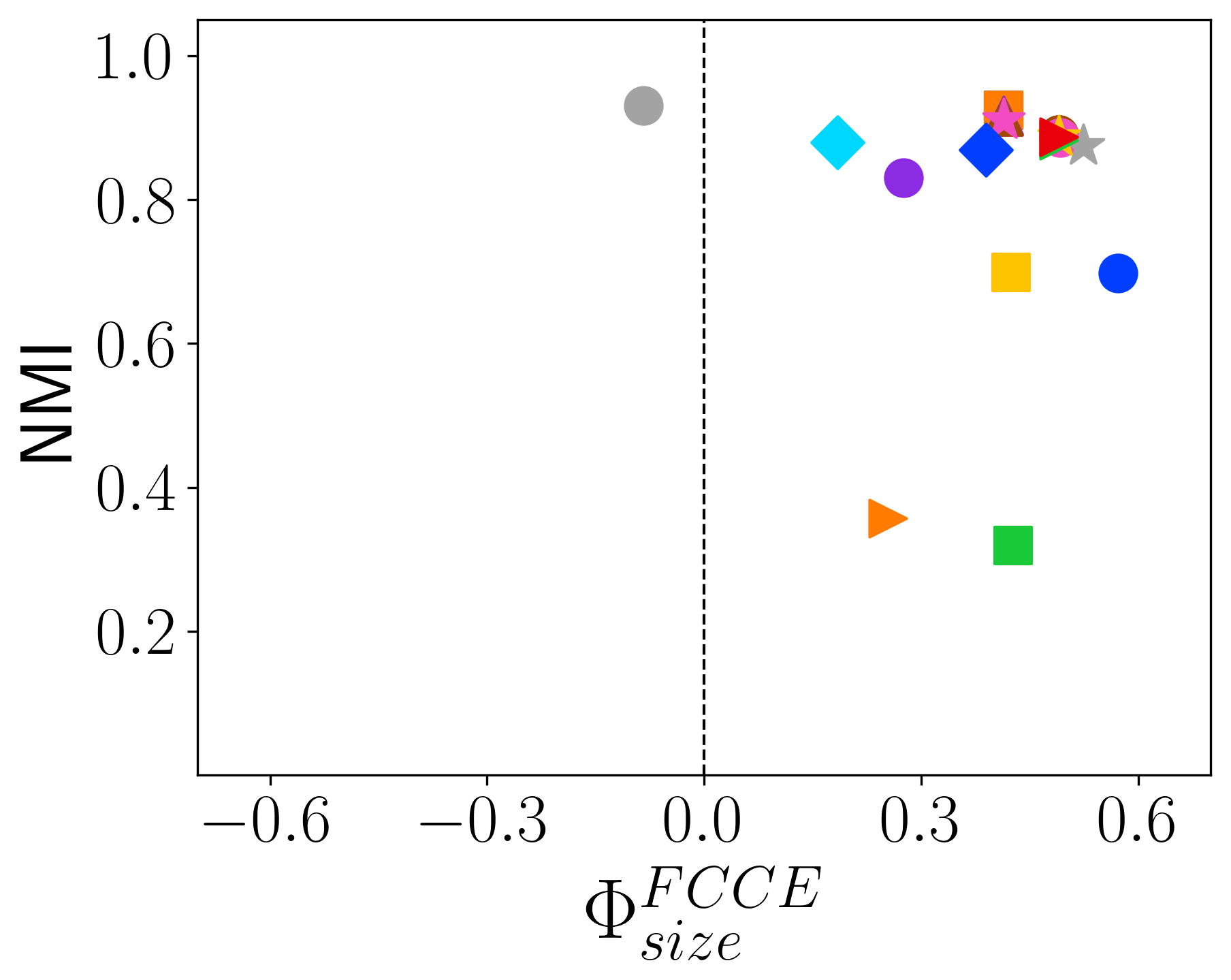}
\end{minipage}
\\
\begin{subfigure}[c]{0.05\textwidth}
\caption*{\rotatebox{90}{Eu-core}}
\end{subfigure}%
\begin{minipage}[c]{0.95\textwidth}
\includegraphics[width=0.3\textwidth]{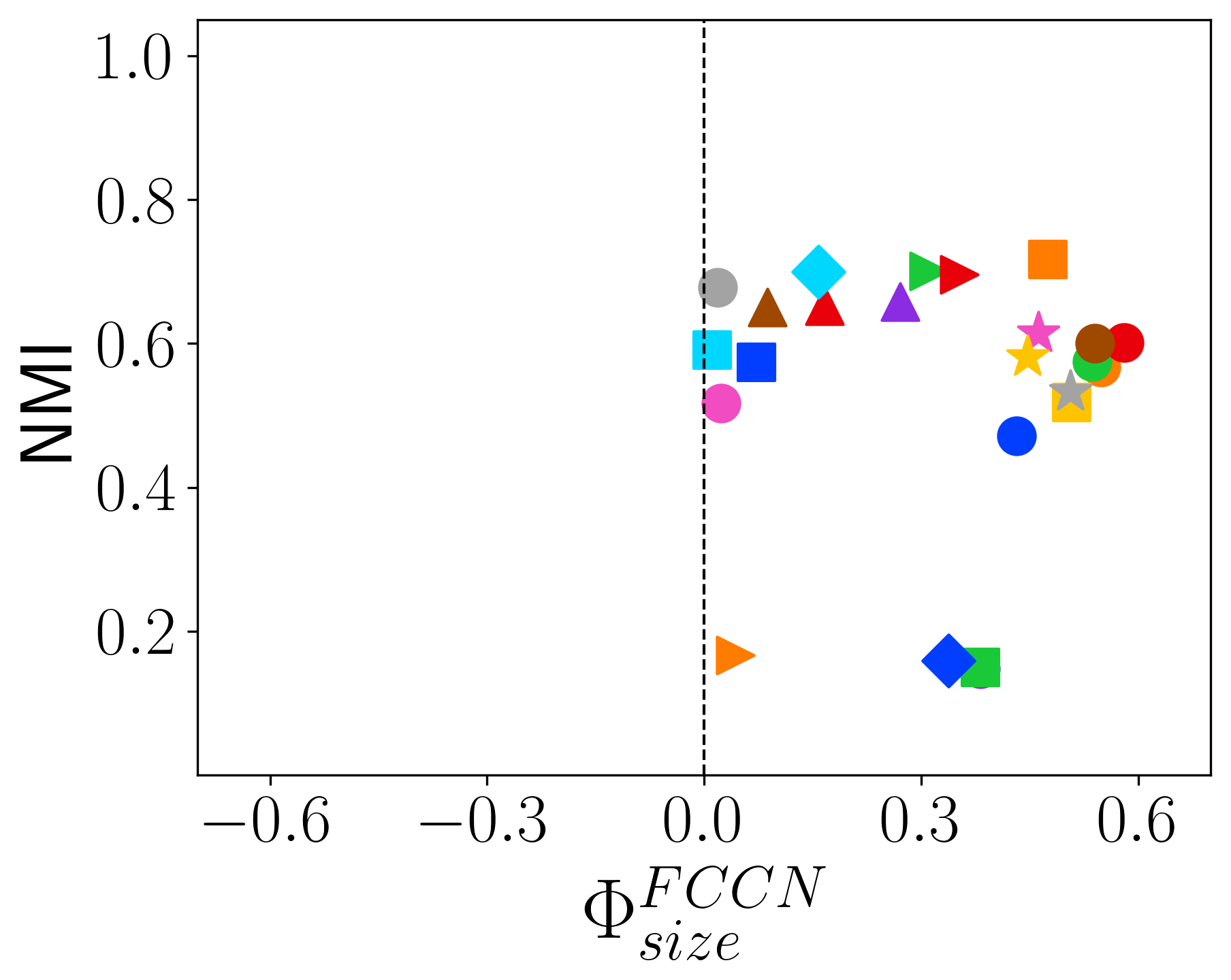}\quad
\includegraphics[width=0.3\textwidth]{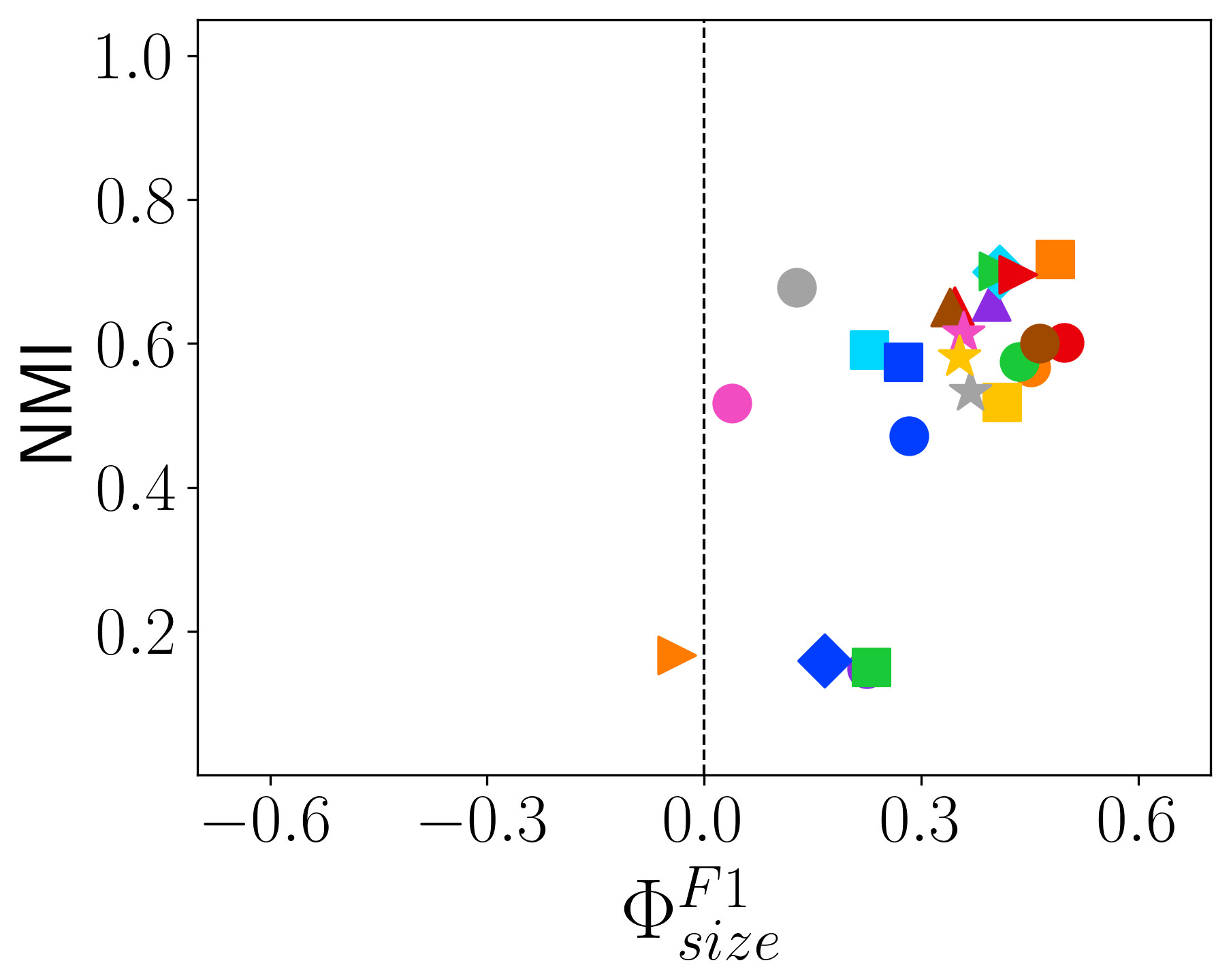}\quad
\includegraphics[width=0.3\textwidth]{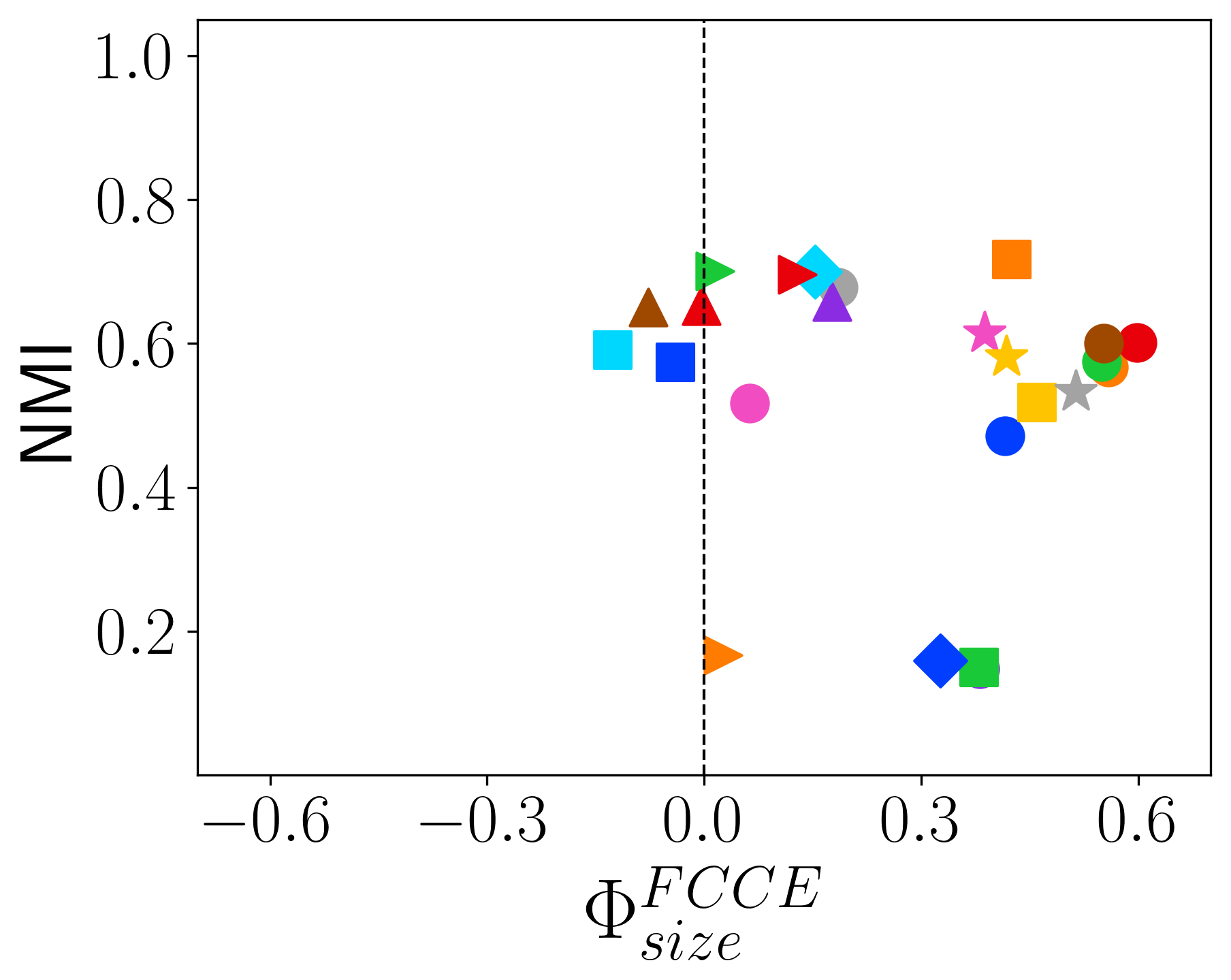}
\end{minipage}
\caption{NMI vs. fairness of community detection methods on real-world networks.}\label{real_world_phi_size}
\end{figure}

We perform experiments on three real-world networks: Polbooks, Football, and Eu-core. Fig.~\ref{real_world_phi_size} presents NMI vs. $\Phi^{F*}_{size}$ for real-world networks. All methods, except EM, tend to favor larger groups, with consistent results across various fairness metrics and community property. It is important to note that no single method is universally fair across all datasets, though Significance consistently performs well. 

\subsection*{Fairness versus Other Evaluation Metrics}\label{sec:results partition quality}

\begin{figure}[t]
\centering
\begin{subfigure}[b]{0.98\textwidth}            
    \includegraphics[width=\textwidth]{figures/legend_ncol6.png}
\end{subfigure}\\
\begin{subfigure}[c]{0.05\textwidth}
\caption*{\rotatebox{90}{RMI}}
\end{subfigure}
\begin{minipage}[c]{0.94\textwidth}
\includegraphics[width=0.31\textwidth]{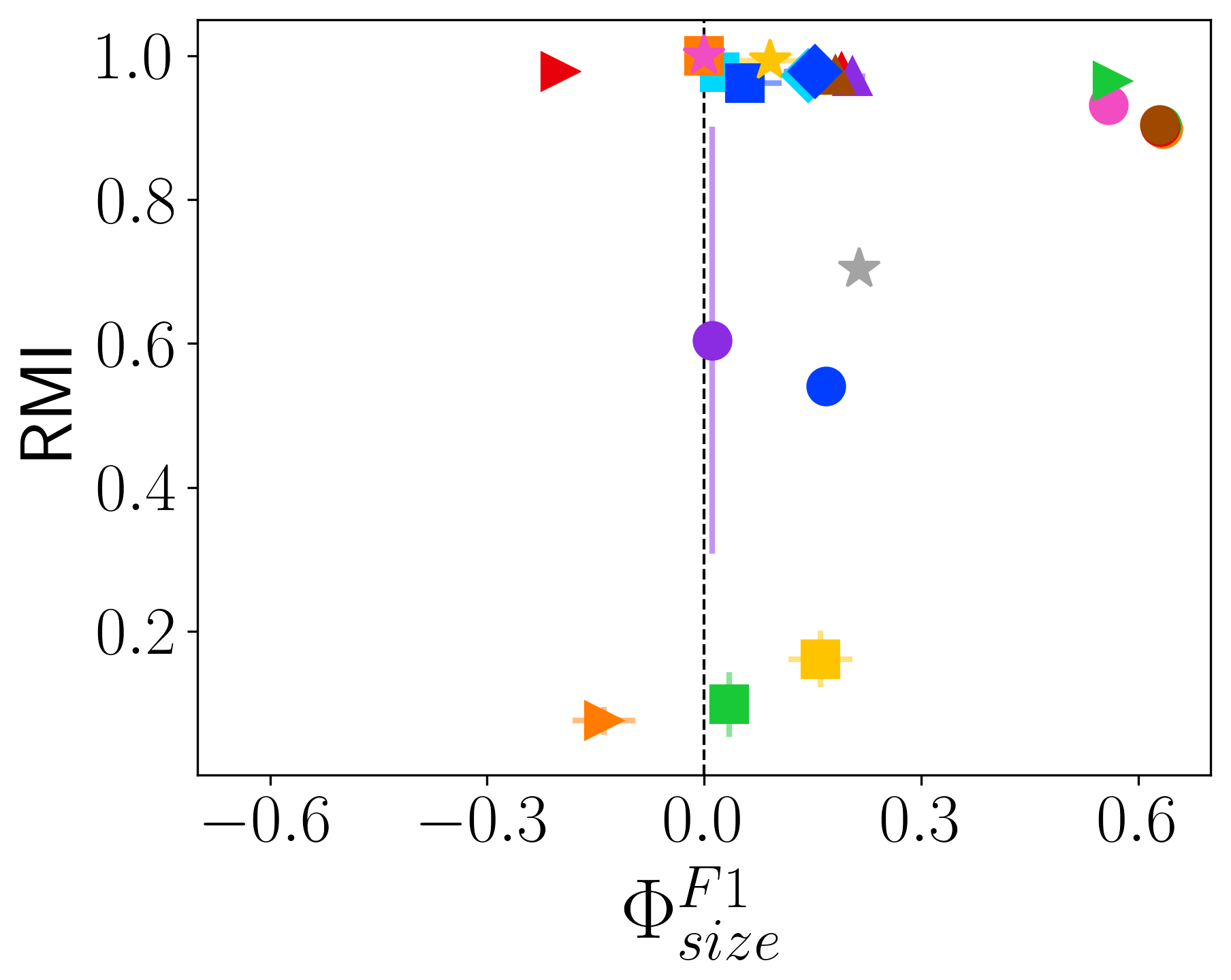}\quad
\includegraphics[width=0.31\textwidth]{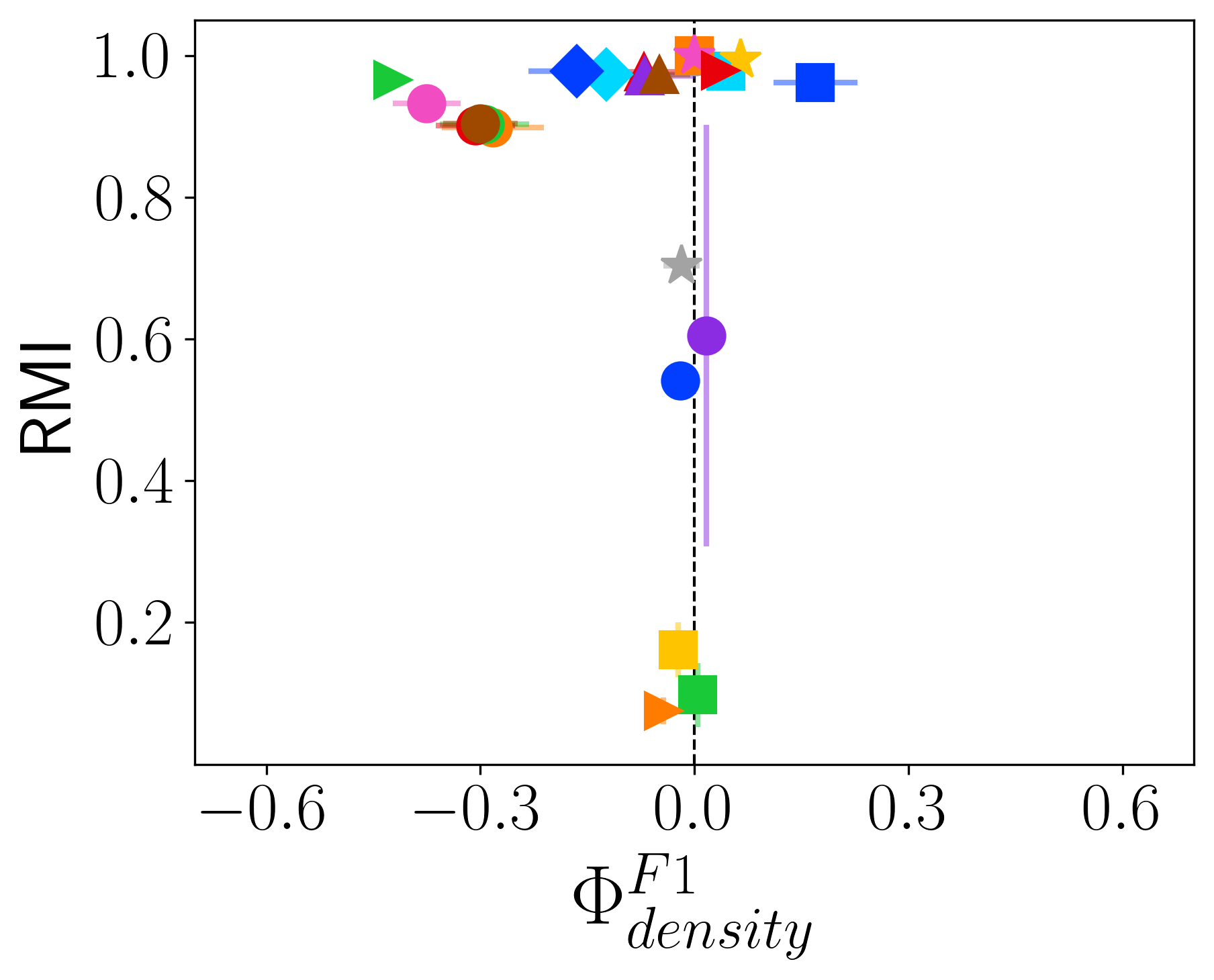}\quad
\includegraphics[width=0.31\textwidth]{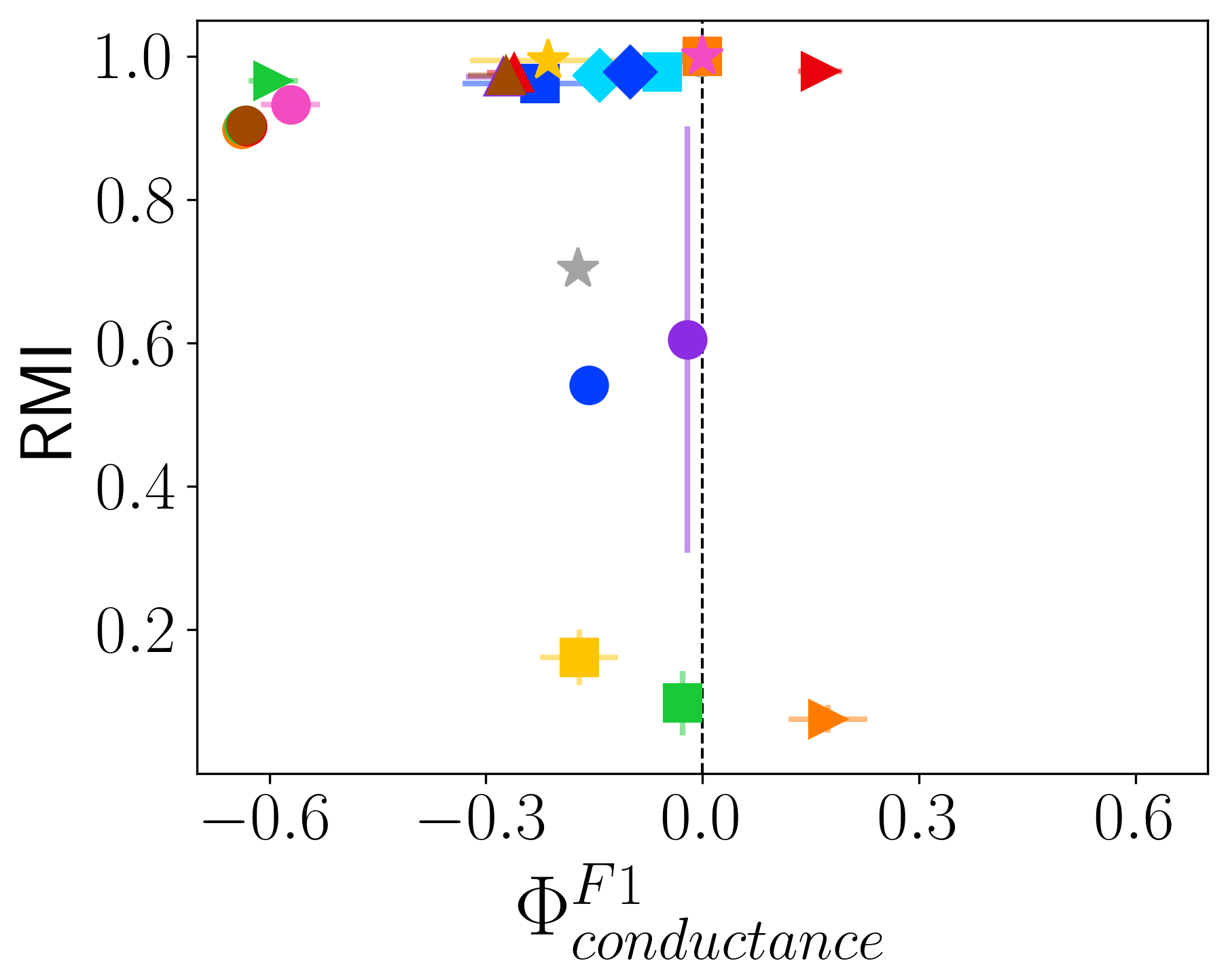}
\end{minipage}
\\
\begin{subfigure}[c]{0.05\textwidth}
\caption*{\rotatebox{90}{ARI}}%
\end{subfigure}%
\begin{minipage}[c]{0.94\textwidth}
\includegraphics[width=0.31\textwidth]{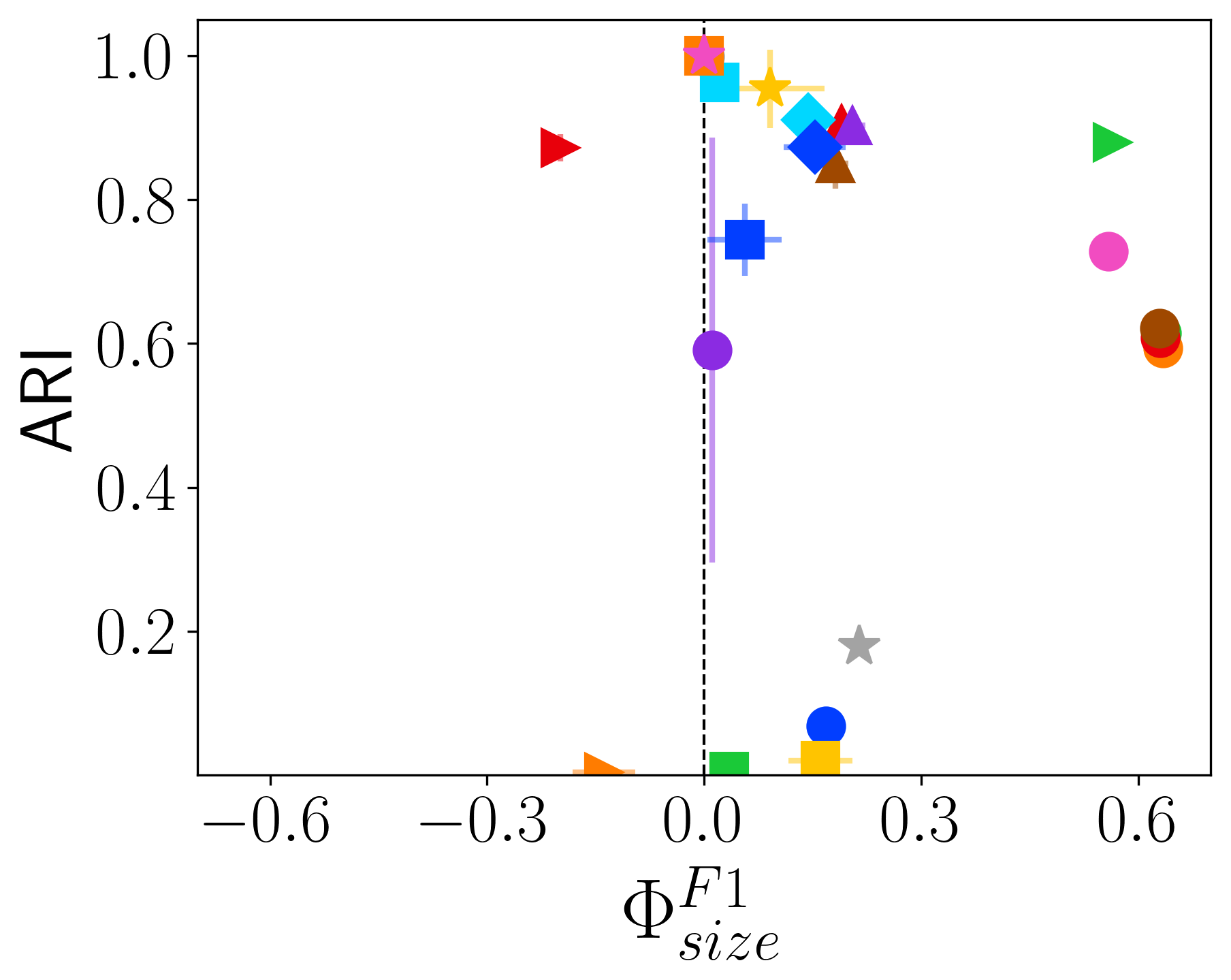}\quad
\includegraphics[width=0.31\textwidth]{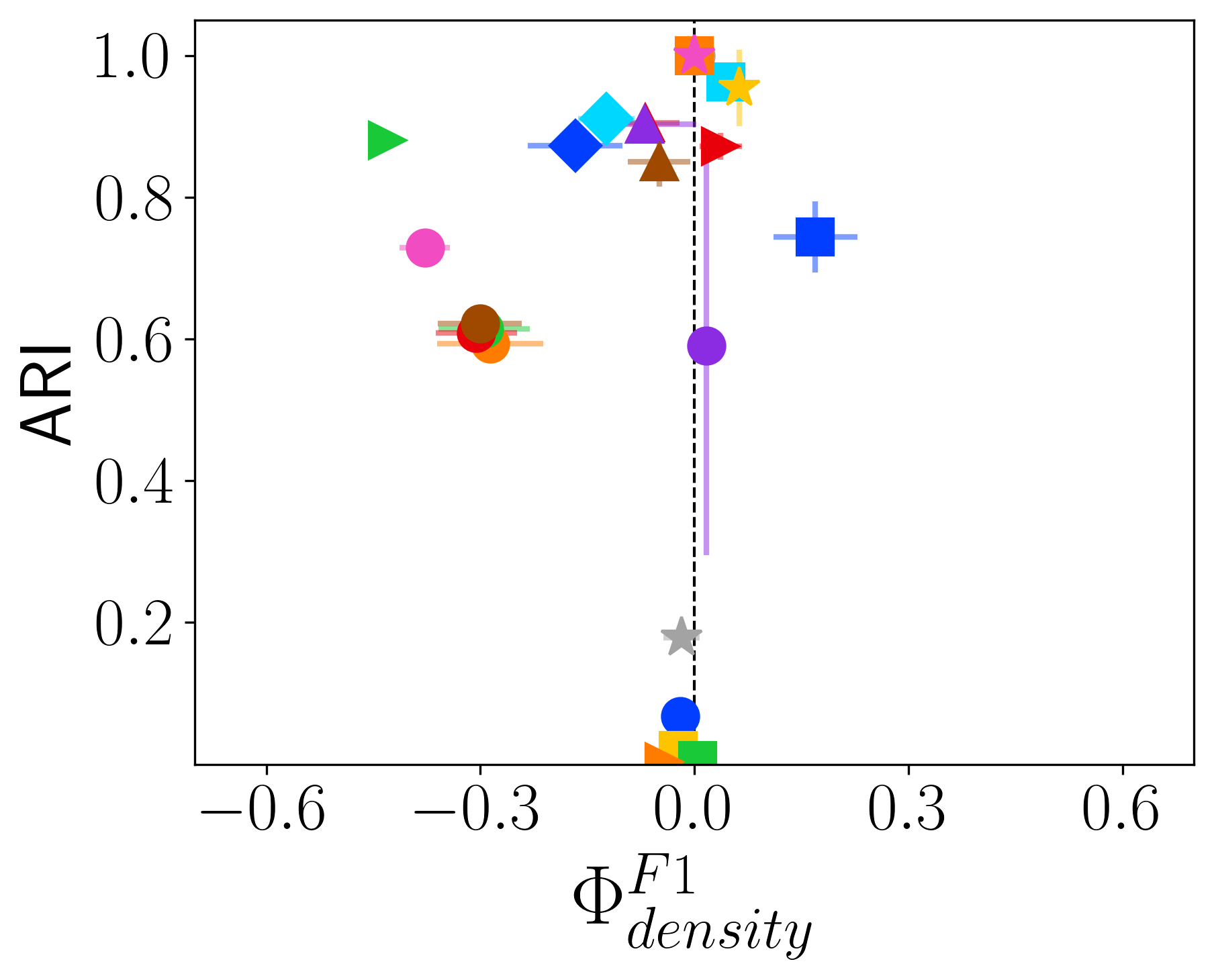}\quad
\includegraphics[width=0.31\textwidth]{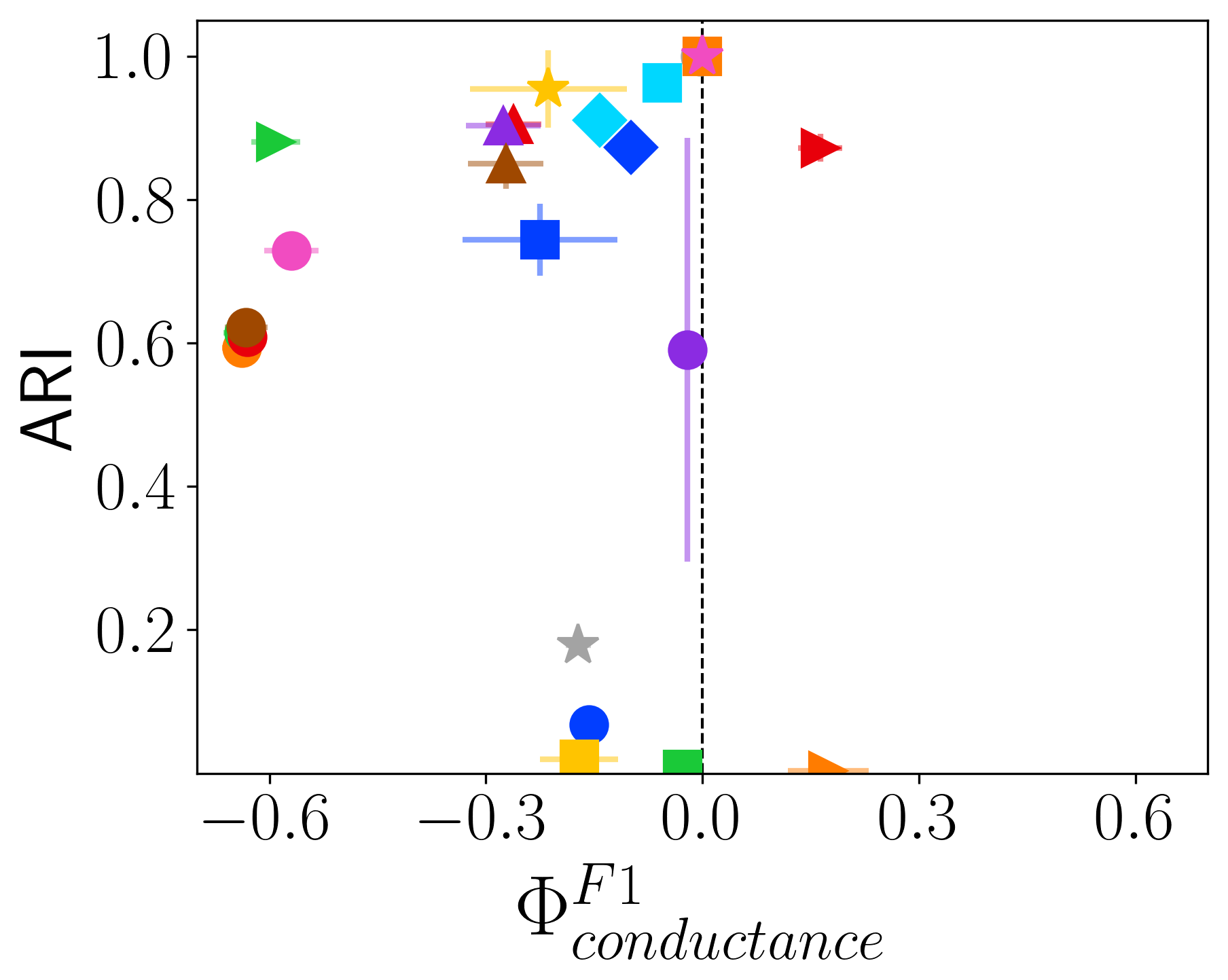}
\end{minipage}
\\
\begin{subfigure}[c]{0.05\textwidth}
\caption*{\rotatebox{90}{PF1}}
\end{subfigure}
\begin{minipage}[c]{0.94\textwidth}
\includegraphics[width=0.31\textwidth]{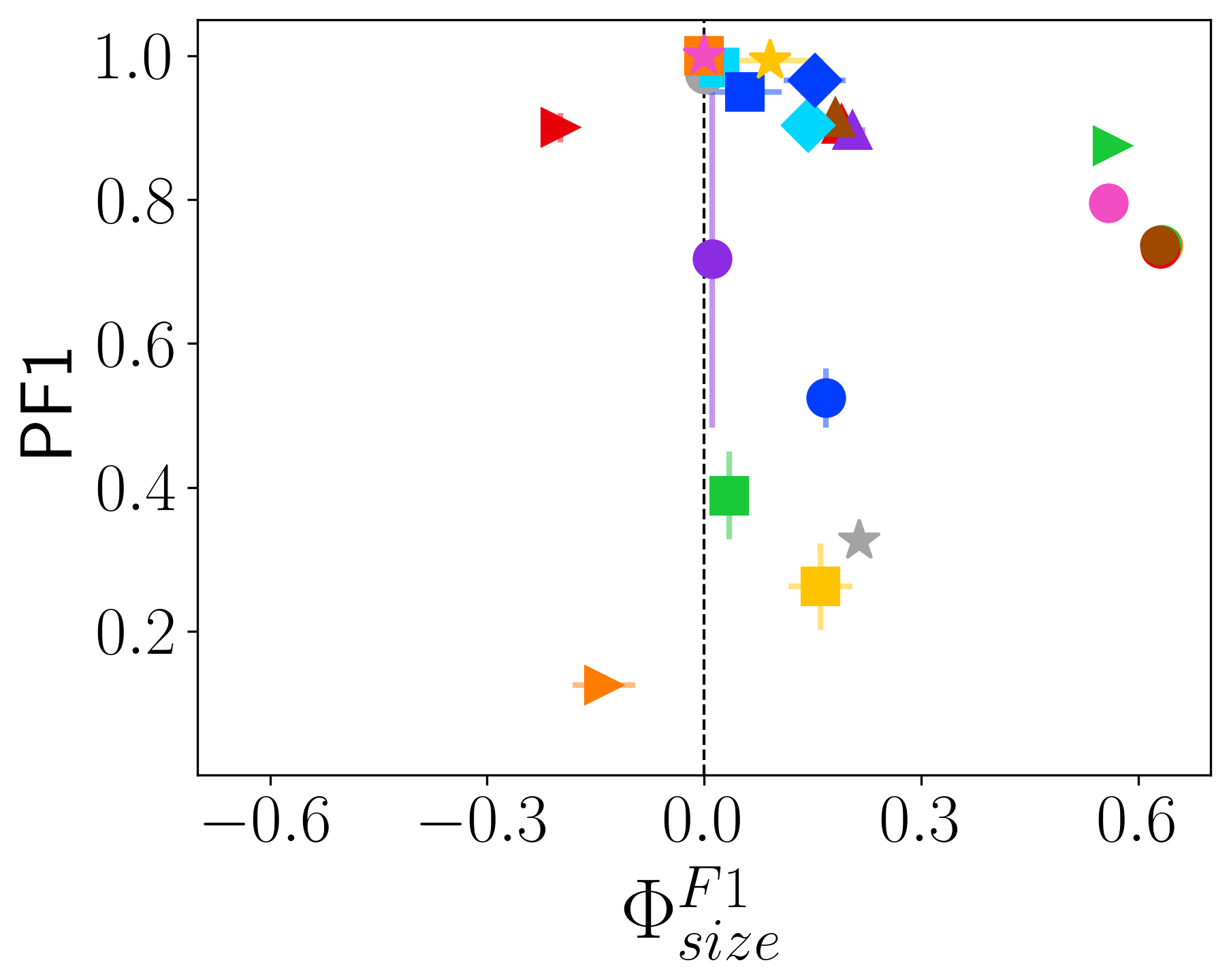}\quad
\includegraphics[width=0.31\textwidth]{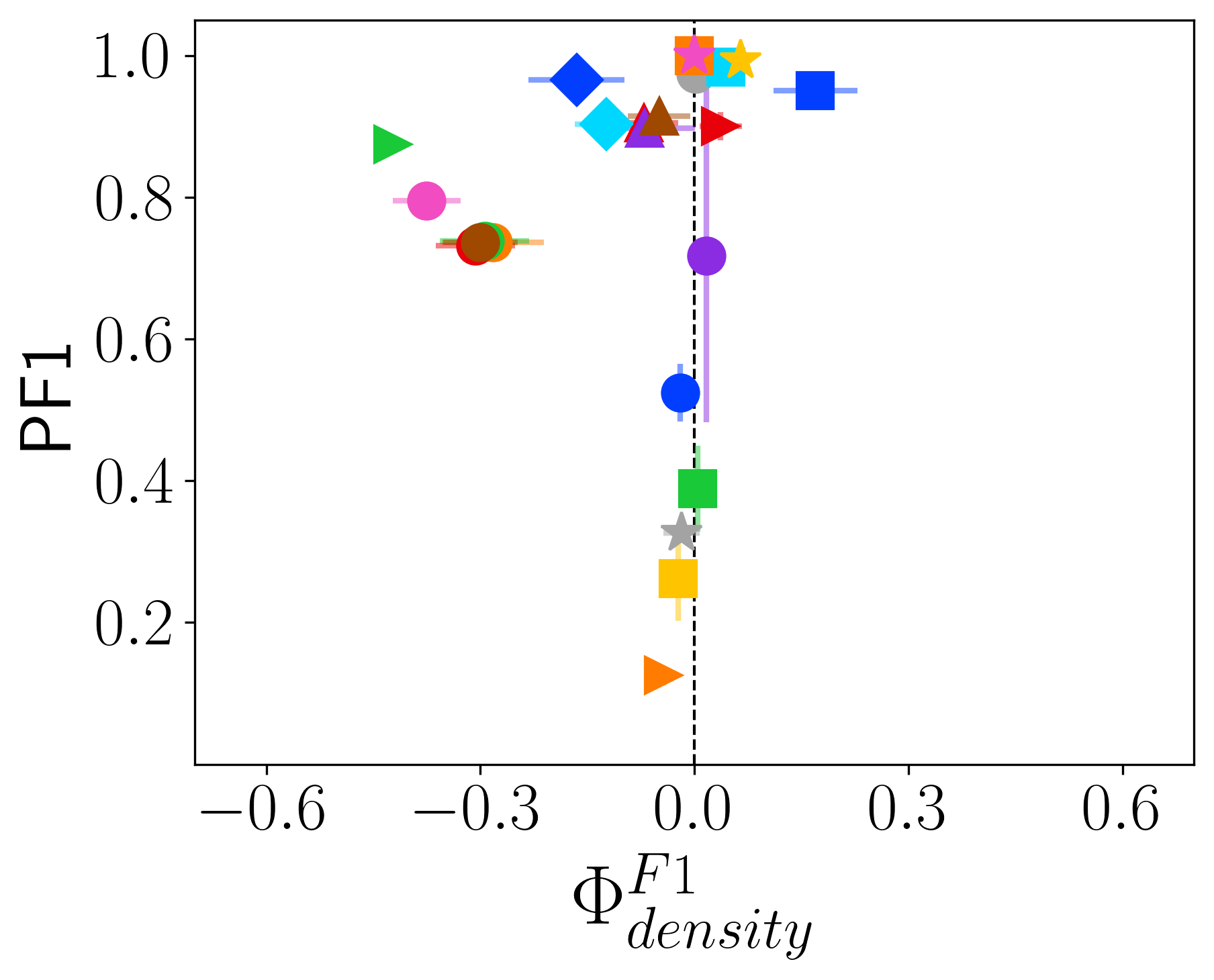}\quad
\includegraphics[width=0.31\textwidth]{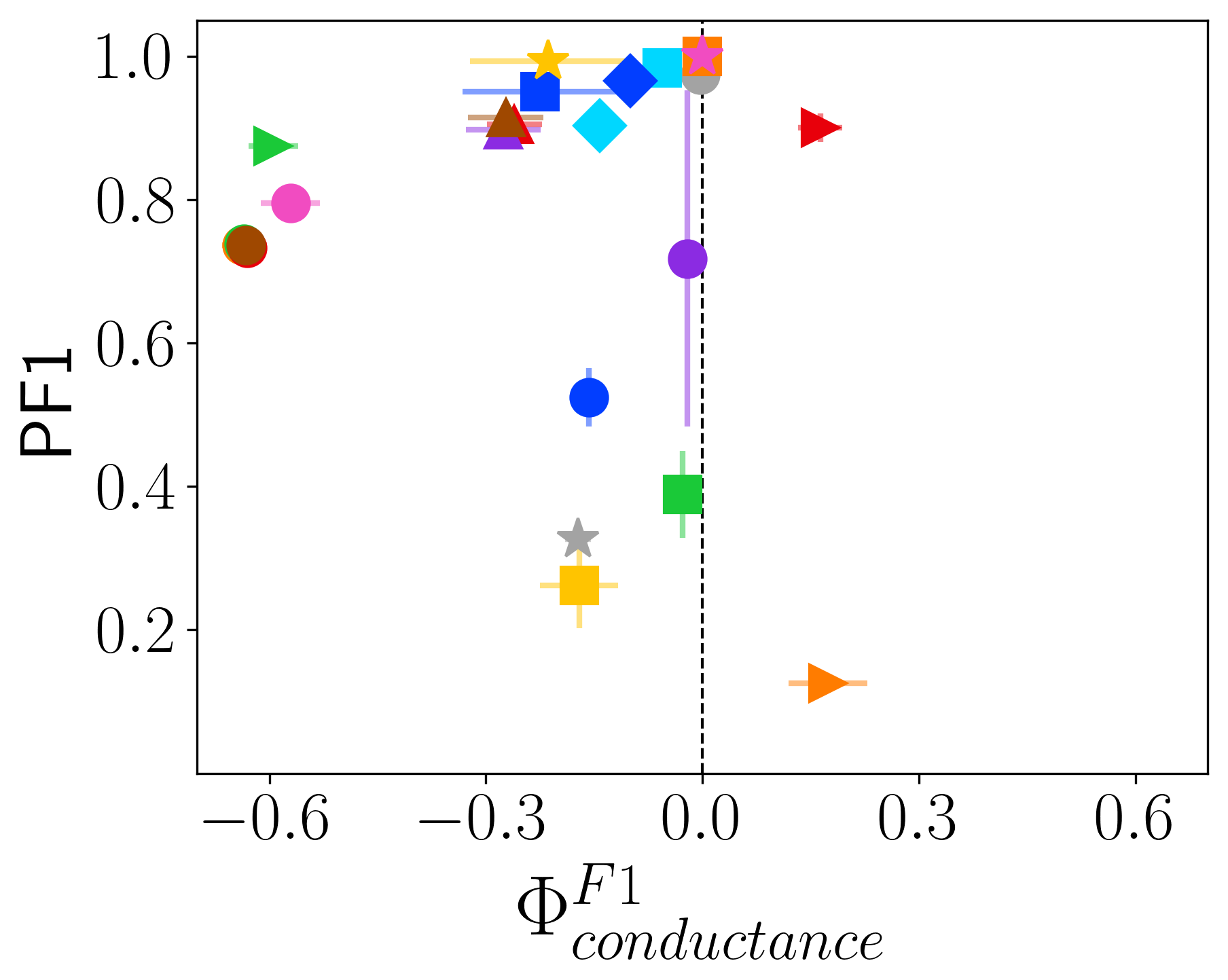}
\end{minipage}
\\
\begin{subfigure}[c]{0.05\textwidth}
\caption*{\rotatebox{90}{NF1}}
\end{subfigure}
\begin{minipage}[c]{0.94\textwidth}
\includegraphics[width=0.31\textwidth]{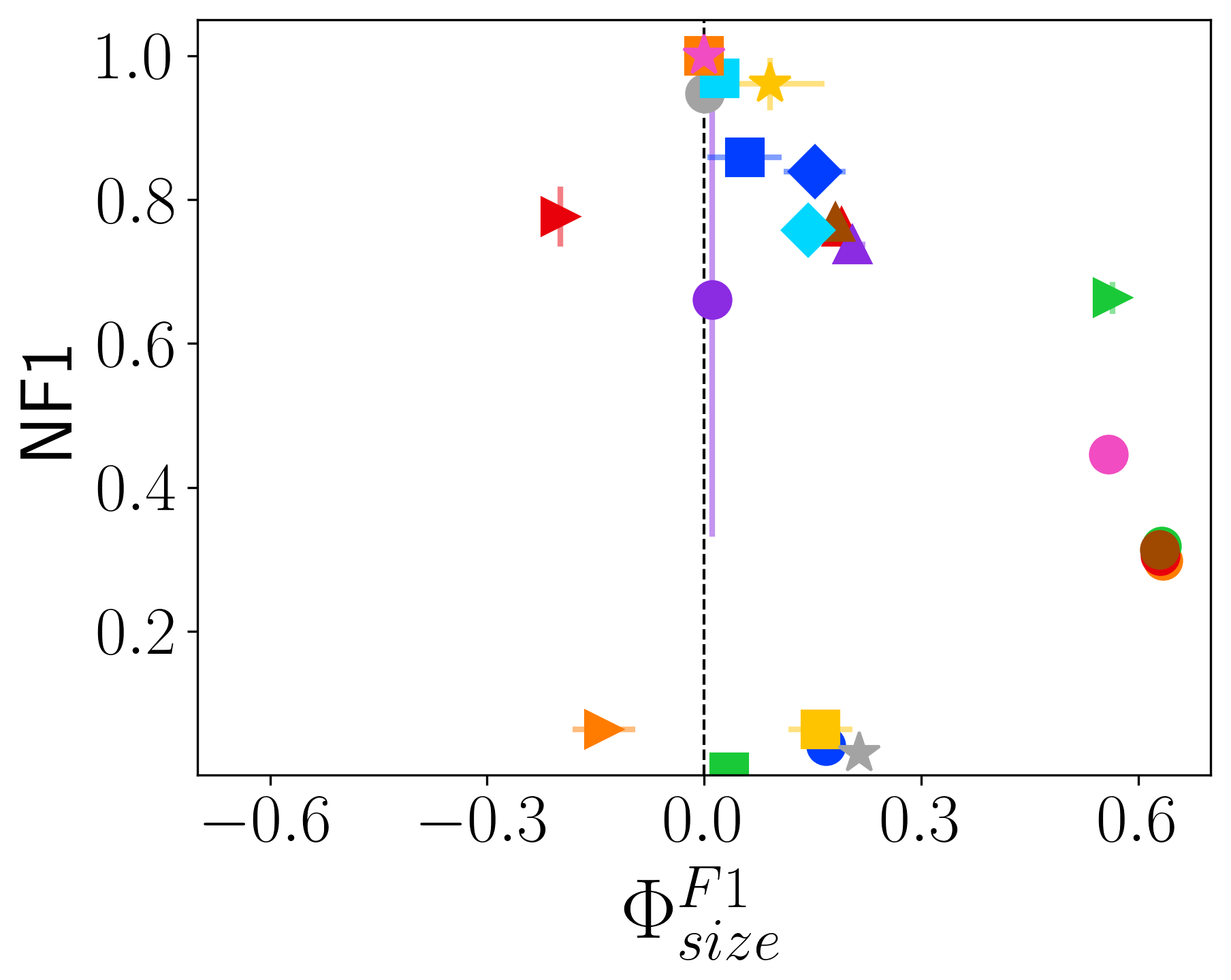}\quad
\includegraphics[width=0.31\textwidth]{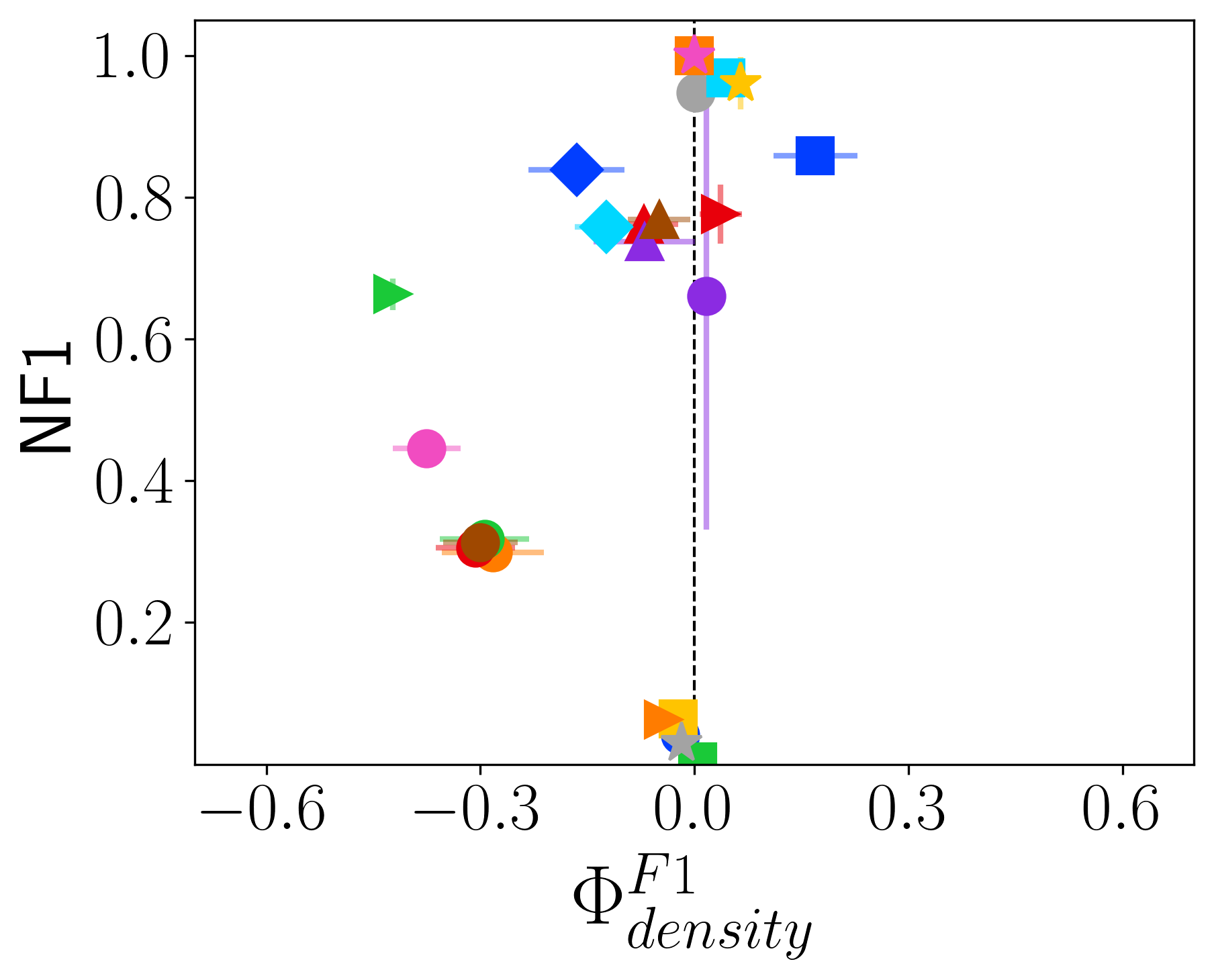}\quad
\includegraphics[width=0.31\textwidth]{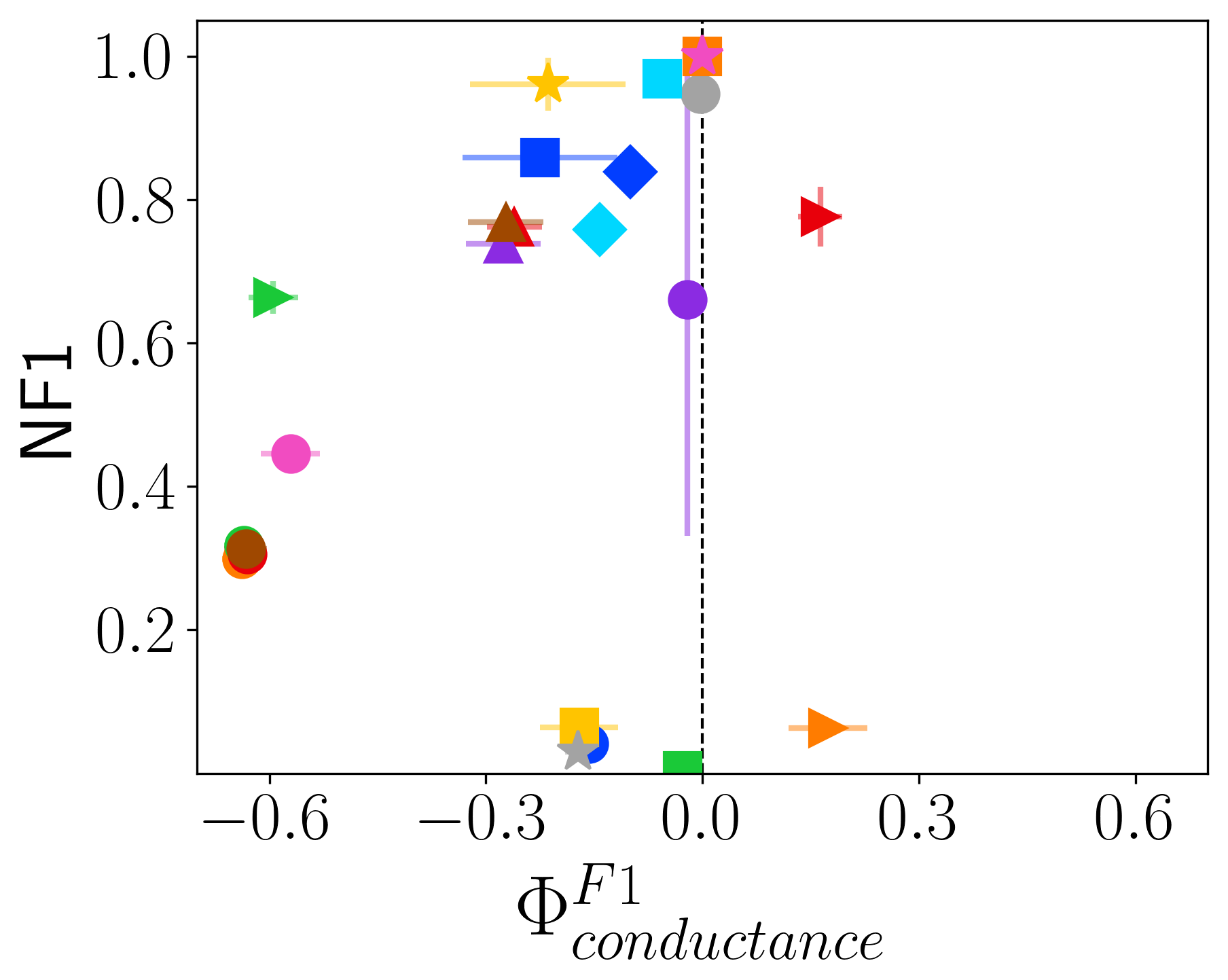}
\end{minipage}
\caption{RMI, ARI, PF1, and NF1 vs. $\Phi^{F1}$ on LFR networks having $\mu=0.2$.}\label{phi_vs_other_metrics} 
\end{figure}

We also assess fairness in relation to four additional metrics that evaluate the quality of the detected communities to ensure the robustness of our findings. These metrics include: (i) Adjusted Rand Index (ARI), (ii) Reduced Mutual Information (RMI), (iii) F1 Score (PF1), and (iv) Normalized F1 Score (NF1) (refer to the Evaluation Metrics Section for details). Figure~\ref{phi_vs_other_metrics} presents the results of $\phi^{F1}$ for the LFR network with $\mu =0.2$, analyzed with respect to size, density, and conductance. This analysis aims to identify community detection methods that achieve both high fairness and high performance across multiple evaluation metrics, making them suitable for broader applications. Additionally, understanding the working dynamics of these methods will provide insights for designing new community detection algorithms with high fairness-performance trade-offs. From this evaluation, the standout community detection methods demonstrating both fairness and high performance include RSC-V, RSC-K, Walktrap, Infomap, and Significance.

\subsection*{Discussion}

Communities within networks vary in size, density, and connectivity. However, most community detection algorithms do not account for structural inequalities among communities and in the network structure, leading to biased detection outcomes and affecting downstream social network analysis tasks \cite{Saxena2024fairsna}. Certain types of communities, particularly smaller in size or having high conductance, are often not properly identified by existing methods. In literature, there exist several metrics to quantify the quality of identified communities \cite{fortunato2016community}; however, the metrics to compute the fairness of a community detection method are underexplored. 

In this work, we introduce group-fairness metrics $\Phi_p^{F*}$, which quantifies bias concerning a given community property $p$. This metric is grounded in the fairness principle that all types of communities should be detected equally well. It first computes the community-wise fairness using the three introduced measures and then uses them to calculate the overall fairness. Next, we evaluate 24 community detection methods on both real-world and synthetic networks. Synthetic networks are generated using the LFR, ABCD, and HICH-BA benchmark models. For LFR and ABCD networks, we analyze different levels of community mixing, while HICH-BA networks are structured based on the number of majority communities and overall community composition. We analyze community detection methods based on their fairness, as measured by $\Phi$, and their performance using Normalized Mutual Information (NMI), Reduced Mutual Information (RMI), Adjusted Rand Index (ARI), Average F1 Score (PF1), and Normalized F1 Score (NF1).

Our analysis of the fairness-performance trade-off in community detection reveals key biases across different community properties and network mixing. If we talk about fairness with respect to community properties, in LFR networks, most methods favor larger communities, except for SBM-Nested, which effectively detects smaller groups due to its hierarchical structure. As the mixing parameter ($\mu$) increases, fairness-oriented methods struggle equally across all sizes, leading to lower NMI. Regarding community density, methods with high NMI, such as Leiden, Louvain, and Spinglass, tend to favor sparse communities at low $\mu$ but shift toward denser ones as $\mu$ increases. The Significance method excels at identifying dense communities but overestimates their number. For community conductance, most methods prefer low-conductance communities, and this bias worsens as $\mu$ increases. At $\mu=0.4$, methods like RSC-V, SBM-Nested, Infomap, and Significance achieve both fairness and high performance. However, at $\mu=0.6$, fairness and accuracy trade-off linearly, where fair methods perform poorly, and high-performing methods show stronger bias.

Community detection methods perform better in ABCD networks compared to LFR, even at high $\xi$ values, but fairness varies with structural mixing. At $\xi=0.2$, Significance, RSC-SSE, Spectral, and Infomap show high fairness, but Spectral becomes less fair as $\xi$ increases, while RSC-V improves. SBM-Nested excels in quality but favors large, low-density, high-conductance communities, reducing fairness. Representation learning-based methods (DeepWalk, Node2Vec, FairWalk) show fairness for conductance but not for other properties, as the likelihood of correctly identifying communities of varying conductance is similar. Regarding density, Paris and Spectral exhibit shifting fairness behavior due to their tendency to merge nodes into larger communities. Most community detection methods favor low-density communities, and unlike LFR networks, fairness does not necessarily improve with increased mixing. For conductance, Significance, RSC-SSE, Infomap, and Walktrap maintain high fairness and performance, while methods like Combo, Leiden, and SBM shift their conductance bias as $\xi$ increases. Additionally, methods requiring predefined community numbers tend to have lower fairness, as they misclassify nodes from low-conductance communities.

In contrast to LFR and ABCD networks, fairness and performance trends in HICH-BA networks differ significantly. While smaller communities in both networks have higher density, their relationship with conductance differs. Across both HICH-BA networks, there is a general bias toward low-conductance communities, consistent with LFR and ABCD results. However, biases concerning community size and density vary. In both networks, methods like RB-ER and Significance predict an excessive number of communities, fragmenting larger ones. In the multi-majority (MMaj) structure, methods such as Combo, Leiden, Louvain, RB-C, Spinglass, SBM, and SBM-Nested perform well, but struggle in MMin, where only Paris, RSC-V, Label Propagation, SBM, and SBM-Nested achieve good results. In the MMin network, overall performance is lower, with only seven methods achieving NMI scores above 0.4. Notably, SBM and SBM-Nested are fair across all community types. While Walktrap and Infomap maintain fairness across different network types, they overestimate the number of communities in HICH-BA networks, leading to lower-quality results. 

Significance, RSC-K, Infomap, Walktrap, and SBM (including SBM-Nested) are among the top-performing community detection methods across various networks. These methods perform well on LFR and ABCD networks, which closely resemble real-world structures. However, Walktrap’s performance declines significantly in highly mixed networks. Fairness and performance often correlate (meaning high-performing methods are more fair), particularly in ABCD networks, where many communities are accurately identified, resulting in high fairness scores even at high mixing levels ($\xi=0.6$). High-performing methods generally tend to show biases toward large, dense, and low-conductance communities. Additionally, modularity-based approaches such as Combo, Leiden, Louvain, RB-C, and RB-ER follow similar performance and fairness patterns. While these methods achieve good results, they tend to favor less dense communities, demonstrating a different bias compared to other community detection methods.

Based on our findings, we recommend Significance and Infomap for general use. These methods consistently achieve high performance and fairness across all networks. These methods do not require users to specify the number of communities, unlike many other high-performing methods. Additionally, both of these methods identify many communities well, often ranking among the fairest methods, and Significance outperforms other methods in real-world networks. While fairness is important, community detection method selection should not be based solely on fairness metrics. A method scoring high in fairness may still detect all communities poorly. Additionally, some methods could achieve better results with optimized parameters rather than the default settings in CDlib. However, Significance and Infomap require no parameter tuning, making their strong fairness-performance trade-off even more notable. Interestingly, Significance does not cluster with modularity-based methods despite optimizing a goodness score similar to those approaches. Unlike modularity-based methods, it optimizes for Significance rather than modularity, resulting in distinct fairness-performance characteristics. We also recommend using the $\Phi^{F1}_{p}$ fairness metric if one needs to be chosen, as it considers multiple factors while computing the fairness for a community, and the overall fairness of the method is captured very well. 

The choice of a community detection method should also align with the specific network analysis task. Different fairness metrics offer insights relevant to practical applications. For instance, in Influence Maximization, when spreading information to remote parts of a network, selecting a community detection method that identifies low-conductance communities well may be advantageous. The proposed fairness measures and our analysis provide valuable insights for designing fairer community detection methods, improving existing algorithms, and optimizing parameter selection. Additionally, our findings highlight that when extreme cases are created using the HICH-BA model, none of the existing community detection methods perform well. This underscores the need to develop new, well-performing methods specifically designed to handle such challenging network structures.

This work focused on undirected, unweighted networks with non-overlapping communities. However, many real-world complex networks contain overlapping communities, and future research could explore fairness metrics tailored for such networks. Moreover, one can propose fairness metrics for hierarchical and hyperedge networks. In this work, the Jaccard similarity-based iterative approach is used to map ground truth and predicted communities. However, other approaches to mapping ground truth communities to predicted communities could be explored, and their impact can be studied. Additionally, our use of linear regression abstracts away specific community-wise performance values, and a deeper examination of these values may provide further insights. Moreover, this work mainly focuses on group fairness, and comparing the community detection methods from the perspective of individual fairness will provide additional insights at the node level. One can propose individual fairness metrics for community detection methods. Finally, the development of fairness metrics that do not depend on ground truth community structures could enhance fairness evaluations in practical applications where ground truth is not readily available and should be explored further.

\section*{Conclusion}

Community detection is crucial for understanding network structures and node behaviors, yet many community detection methods overlook structural inequalities, leading to biased results. In this work, we introduced group-fairness metrics $\Phi^{F*}_{p}$ to evaluate the fairness of community detection methods concerning community properties. Our approach maps ground truth to detected communities and evaluates fairness using three community-wise measures (FCCN, F1, FCCE), which are then aggregated into an overall fairness score $\Phi$. We analyzed the performance-fairness trade-off analysis of 24 community detection methods from six different classes across real-world and synthetic networks (LFR, ABCD, and HICH-BA network generating models). Our analysis reveals that no single class of methods consistently outperforms others, but certain patterns emerge, as explained in the Discussion section. We also observe that the community detection methods that take the number of communities as input have the upper hand over other methods, though they still do not perform well compared to other methods. Our analysis highlights that Significance and Infomap are both high-performing and fair across various community properties, making them practical choices without requiring extensive knowledge of the network structure. However, in networks with extreme structures like MMaj and MMin, no method effectively identifies high-quality communities of all types. In cases with multiple majority groups, Spinglass could be used, while for networks with multiple minority groups, SBM, SBM-Nested, and Label Propagation are suitable options. 

In future work, we will explore individual and group fairness metrics for networks with overlapping communities and hyperedge networks.

\bibliography{bibliography}
\clearpage

\end{document}